\documentclass{article}
\pdfoutput=1
\usepackage{pdfpages}
\usepackage{icrc2011}

\title{The JEM-EUSO Mission: Status and Prospects in 2011}

\newcommand{\etal}{\MakeLowercase{\textit{et al. }}} 
\shorttitle{JEM-EUSO}

\authors{The JEM-Euso Collaboration}
\afiliations{}

\abstract{Contributions of the JEM-EUSO Collaboration to the 32nd International Cosmic Ray Conference, Beijing, August, 2011.
}

\begin{document}
\maketitle

\onecolumn{
{\bf The Collaboration:} \\
J.H.~Adams Jr.$^{md}$, 
S.~Ahmad$^{ba}$, 
J.-N.~Albert$^{ba}$, 
D.~Allard$^{bb}$, 
M.~Ambrosio$^{df}$, 
L.~Anchordoqui$^{me}$,
A.~Anzalone$^{dh}$, 
Y.~Arai$^{eu}$, 
C.~Aramo$^{df}$, 
K.~Asano$^{es}$
% ...
\\


{\center
% ...
$^{ba}$ Laboratoire de l'Acc\'el\'erateur Lin\'eaire, Univ Paris Sud-11, CNES/IN2P3, Orsay, France\\
$^{bb}$ APC, Univ Paris Diderot, CNRS/IN2P3, CEA/Irfu, Obs de Paris, Sorbonne Paris Cit\'e, France\\
% ...
$^{df}$ Istituto Nazionale di Fisica Nucleare - Sezione di Napoli, Italy\\
% ...
$^{dh}$ INAF - Istituto di Astrofisica Spaziale e Fisica Cosmica di Palermo, Italy\\
% ...
$^{es}$ Interactive Research Center of Science, Tokyo Institute of Technology, Tokyo, Japan\\ 
% ...
$^{eu}$ High Energy Accelerator Research Organization (KEK), Tsukuba, Japan\\ 
% ...
$^{md}$ University of Alabama in Huntsville, Huntsville, USA\\
$^{me}$ University of Wisconsin-Milwaukee, Milwaukee, USA\\
% ...
}
}

\newpage

\onecolumn{
{\bf Contents} \\

\begin{tabular}{lllll}
1. 	&page 7  	&The JEM-EUSO mission 	&T. Ebisuzaki &ID1628 \\
& & & &  \\
2. 	&page 11		&Science objectives of the JEM-EUSO mission 	&G.A. Medina-Tanco 	&ID0956 \\
& & & &  \\
3. 	&page 15	&Overview of the JEM-EUSO Instruments			&F. Kajino	&ID1216 \\
& & & &  \\
4. 	&page 19	&Requirements and Expected Performances &A. Santangelo	&ID0991 \\
& &of the JEM-EUSO mission	 & &  \\
& & & &  \\
5. 	&page 23	&The potential of the JEM-EUSO telescope for  	&G. Medina Tanco	&ID0930 \\
& &the astrophysics of extreme energy photons & &  \\
& & & &  \\
6. 	&page 27	&Neutrino astrophysics with JEM-EUSO			&G. Medina Tanco	&ID0958 \\
& & & &  \\
7. 	&page 31	&The Focal Surface Detector of the JEM-EUSO Telescope	& Y. Kawasaki	& ID0472 \\
& & & &  \\
8. 	&page 35	&The JEM-EUSO Focal Surface Mechanical Structure	&M. Ricci	&ID0335 \\
& & & &  \\
9. 	&page 39	&SPACIROC: A Front-End Readout ASIC for spatial        & S. Ahmad        & ID0236  \\
& &cosmic ray observatory  & &  \\
& & & &  \\
10. &page 43	&Performance of a front-end ASIC for JEM-EUSO        & H. Miyamoto        & ID0775  \\
& & & &  \\
11. &page 47	&High Voltage system for the JEM-EUSO Photomultipliers        & J. Szabelski        & ID0216  \\
& & & &  \\
12. &page 51	&The Cluster Control Board of the JEM-EUSO mission        & J. Bayer        & ID0836  \\
& & & &  \\
13. &page 55	&The Housekeeping subsystem of the JEM-EUSO instrument        & G. Medina Tanco        & ID0961  \\
& & & &  \\
14. &page 59	&Data Acquisition System of the JEM-EUSO project        & M. Casolino        & ID1219  \\
& & & &  \\
15. &page 63	&The Development of Photo-Detector Module Electronics      & I. Park        & ID1246  \\
& &for the JEM-EUSO Experiment    & &  \\
& & & &  \\
16. &page 67	&The JEM-EUSO time synchronization system        & G. Osteria        & ID1131  \\
& & & &  \\
17. &page 71	&Technological developments in Russia for the        & B. Khrenov        & ID1261  \\
& &JEM-EUSO mission  & &  \\
& & & &  \\
18. &page 75	&Calibration of JEM-EUSO photodetectors        & P. Gorodetzky        & ID0218  \\
& & & &  \\
19. &page 79	&The JEM-EUSO optics design        & A. Zuccaro Marchi        & ID0852  \\
& & & &  \\
20. &page 83	&JEM-EUSO lens manufacturing        & Y. Takizawa        & ID0874  \\
& & & &  \\
21. &page 87	&Testing of Large Diameter Fresnel Optics for        & J. Adams        & ID1100  \\
& &Space Based Observations of Extensive Air Showers  & &  \\
& & & &  \\
\end{tabular}
}

\newpage

\onecolumn{
\begin{tabular}{lllll}
22. &page 91	&Atmospheric Monitoring System of JEM-EUSO        & A. Neronov        & ID0301  \\
& & & &  \\
23. &page 95	&The IR-Camera of the JEM-EUSO (JAXA) Space Observatory        & J.A. Morales        & ID1031  \\
& & & &  \\
24. &page 99	&Cloud Coverage and its Implications for Cosmic Ray        & M. Bertaina        & ID0398  \\
& &Observation from Space  & &  \\
& & & &  \\
25. &page 103	&A Comparison of Different Cloud Detection Methods         & A. Anzalone        & ID1152  \\
& &for the JEM-EUSO Atmospheric Monitoring System & &  \\
& & & &  \\
26. &page 107	&Estimation of JEM-EUSO experiment duty cycle based        & P. Bobik        & ID0886  \\
& &on Universitetsky Tatiana measurements  & &  \\
& & & &  \\
27. &page 111	&ESAF Simulation of Ultra-High Energy Cosmic Rays in  & G. Saez Cano        & ID1034  \\
& &cloudy conditions for the JEM-EUSO (JAXA) Space Observatory  & &  \\
& & & &  \\
28. &page 115	&The ESAF Simulation Framework for the JEM-EUSO mission        & F. Fenu        & ID0829  \\
& & & &  \\
29. &page 119	&The ESAF reconstruction framework for the JEM-EUSO mission        & F. Fenu        & ID0633  \\
& & & &  \\
30. &page 123	&Simulation framework of STM code for development of        & K. Higashide        & ID1240  \\
& &JEM-EUSO instrument  & &  \\
& & & &  \\
31. &page 127	&Estimation of effective aperture for extreme energy          & K. Shinozaki        & ID0979  \\
& &cosmic rays by space-based JEM-EUSO Mission & &  \\
& & & &  \\
32. &page 131	&Very precise Fluorescence Yield measurement        & D. Ragaigne-Monnier        & ID0212  \\
& &using a MeV electron beam  & &  \\
& & & &  \\
33. &page 135	&Fluorescence yield by electron in moist air and its application  & N. Sakaki &ID0520 \\
& & to the observation of ultra high energy cosmic rays from space& &  \\
\end{tabular}
}

\clearpage
\newpage

\vspace*{2cm}
\clearpage
\newpage

	\includepdf[pages=-]{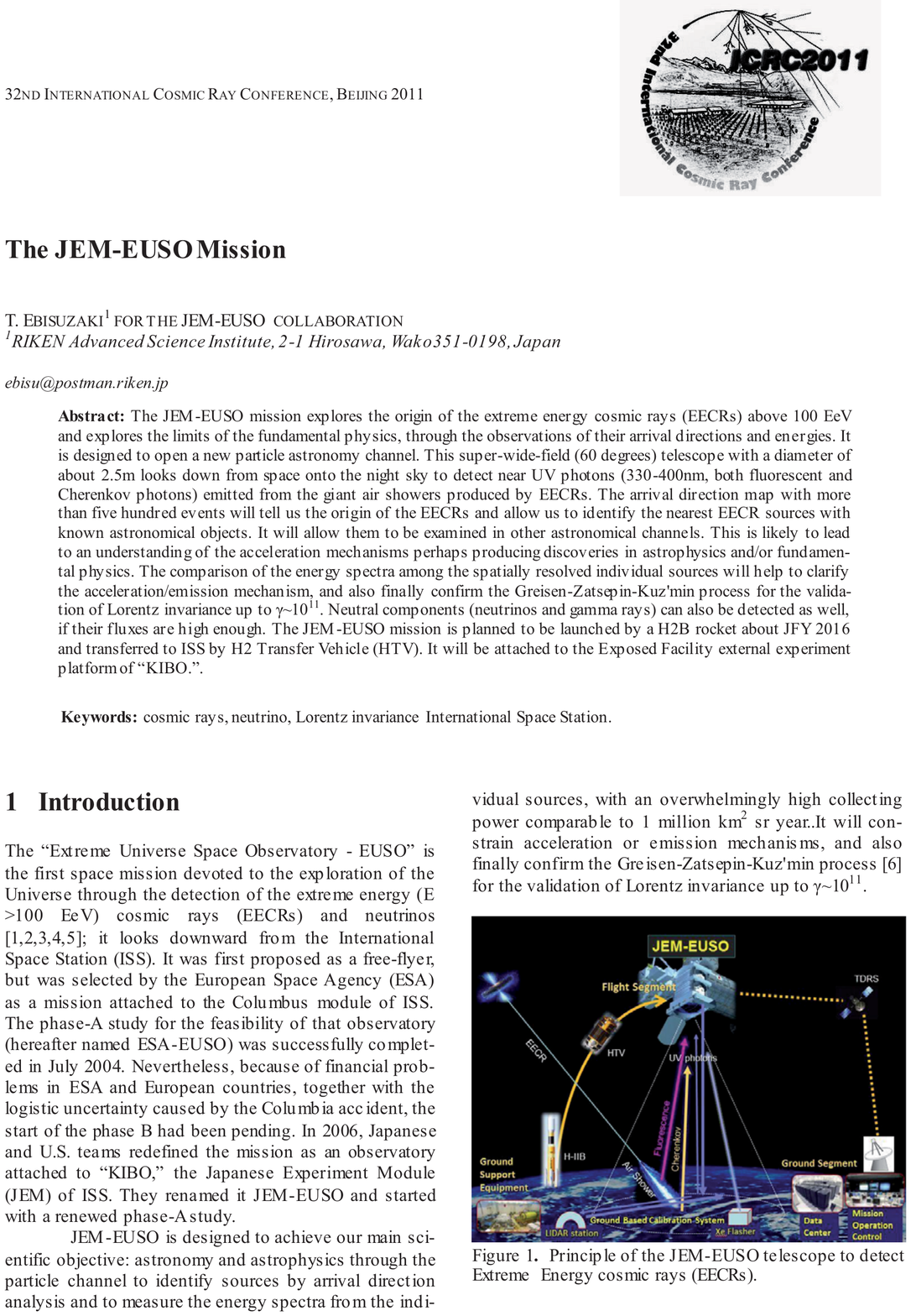}
\hspace{4cm}
\newpage

\newpage
\normalsize
\setcounter{section}{0}
\setcounter{figure}{0}
\setcounter{table}{0}
\setcounter{equation}{0}



\title{Science objectives of the JEM-EUSO mission}

\shorttitle{G. Medina-Tanco \etal JEM-EUSO Science objectives}

\authors{
G.~Medina-Tanco$^{1}$, 
T.J.~Weiler$^{2}$, 
M.~Teshima$^{3}$, 
T.~Ebisuzaki$^{4}$, 
P.~Picozza$^{5}$, 
A.~Santangelo$^{6}$, 
E.~Parizot$^{7}$,
M.~Bertaina$^{8}$ 
for the JEM-EUSO Collaboration}
\afiliations{
$^{1}$Instituto de Ciencias Nucleares, UNAM, Ciudad Universitaria, M\'exico D. F. 04510. (M\'exico).\\
$^{2}$Department of Physics and Astronomy, Vanderbilt University Nashville, TN 37235 USA.\\
$^{3}$Max-Planck-Institut fuer Physik, F\"ohringer Ring 6, D-80805 Munchen, Germany.\\
$^{4}$RIKEN Advanced Science Institute, Japan.\\
$^{5}$INFN, Sezione di Rome Tor Vergata, I-00133 Rome, Italy \& University of Rome Tor Vergata.\\
$^{6}$Institute fuer Astronomie und Astrophysik Kepler Center for Astro and Particle Physics
Eberhard Karls University Tuebingen Germany.\\
$^{7}$APC - Universit\'e Denis Diderot - Paris VII, France.\\
$^{8}$Dipartimento di Fisica Generale dell' Universit\`a di Torino, Torino, Italy\\
}
\email{gmtanco@nucleares.unam.mx}

\abstract{JEM-EUSO is a space telescope proposal, devoted to the observation of the ultraviolet fluorescence light emitted by extreme energy cosmic ray (EECR) atmospheric cascades \cite{JEMEUSOMission}. The fluorescence technique has proved to be extremely successful from the ground and JEM-EUSO will be the first detector to use it from space. The telescope possesses an innovative wide field of view Fresnel optics which, combined with a highly sensitive focal surface and an observation altitude in excess of 360 km, will allow it to reach an unprecedented exposure of 10$^6$ km$^2$ sr yr at $3 \times 10^{20}$ eV. These capabilities go far beyond what can be practically achieved by ground observatories. The large number of expected events will allow the identification of relatively nearby individual sources of EECR and determine their spectra. Point spread function analysis will also be used to study the Galactic magnetic field. Furthermore, baryons, photons and neutrino primaries can be discriminated with considerable accuracy, and upper limits to the fluxes of the last two will be improved by at least a factor of 10 beyond present experiments. Moreover, the mass target inside the field of view is $\sim 10^{12}$ ton which, depending on the actual astrophysics scenario, makes very likely the observation of up to a few cosmogenic neutrinos per year. Other exploratory objectives include the observation of atmospheric phenomena, like night-glow, high altitude plasma discharges and meteors.}


\maketitle

\section{Introduction} \label{sec:intro}

Cosmic rays (CR) at the highest energies may be messengers of the most extreme environments in the universe. This challenging extreme energy region, at the frontier of present scientific knowledge, is the scope of the JEM-EUSO mission. JEM-EUSO is intended to address basic problems of fundamental physics and high energy astrophysics by investigating the nature and origin of extreme energy cosmic rays (EECR). JEM-EUSO will pioneer the observation from space of EECR-induced extensive air showers (EAS), making accurate measurements of the energy, arrival direction and identity of the primary particle using a target volume far greater than which is possible from the ground. The corresponding quantitative jump in statistics will clarify the origin (sources) of the EECR and, possibly, the particle physics mechanisms operating at energies well beyond those achievable by man-made accelerators. Furthermore, the spectrum of scientific goals of the JEM-EUSO mission also includes as exploratory objectives the detection of high energy gamma rays and neutrinos, the study of cosmic magnetic fields, and testing relativity and quantum gravity effects at extreme energies. In parallel, all along the mission, JEM-EUSO will systematically survey atmospheric phenomena over the Earth surface. \\
 
\section{Main objectives} \label{sec:MainObjectives}

The CR can be considered as the ÒParticleÓ channel complementing the ÒElectromagneticÓ one of conventional astronomy. The main objective of JEM-EUSO is to initiate a new field of astronomy and astrophysics that uses the extreme energy particle channel ($10^{19.5} eV < E < 10^{21}$ eV). JEM-EUSO is designed to achieve more than $10^5$ km$^{2}$ sr yr above $7 \times 10^{19}$ eV during its first three years of operation which, given current uncertainties, amounts to the detection of between 500 and 800 events with energy above $5.5 \times 10^{19}$ eV \cite{JEMEUSOPerformance}. Such a number of events makes possible the following targets: (a) identification of sources by high-statistics arrival direction analysis; 
(b) measurement of the energy spectra from individual sources to constrain acceleration or emission mechanisms.

A remarkable characteristic of the EECR flux is that few
astrophysical candidates are known which can attain such energies with the acceleration
mechanisms we are presently aware of \cite{HillasPlot,StanevSummaryAccelSites2003}. 
This fact makes imperative the identification of both, those sources and of the powering 
mechanisms at play.

\begin{figure}[!bt] 
\centering
\includegraphics[width=8.5cm]{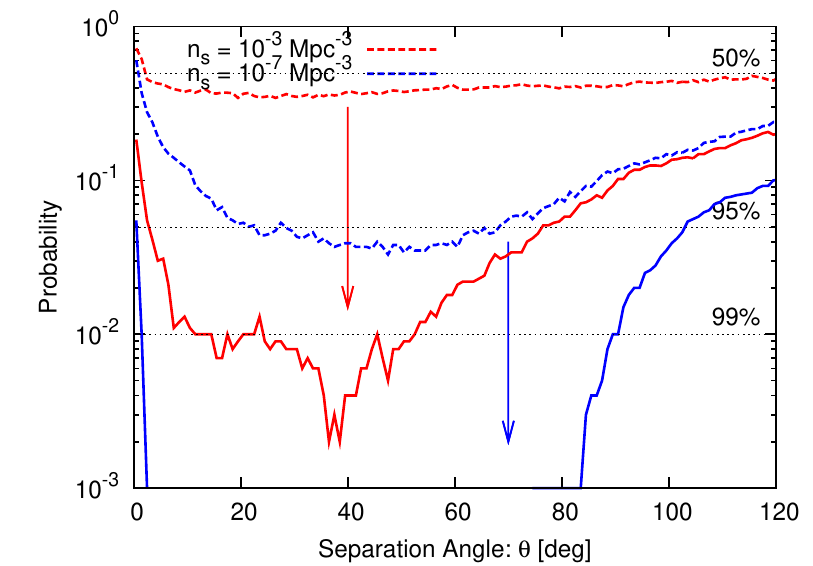}\label{figSourceDensityTakami}
\caption{Probability that positive excess in the arrival distribution of EECRs (for Fe injection at LSMD from IRAS PSCz), compared to an isotropic distribution, is NOT realized (p). The probability that positive excess is realized (1-p) is indicated by numbers in the figure \cite{Takami}.}
\end{figure}

Given that a correlation between the arrival directions of EECR and the Galactic plane has never been 
observed, not to mention the relative calmness of the Milky Way, it is broadly accepted that
the particles have an extragalactic origin. Furthermore, in all the most conservative models,
the sources either follow the distribution of luminous matter or that of the associated dark 
matter. In either case, at large enough energies, anisotropy in the arrival directions 
is expected in the form of an enhanced correlation with nearby luminous matter, as data from 
the Pierre Auger Observatory presently implies \cite{AugerAGN}. To complicate the picture 
even further, the particles are widely thought to be predominantly baryons and, therefore, to posses
charge during at least a significant portion of their transit through the intergalactic medium. 
Magnetic fields of poorly known intensity and topology are likely widespread throughout the 
universe, blurring any correlation between arrival directions and source position on the sky. 

Particles also interact with the CMBR and the IR background. At the energies of 
JEM-EUSO the dominant target is the CMB which leads, in the case of HE protons, to 
photo-pair and photo-pion production. Above $\sim 10^{19.6}$ eV 
the latter dominates and can effectively decelerate particles to below the threshold for photo-pion
production in few tens of Mpc, strongly suppressing the EECR energy spectrum (the GZK cut-off), 
and effectively setting a horizon at $\sim 100$ Mpc. Nuclei, on the other hand, lose 
energy mainly by photo-disintegration. The end result is a similar 
attenuation length for Fe, but shorter for intermediate nuclei. Therefore, the volume of 
universe sampled by EECR, regardless of their mass, is local in cosmic terms and 
encompasses a region where the large scale matter distribution (LSMD) is inhomogeneous. 
Thus, under general assumptions and given enough statistics, the footprint of the source 
distribution should emerge from the EECR flux.

\begin{figure}[!h]
\centering
\includegraphics[width=0.5\textwidth]{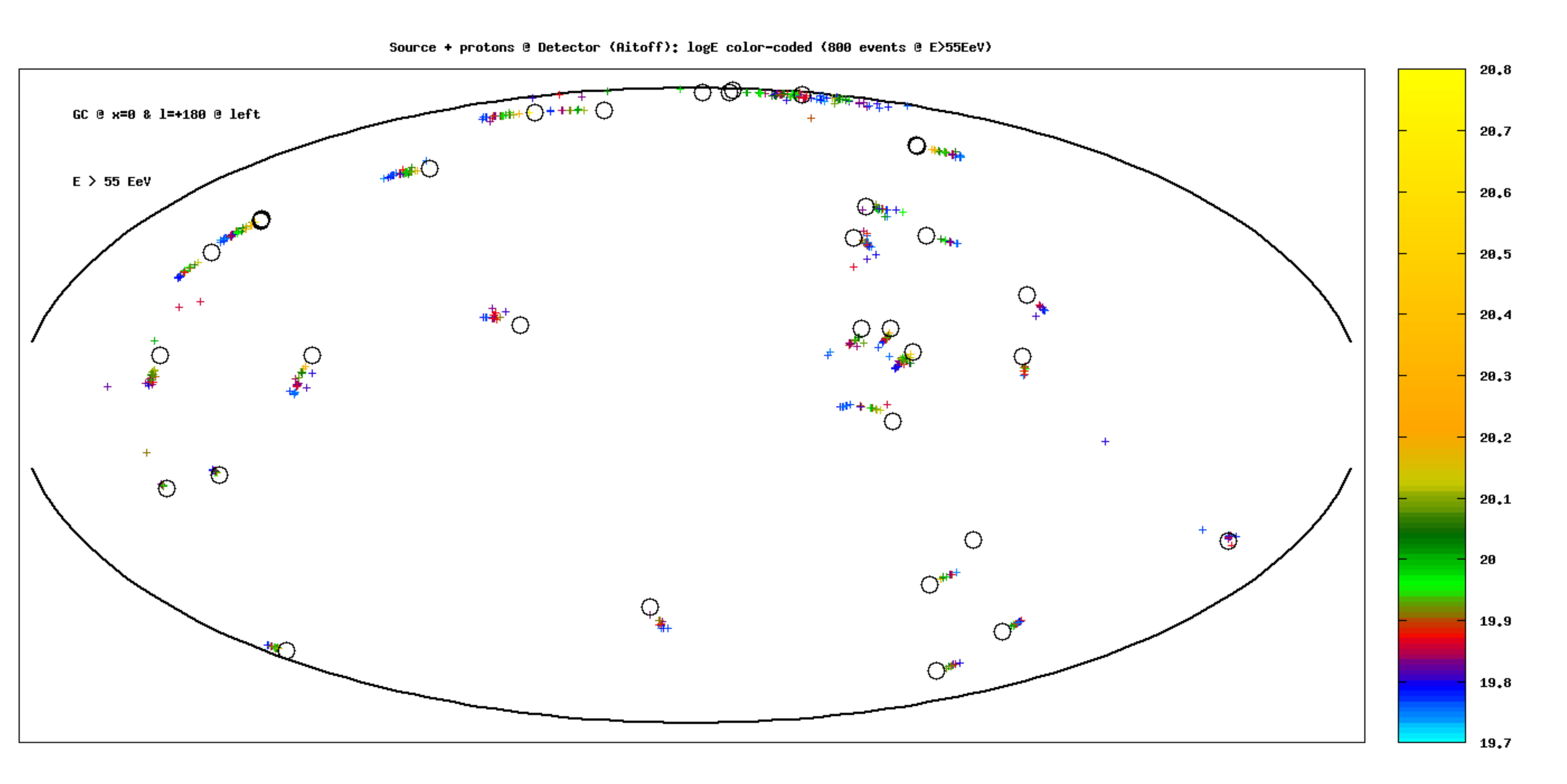}\label{figSkyIndivSrc}
\includegraphics[width=0.5\textwidth]{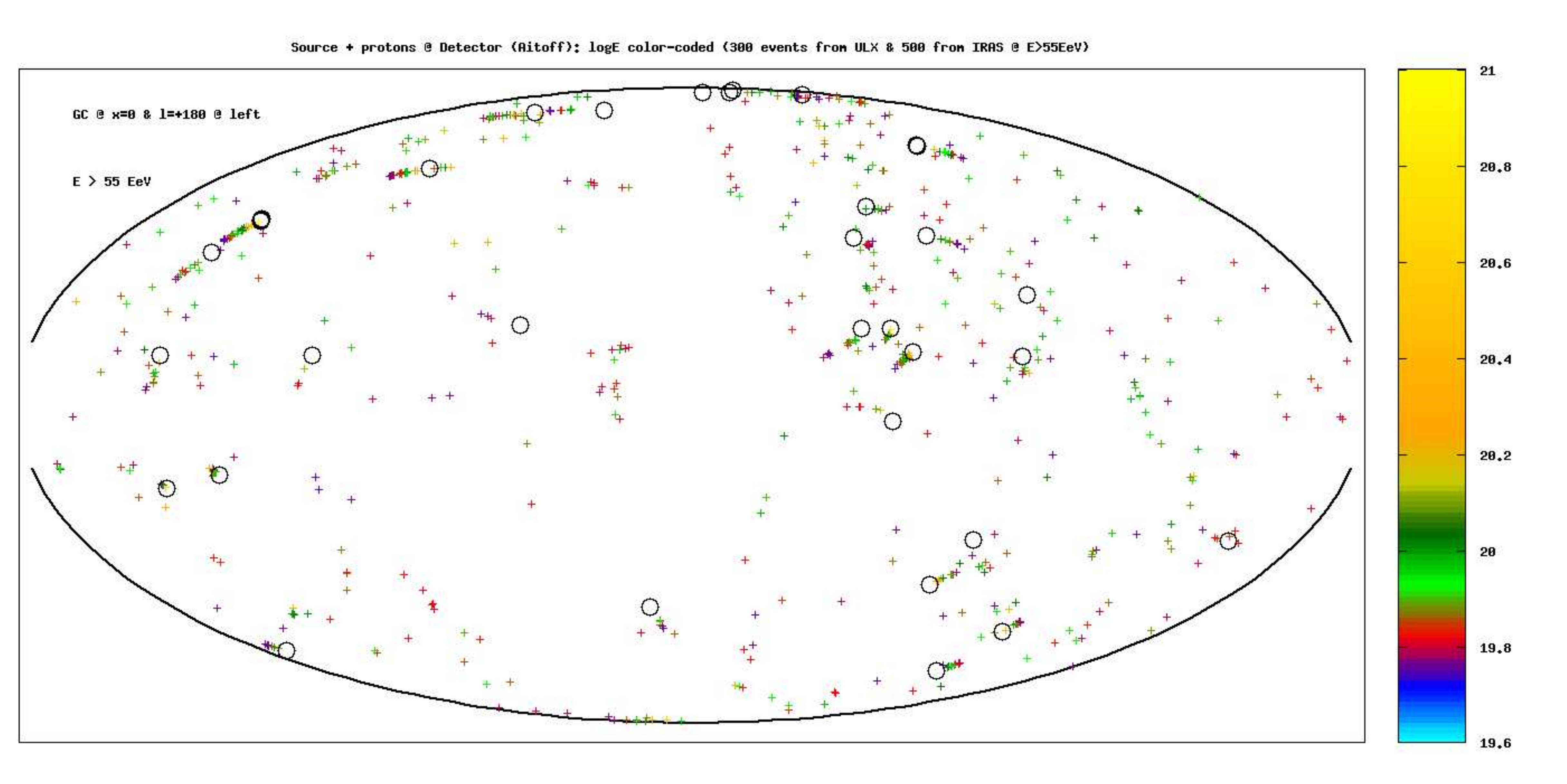}
\caption{Simulated distribution of arrival direction of EECRs protons with $E>55$ EeV, for (top) ULX sources and (bottom) a combination of ULX sources which contribute 37\% of the events to a background originated in LSMD IRAS galaxies contributing 63\%  of events. Black circles denote ULX positions. The energy of events is color-coded in a log-scale. The intergalactic field (IGMF) is modeled following Faraday Rotation constraints and the LSMD traced by de IRAS PSCz. The GMF corresponds to a BSS-S disk with vertical component and $h_{z} \sim 250$ pc, immersed in an ASS-A magnetized halo extending up to 20 kpc. The injection at the sources follows $dN_{inj}/dE \propto E^{-2.7}$ \cite{GMT&Toshi}.}
\end{figure}

The identification of the sources can follow different paths. First, a statistical identification 
can be attempted. In this case, arrival directions and source positions from candidate 
astrophysical catalogues are globally compared and the corresponding correlation is quantified.
This has been attempted many times 
in the literature for the various experiments for a variety of astronomical catalogues and, 
most notably, recently for Auger \cite{AugerAGN} and HiRes \cite{HiResAGN} data. However, 
the results are always severely bounded by the low available statistics at the highest
energies and, to a lesser extent, by the small observed fraction of the sky and the strong 
exposure dependence on declination. JEM-EUSO, with its full sky coverage, low declination
dependence of the exposure and large aperture, can significantly improve this kind of analysis.

There are several approaches to infer the density of nearby sources of EECR. 
If magnetic deflections are not too large, a low density of sources implies a 
relatively high EECR luminosity per source and, therefore, a smaller number of large multiplicity
clusters of events is expected, while the opposite should occur in a large density scenario.
The degree of clustering over the celestial sphere should also be dependent on the large scale 
spatial distribution of the sources. However, in practice, the number of parameters involved 
when trying to explore this avenue leads to ambiguous results due to the present limited data
set \cite{Couco2009SrcDensity}. Again, JEM-EUSO will have a strong impact in this arena,
since its increased statistics will allow the discrimination of source densities in the interval
$n_{s} \sim 10^{-7} - 10^{-3}$ Mpc$^{-3}$ at more than 99\% confidence level, as it is shown 
in Figure \ref{figSourceDensityTakami} in comparison to the present statistics of Auger above
55 EeV.

\begin{figure}[!h]
\centering
\includegraphics[width=0.5\textwidth]{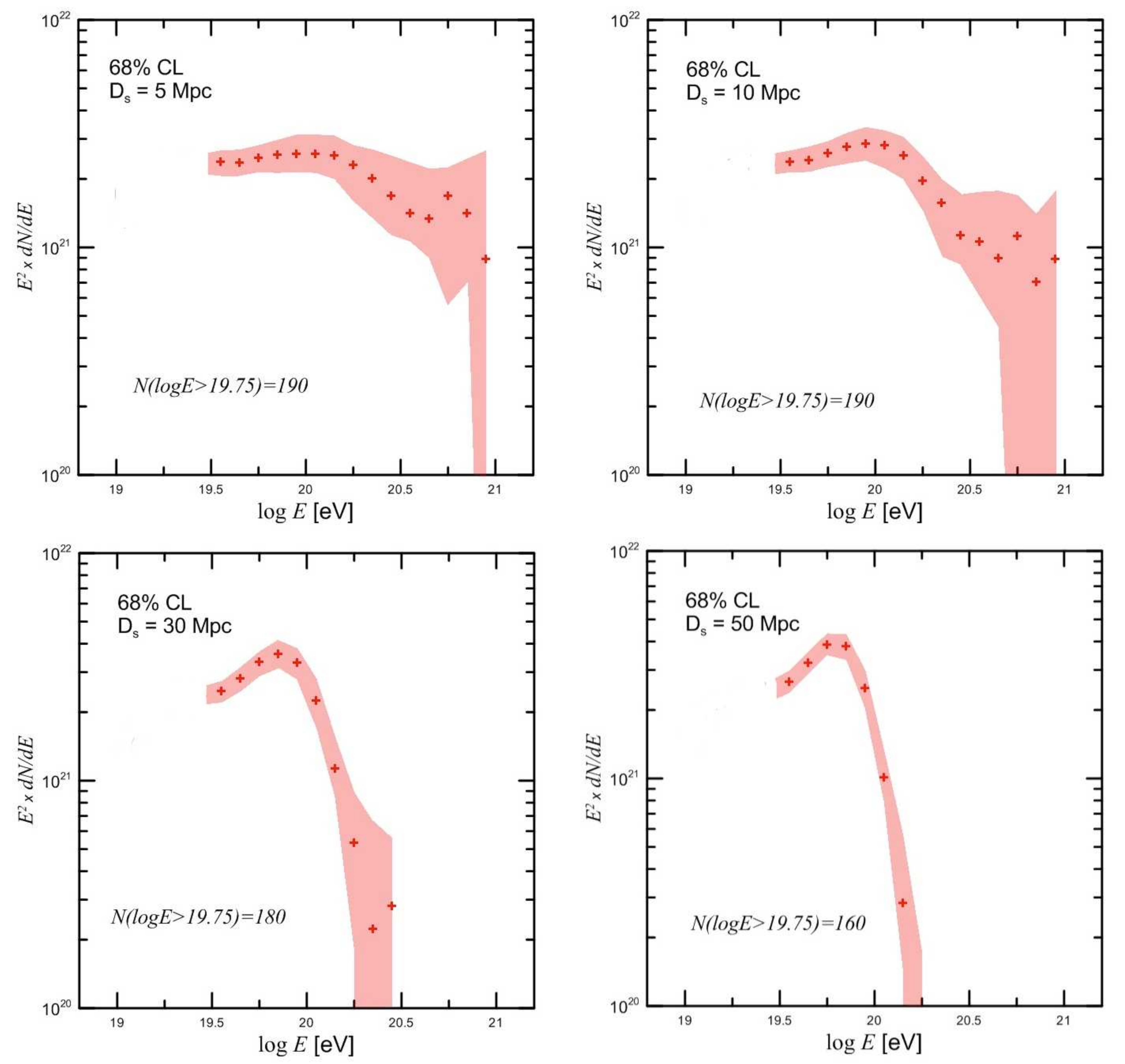}\label{figIndivSpectra}
\caption{
Simulated observed spectra of a point sources as a function of distance. The median and the upper and lower 68\% CL are shown for each spectrum. The hypothetical sources have the same flux at Earth, which amounts to  $\sim 160-190$ events above 55 EeV. If achieved in 5 yrs of operation of JEM-EUSO, it corresponds to a collection rate at Auger of $<4$ yr$^{-1}$ per source. $dN_{inj}/dE \propto E^{-2}$, IGMF $\sim 1$ nG and $L_{c} \sim1$ Mpc. Incoming events are selected with an appropriate trigger probability and their energies are convoluted with an energy and azimuth dependent error \cite{GMT&Toshi}.
}
\end{figure}

Another novel possibility is to directly observe individual sources. 
In  this context, an individual source is a very high multiplicity cluster whose 
events are genetically related. Indications of such a cluster may be already popping up in 
the Auger data in the general direction of Cen A. Whether this enhancement is the product 
of a single astrophysical object or the combined effect of a compact more distant region of 
individual sources, e.g., the huge Shapley supercluster behind Cen A, is impossible to tell 
at the present level of statistics. Other relatively nearby sources may be also contributing significantly 
to the EECR flux, although masked at present by the limited fraction of the sky available to Auger 
and the strong declination dependence of its exposure. In fact, M87 and the Virgo cluster 
may be just such an example. JEM-EUSO, on the other hand, will be able to detect those 
sources if they exist. Figure \ref{figSkyIndivSrc} shows how the JEM-EUSO sky after 3 yrs 
of exposure could look like if some particular class of object, ultra-luminous X-ray Galaxies (ULX), 
were sources of EECR contributing 37 \% of the total flux originated from the LSMD as traced by 
the IRAS catalog (bottom).  Furthermore, if several sources are found with at least dozens of 
observed EECR events, then
the observed differences in spectral features among those sources Figure \ref{figIndivSpectra}, 
combined with a multi-wavelength approach, will provide direct clues on the identity of the 
sources and the acceleration mechanism involved.

The energy dependent distortions of the sources' point spread functions as a result of the 
Galactic magnetic field can be clearly seen as a function of the position on the sky (top 
panel in Fig.\ref{figSkyIndivSrc}). This pattern of distortions, over the celestial sphere can 
be used to infer the large scale structure of the Galactic magnetic field (GMF).

\section{Exploratory objectives} \label{sec:SecondaryObjectives}

\begin{figure}[!h] 
\centering
\includegraphics[width=0.5\textwidth]{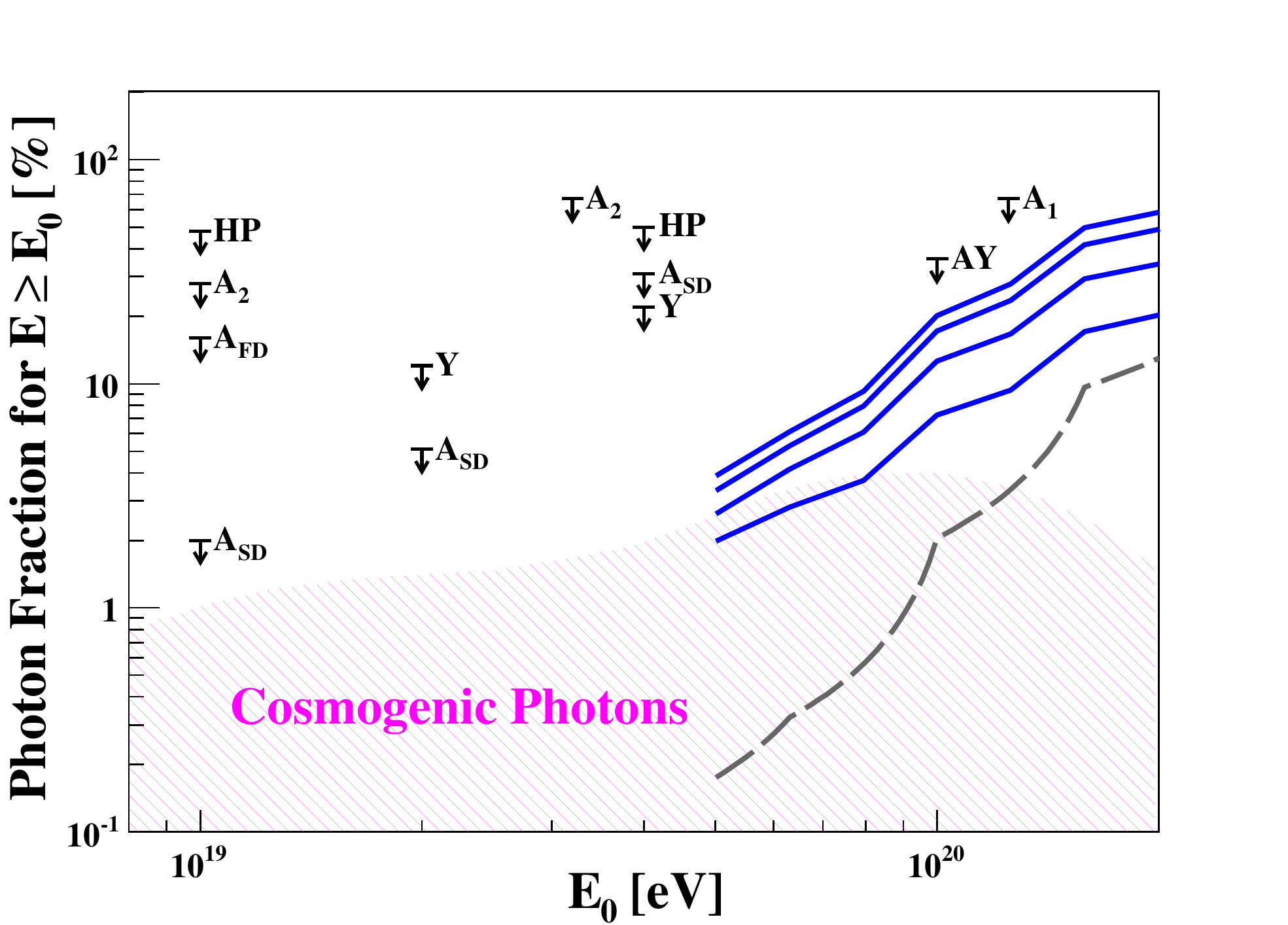}\label{figUpperLimitsXmax}
\caption{Upper limits on the fraction of photons in the integral cosmic ray flux at 95\% of confidence level.
Dashed line corresponds to the ideal case in which it is known that there is no photon in the data. Blue solid lines are the
upper limits obtained by using $X_{max}$; from bottom to top, different $X_{max}$ reconstruction uncertainties of 0, 70, 120 and 150 g cm$^{-2}$ are considered. See \cite{Supa&GMTgammas} for detils.}
\end{figure}

Gamma rays at extreme energies are a natural consequence of $\pi_{0}$ production during EECR proton propagation
through the CMB. A gamma-ray flux higher than expected from this secondary production would signify a new production mechanism, such as top-down decay/annihilation, or a breaking of Lorentz symmetry. Nuclei, on the other, would produce a much smaller gamma background. Therefore, the flux of gamma rays in extreme energy is a key parameter to discriminate origin models. Figure \ref{figUpperLimitsXmax} summarizes existing limits on the gamma-ray flux and shows the sensitivity of gamma rays by five years operation of the JEM-EUSO Mission. The Auger Observatory reported the upper limit on gamma ray flux as a few percent of EECR flux above 10 EeV \cite{AugerPhotonLimit}. Under the null gamma ray assumption, JEM-EUSO is capable of putting more stringent upper limit by an order of magnitude at overlapping energies.  To give the constraint on origin models or their parameters, the gamma ray flux above 100 EeV is essential and will be constrained in an unprecedented way after five years operation of JEM-EUSO.

\begin{figure}[!t] 
\centering
\includegraphics[width=0.5\textwidth]{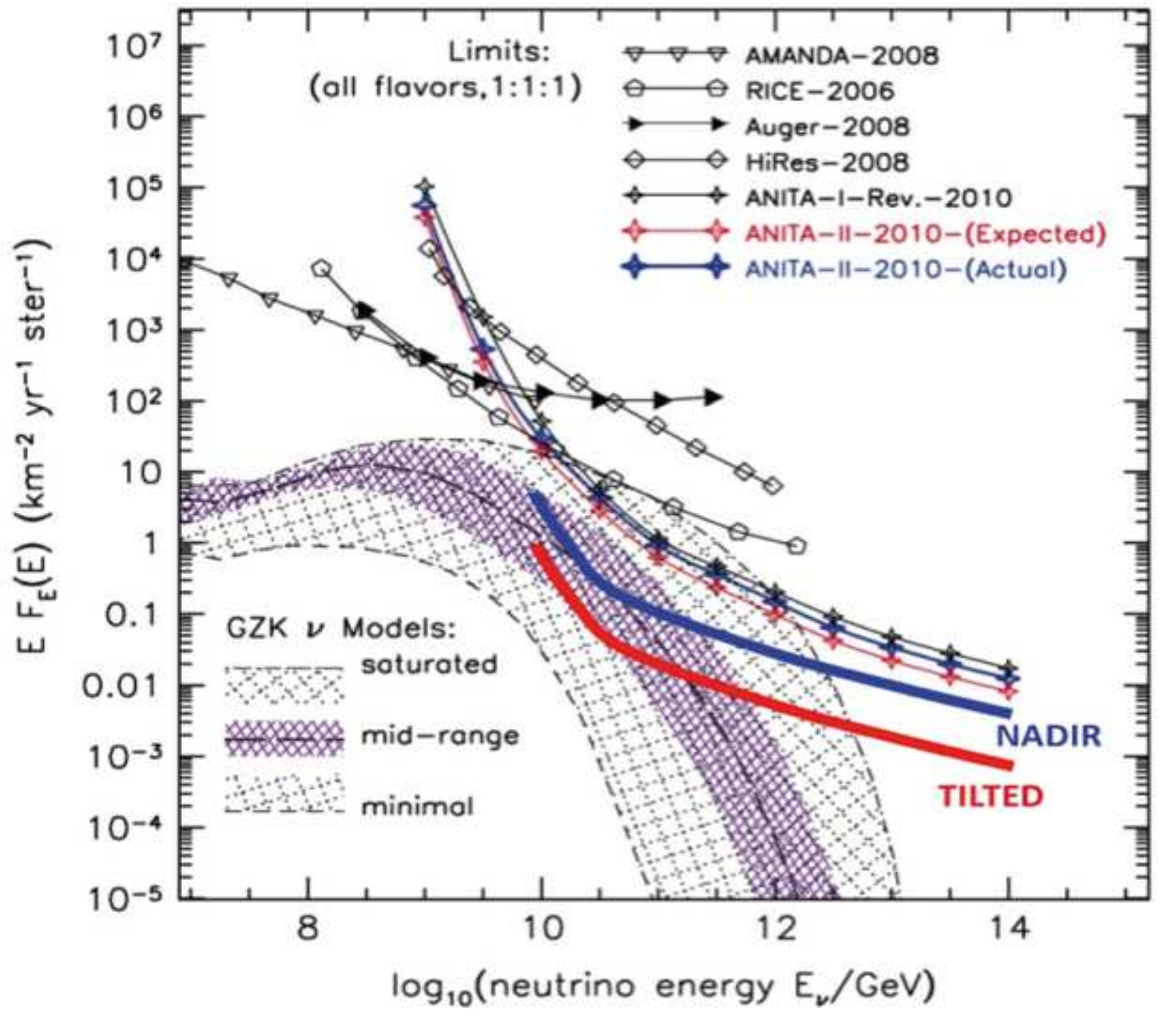}\label{figNeutrinoLimits}
\caption{Flux sensitivity of JEM-EUSO detecting 1 event/energy-decade/yr. An observational efficiency of 25\% is assumed. Thick blue and red curves show the case of nadir and tilted modes, respectively. Adapted from \cite{ANITA}}
\end{figure}

During proton propagation through the CMBR, $\nu$ are produced. These {\it cosmogenic neutrinos\/} constitute a guaranteed flux at Earth and contain extremely valuable information on the redshift evolution of the sources. Besides the cosmogenic flux there may also be contributions from hadronic interactions at the acceleration sites and from top-down processes. 
JEM-EUSO can detect neutrinos evolving deep in the atmosphere or, in the case of bursts of upward going neutrinos interacting inside the outermost layers of the crust, as expected form GRB, through direct Cherenkov. 
Figure \ref{figNeutrinoLimits} shows the flux sensitivity of JEM-EUSO for several neutrino production 
models for both nadir and tilted mode operation of the telescope.  The discovery of EE $\nu$ beyond 100 EeV has profound implications on our understanding of production mechanisms, since protons of energy $>1$ ZeV at the source are required to create such energetic $\nu$ via the pion chain. Higher energy neutrinos should originate either by top-down mechanisms or by less understood bottom-up channels, like exotic plasma phenomena or unipolar induction in extreme environments.

Furthermore, the $\nu$ cross-section is uncertain and highly model-dependent. Extra-dimension 
models \cite{NuExtraDim} in which the Universe is supposed to consist of 10 or 11 dimensions 
are among the favored models to unify quantum mechanics and gravitation theory. In these models, 
the predicted neutrino cross-section is $10^{2}$ times larger than the Standard Model prediction. 
Under these conditions, JEM-EUSO should observe $100$s of $\nu$ events, which would immediately validate experimentally low-scale unification. In addition, the ratio of horizontal to upward $\nu$-originated EAS gives a quantitative estimation of $\nu$ cross-section around $10^{14}$ eV center of mass energies \cite{NuCrossSectionEstimate}.

Additionally, a stringent test of relativity could be made from high multiplicity sources at known distances. If the GZK steepening functions consistently deviate at some directions in the sky, external vector fields might be emerging which are not unidirectionally Lorentz Invariant. On the other hand, verification of LI at EHE would disfavor such vector fields. \cite{N_Myu}. 

%
%
%

%
%
%

\clearpage


  \includepdf[pages=-]{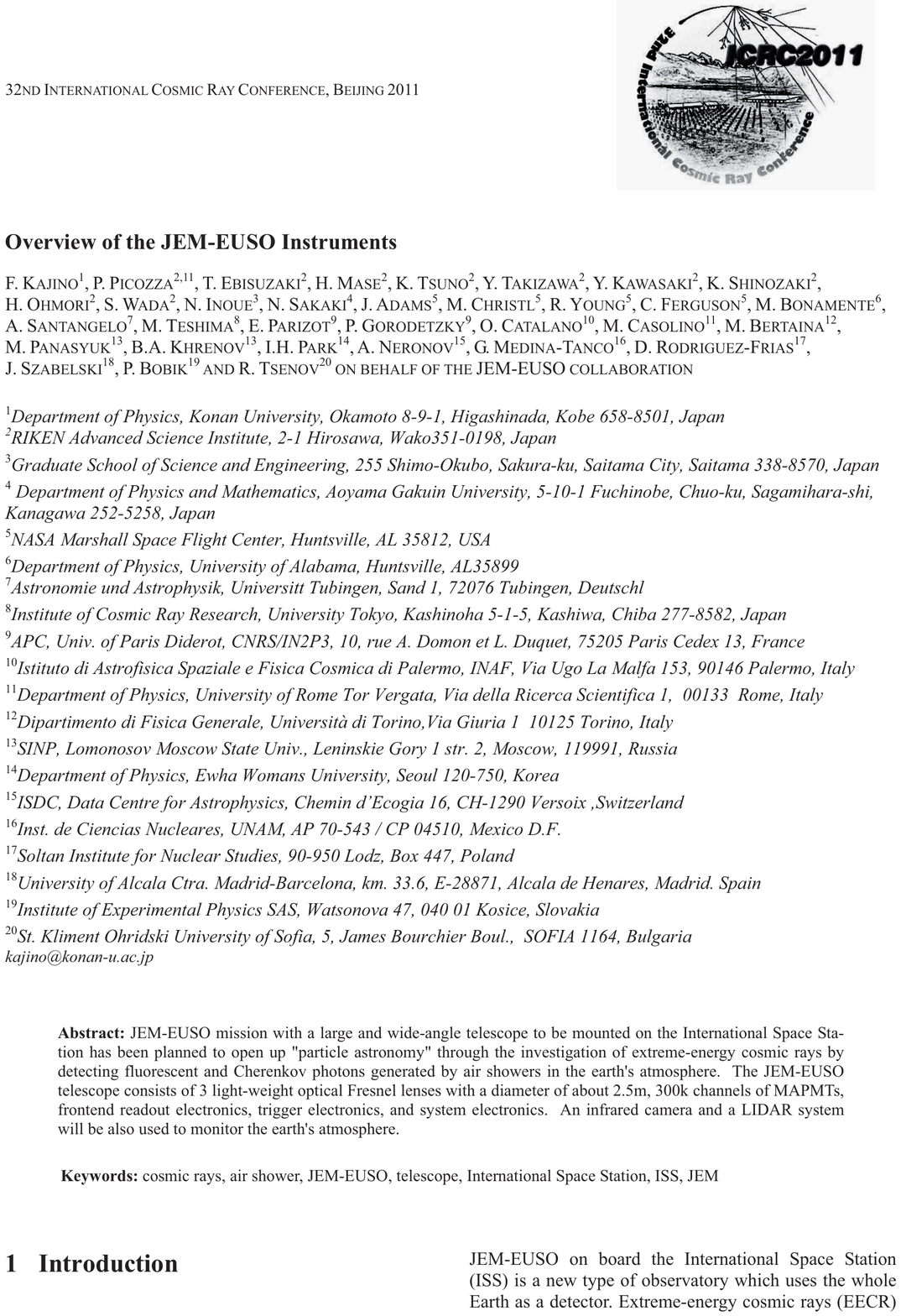}

\newpage
\normalsize
\setcounter{section}{0}
\setcounter{figure}{0}
\setcounter{table}{0}
\setcounter{equation}{0}



\title{Requirements and Expected Performances of the JEM-EUSO mission}

\shorttitle{A. Santangelo, M. Bertaina \etal Requirements and Expected
Performances of the JEM-EUSO mission}

\authors{
M.~Bertaina$^{1}$,
A.~Santangelo$^{2,3}$,
K.~Shinozaki$^{3}$, 
F.~Fenu$^{2,3}$,
T.~Mernik$^{2}$,
P.~Bobik$^{5}$,
F.~Garino$^{1}$,
K.~Higashide$^{3,4}$,
G.~Medina Tanco$^{6}$,
G.~Saez Cano$^{7}$,
on behalf of the JEM-EUSO Collaboration}
\afiliations{
$^1$ Dipartimento di Fisica Generale dell' Universit\`a di Torino, Torino, Italy\\
$^2$ IAAT, Kepler Center f\"ur Astro- und Teilchenphysik, Universit\"at
T\"ubingen, T\"ubingen, Germany\\
$^3$ Computational Astrophysics Laboratory, RIKEN, Wako, Japan\\
$^4$ Graduate School of Science and Engineering, Saitama University, Japan\\
$^5$ Instutute of Experimental Physics SAS, Kosice, Slovakia\\
$^6$ ICN-UNAM, Mexico City, Mexico\\
$^7$ SPace \& AStroparticle (SPAS) Group, University of Alcal\'a, Madrid, Spain\\
}
\email{Andrea.Santangelo@uni-tuebingen.de}

\abstract{
In this paper we describe the requirements and the expected performances of the Extreme 
Universe Space Observatory (EUSO) onboard the Japanese Experiment Module 
(JEM) of the International Space Station. Designed as the first mission to 
explore the Ultra High Energy (UHE) Universe from space, JEM-EUSO will monitor, 
night-time, the earth's atmosphere to record the UV (300-400 nm) tracks 
generated by the Extensive Air Showers produced by UHE primaries 
propagating in the atmosphere. After briefing summarizing the main aspects 
of the JEM-EUSO Instrument and mission baseline, we will present in details
 our studies on the expected trigger rate, the estimated exposure, as well 
as on the expected angular, energy, and X$_{max}$ resolution. Eventually, the 
obtained results will be discussed in the context of the scientific 
requirements of the mission.}

\maketitle

\section{Introduction}
JEM-EUSO \cite{Yoshi,Toshi} is an innovative space-based mission with the aim of detecting 
Ultra High Energy Cosmic Rays (UHECR) from the International Space Station (ISS), by using the earth's atmosphere as a fluorescence detector. JEM-EUSO consists of an UV telescope and of an atmosphere monitoring system. Orbiting the earth every $\sim$90 minutes, JEM-EUSO is designed to detect, from an 
altitude of 350-400 km, the moving track of the UV (300-400 nm) fluorescence photons produced during the development of Extensive Air Showers (EAS) in the atmosphere.
The telescope, which contains a wide Field-of-View ($\pm$30$^\circ$, FOV) optics composed by Fresnel lenses \cite{Takky}, records the EAS-induced tracks with a time resolution of 2.5$\mu$s and a spatial resolution of about 0.5 km ($\sim0.07^\circ$) in nadir mode by using a highly pixellised focal surface ($\sim$3$\times$10$^5$ pixels) \cite{Kajino}. These time-segmented images allow an accurate measurement of the energy and arrival direction of the primary particles.\\
Since the ISS orbits the earth in the latitude range $\pm 51^{\circ}$,
moving at a speed of $\sim 7$ km/s, the variability of the FOV observed by JEM-EUSO is much higher than that observed by ground-based experiments. In particular the atmospheric conditions, which eventually determine the acceptance, must be carefully monitored via an atmosphere monitoring system consisting of an infrared camera \cite{Spain} and a LIDAR \cite{Andrii}.\\ 
Thanks to the ISS orbit, JEM-EUSO will monitor, with a rather uniform exposure, both hemispheres minimizing the systematic uncertainties that strongly affect any comparison between different observatories exploring, from ground, different hemispheres.\\
The other great advantage of JEM-EUSO, in comparison to any existing or studied ground-based observatory, is the significant increase of aperture (see Section ~\ref{sec:aperture}). 
There are however other relevant advantages in using space-based UHE observatories. First, the non-proximity of the detector to the EAS considerably reduces all problems associated with the determination of the solid angle and with the different attenuation suffered by the UV light in the atmosphere. Second, the near-constant fluorescence emission rate at different heights below the stratosphere simplifies all assumptions on the energy-fluorescence yield relation at the EAS maximum as well as on the dependence of the EAS time structure on the production height \cite{Kenji}. Third, the observation from space minimize uncertainties due to scattering by aerosols limited to altitudes below the atmospheric boundary layer. Finally, as the EAS maximum develops, for most zenith angles, at altitudes higher than 3-5 km from ground, space measurements are also possible in cloudy sky
conditions. Compared to ground-based detectors, the duty cycle is therefore mainly limited by the moon phases, while the cloud impact is less relevant.\\
The JEM-EUSO observational approach mainly relies on the fact that a substantial fraction of the UV fluorescence light generated by the EAS can reach a light-collecting device of several square meters: typically a few thousand photons reach the JEM-EUSO detector for a shower produced by a $10^{20}$ eV particle.
JEM-EUSO is designed to record not only the number of photons but also their direction and arrival time.
It is the observation of the specific space-time correlation that allows to very precisely identify EAS tracks in the night glow background.\\
We wish also to observe that JEM-EUSO has considerably improved with respect to the original Extreme Universe Space Observatory \cite{EUSO} studied by the European Space Agency. Main improvements have to be ascribed to the new optics \cite{Takky} (with $\sim$1.5 better throughput and $\sim$1.5 better 
focusing capability), to the photo-detector \cite{Kawasaki} ($\sim$1.6 higher detection efficiency), to the better geometrical layout of the focal surface that maximizes the filling factor \cite{Ricci}, and to the improved performance of the electronics \cite{Il,Joerg}, which allows to 
exploit more complex trigger algorithms \cite{Bertaina}.\\
The key element to estimate the science potential of JEM-EUSO is its exposure. This is determined by three main contributions: the trigger aperture, the observational duty cycle and the cloud impact. In the following sections the three terms are discussed in details.

\section{Night-glow background and estimation of the observational duty cycle}
\label{sec:duty}
The UV tracks of EAS must be discriminated in the night-glow background. One key parameter is therefore the fraction of time in which EAS observations are not hampered by the brightness of the sky. We define \textit{observational duty-cycle} the fraction of time in which the sky is dark enough to measure EAS. 
Pavol et al. \cite{Pavol} have conducted an analysis of the duty-cycle using measurements performed by the 
Tatiana satellite rescaling them to the ISS orbit. In this estimate all major atmospheric effects, such as lightnings, meteors and anthropic lights (e.g. city lights) have been included.
Results indicate that for a zenith angle position of the sun higher than 108$^\circ$ (120$^\circ$), the fraction of time in which the night-glow 
background is less than 1500 ph/m$^2$/ns/sr is 22\% (18\%). In fact the mean of all background levels 
less than 1500 ph/m$^2$/ns/sr, weighted according to their relative occurrence, is equivalent to an average 
background of 500 ph/m$^2$/ns/sr: the so-called standard background actually 
measured by different balloon experiments. This is a conservative estimate for the highest energies where measurement
could be performed even in a higher background condition. These recent studies confirms previous estimates of 18\%--22\% performed in the context of the EUSO studies, based on a combined analytical and simulation approach \cite{Berat2}. We therefore assume a value of 20\% as the most probable value for the observational duty-cycle of the mission. 

\section{The cloud impact}
\label{sec:clouds}
Space based UHE observatories can observe EAS induced tracks also in cloudy conditions: this is typically not the 
case for ground-based observatories. In fact if the maximum of the shower is above the cloud top layer the reconstruction 
of the shower's parameters is still possible. It is clear that the same cloud top layer will affect in different ways showers of various inclination 
or originating from different type of primary particles (e.g. neutrino will develop much deeper in the atmosphere compared to protons).
Thin clouds ($\tau <$1, typical of cirrus) might affect the measurement of the energy but arrival direction will still be 
nicely measurable. Thick clouds ($\tau >$1) will strongly impact the measurement only if located at high altitudes.
As an example, a 60$^\circ$ zenith-angle inclined shower will reach the 
shower maximum at an altitude of 6--7 km, much higher than the typical range of stratus.
In order to quantify the effective observational time, a study on the
distribution of clouds as a function of altitude, optical depth and 
geographical location has been performed using different meteorological
data sets \cite{ggarino}. 
Table~\ref{tab1} reports the results of the occurrence of each cloud typology
for oceans during daytime using visible and IR information. 
 
\begin{figure}[th!]
  \centering
  \includegraphics[width=3.0in]{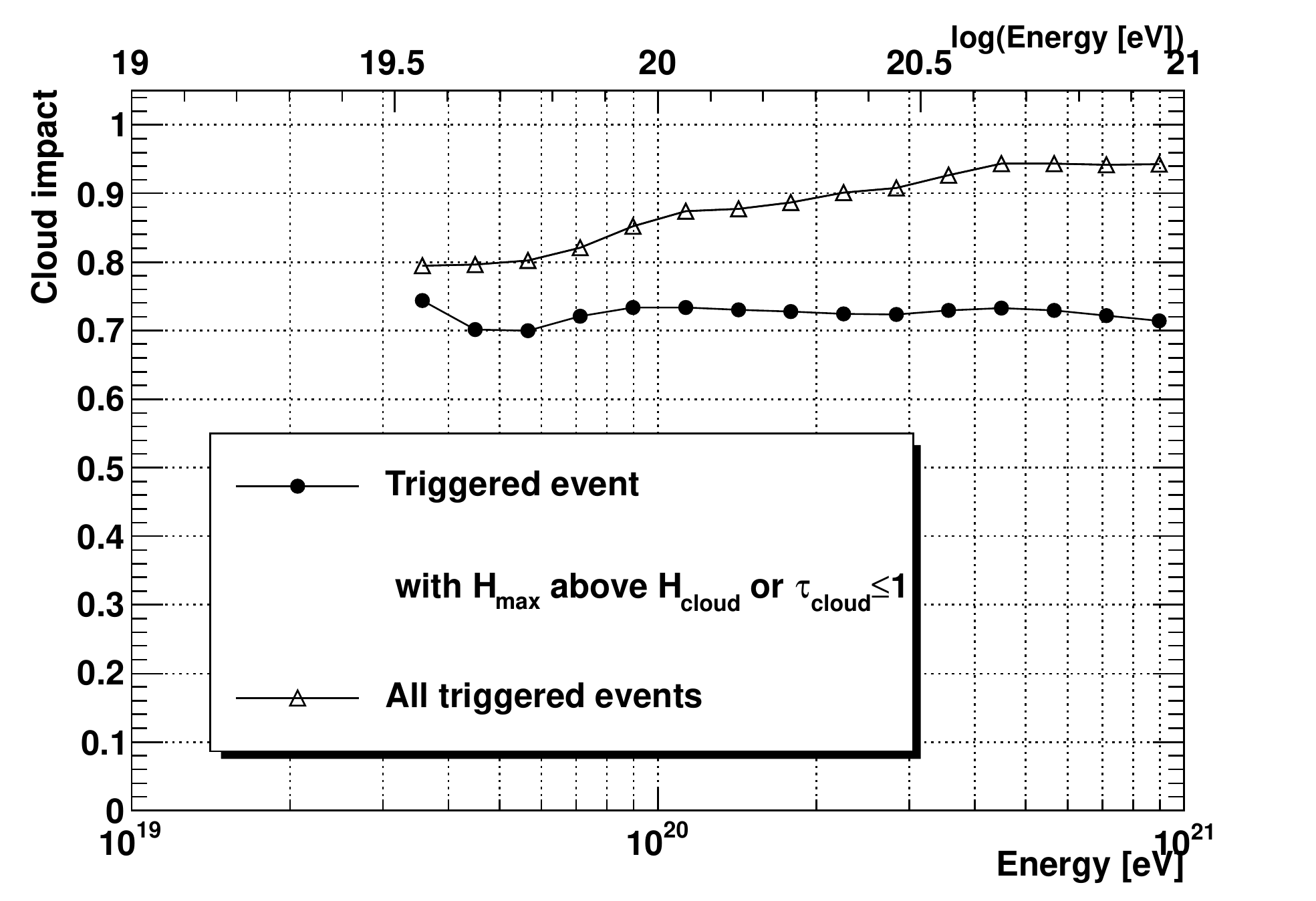}
   \vspace{-5mm}
  \caption{Ratio in the trigger efficiency for clear-sky and cloudy
conditions. }
  \label{fig:ratio}
 \end{figure}

 \begin{table}[th!]
  \caption{Relative occurrence (\%) of clouds between 50$^\circ$N and 50$^\circ$S
latitudes on TOVS database in the matrix of cloud-top altitude vs optical
depth. Daytime and ocean data are used for the better accuracy of the
measurements.\\}
   {
  \centering
  \begin{tabular}{|l|c|c|c|c|c|}
  \hline
Optical Depth&\multicolumn{4}{c|}{Cloud-top altitude}\\ \hline
 &$<$3km&3-7km&7-10km&$>$10km\\ \hline
$>$2 & 17.2 & 5.2 & 6.4 & 6.1\\ \hline
1-2 & 5.9 & 2.9 & 3.5 & 3.1\\ \hline
0.1-1 & 6.4 & 2.4 & 3.7 & 6.8\\ \hline
$<$0.1 & 29.2 & $<$0.1 & $<$0.1 & 1.2\\ \hline
  \end{tabular}
  \label{tab1}
   }
 \end{table}
In Table~\ref{tab1} cloud coverage data taken during daytime are chosen since they are in 
general more precise.  The same applies to data of clouds above the oceans, more reliable than the ones taken
above land. A comparison between day and night cloud coverage has been then performed 
for data above land as higher variations are expected in comparison with day/night variation above the oceans. 
Differences however resulted to be of only a few
percents. 
The results of Table~\ref{tab1} can be understood as it follows.
Clear sky corresponds to $\tau<0.1$ and this accounts for $\sim$30\% of the observation time. Clouds 
below 3 km height do not hamper the measurements as the shower maximum 
will develop at higher altitudes, regardless of their $\tau$ and they account 
for another $\sim$30\%, which gives a total of $\sim$60\% of the time when 
the measurement is possible with no major correction. Thick optically depth ($\tau>$1) high clouds (h$>$7km) 
will prevent the possibility of any measurements, and they account for 
$\sim$19\%. For the remaining $\sim$21\% angular and energy measurements will be possible for very 
inclined showers (zenith angle $>$60$^\circ$) which correspond 
to the best set of showers characterized by long tracks. 
For the non inclined showers of this last sample arrival direction analysis will still be possible while the energy estimation will be severely hampered by the 
shower attenuation in the atmosphere.\\
More quantitative results have been obtained by simulating showers according to the conditions of Table \ref{tab1}, determining 
the trigger efficiency in the different conditions, and by convoluting it with the corresponding aperture.
Fig.~\ref{fig:ratio} shows the ratio between the aperture in cloudy 
conditions compared to clear sky, for all events and for those
events which have 'good quality' characteristics (clouds with $\tau<1$, and
shower maximum well above the cloud top height). 

More details can be found in ~\cite{Lupe}. 
From these results we conclude that 70\% is a conservative estimate of the fraction of observing time in which
the measurement will not be hampered by atmospheric factors. This number
convoluted with the 20\% duty-cycle, provides a final
14\% factor to be applied to the aperture to determine the
exposure Fig.~\ref{fig:aperture}.

\section{Trigger rate and estimated exposure}
\label{sec:aperture}
\begin{figure}[th!]
  \centering
  \includegraphics[width=3.5in]{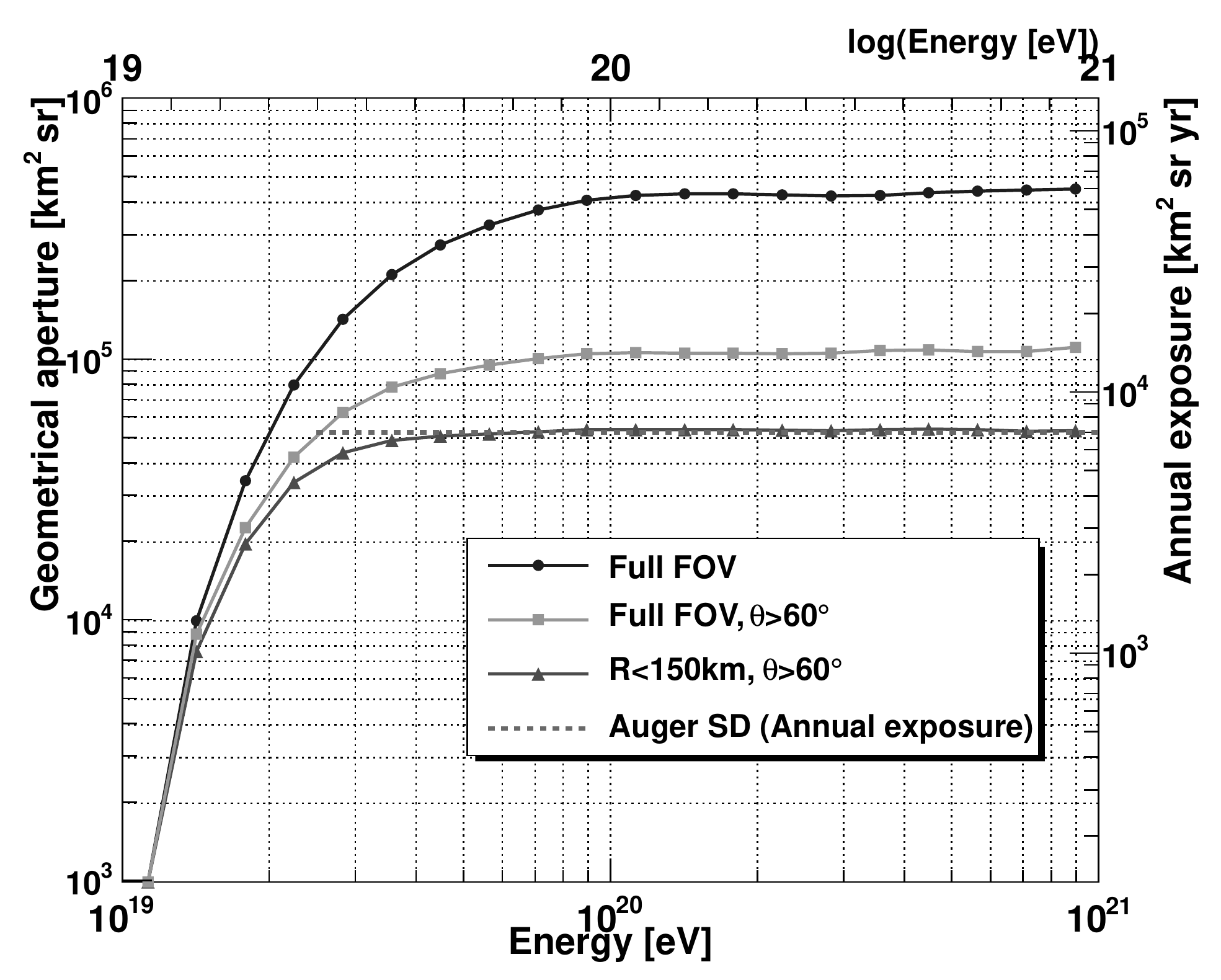}
   \vspace{-5mm}
  \caption{Aperture and annual exposure of JEM-EUSO for different quality cuts.}
  \label{fig:aperture}
 \end{figure}

The last parameter needed to estimate the aperture and the exposure is the trigger
efficiency. Main objective of the trigger system is to reduce the rate of UHECR 
candidates to $\sim$0.1 Hz, limit imposed by downlink telemetry
capabilities. The rejection level of the trigger algorithm 
determines the aperture of the instrument as a function of the energy. The rejection power depends also on the average night-glow 
background. In the following, the background has been assumed to be 500 ph/m$^2$/ns/sr.\\
Fig.~\ref{fig:aperture} shows the full aperture, and annual exposure of
JEM-EUSO in nadir mode for the entire FOV of the detector and for a few 
high quality conditions corresponding to "quality cuts". Quality cuts are defined by the better performance of the optics
in the center of the FOV or for showers with inclined zenith angles 
($\theta>60^\circ$ from nadir), which produce longer and less attenuated tracks.\\
Fig.\ref{fig:aperture} shows that 80-90\% of the full aperture is already reached at
energies $\sim$2-3$\times$10$^{19}$~eV when the foot print of the shower is located
in the central part of the FOV ( R$<$150 km from nadir) and for showers with zenith angles 
$\theta>60^\circ$ (more details in \cite{Kenji2}). The 80-90\% is reached at 
$\sim$5$\times$10$^{19}$~eV if showers distributed in the entire FOV are considered.

The convolution of the trigger aperture with the observational duty cycle and the cloud
impact gives the annual exposure. In the most stringent conditions JEM-EUSO has an annual exposure equivalent to Auger
($\sim 7\times 10^{3}$ km$^2$~sr~y) while it reaches $\sim 60 \times 10^{3}$ km$^2$~sr~y at 10$^{20}$~eV that is
9 times Auger. JEM-EUSO will well overlap (about one order of magnitude, starting from 2-3$\times$10$^{19}$~eV) with ground-based experiments to cross-check systematics and performances. At higher energies JEM-EUSO will be able to accumulate statistics at a pace per year at about one order or magnitude higher than currently existing ground based detectors. JEM-EUSO will also be operates in tilt mode to further increase the exposure at the highest energies ($E> 3-5 \times 10^{20}$ eV) by a factor of $\sim$3 compared to nadir mode. The optimization of the tilt parameters is still under evaluation.
 
 \begin{table*}[ht!]
  \caption{Relative comparison of apertures and exposures of current and planned
UHECR observatories.}
   {  \vspace{+2mm}
   \scriptsize
  \centering
  \begin{tabular}{|l|c|c|c|c|c|c|c|c|}
  \hline
Observatory&Aperture&Status&Start&Lifetime&Duty Cycle $\times$ &Annual Exposure &Relative Exposure\\ 
&(km$^2$sr)& & & & Cloud Impact& (km$^2$ sr yr) & Auger = 1 \\ \hline
Auger&7000&Operations&2006&4(16)&1&7000&1\\ \hline
TA&1200&Operations&2008&2(14)&1&1200&0.2\\ \hline
TUS&30000&Developed&2012&5&0.14&4200&0.6\\ \hline
JEM-EUSO (E$\sim10^{20}$eV)&430000&Design&2017&5&0.14&60000&9\\ \hline
JEM-EUSO (highest E) tilt mode $^\circ$&1500000&Design&2017&5&0.14&200000&28\\ \hline
  \end{tabular}
  \label{tab2}
   }
 \end{table*}

\section{Reconstruction capabilities}
\label{requirements}

The JEM-EUSO reconstruction capabilities have been estimated using the ESAF code
\cite{Berat}, a software for the simulation of space based UHECR 
detectors developed in the context of the EUSO ESA mission. 
Currently the ESAF code is being updated to the most recent 
JEM-EUSO configuration \cite{Fenu}. The technique to reconstruct the
different shower parameters is extensively discussed in \cite{Mernik}.
Regarding the energy reconstruction, at the current status of 
development of the instrument and of the reconstruction algorithms, 
proton showers with zenith angle $\theta> 60^{\circ}$
are reconstructed in clear-sky conditions with a typical energy resolution 
$\Delta E / E$ of $\sim 25\% (20\%)$ at energies around $4\times 10^{19}$ ($10^{20}$)~eV. The 
energy resolution slightly worsen for more vertical showers where it is of the order of 30\% around $10^{20}$~eV.
This result indicates that the reconstruction of events with $E<5\times10^{19}$~eV is still possible
confirming the possibility of overlapping with ground based experiments over a sufficient wide energy range.
Regarding the arrival direction analysis, our current results (\cite{Mernik}) indicate that showers of energy 
$E\sim 7\times10^{19}$~eV and zenith angle $\theta$ $>$ 60$^{\circ}$
can be reconstructed with a 68\% separation angle less than 2.5$^{\circ}$. Eventually our still preliminary results indicate that the $X_{max}$ uncertainties 
the $\sigma_{X_{max}}$ are better than 70~g/cm$^2$ for $E\sim10^{20}$~eV

\section{Meeting the Scientific Requirements}
The scientific requirements of the mission are described in detail in 
\cite{Toshi}. They can be summarized as: Observation area greater than $1.3\times10^5$km$^2$; Arrival direction determination accuracy
better than 2.5$^{\circ}$ for 60$^{\circ}$ inclined showers at
E $>$1$\times$10$^{20}$ eV (standard showers); Energy determination accuracy better than $30\%$  for standard showers; $\sigma_{X_{max}}$ $<$ 120~g/cm$^2$.

Results of simulations shown in the previous section confirm that the requirements can be already achieved with the current configuration. 

The number of events that JEM-EUSO will observe depends of course on the UHE flux, which is uncertain especially at the highest energies. 
The apertures shown in fig.~\ref{fig:aperture} can be however converted into number of events, assuming fluxes reported in literature by the Pierre Auger and the HiRes observatories 
\cite{aauger,hhires}, and a conversion factor 0.14 between aperture and exposure (cloud impact included). 
We obtain more than 500 events with energy E $>$5.5$\times$10$^{19}$ eV for the flux measured by Auger and more than 1200 in the case of the HiRes spectrum.\\
A synthetic comparison between the JEM-EUSO aperture and exposure
and the ones of other observatories is reported in Table~\ref{tab2}. 
\section{Conclusions}
The expected performance of the JEM-EUSO mission has been reviewed. Simulations show that 
JEM-EUSO can reach almost full efficiency already at 
energies around 2.5-3$\times$10$^{19}$ for a restricted subset of 
events, and full aperture at energies E$>$ 5.5$\times$10$^{19}$~eV. The 
expected annual exposure of JEM-EUSO at 10$^{20}$ is equivalent 
to about 9 years exposure of Auger. 
The duty cycle and the impact of clouds has been assessed. Results
indicate that the assumptions of 20\% operational time and 70\%
cloud impact are matched by the present analysis.
The angular, energy and X$_{max}$ 
resolutions satisfy the requirements. The total number of events
expected at energies E$>$ 5.5$\times$10$^{19}$ in 5 years of operations ranges between 500 and 1200 events.

\vspace*{2mm} \footnotesize{
{\bf Acknowledgement:} 
This work has been partially supported by Deutsches Zentrum fuer Luft- und Raumfahrt, and by the Italian Ministry of Foreign Affairs, General Direction for the
Cultural Promotion and Cooperation.}

\vspace{-2mm}
\

\clearpage


\newpage
\normalsize
\setcounter{section}{0}
\setcounter{figure}{0}
\setcounter{table}{0}
\setcounter{equation}{0}



\title{The potential of the JEM-EUSO telescope for the astrophysics of extreme energy photons}

\shorttitle{A.D. Supanitsky \etal High energy photons in JEM-EUSO}

\authors{A.D. Supanitsky$^{1,2}$, G. Medina-Tanco$^{2}$ for the JEM-EUSO Collaboration.}
\afiliations{$^1$Instituto de Astronom\'ia y F\'isica del Espacio (IAFE), UBA-CONICET, Argentina.
\\ $^2$Instituto de Ciencias Nucleares, UNAM, Circuito Exteriror S/N, Ciudad Universitaria, M\'exico D. F. 04510, M\'exico.}

\email{supanitsky@iafe.uba.ar}

\abstract{
Extreme energy photons are expected to be a minor component of 
the ultra high energy cosmic rays. Nevertheless, they are the carriers 
of very important astrophysical information related to the origin and 
propagation of such ultra energetic particles. JEM-EUSO is an orbital 
fluorescence telescope intended to observe the highest energy component 
of the cosmic rays, including photons and neutrinos. In this work we 
study several techniques to improve the discrimination between photon 
and proton showers in the context of the JEM-EUSO telescope. The most 
important parameter used to discriminate between protons and gammas is 
the atmospheric depth of the maximum of the showers, $X_{max}$. However,
it can be demonstrated that, for a given available statistics, additional
information is needed in order to take advantage of the full potential of 
the instrument. We propose and study additional parameters, related to 
the shape of the longitudinal profile, in order to obtain a better 
discrimination than the one given by $X_{max}$ alone.   
}

\maketitle

\section{Introduction}

Extreme high energy photons can be generated as a consequence of the interaction of the cosmic rays during their 
propagation through the intergalactic medium \cite{aGelmini:08}. They can also be produced as by-products of the
cosmic rays interactions in the acceleration sites \cite{aOstapchenko:08} and, although disfavored by present 
data, they can be generated in top-down scenarios involving the decay of super heavy relic particles 
or topological defects \cite{aAloisio:04}. At present there is no ultra high energy photon unambiguously 
identified.

High energy photons initiate air showers when they interact with the molecules of the atmosphere. In the high 
energy region the characteristics of such air showers are dominated by the LPM effect and pre-showering (i.e., 
photon splitting) in the geomagnetic field (see Ref. \cite{aRisse:07} for a review). In this work we present an
improved version of the methods, recently proposed in Ref. \cite{aSupanitsky:11}, to calculate the upper limits 
on the photon fraction in the integral flux by using the $X_{max}$ parameter, the atmospheric depth of the maximum 
development of the showers, in the context of the JEM-EUSO mission \cite{aTakahashi:09}. We also study and propose 
new parameters in order to improve the proton-photon separation in the energy range relevant to JEM-EUSO.

\section{Upper limit calculation: $X_{max}$}

Let us consider the ideal situation in which it is known that there is no photons in a given sample of $N$ events.
For this case, the expression for the upper limit to the photon fraction is given by \cite{aRisse:07},
\begin{equation}
\mathcal{F}_{\gamma}^{min} = 1-(1-\alpha)^{1/N}
\label{Fuplideal}
\end{equation} 
where $\alpha$ is the confidence level of rejection. However, in practice, the probability of the existence of 
photons must be realistically assessed through some observational technique which involves the determination of 
experimental parameters, which leads unavoidably to less restrictive upper limits than the previous one.

The method used to calculate the upper limit by using $X_{max}$ parameter is based on the abundance estimator first 
introduced in \cite{aSupanitsky:08a},
\begin{equation}
\xi_{X_{max}} = \frac{1}{N} \sum_{i=0}^{N} \frac{f_{\gamma}(X_{max}^i)}{f_{\gamma}(X_{max}^i)+f_{pr}(X_{max}^i)}
\label{eatdef}
\end{equation}
where $f_\gamma(X_{max})$ and $f_{pr}(X_{max})$ are the photon and proton distribution functions, $X_{max}^{i}$ are 
experimental values of $X_{max}$ and $N$ is the sample size. $\xi_{X_{max}}$ is an estimator of the photon abundance, 
$c_{\gamma}=N_\gamma/N$ where $N_\gamma$ is the number of photons in the sample. The mean value and the variance of 
$\xi_{X_{max}}$ are given by,
\begin{eqnarray}
E[\xi_{X_{max}}] &=& u_1 c_\gamma + u_2, \\
Var[\xi_{X_{max}}] &=& \frac{1}{N} \left[v_1 c_\gamma + v_2 + u_1^2 c_\gamma (1-c_\gamma) \right],
\end{eqnarray}
where $u_1=\alpha_1-\alpha_2$, $u_2=\alpha_2$, $v_1=\alpha_3-\alpha_4+\alpha_2^2-\alpha_1^2$ and $v_2=\alpha_4-\alpha_2^2$. 
Here 
\begin{eqnarray}
\alpha_1 &=& \int dX_{max} \frac{f_{\gamma}(X_{max})^2}{f_{\gamma}(X_{max})+f_{pr}(X_{max})}, \\
\alpha_2 &=& \int dX_{max} \frac{f_{\gamma}(X_{max}) f_{pr}(X_{max})}{f_{\gamma}(X_{max})+f_{pr}(X_{max})}, \\
\alpha_3 &=& \int dX_{max} \frac{f_{\gamma}(X_{max})^3}{[f_{\gamma}(X_{max})+f_{pr}(X_{max})]^2}, \\
\alpha_3 &=& \int dX_{max} \frac{f_{\gamma}(X_{max})^2 f_{pr}(X_{max})}{[f_{\gamma}(X_{max})+f_{pr}(X_{max})]^2}. 
\end{eqnarray}
Note that the last term in the expression of the variance has to do with the binomial fluctuations of the process.
It is assumed, for the present calculation, that the distribution functions of $\xi_{X_{max}}$ is Gaussian,
which is valid for large enough values of $N$ (due to the central limit theorem).

The upper limit to the photon fraction, for the case in which there is no photons in the sample, is given by,
\begin{eqnarray}
\mathcal{F}_{\gamma} &=& \frac{1}{2 u_1^2 (1+s_\alpha^2/N)} \left[ \frac{s_\alpha^2}{N} (v_1+u_1^2) + \right. \nonumber \\
&& \left. \sqrt{\frac{s_\alpha^4}{N^2} (v_1+u_1^2)^2+4 \frac{u_1^2 v_2}{N} (1+\frac{s_\alpha^2}{N})} \right],
\label{Fuplxi}
\end{eqnarray}
where, $s_\alpha = \sqrt{2}\ \textrm{Erf}^{-1}(2 \alpha-1)$ and 
\begin{equation}
\textrm{Erf}(x) = \frac{2}{\sqrt{\pi}} \int_0^x dt \exp(-t^2). 
\end{equation}

A shower library was generated by using the last version of CONEX \cite{aconex} (v2r2.3) which consist of 
$1.1 \times 10^{5}$ proton showers following a power law energy spectrum of spectral index $\gamma = -1$ in the interval 
[$10^{19.7}, 10^{21}$] eV and with uniformly distributed arrival directions. Also $1.5 \times 10^{5}$ photon showers 
were generated under the same conditions but in this case cores were also uniformly distributed on the surface of 
the Earth in order to properly take into account pre-showering effect in the geomagnetic field. The distribution 
functions needed to calculate $\mathcal{F}_{\gamma} $ are obtained from the simulated data by using the non-parametric 
method of kernel superposition with adaptive bandwidth \cite{aSilvermann:86,aSupanitsky:08a}.

\begin{figure}[!ht]
\centering
\includegraphics[width=8.5cm]{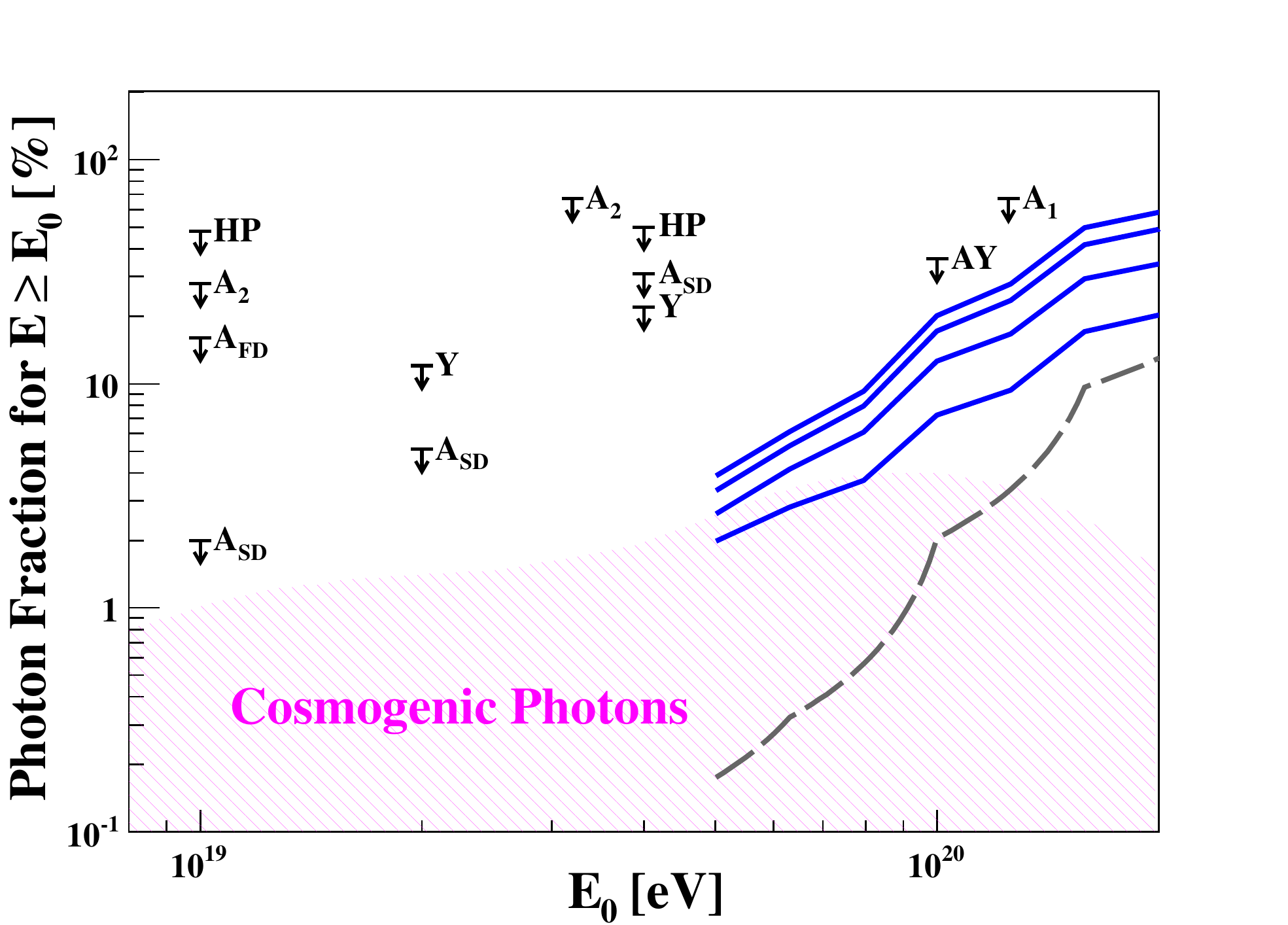}
\caption{The upper limits on the fraction of photons in the integral cosmic ray flux at 95\% of confidence level.
Dashed line corresponds to the ideal case in which it is known that there is no photon in the data. Blue lines 
are the upper limits obtained by using $\xi_{X_{max}}$ method, they correspond, from bottom to top, to a Gaussian 
uncertainty of 0, 70, 120 and 150 g cm$^{-2}$. Shadow region is the prediction for the GZK photons \cite{aGelmini:08}. 
Black arrows are experimental limits, HP: Haverah Park \cite{aAve}; A$_1$, A$_2$: AGASA \cite{aRisse:05,aShinozaki:02}; 
A$_{\textrm{FD}}$, A$_{\textrm{SD}}$: Auger \cite{aaugerFD,aaugerSD}; AY: AGASA-Yakutsk \cite{aRubtsov:06}; Y: Yakutsk 
\cite{aGlushkov:07}.}
\label{UpperL}
\end{figure}

Fig. \ref{UpperL} shows the upper limits on the fraction of photons in the integral cosmic ray flux, at 95\% of 
confidence level, obtained in the ideal case $\mathcal{F}_{\gamma}^{min}$ (dashed line), by using the $\xi_{X_{max}}$ 
method (blue lines), and also the upper limits obtained by different experiments. The calculation is done for 
$E\geq 5\times10^{19}$ eV and $\theta \in [30^\circ, 80^\circ]$. For each method, the lines from bottom to top 
correspond to a Gaussian uncertainty on the determination of $X_{max}$ of 0, 70, 120 and 150 g cm$^{-2}$.  
The number of events above a given energy, $E_0$, and the spectrum are obtained from Ref. \cite{aInoue:09}. The number 
of events corresponds to two years in nadir mode plus three years in tilt ($\alpha_{Tilt}=38^{\circ}$) mode for the 
JEM-EUSO mission. Also, a reconstruction efficiency, taking into account the presence of clouds, of $\epsilon_{R}=50\%$ 
is assumed (see Ref. \cite{aBerat:10}). 

Note that, at the highest energies, the upper limit curves can be underestimated due to the decrease in the number
of events. In this energy region the Gaussian approximation of the distribution function of $\xi_{X_{max}}$ could
be not so good. 

\section{Photon-proton separation with skewness}

\begin{figure}[!ht]
\centering
\includegraphics[width=8cm]{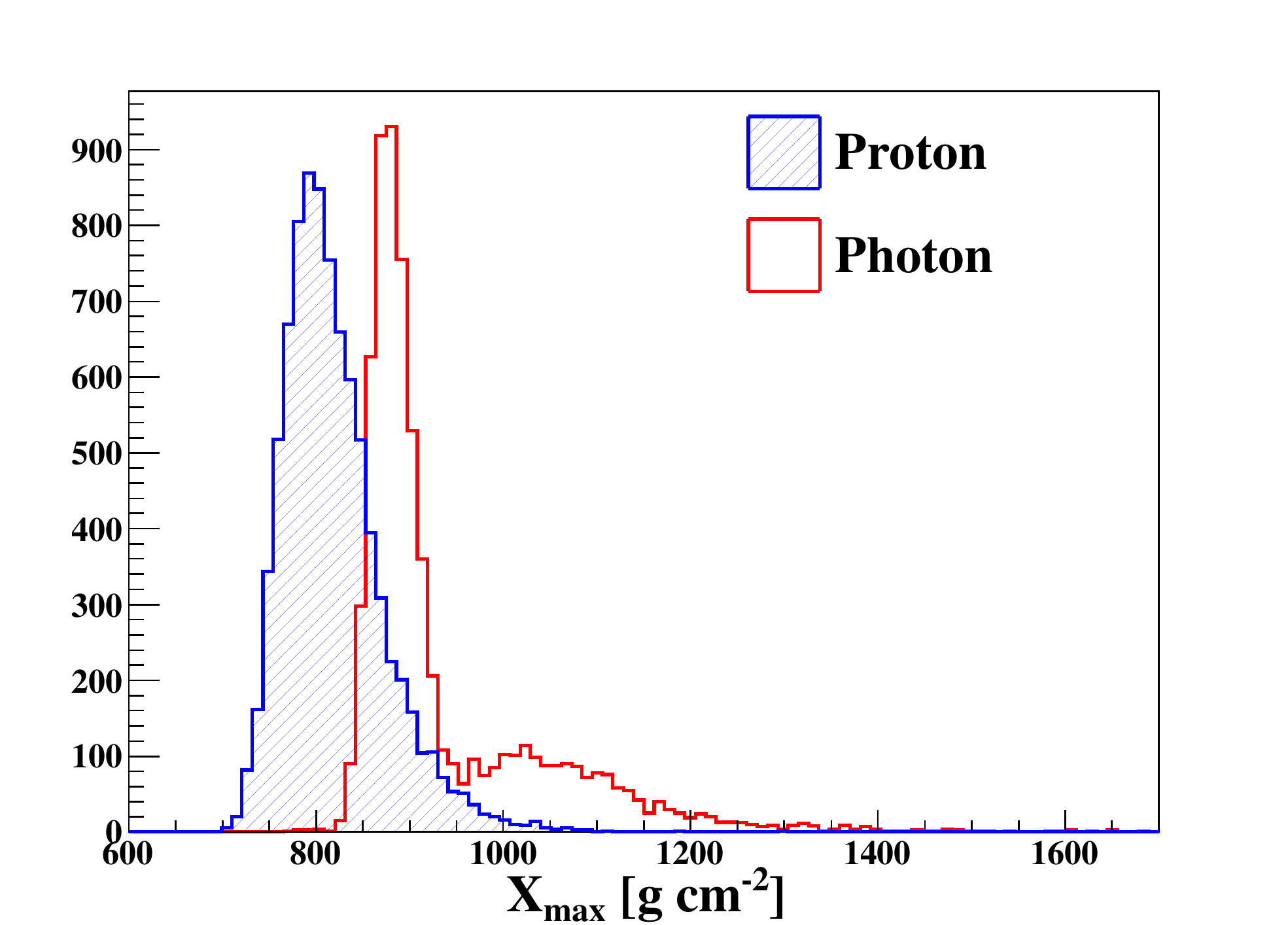}
\includegraphics[width=8cm]{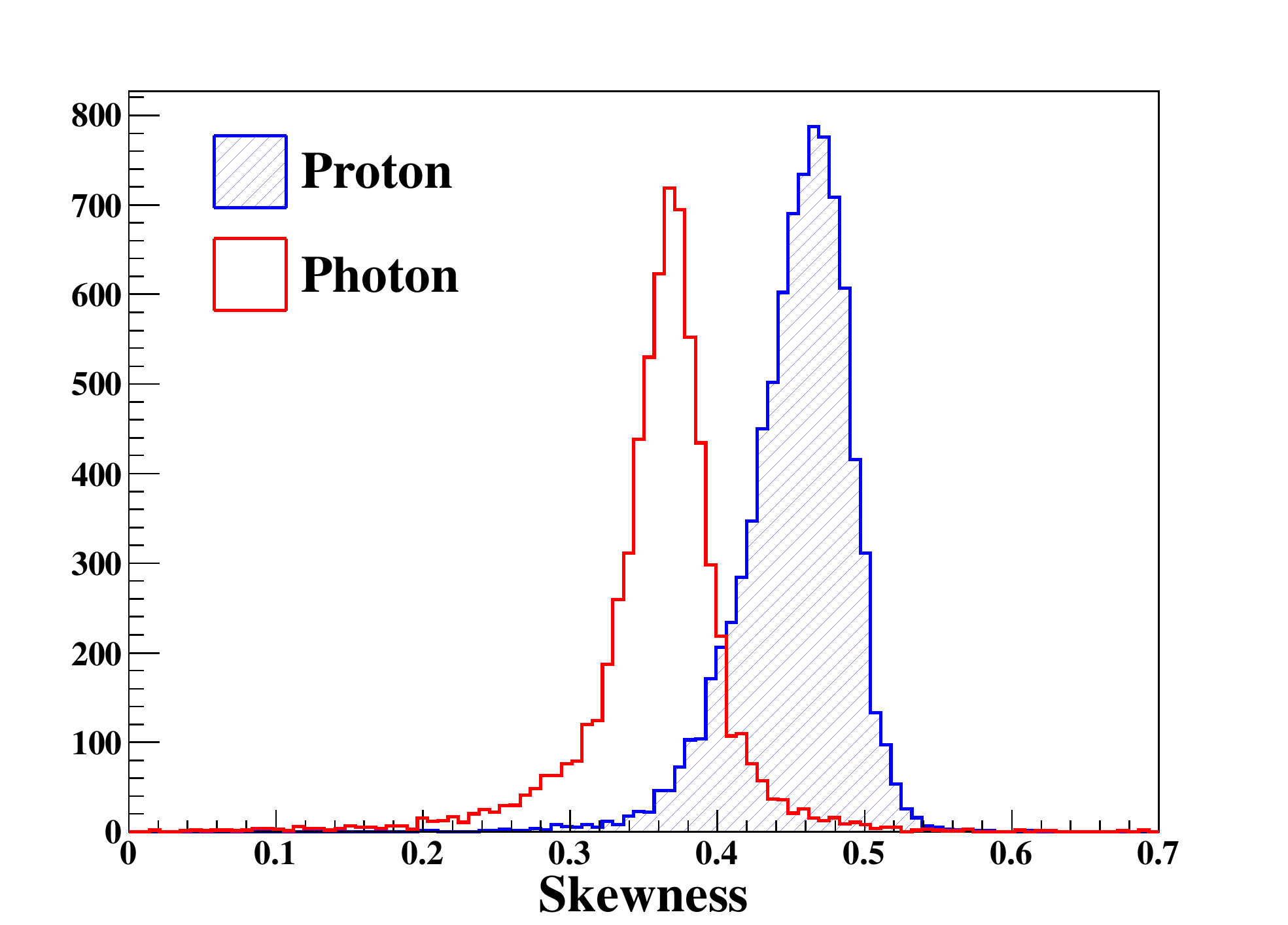}
\caption{$X_{max}$ and skewness distributions for $\theta \in [30^\circ, 80^\circ]$ and $\log(E/eV) \in [ 19.7, 20 ]$.
Skewness is calculated with the part of the longitudinal profile between $X_{max} \pm $ 1000 g cm$^{-2}$.}
\label{XmaxSkewMono}
\end{figure}

In Ref. \cite{aSouza:07} it is shown that the skewness of the longitudinal profile of the showers is one of the best 
parameters to discriminate between primaries. However, its discrimination power depends on the part of the track of 
the cascade observed. Figure \ref{XmaxSkewMono} shows the distributions of $X_{max}$ and skewness for 
$\theta \in [30^\circ, 80^\circ]$ and $\log(E/eV) \in [ 19.7, 20 ]$. Skewness is calculated by using the part of the 
profiles between $X_{max}$-1000 g cm$^{-2}$ and $X_{max}$+1000 g cm$^{-2}$. As discussed in \cite{aSupanitsky:11} 
the $X_{max}$ distribution presents two peaks, the one at lower values corresponds to the photons that suffered photon 
splitting in the geomagnetic field and the one at higher values corresponds to the ones that do not. From the figure
it can be seen that skewness separates better photons from protons than $X_{max}$. The merit factor measures how good
is a given parameter to discriminate between two species, it is defined as 
$MF=(E[x_{pr}]-E[x_{ph}])/\sqrt{Var[x_{pr}]+Var[x_{ph}]^2}$, where $E[x_{A}]$ and $Var[x_{A}]$ are the mean value and the 
variance of $x_A$ with $A=\{pr,\ ph\}$ and $pr=$ proton and $ph =$ photon. The merit factors of $X_{max}$ and skewness 
are $\sim 1$ and $\sim 1.5$, respectively, i.e., as mentioned before, the discrimination power is larger for skewness.  

\begin{figure}[!ht]
\centering
\includegraphics[width=8cm]{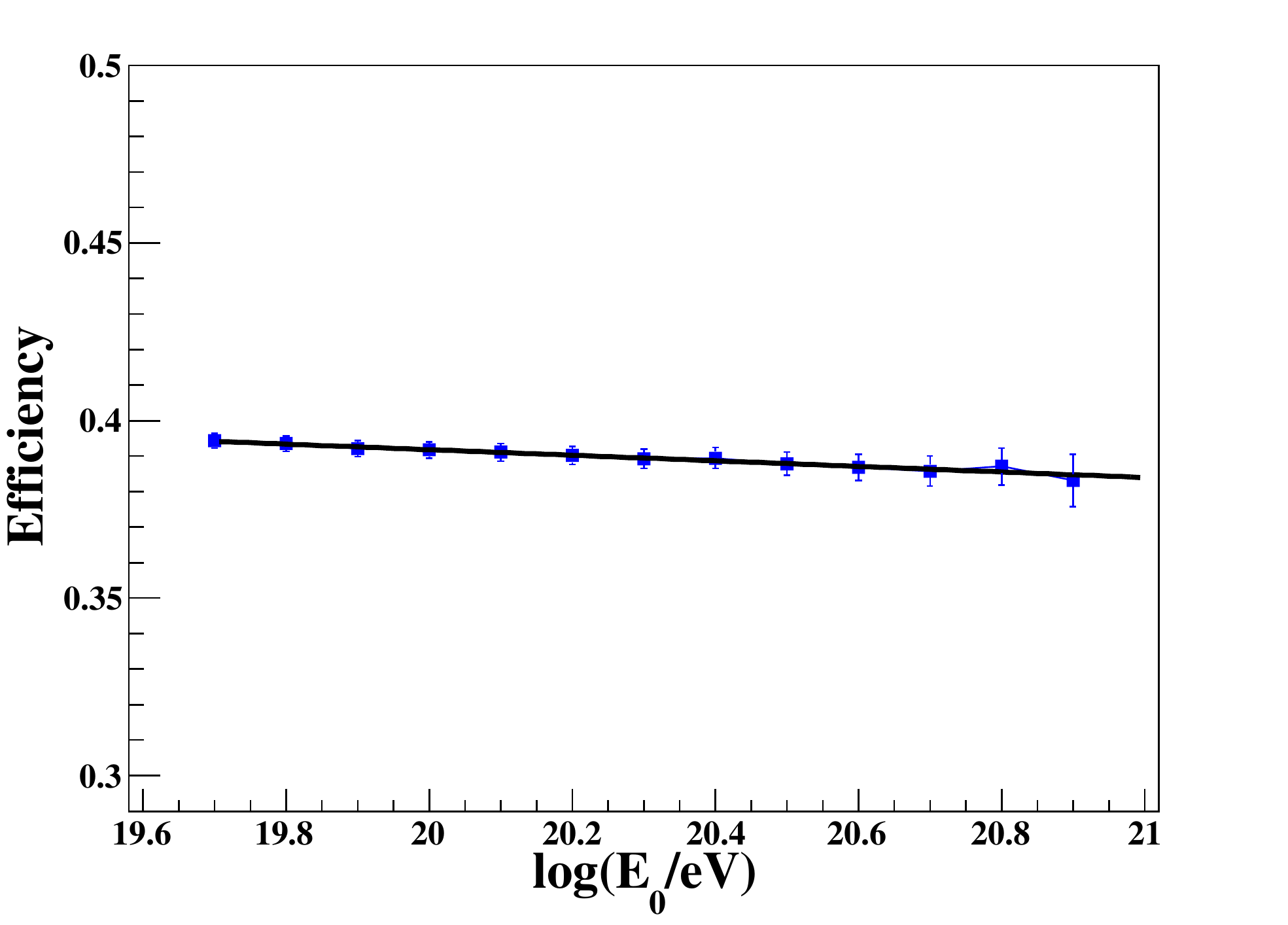}
\caption{Efficiency of the cut imposed to the events to calculate skewness for $E\geq E_0$. Solid line:
linear fit of the points.}
\label{EffSkewCut}
\end{figure}

The cut imposed to the events to calculate skewness has an efficiency. In particular, the profiles that passes the cut 
are such that they must hit the ground after propagating throughout an atmospheric depth larger than $X_{max}$+1000 g cm$^{-2}$.
Figure \ref{EffSkewCut} shows the efficiency of such cut for protons and for events with energy $\geq E_0$. For photons 
the efficiency is about a 10\% smaller. Note that this cut favors showers of larger zenith angles, in particular, for
protons, all the showers with $\theta > 55^\circ$ pass the cut.  

\begin{figure}[!ht]
\centering
\includegraphics[width=8cm]{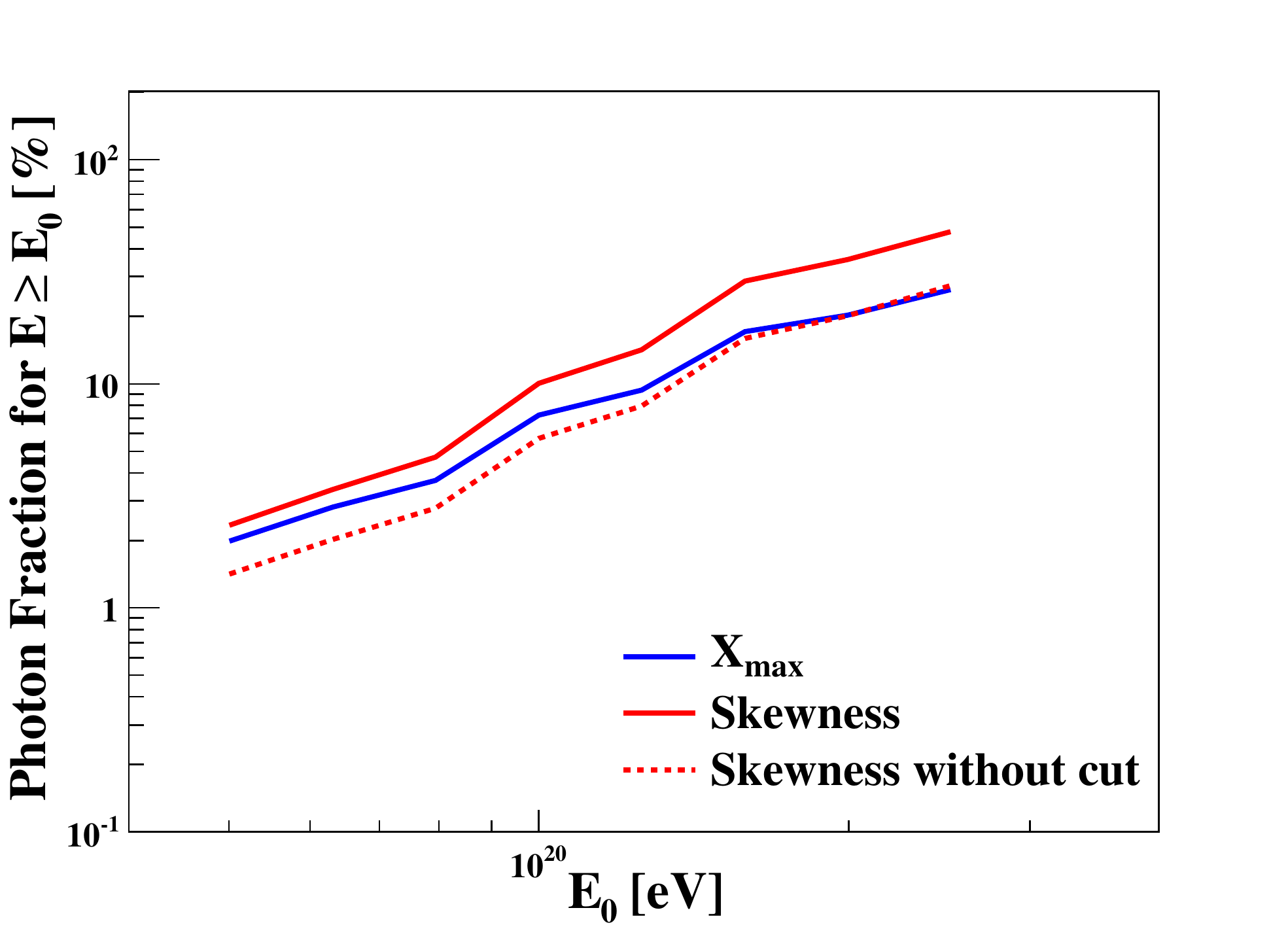}
\caption{Upper limits on the fraction of photons in the integral cosmic ray flux at 95\% of confidence level. 
Blue line is the upper limit obtained by using $X_{max}$ parameter without including any reconstruction uncertainty. 
Red line correspond to skewness with (solid) and without (dotted) including the efficiency of the cut.}
\label{UpperLskew}
\end{figure}

The upper limit on the photon fraction in the integral flux is obtained by using the method described above but in
this case considering skewness instead of $X_{max}$. Figure \ref{UpperLskew} shows the results obtained, the blue 
curve corresponds to the result obtained for $X_{max}$ without reconstruction uncertainty, the red curves correspond 
to skewness with (solid line) and without (dotted line) including the efficiency of the cut. As expected, the upper 
limit obtained by using skewness without including the efficiency of the cut is smaller, in almost the entire energy 
range, than the one obtained by using $X_{max}$ without considering reconstruction uncertainties. However, when the 
efficiency of the cut is included the upper limit become larger than the corresponding to $X_{max}$. Note that the 
upper limit curves corresponding to skewness increase faster with primary energy than the corresponding to $X_{max}$. 
This is due to the fact that the separation given by skewness is better for photons that do not convert in the 
geomagnetic field and the fraction of non converted photons decreases with primary energy.       

\section{Two dimensional analysis}

In order to improve the separation between photons and protons we introduce a new parameter intended to increase
the separation power of profiles corresponding to converted photons. There is a population of converted photons 
for which the longitudinal profile is wider and with smaller $N_{max}$ (number of charged particles at $X_{max}$). 
Therefore, a good parameter to separate this population should be the second derivative of the profile evaluated at 
the $X_{max}$ position. For that purpose, following Ref. \cite{aScherini:07}, a Gaussian fit of the profiles as a 
function of the age parameter $s=3 X/(X+2 X_{max})$ is performed. The second derivative at the maximum is given by,
\begin{equation}
a_s = \frac{N_{max}}{\sigma^2},
\end{equation}     
where $\sigma^2$ is the variance obtained form the fit. Note that $N_{max}$ and $\sigma$ are parameters sensitive 
to mass composition (see references \cite{aSouza:07} and \cite{aScherini:07}, respectively). 

\begin{figure}[!ht]
\centering
\includegraphics[width=8cm]{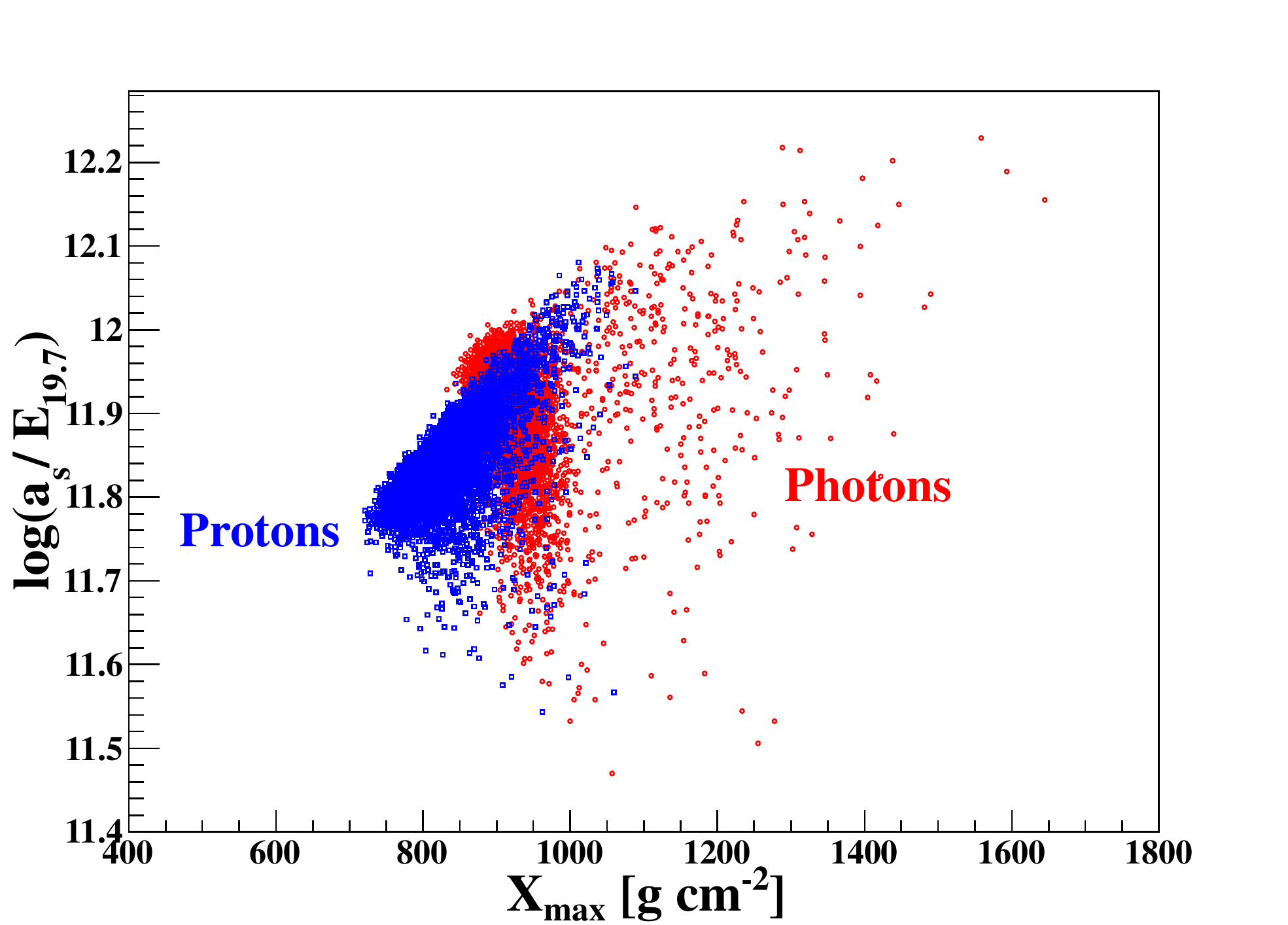}
\caption{$la_s=\log(a_s/E_{19.7})$ as a function of $X_{max}$ for $\log(E/eV) \in [20.1, 20.2]$ and 
$\theta \in [30^\circ, 80^\circ]$.} 
\label{AsXmax}
\end{figure}

Figure \ref{AsXmax} shows $la_s=\log(a_s/E_{19.7})$, with $E_{19.7}=E/10^{19.7} \textrm{eV})$, as a function of $X_{max}$ 
for $\log(E/eV) \in [20.1, 20.2]$. The figure shows that, at these energies, most of the photons suffered photon 
splitting (the majority of the photon showers have $X_{max}< 1000$ g cm$^{-2}$). It also shows that $la_s$, in 
combination with $X_{max}$, also helps to the separation of narrower profiles with large $N_{max}$.  

A two dimensional analysis is developed in order to assess the improvement on the separation between protons and photons 
combining different pairs of parameters. The method described in Ref. \cite{aSupanitsky:08b} is used to obtain the 
classification probability corresponding to different sets of parameters. The non-parametric method of kernel superposition 
with adaptive bandwidth \cite{aSilvermann:86,aSupanitsky:08a} is used to estimate the distribution functions corresponding to each 
primary and the leave-one-out technique \cite{aChilingarian:89} is used to estimate the classification probabilities for each 
set of parameters considered.

\begin{figure}[!ht]
\centering
\includegraphics[width=8cm]{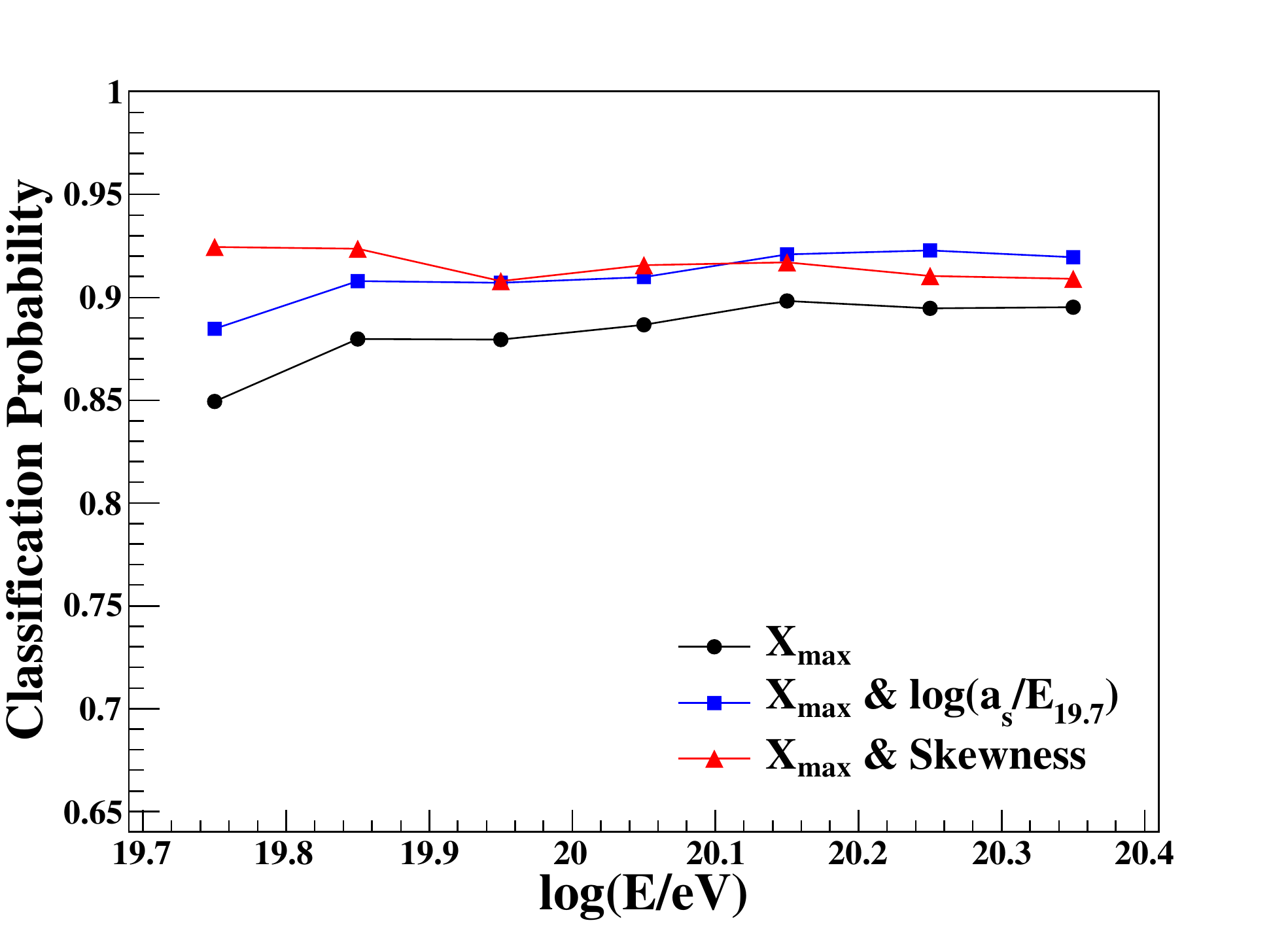}
\caption{Classification probability as a function of primary energy for $\theta \in [30^\circ, 80^\circ]$.}
\label{Pcl}
\end{figure}

Figure \ref{Pcl} shows the classification probability as a function of primary energy, for $\theta \in [30^\circ, 80^\circ]$
and for the set of parameters: $\{X_{max}\}$, $\{X_{max}, \textrm{skewness}\}$ and $\{X_{max}, la_s \}$. From the figure 
it can be seen that adding skewness or $la_s$ to the calculation with $X_{max}$ alone, the classification probability 
increases. For energies of order of $10^{19.8}$ eV the skewness parameter increases the classification probability of 
$X_{max}$ alone about 8\% whereas $la_s$ about 4\%. This is due to the fact that skewness is good for the separation 
between protons and non converted photons, which at these energies are still quite abundant. As the energy increases 
the improvement due to the addition of skewness decreases and the corresponding to $la_s$ increases. This is due to the 
fact that the fraction of converted photons increases with primary energy and $la_s$ is good for the separation of this 
population whereas skewness do not.  

The use of more than one parameter to calculate an upper limit to the photon fraction is a work in progress and will 
be presented in a forthcoming publication.

\clearpage


\newpage
\normalsize
\setcounter{section}{0}
\setcounter{figure}{0}
\setcounter{table}{0}
\setcounter{equation}{0}



\title{Neutrino astrophysics with JEM-EUSO}

\shorttitle{A.D. Supanitsky \etal Neutrino astrophysics with JEM-EUSO}

\authors{A.D. Supanitsky$^{1,2}$, G. Medina-Tanco$^{2}$ for the JEM-EUSO Collaboration.}
\afiliations{$^1$Instituto de Astronom\'ia y F\'isica del Espacio (IAFE), UBA-CONICET, Argentina.
\\ $^2$Instituto de Ciencias Nucleares, UNAM, Circuito Exteriror S/N, Ciudad Universitaria, M\'exico D. F. 04510, M\'exico.}

\email{supanitsky@iafe.uba.ar}

\abstract{
High energy neutrinos play a fundamental role on the understanding of
several astrophysical phenomena and, in particular, on the origin and propagation
of extreme energy cosmic rays. JEM-EUSO is a proposed orbital detector to be installed
onboard the International Space Station. It is designed to observe the fluorescence 
light produced by the air showers initiated by the extreme energy component of the 
cosmic rays, including gamma rays and neutrinos. In this work we study the discrimination
capability of the mission between nearly horizontal neutrino and proton showers, at the 
highest energies, by using the atmospheric depth of maximum development. We propose a new 
method to discriminate between electron neutrino and tau horizontal showers, developing very 
deep in the atmosphere, by using the multi-peak structure that they present. We also study the 
flux of tau leptons emerging from the Earth, including the case of the presence of oceans, 
produced by the interaction of tau neutrinos inside the Earth for a given model of gamma ray 
bursts and in the context of the JEM-EUSO mission.
}

\maketitle

\section{Introduction}

The astrophysical information carried by very high energy neutrinos is very important for the 
understanding of the origin and propagation of the cosmic rays. Such particles can be produced during 
the propagation of the cosmic rays in the interstellar medium \cite{Berezinsky:69}, as by-products of 
the hadronic interaction in the sources \cite{Ostapchenko:08} and as the main product of the decay of 
superheavy relic particles \cite{Aloisio:04,Battacharjee:00}.

In this work we study the characteristics of inclined tau and electron neutrino showers and its 
discrimination from the proton component in the context of the JEM-EUSO mission \cite{Takahashi:09}. 
The main parameter used to separate the different species is $X_{max}$, the atmospheric depth of the 
maximum development of the showers. 

Also, an extension of the study presented in Ref. \cite{Supanitsky:11} about showers initiated by 
neutrinos that interact in the central region of the field of view of the JEM-EUSO telescope (in nadir
 mode) is developed. A possible technique to identify the presence of both tau and electron neutrinos 
in a given sample is proposed. 

Finally, the propagation of tau neutrinos inside the Earth is studied. In particular, neutrinos originated 
in gamma ray bursts are considered, for which the propagation in the presence of oceans is compared with 
the one in the mantle of the Earth.        

\section{Inclined neutrino showers}

Neutrinos can initiate atmospheric air showers when they interact with the nucleons of the air molecules. 
The probability that a neutrino interact in the atmosphere increases with zenith angle because of the 
increase of the number of target nucleons. High energy neutrinos, propagating through the atmosphere,
can suffer charge (CC) and neutral (NC) current interactions. The CC interactions are the most important 
for the space observations because in the NC interactions most of the energy is take by a secondary 
neutrino which could produce an observable air shower just in the case it suffers a subsequent CC 
interaction. The shower produced by the hadronic component resultant from the NC interaction is difficult 
to observe from the space due to the high energy threshold of the telescope. 

As a result of a CC interaction, a very high energy lepton, which takes most of the energy of the incident 
neutrino, is generated. Typically, it takes $\sim 80\%$ of the neutrino energy at $E_{\nu} \cong 10^{20}$ eV, 
the rest of the energy goes into the hadronic component.

Proton and neutrino showers of $E = 10^{20}$ eV and $\theta = 85^\circ$ are simulated in order to study their 
characteristics and its possible identification. The last version of CONEX \cite{conex} (v2r2.3) with QGSJET-II 
\cite{QGSJETII} is used to generate the proton and neutrino showers. Electron and tau neutrino showers are 
considered. The program PYTHIA \cite{pythia}, linked with LHAPDF \cite{lhapdf}, is used to simulate the 
electron neutrino-nucleon interactions. The CTEQ6 \cite{cteq6} set of parton distribution functions are used. 
The air showers are generated injecting the produced particles in CONEX \cite{Supanitsky:11}. For the case of 
tau neutrinos it is just consider the decay of tau leptons of $E = 10^{20}$ eV, for which the 
simulation program TAUOLA \cite{tauola} is used. 

The interaction points of the neutrino showers are simulated by taking at random values of the
atmospheric depth from an exponential distribution, $P(X) \propto \exp(-X/\lambda_\nu(E_\nu))$ with 
$\lambda_\nu (10^{20} eV) = 3.2\times10^7 $ g cm$^{-2}$, in the interval $[0, X_{end}]$ where $X_{end}$ is 
the atmospheric depth from the top of the atmosphere to the core position, which for $\theta=85^\circ$ is
$\sim 10573$ g cm$^{-2}$. Note that due to the large mean free path of the neutrinos the exponential distribution 
can be approximated by the uniform distribution in the interval $[0, X_{end}]$. 

\begin{figure}[!ht]
\centering
\includegraphics[width=8.5cm]{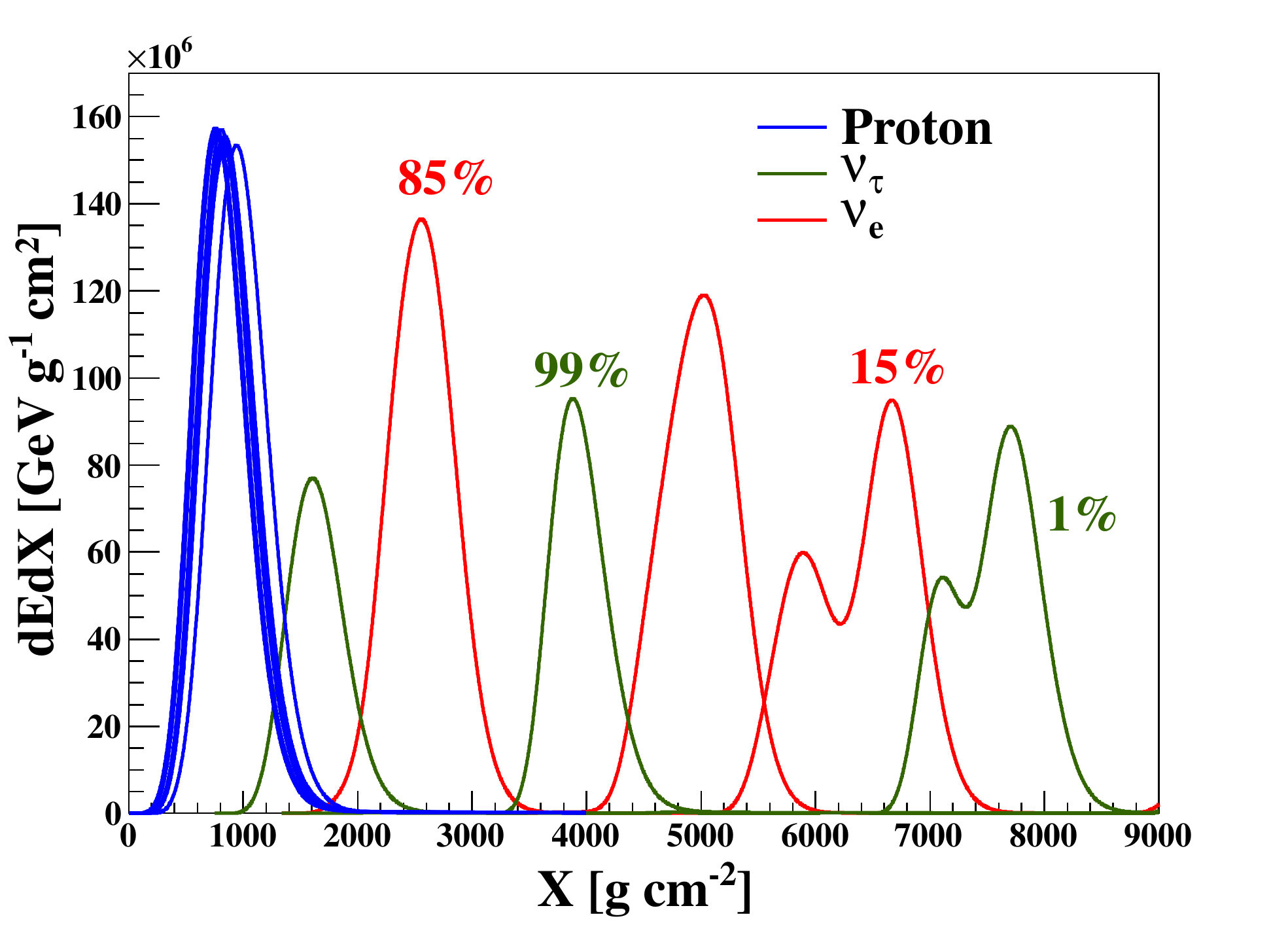}
\caption{Simulated proton and neutrino showers of $E = 10^{20}$ eV and $\theta = 85^\circ$.}
\label{Prof}
\end{figure}

Figure \ref{Prof} shows the profiles for some simulated events. The neutrino showers that develop deeper in the 
atmosphere can present more than one peak, this is due to the LPM fluctuations suffered by showers dominated by the
electromagnetic component. For the case of tau leptons, just in $\sim 18\%$ of the decays a high energy electron 
or positron is produced, $\tau^\pm \rightarrow e^\pm \nu_\tau \nu_e$. The showers produced by this channel that
develop deep in the atmosphere can have more than one peak. On the other hand, in every electron neutrino 
interaction a high energy electron or positron is produced, increasing the probability of finding a shower
with more than one peak. This is the reason why 15\% of electron neutrino showers present more than on peak 
whereas the same happens with just 1\% of the tau neutrino showers.  

Note that about 17.51\% of the taus decay into a muon and two neutrinos, $\tau^\pm \rightarrow \mu^\pm \nu_\tau \nu_\mu$.
The showers initiated by the muons are quite difficult to observe from the space because the deposited energy
of these kind of showers is much smaller than the regular ones. Therefore, these type of showers are excluded
from the subsequent analyses. 

\begin{figure}[!ht]
\centering
\includegraphics[width=8.5cm]{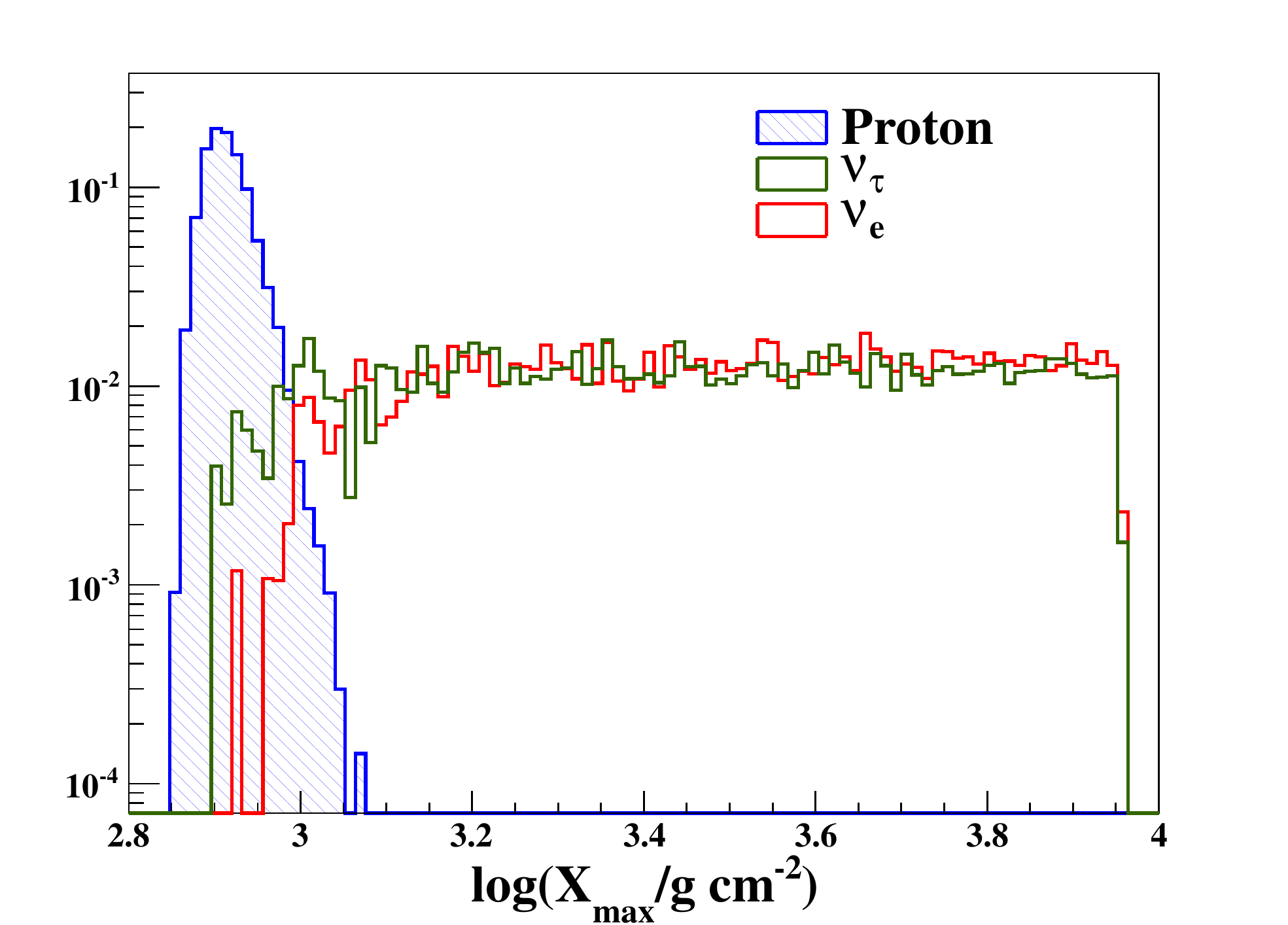}
\caption{Distribution of the first peak of the profiles for proton and neutrinos of $E_\nu = 10^{20}$ eV and 
$\theta = 85^\circ$.}
\label{XmaxI85}
\end{figure}

Figure \ref{XmaxI85} shows the distributions of the first peak of the simulated showers. Just the events with the first 
peak above 1 km of altitude are taken into account, which is equivalent to consider the ones whose first peak has an 
atmospheric depth less than $\sim 9000$ g cm$^{-2}$. It can be seen that, above $\sim 1600$ g cm$^{-2}$ the distributions 
corresponding to tau and electron neutrinos are flat and extended over a huge interval of atmospheric depth, which allows 
a very efficient discrimination of the neutrino showers from the proton ones. Note that, as expected, the overlap between
the $X_{max}$ distributions of protons and taus is larger than the corresponding one to protons and electron neutrinos.  

\section{Horizontal neutrino showers}

\begin{figure}[!ht]
\centering
\includegraphics[width=8.5cm]{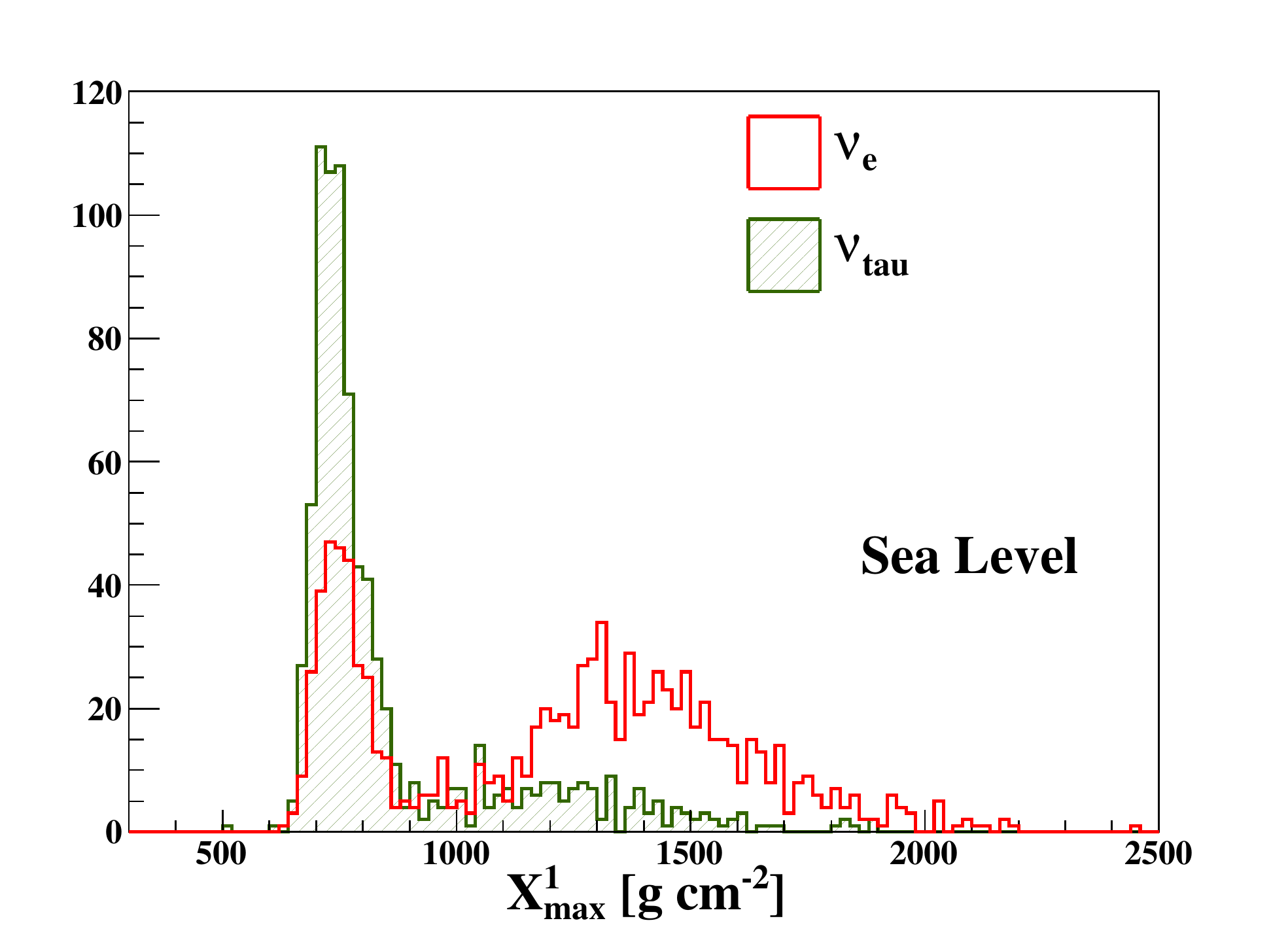}
\caption{Horizontal tau and electron neutrino showers of $E=10^{20}$ eV injected
in the center of the field of view of the JEM-EUSO telescope and at sea level.}
\label{XMaxI90}
\end{figure}

In this section the showers generated by horizontal neutrinos that interact in the central 
region of the field of view of the JEM-EUSO telescope, in nadir mode, are considered. 
For the case of electron neutrinos, this showers are dominated by the LPM effect and can
present more than one peak \cite{Supanitsky:11}. As mentioned before, just $\sim 18\%$ of 
the tau showers are initiated by electrons or positrons, diminishing in this way the 
probability to find showers with more than one peak. Figure \ref{XMaxI90} shows the 
distributions of the first peak for tau and electron neutrino showers of $E=10^{20}$ eV
and $\theta=90^\circ$ injected in the center of the field of view (fov) of JEM-EUSO at sea level. 
These distributions present two populations. The population with smaller values of $X_{max}^1$ 
corresponds to showers dominated by the hadronic component and the other one corresponds
to showers dominated by the electromagnetic component. As expected, the hadronic population 
is more important for tau showers. 

\begin{figure}[!ht]
\centering
\includegraphics[width=8.5cm]{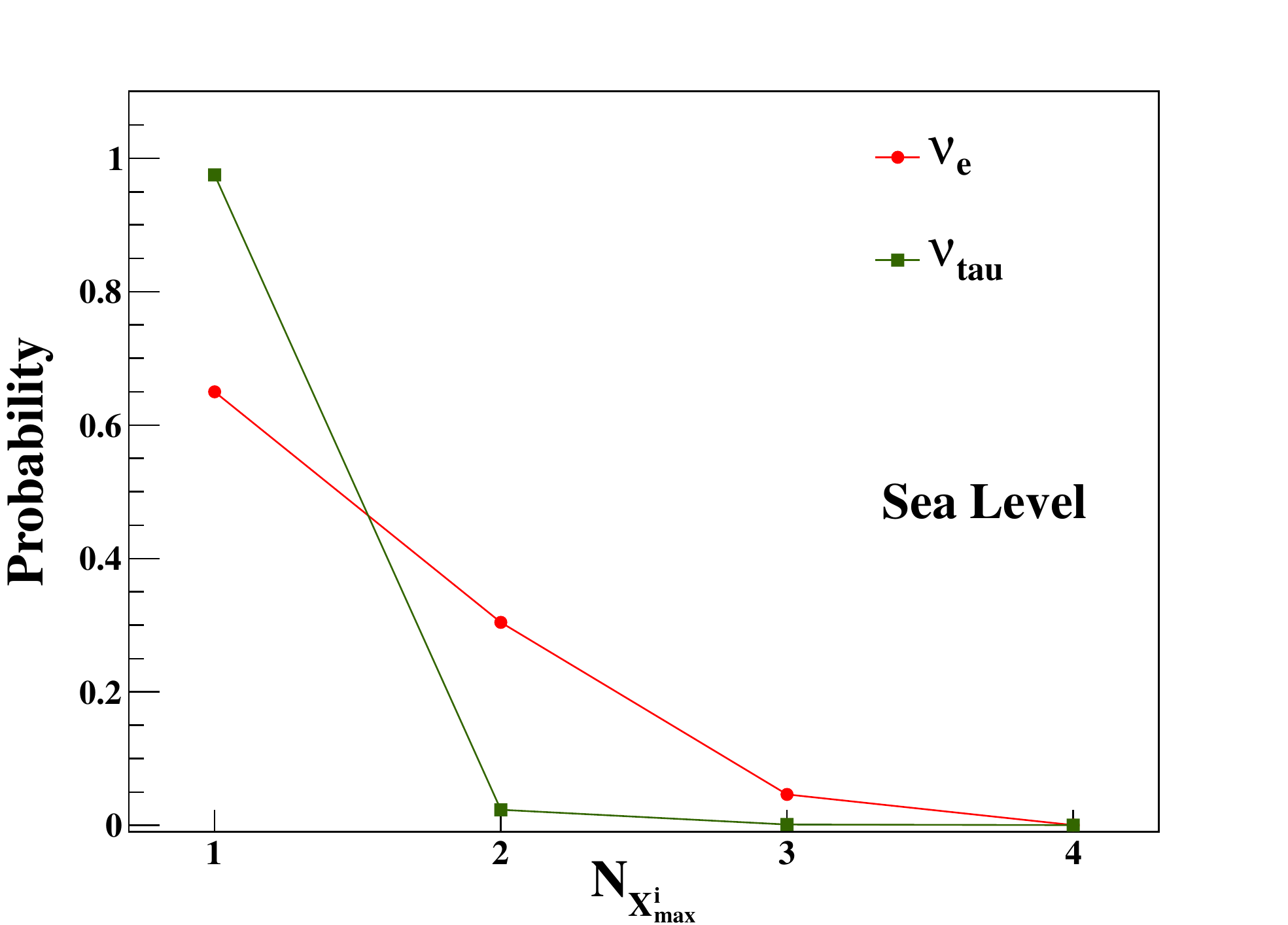}
\caption{Probability to find showers with exactly $N_{X_{max}^i}$ peaks for horizontal tau and 
electron neutrino showers of $E=10^{20}$ eV, injected in the center of the field of view of the 
JEM-EUSO telescope in nadir mode.}
\label{PXmaxI}
\end{figure}

Figure \ref{PXmaxI} shows the probability to find showers with exactly $N_{X_{max}^i}$ peaks
for electron neutrino and tau showers. At sea level, the probability to find a tau shower with 
just one peak is $\sim98\%$ whereas for electron neutrinos is $\sim 65\%$. At an altitude of 
5 km, the probability to find a tau shower with just one peak is $\sim99\%$ whereas for 
electron neutrinos is $\sim 76\%$. The reduction of the probability to find more than one peak 
with increasing altitudes has to do with the fact that the development of the showers takes place 
in regions with smaller values of air density, therefore, the influence of the LPM effect is also 
reduced.   

\begin{figure}[!ht]
\centering
\includegraphics[width=8cm]{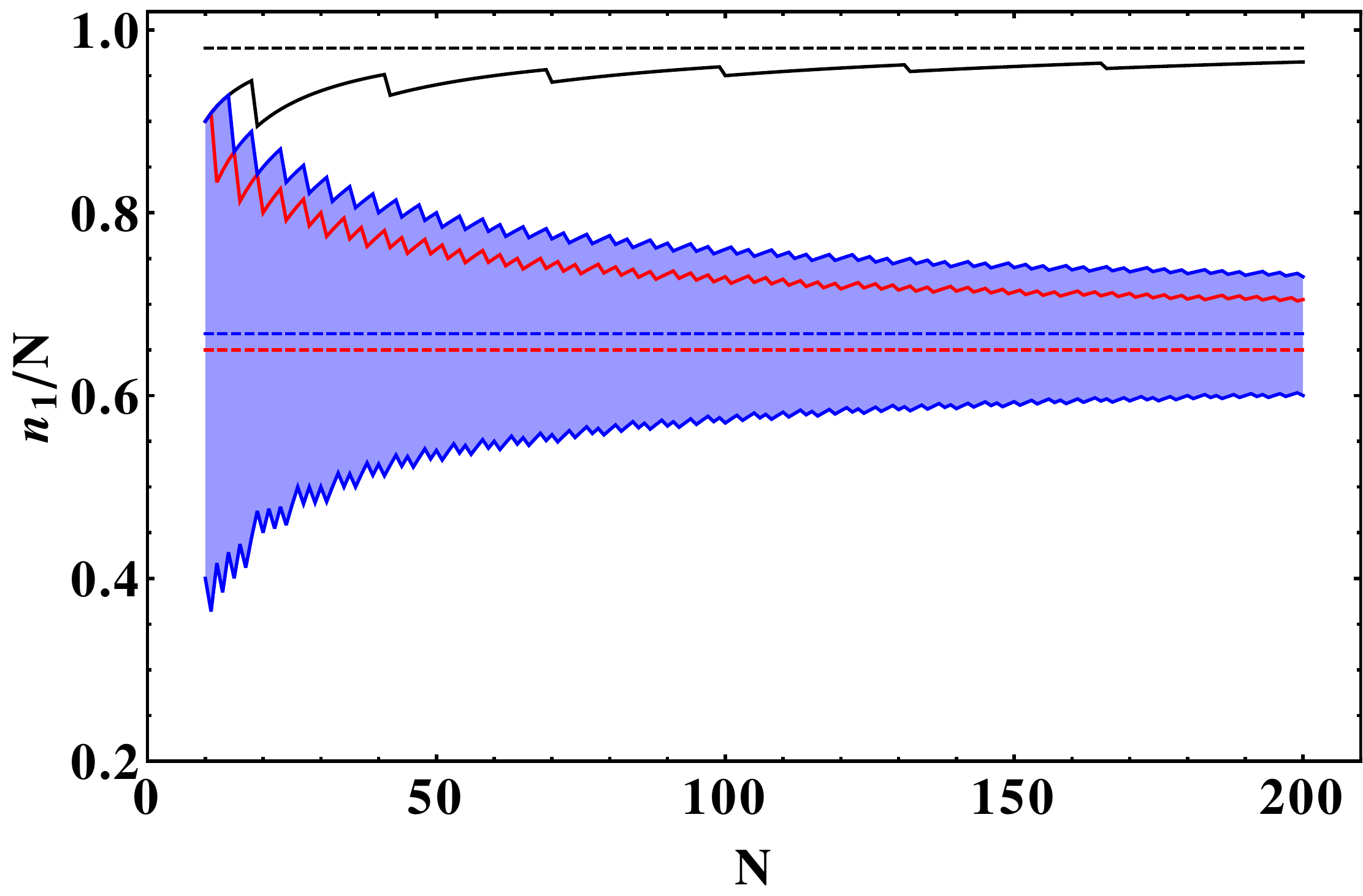}
\includegraphics[width=8cm]{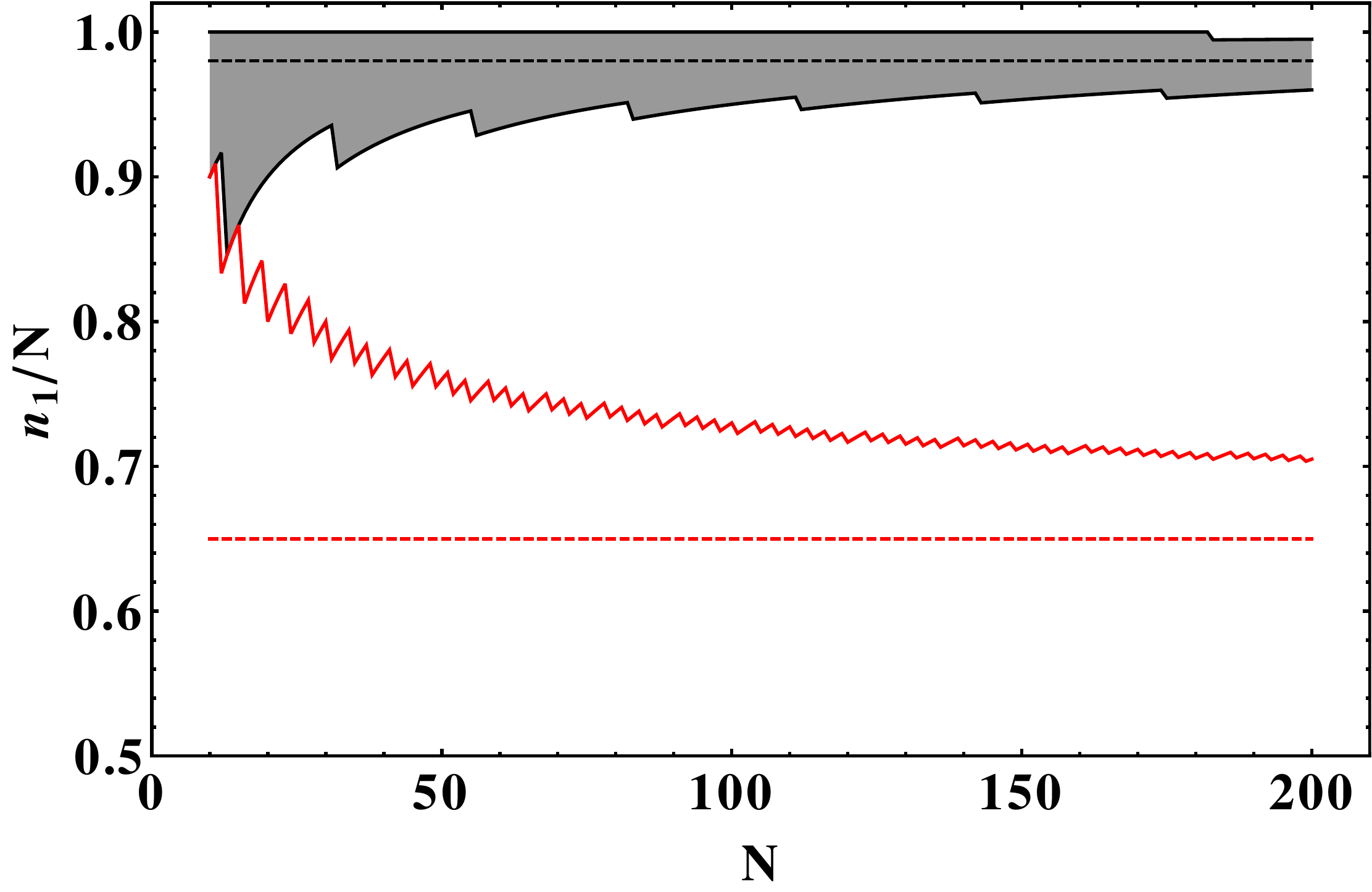}
\caption{Shadowed region corresponds to 95\% of probability to find a fraction $n_1/N$ of showers with just 
one peak, for a mixture of equal number of incident tau and electron neutrinos (upper panel) and for the case 
in which the incident flux contains just tau neutrinos (bottom panel), as a function of the sample size $N$.}
\label{Proba}
\end{figure}

The difference between the number of showers with just one peak can be used to study the relative
abundance of tau and electron neutrino showers. Due to the large decay length of the taus at 
$10^{20}$ eV ($\sim 5000\ \textrm{km}$), the ratio between the number of tau neutrino and electron 
neutrino horizontal showers (starting in the center of the fov of the telescope) is of order of 
$N_{sh}(\nu_\tau)/N_{sh}(\nu_e) \cong 0.07$, assuming that the relative abundances of the incident 
flux are equal to one. Upper panel of figure \ref{Proba} shows the region (in blue) of 95\% of
probability to find a fraction of $n_1/N$ showers with just one peak, as a function of the sample 
size $N$, obtained from simulations, for a mixture of equal number of incident tau and electron 
neutrinos. Note that showers corresponding to the muonic decay channel of the tau are not included 
in the analysis. Observed values of $n_1/N$ smaller than the black solid line reject the hypothesis 
that the sample is composed by tau showers alone, with probability of rejection larger or equal to 
0.95, depending of the particular value of $n_1/N$. Also, observed values of $n_1/N$ larger than the 
red solid line reject the hypothesis that the sample is composed by just electron neutrino showers 
with probability of rejection larger or equal to 0.95, again depending on the particular value of 
$n_1/N$. From the figure it can be seen that for 95\% of the cases, samples of more than 15 showers 
are needed to be able to reject the hypothesis of having tau showers alone. Although not shown in 
the figure, the number of events needed to reject the hypothesis of having a sample with electron 
neutrinos alone has to be grater than $\sim 9300$.   

Bottom panel of figure \ref{Proba} shows the region of 95\% of probability to find a fraction of $n_1/N$ showers 
with just one peak obtained from a binomial distribution, for the case in which the samples have just tau showers,
as a function of the sample size $N$. It can be seen that for 95\% of the cases, samples of more than $\sim 17$ 
events are needed to be able to reject the hypothesis of having electron neutrino showers alone. 

It is important to note that as the energy decreases the ratio $N_{sh}(\nu_\tau)/N_{sh}(\nu_e)$ goes to one, 
as a consequence the number of events needed to reject the hypothesis of having a sample with electron neutrinos 
alone decreases drastically.   

\section{Earth skimming tau neutrinos}

Gamma ray bursts are potential sources of high energy cosmic rays \cite{Waxman:95,Vietri:95}. If the cosmic 
rays are efficiently accelerated in GRBs a neutrino flux is expected as a result of the photo-hadronic 
interactions of protons with the photons present in the acceleration site \cite{Waxman:98}. The detection 
of high energy neutrinos in coincidence with GRBs should be a proof of the acceleration of cosmic rays in 
this kind of events. 

Depending on the redshift of the GRB, the JEM-EUSO telescope will be able to observe Earth skimming tau 
neutrinos, detecting the Cherenkov flashes originated by the showers produced by the decay of the taus 
after propagation inside the Earth \cite{Asano:09}.  

A modified version of the ANIS \cite{anis} program is used to propagate tau neutrinos inside of the Earth. 
We have improved the propagation and energy lose of the taus in order to study the case in which the taus 
traverse interfaces between rock and water which is the case of taus emerging from or entering to the oceans.  

Two cases are considered, for the first one, the last or external layer of the Earth, of 3 km of thickness, 
is composed by standard rock of density $2.6$ g cm$^{-3}$. For the second case, this last layer is composed 
by water, i.e. of density $1$ g cm$^{-3}$. Following Ref. \cite{Asano:09} tau neutrinos of $70^\circ$ of 
nadir angle are considered. The energy spectrum of the tau neutrinos, used in the simulations, is the one 
corresponding to figure 2 of Ref. \cite{Asano:09}. 

\begin{figure}[!ht]
\centering
\includegraphics[width=8cm]{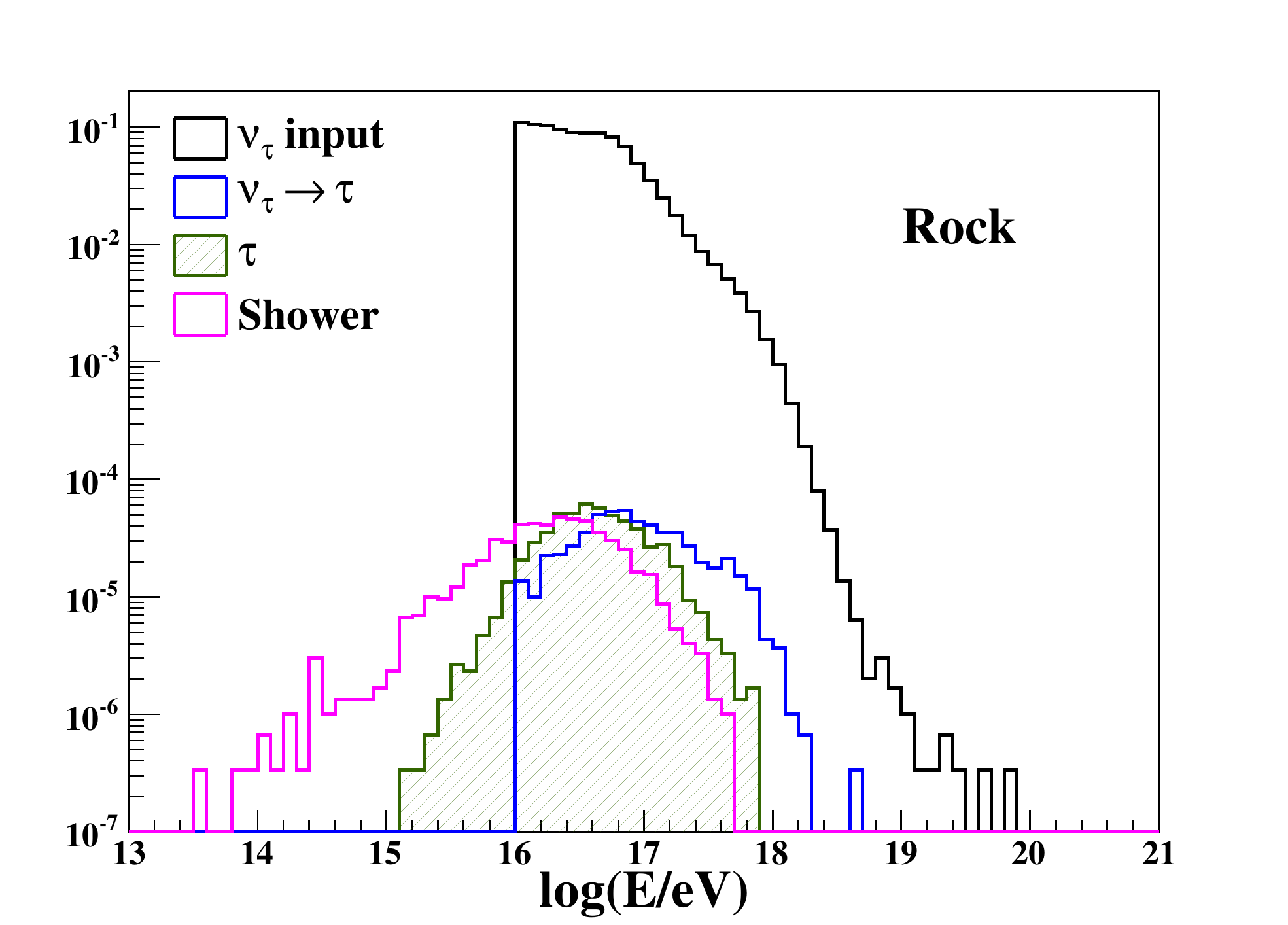}
\caption{Energy distributions corresponding to the propagation of tau neutrinos of $70^\circ$ of nadir 
angle following the energy spectrum of the GRB model of figure 2 of Ref. \cite{Asano:09}. Black lines: 
input spectra; blue lines: neutrinos that produced an emerging tau; green lines: emerging taus; magenta 
lines: energy that goes to the showers (see the text for details).}
\label{NuTauGRB}
\end{figure}

Figure \ref{NuTauGRB} shows the energy distributions of the tau neutrinos injected into the simulation 
(black lines), the ones that produced an emerging tau lepton (blue lines), the emerging taus (green lines)
and the energy that effectively goes to the shower (magenta lines). This last distribution is obtained 
by simulating the tau decay with TAUOLA and summing the energy of the particles that contribute to the 
shower, i.e. all particles excepting neutrinos.   

In the case of rock the probability of a tau to emerge from the Earth is 
$P_R(\nu_{\tau}\rightarrow \tau) = 5.7\times10^{-4}$ and the median of the energy distribution
of the taus is $med(E_\tau^R) \cong 4\times 10^{16}$ eV. For the case in which the last layer of the 
Earth is composed by water $P_W(\nu_{\tau}\rightarrow \tau) = 2.9\times10^{-4}$ and 
$med(E_\tau^W) \cong 6\times 10^{16}$ eV. Therefore, the number of emerging taus for the case where 
the last layer of the Earth is made of water is about a factor two smaller than the corresponding 
to rock, whereas, on average, the energy of the emerging taus is larger for the case of water. 
This is due to the fact that the energy lose of taus is smaller in the presence of water because 
the density of water is smaller. In principle, the presence of oceans could deteriorate the 
detectability of tau showers because of the reduction of the number of emerging taus.

The simulation of the Cherenkov photons that reach the JEM-EUSO telescope is the last step to complete
the simulation chain for Earth skimming tau neutrinos (without considering the detector which is simulated
with the ESAF \cite{Berat:10} software). These simulations are under development and will allow us to study 
in detail the influence of the presence of oceans on the detectability of tau neutrinos from GRBs.

\clearpage


  \includepdf[pages=-]{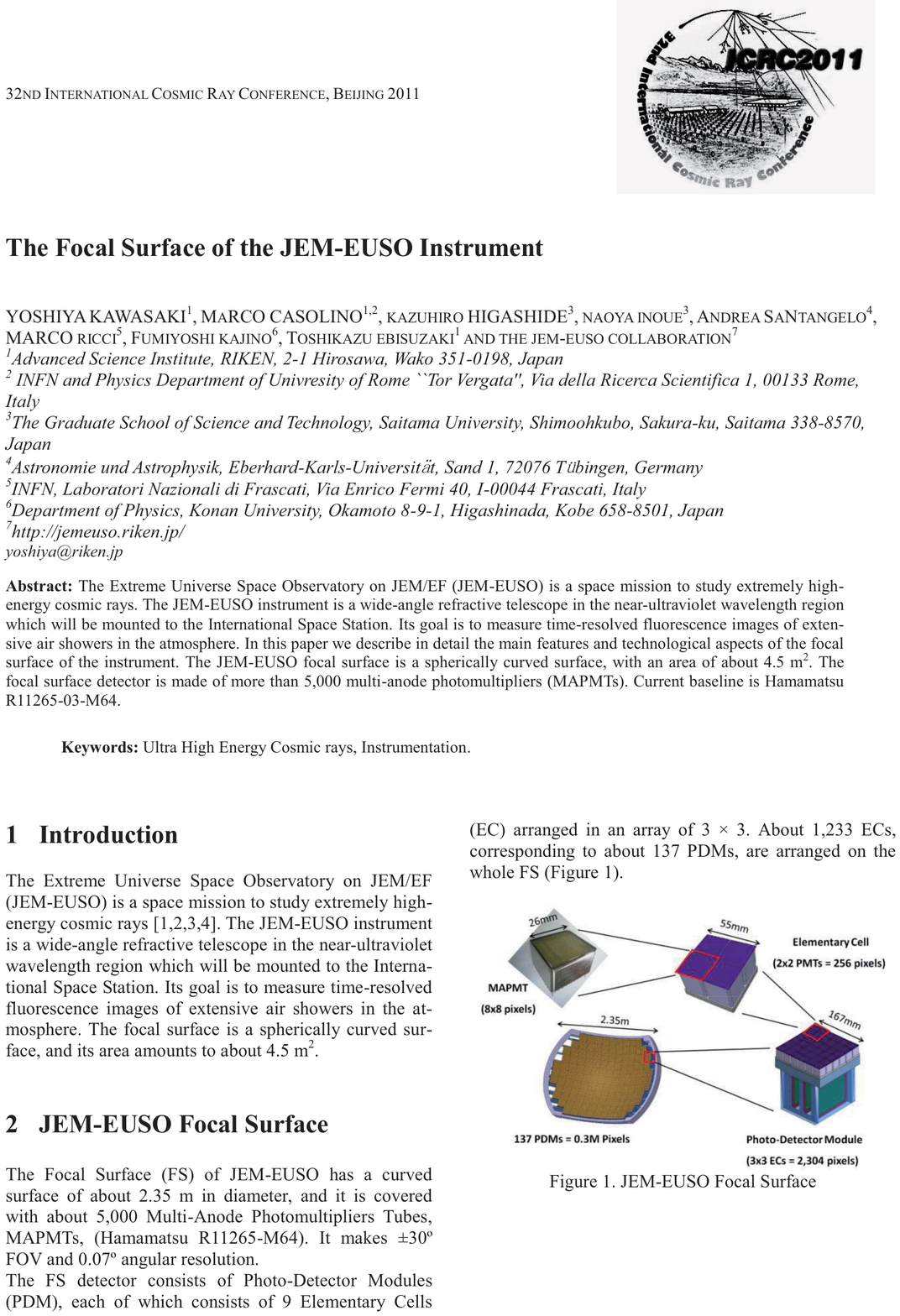}
 \hspace{4cm}
 \newpage

  \includepdf[pages=-]{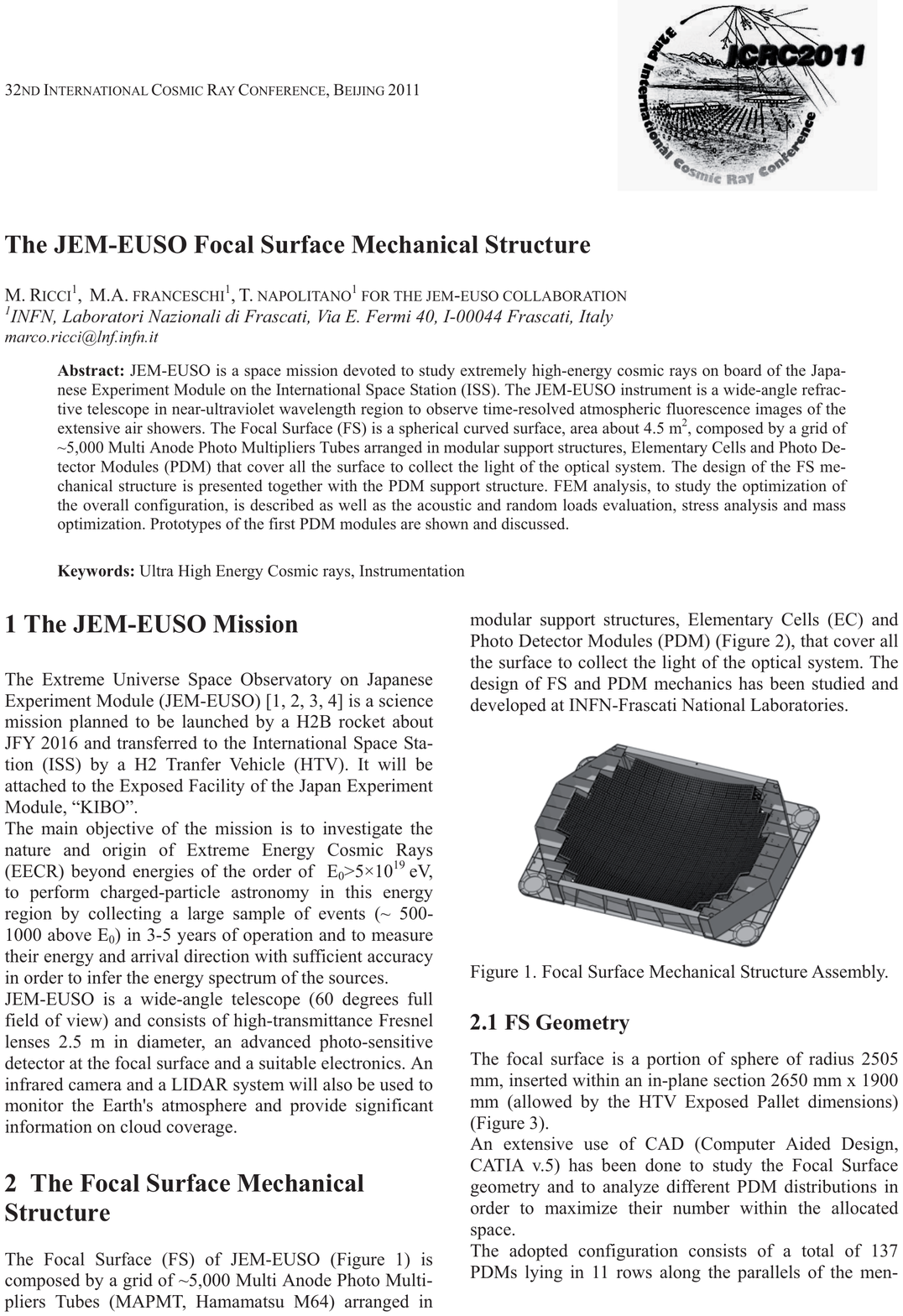}

  \includepdf[pages=-]{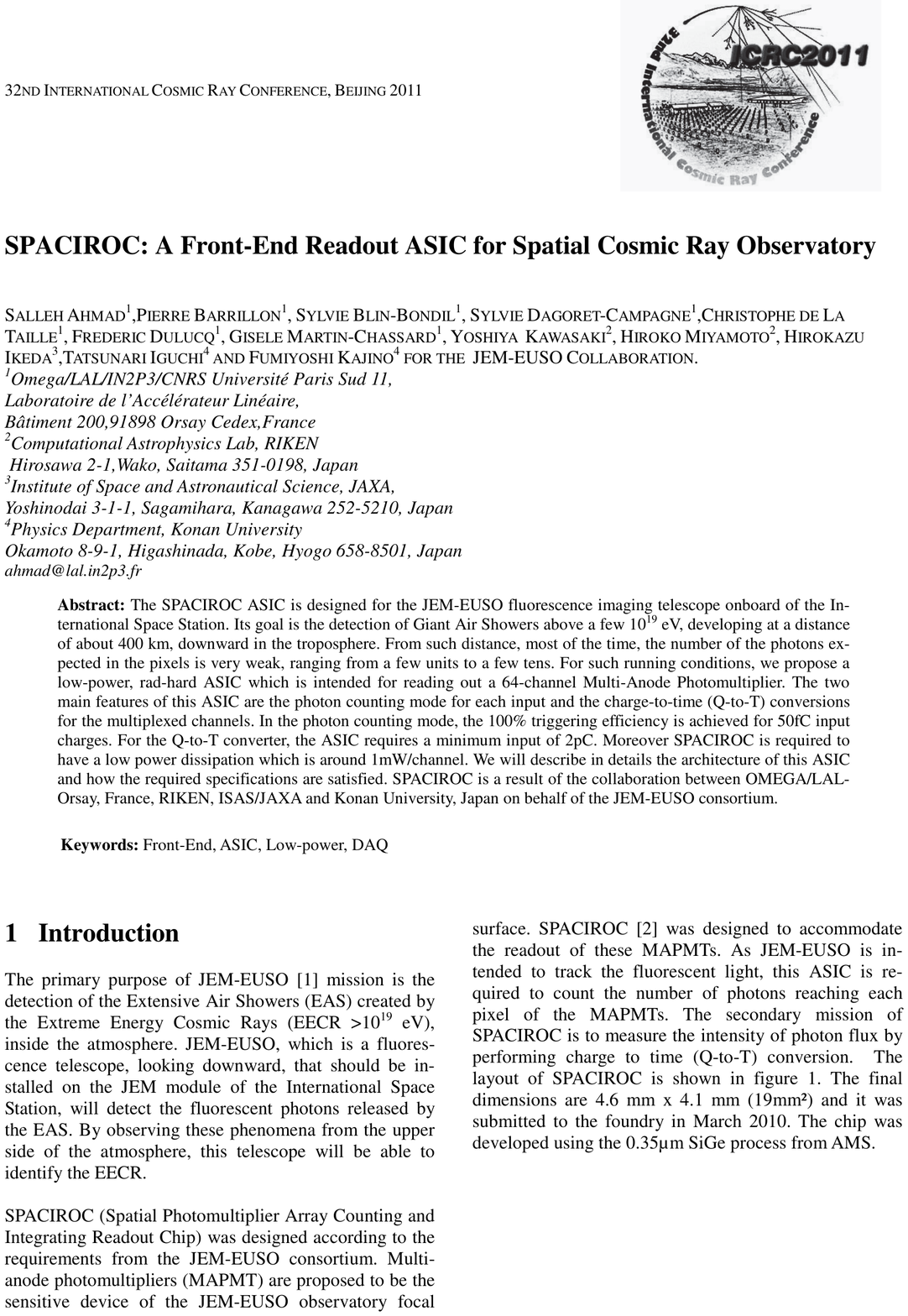}

\newpage
\normalsize
\setcounter{section}{0}
\setcounter{figure}{0}
\setcounter{table}{0}
\setcounter{equation}{0}



\title{Performance of a front-end ASIC for JEM-EUSO}

\shorttitle{H. Miyamoto \etal Performmance of front-end ASIC for JEM-EUSO}

\authors{Hiroko Miyamoto$^{1}$, Salleh Ahmmad$^{2}$, Pierre Barrillon$^{2}$,
Sylvie Blin-Bondil$^{2}$, Sylvie Dagoret-Campagne$^{2}$,
Christophe De La Taille$^{2}$, Frederic Dulucq$^{2}$, Tatsunari Iguchi$^{3}$,
Hirokazu Ikeda$^{4}$, Fumiyoshi Kajino$^{3}$, Yoshiya Kawasaki$^{1}$,
Gisele Martin-Chassard$^{2}$, Kenji Yoshida$^{3}$
on behalf of the JEM-EUSO collaboration}
\afiliations{$^1$Advanced Science Institute, RIKEN\\
Hirokosawa 2-1, Wako, Saitama 351-0198, Japan\\
$^2$OMEGA/LAL/IN2P3/CNRS/Universit$\grave{e}$ Paris-Sud, Laboratoire de 
l'Acc$\grave{e}$l$\grave{e}$rateur Lin$\grave{e}$aire\\
B$\hat{a}$timent 200, 91898 Orsay Cedex, France\\
$^3$Physics Department, Konan University\\
Okamoto 8-9-1, Higashinada, Kobe, Hyogo 658-8501, Japan\\
$^4$Institute of Space and Astronautical Science, JAXA\\
Yoshinodai 3-1-1, Sagamihara, Kanagawa 252-5210, Japan
}
\email{hirokom@riken.jp}

\abstract{
The SPACIROC (Spatial Photomultiplier Array Counting and Integrating
ReadOut Chip) is a Front-End ASIC designed for the space-borne
fluorescence telescope JEM-EUSO\cite{jemeuso1}\cite{jemeuso2}.
The device is designed for features of single photon counting, dynamic
range of 1 photoelectron (PE) to 1500 PEs, double pulse resolution of
10 ns, and low power consumption ($<$1 mW/ch).
SPACIROC reads output signals from a 64-channel Multi-Anode Photomultiplier
Tube (MAPMT). Input photons are measured in the two features as following:
photon counting mode for each input and charge-to-time (Q-to-T) conversion
mode for the multiplexed channels. The combination of these two features
enables the large dynamic range as described above.
We will report the performance of the ASIC such as power consumption,
double pulse resolution, dynamic range and linearity.
}

\maketitle

\section{Introduction}
%
%
JEM-EUSO (Extreme Universe Space Observatory on board Japanese
Experiment Module) is a mission which
aims the observation of Extreme Energy Cosmic Rays (EECRs) with a
space-borne fluorescence telescope on the International Space Station
(ISS). 
The detector will consist of five thousand 1-inch-square MAPMTs, and
will allow an area of about 400 kilometers in diameter of
Earth's atmosphere to be imaged in the field of view.
Since 2006 the Phase A study of JEM-EUSO has been continued with
extensive simulations, design, and prototype hardware developments
that have significantly improved the JEM-EUSO mission profile,
targeting the launch of 2016 in the framework of the second phase of 
JEM/EF (Japanese Experiment Module/Exposure Facility) utilization.\\
The main physical sources of interest of JEM-EUSO are the fluorescent
and Cherenkov UV photons induced by cosmic rays with energies
higher than $10^{19}eV$ impinging on the atmosphere leading the
development of an Extensive Air Showers (EAS) in the troposphere.
The JEM-EUSO telescope will determine the energies and directions
of extreme energy primary particles by recording the tracks of EAS
with a time resolution of about $1{\mu}s$ and an angular resolution
of about $0.1^\circ$.\\
About the JEM-EUSO status and general project information, see also
\cite{jemeuso3} and \cite{jemeuso4} in this conference.
\section{JEM-EUSO Focal Surface}
 \begin{figure}[h!]
  \centering
  \includegraphics[width=\hsize]{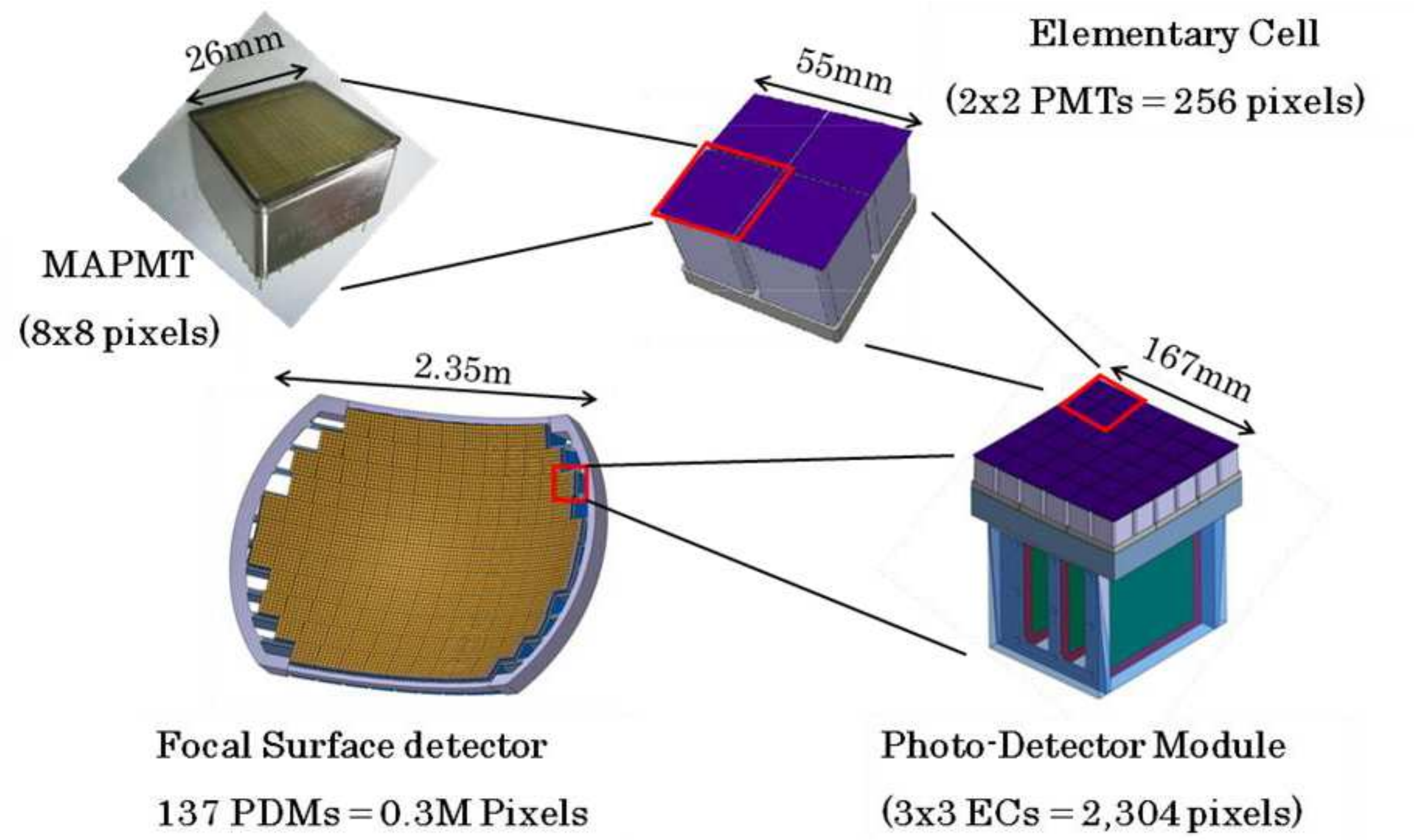}
  \caption{FS detector modules.}
  \label{FS_DM}
 \end{figure}
The Focal Surface (FS) of the JEM-EUSO telescope consists of a curved
surface of about 2.35 m in diameter which is covered with 5,000 of
64-channel MAPMTs (Hamamatsu R11265-M64).
A JEM-EUSO FS Photo-Detector Module (PDM) consists of an array of
3$\times$3 Elementary Cells (ECs), each of which consist of 2$\times$2
MAPMTs.
About 1,233 ECs, corresponding to about 137 PDMs, are arranged on the
on the whole FS (See the Fig.\ref{FS_DM}).
MAPMTs capture the photons from the Earth atmosphere, convert them in
its photocathode into photoelectrons and induce pulses from the charges
on their anodes and dynode output.
The Front-End ASIC transforms the charges from MAPMTs into digital
numbers which can be processed in next stages by digital electronics.
Similarly the trigger stages process digitally those charges which have
been previously converted into numbers.\\
About the JEM-EUSO focal surface, see also the
contributions\cite{jemeuso5}\cite{jemeuso6}\cite{jemeuso7}
in this conference.
\section{Front-End ASIC : SPACIROC}

\begin{figure}[!h]
  \centering
  \includegraphics[width=3.2in]{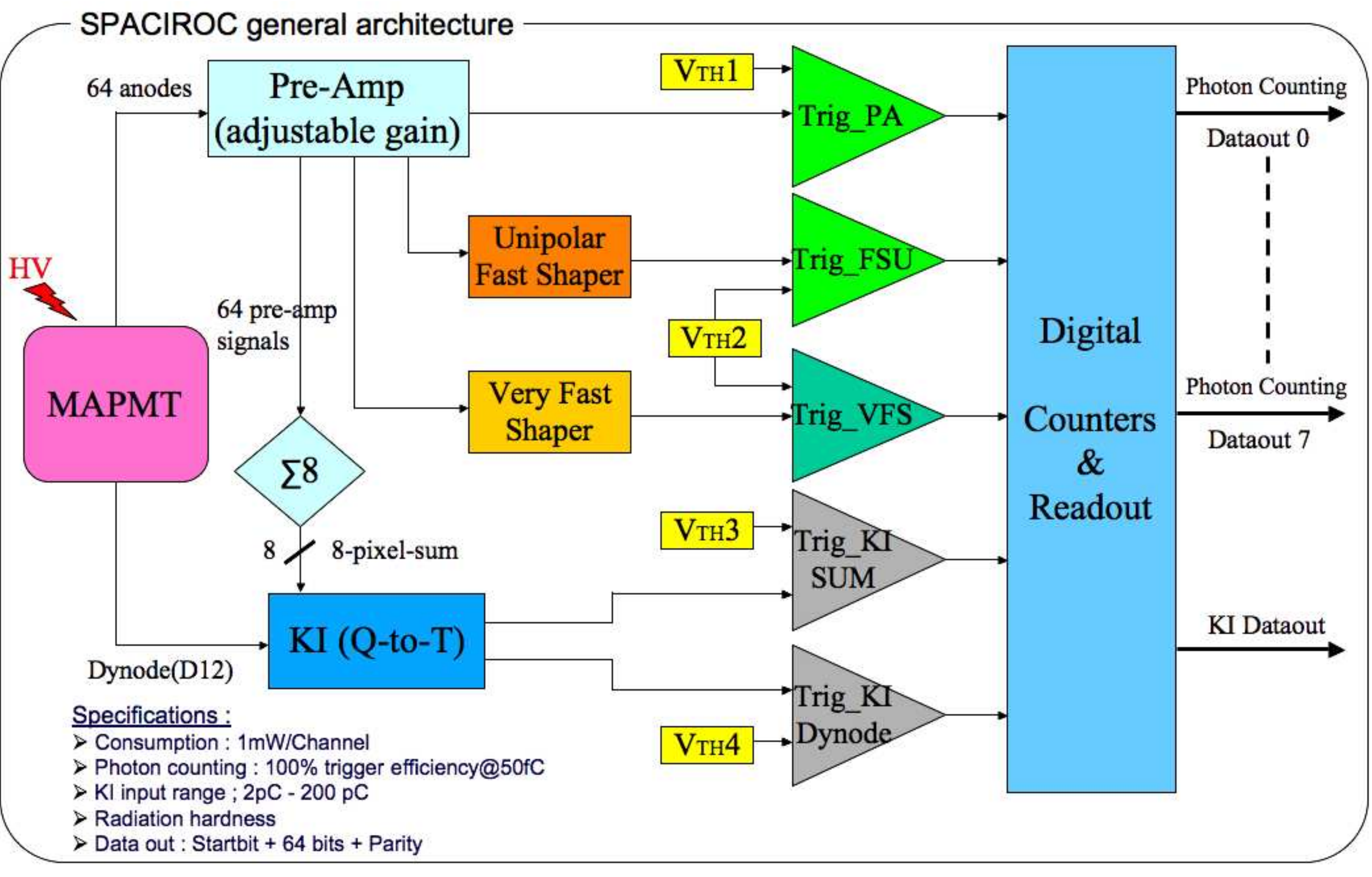}
  \caption{SPACIROC general architecture.}
  \label{SPACIROCschem}
\end{figure}
Fig.\ref{SPACIROCschem} shows the general architecture of the SPACIROC.
SPACIROC consists of two analog blocks and one digital part.
One of the analog parts is dedicated for Photon Counting, and another
is for so called ``KI" Charge to Time (Q-to-T) converter.
The digital part is build to count the detected photons. 
The 64-channel Photon Counting block discriminate the preamplifier
signal into trigger pulses.
Each of 64 channels of photon counting block consist of a preamplifier,
two shapers and three compartors (trigger discriminators) as below:
\begin{itemize}
\setlength{\itemsep}{-0.5mm}
\setlength{\itemindent}{-3mm}
\item Trig$\_$pa $\ \ \ :$ Trigger of the signal coming directly from
preamplifiers
\item Trig$\_$FSU : Trigger of signal from Unipolar Fast Shaper (FSU)
\item Trig$\_$VFS : Trigger for Very Fast Shaper (VFS).
\end{itemize}
Charge signals from 64 anodes of a PMT are first fed into preamplifiers
before sent to various shapers and discriminators in the latter part
of ASIC.
Then divided to 3 photon counting (PC) and outputs : preamp, FSU, VFS.
At the end of each acquisition window, so called Gate Time Unit
(GTU=2.5$\mu$s), the counter values are readout through 8 serial links
in order to reduce overhead.
The first 8 inputs of KI takes the pre-amplified signals from the
photon counting (sum of every 8 channels),
while 9th input takes a signal coming directly from the last dynode of
the MAPMT.
In a similar manner to the photon counting readout, the counter data are
sent through a serial link at the end of each GTU.\\
For more details of design and specification of SPACIROC, see
also\cite{jemeuso8} in this conference.
\subsection{Requirement}
The electronics system is required to keep a high trigger efficiency
with a flexible trigger algorithm as well as a reasonable linearity
in the energy range of 4$\times10^{19}$ to $10^{21}eV$ for EECRs. \\
The Front-End ASIC is required to count single photoelectrons, i.e.,
with considering the MAPMT gain of 5$\times$$10^5$, 0.08 $pC$.\\
Number of photoelectrons per GTU per pixel by the fluorescent light
from EAS generated by EECR with $10^{20}$eV is obtained by simulations
to be roughly 250 at around the shower maximum.
By multiplying a safety factor of 2, it becomes 500 PEs$/$GTU$/$pixel
which correspond to 40 pC$/$GTU$/$pixel.\\
Also, to operate JEM-EUSO under a quite limited power in the space,
a power consumption of less than 1mW/ch is essential.\\
SPACIROC achieves 100 \% trigger efficiency for charge greater than 1/3 PEs,
a dynamic range of over 1000 with having two analog processing modes as
described below, low power consumption of 1 mW/channel
and thus fulfills the requirement for the JEM-EUSO electronics.\\
\subsection{Test Method}
\begin{figure}[!h]
  \centering
  \includegraphics[width=3.2in]{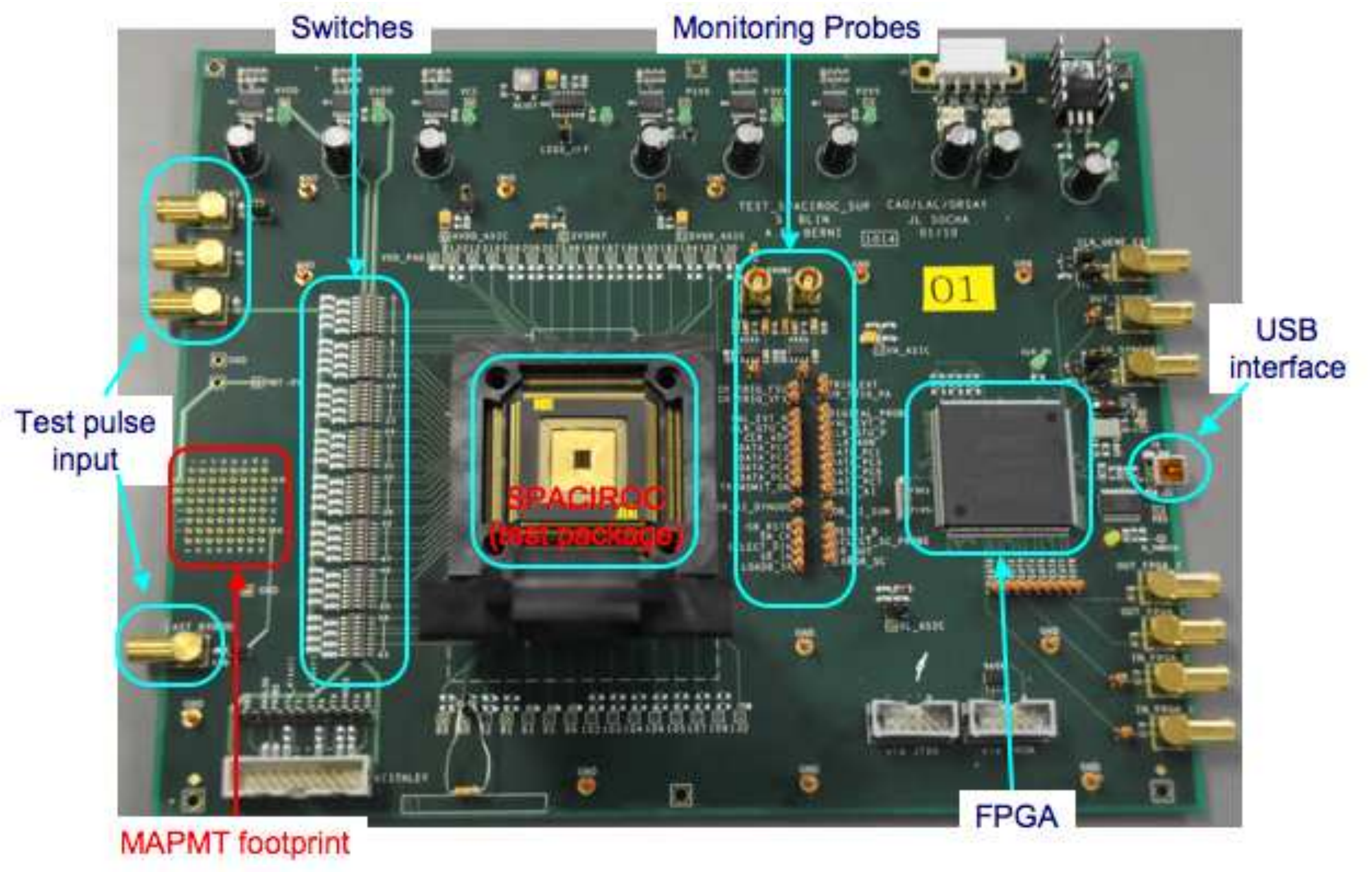}
  \caption{Test board for SPACIROC.}
  \label{testBoard}
 \end{figure}
Various tests have been done by using a test board shown in
Fig.\ref{testBoard}.
The test board consists of a socket for ASIC, FPGA and USB connection
to PC, MAPMT 64 anodes and a dynode footprint, and various test points.
Test pulses are fed into the board from a pulse generator.
PMT signals are also fed into the board after the test pulse measurement.
The registers inside the ASIC are controlled by a PC using a LabView
software via an FPGA and USB connection.
\subsection{Power Consumption}
Due to the limited electrical power available from the ISS for the
JEM-EUSO experiment, a maximum of 0.8 mW is allowed per detection
channel in the ASIC.
For the first prototype of SPACIROC, the measured power consumption
is 1.1 mW/channel.
This is partially because of the design bugs which makes some unused
componet always on and non negligible power dissipation is occuring.
Therefore, the next version of SPACIROC is expected to reduce the
power consumption and fulfill the requirement.
\subsection{Radiation Hardness}
For the experiments in the space, the radiation hardness is essential.
Two effects have been clearly identified as harmful to the ASIC.
One is ``Single Event Latchup'' which can destroy the circuits, another
is ``Single Event Upset'' which may affect the SPACIROC functionalities.
In case of the analogue part of SPACIROC, it has been designed to take
into account the radiation effects on electronic systems.
For example, the layout is done carefully in order to minimize the
single event latchup effect.
Also, a mechanism to detect single event upset is added.
For the total ionizing doze effect, we exposed the ASIC chip against the
radiation of 70 MeV proton beam with the ASIC running with maximum gain
and all the capacitors and registers on to see if it causes any effects.
As a result, we confirmed no significant effect or difference in
configuration parameters such as threshold and preamplifier gains before
and after the test in the operation.
For other effects, we are also planning to test the ASIC with a heavy
ion beam near future.
\subsection{Single Photoelectron response}
\begin{figure}[!h]
  \centering
  \includegraphics[width=2in]{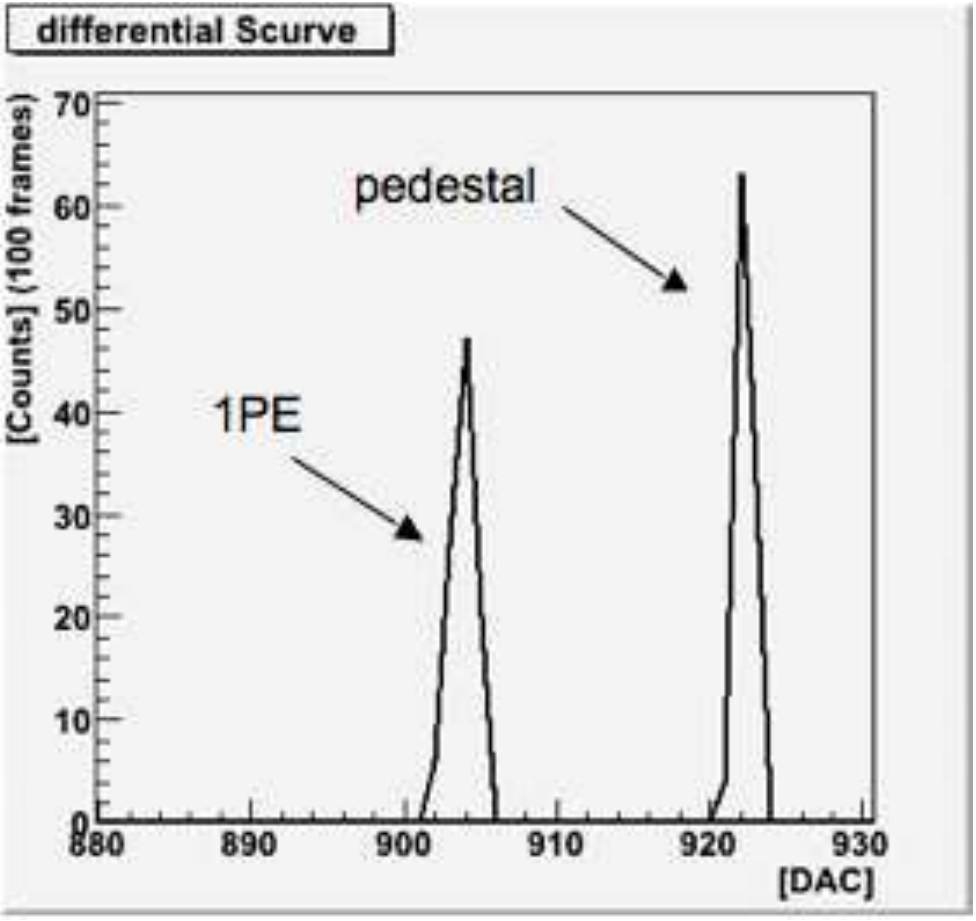}
  \caption{
    Differential S-curve from preamplifier output with
    the input charge of 80fC which is corresponding to 1PE for the
    gain of 5$\times$$10^5$.
  }
  \label{PA_80fC}
 \end{figure}
\begin{figure}[!h]
  \centering
  \includegraphics[width=2.in]{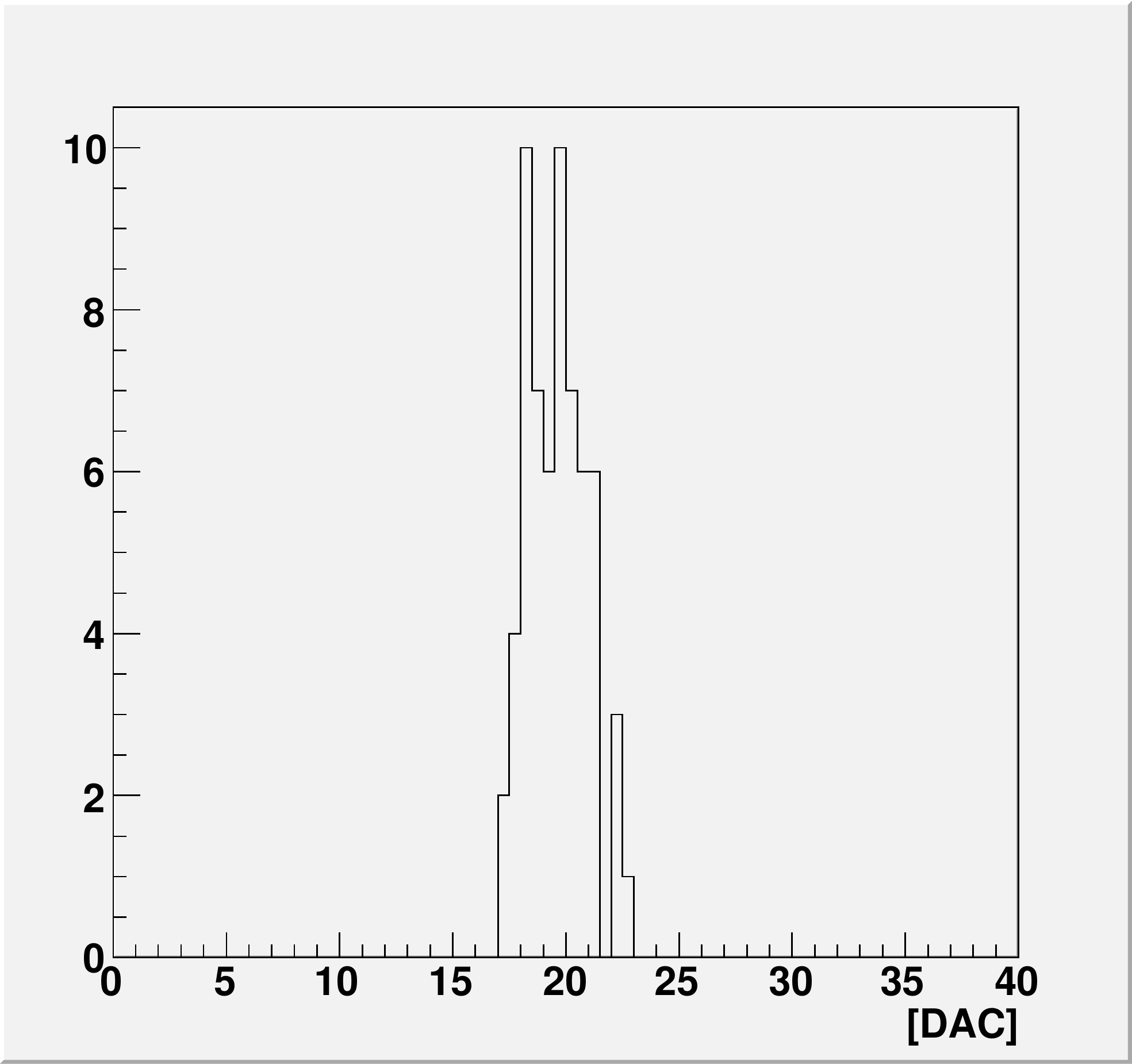}
  \caption{
    Amplitude of 64 ch preamp. RMS is 1.32 DAC unit.
  }
\label{ampl_PA_80fC}
\end{figure}

Fig.\ref{PA_80fC}
shows the differential of a Trig$\_$PA S-curve.
The input charge is 80$fC$ which corresponds to 1 photoelectron for
a PMT gain of 5$\times$$10^5$ generated by a pulse generator Tektronix
AFG 3102.
The single photoelectron peak is clearly seen thus we obtained the
minimum threshold of 0.13 PEs on the average of 64 channel outputs.
Fig.\ref{ampl_PA_80fC}
shows the amplitude, i.e., subtract pedestal from 1 PE peak,
of all 64 channel outputs of preamplifier and FSU respectively. 
The fluctuation in an ASIC is $\sim$1.3 for preamplifier and
$\sim$3.7 DAC unit for FSU. 
\begin{figure}[!h]
  \centering
  \includegraphics[width=2.in]{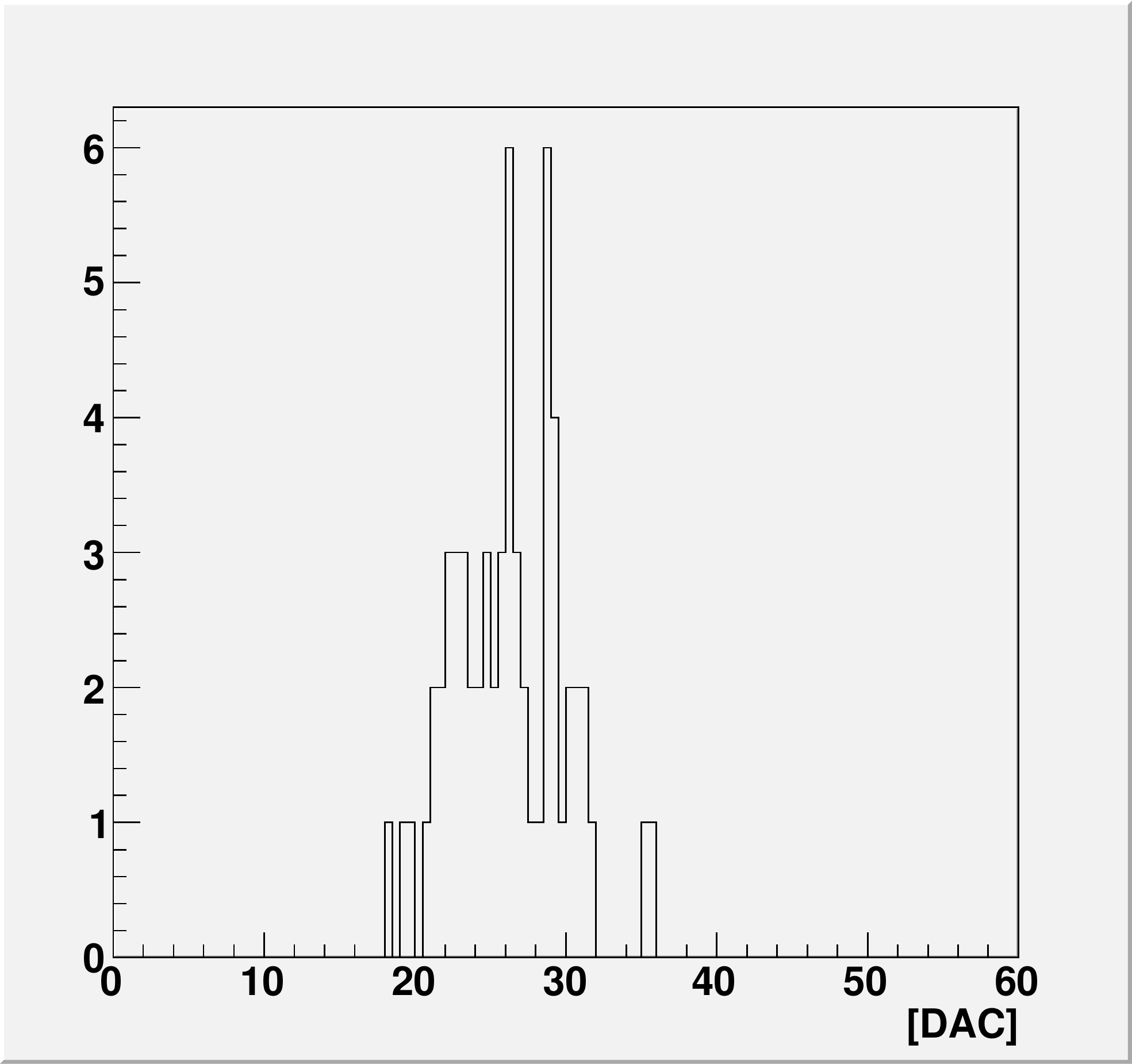}
  \caption{
    Amplitude of 64 ch FSU. RMS is 3.67 DAC unit.
  }
\label{ampl_FSU_80fC}
\end{figure}
\subsection{Linearity and dynamic range}
\begin{figure}[!h]
  \centering
  \includegraphics[width=3.2in]{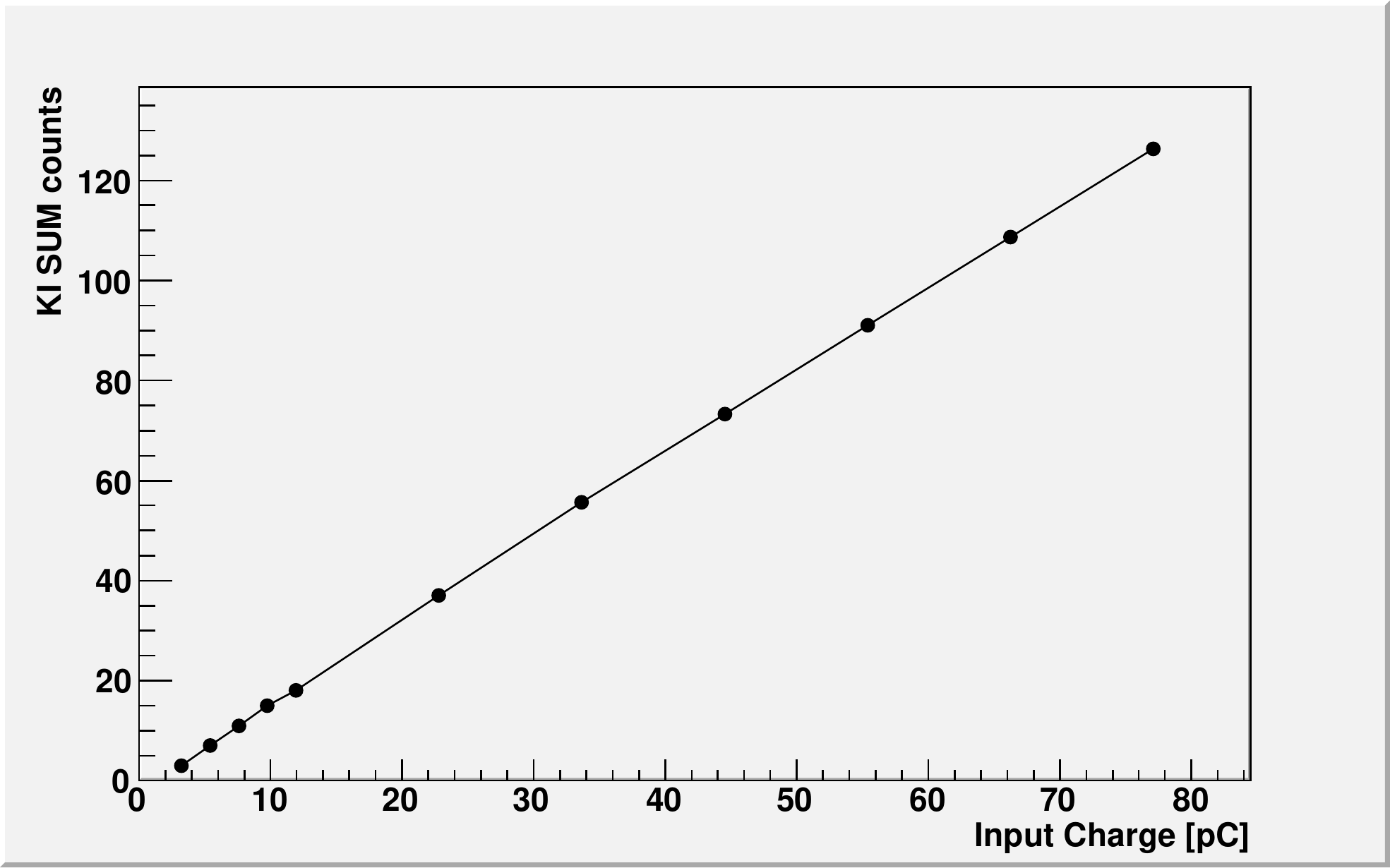}
  \caption{example of KI$\_$SUM digital output.}
  \label{KI_SUM_linearity}
 \end{figure}

In photon counting mode, the measured double pulse resolutions
for Trig$\_$PA is 36ns, and 30ns for Trig$\_$FSU.
Currently, further tests with input pulses of random timing are
ongoing to estimate the actual linearity and dynamic range in photon
counting mode for the PMT signal readout.\\
Fig.\ref{KI_SUM_linearity} shows the KI$\_$SUM counts, which
corresponds to the measured width of KI$\_$SUM output pulse
as a function of input charge ($pC$).
It is shown that there is a sufficient linearity between the
input charge and output pulse width within a dynamic range of
the input charge of 0.3 PEs to 80 PEs. In this case KI reaches
the maximum counter bit (7 bit dedicated for KI$\_$SUM) while
the linearity of KI itself still continues.
The study on the optimum KI parameters such as of KI pulse
width and dynamic range adjust for the Q-to-T conversion
is still ongoing.
\subsection{PMT response}
\begin{figure}[!hb]
  \centering
  \includegraphics[width=3.4in]{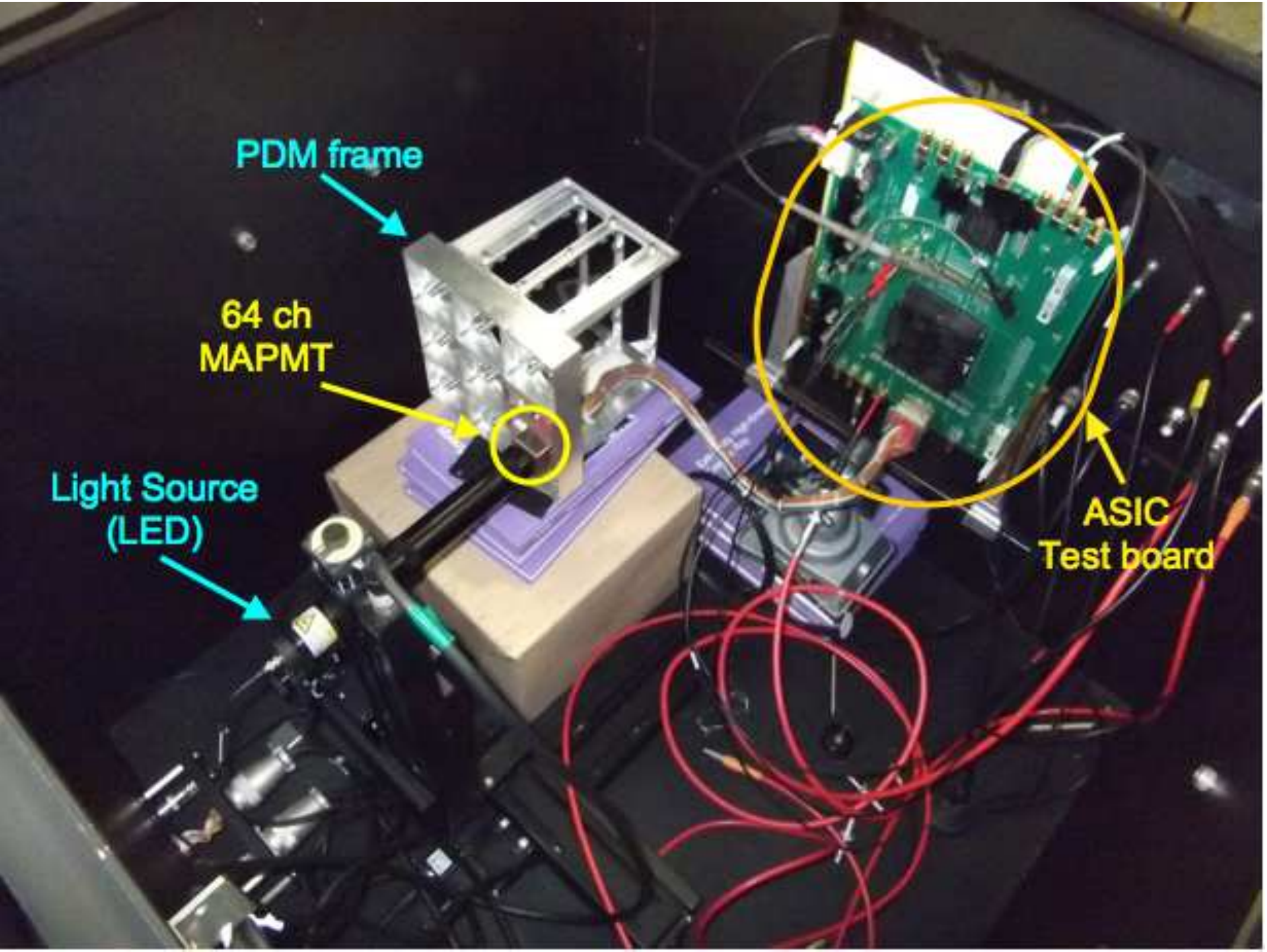}
  \caption{Integration test of MAPMT on a PDM frame connected to SPACIROC.}
  \label{integration}
 \end{figure}

\begin{figure}[!h]
  \centering
  \includegraphics[width=2.4in]{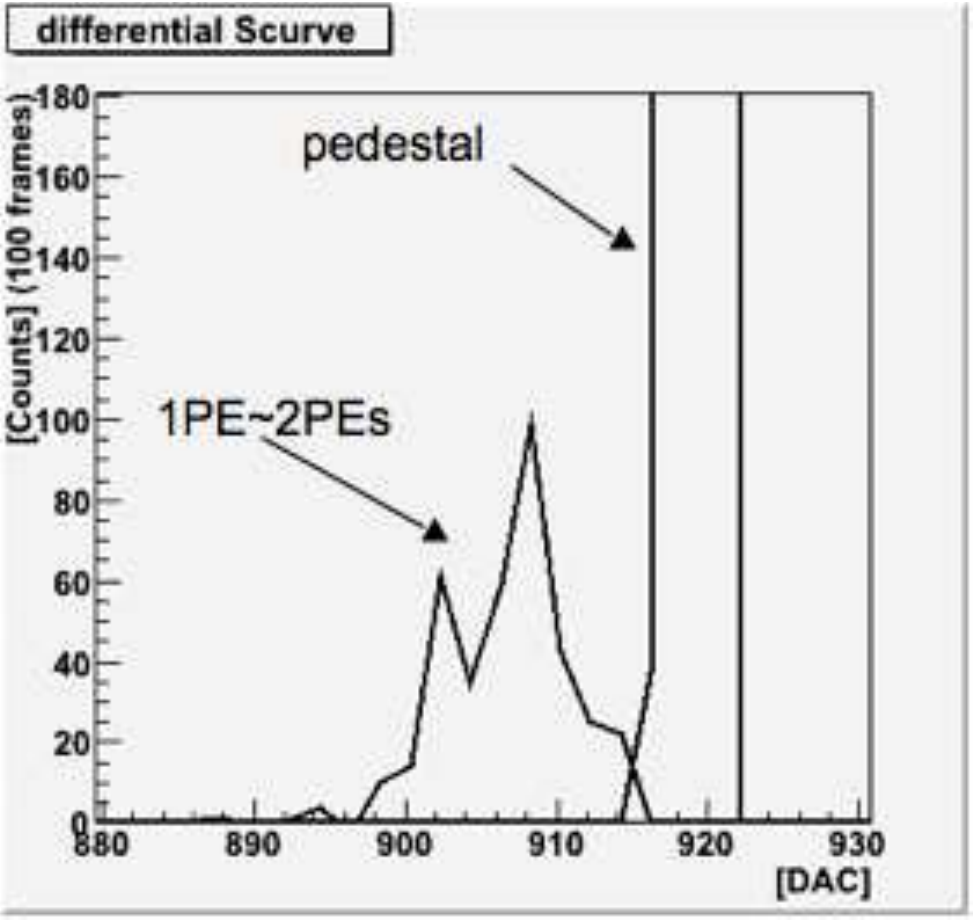}
  \caption{Distribution of number of input pulses using Trig$\_$PA.}
  \label{PMTresponse}
 \end{figure}

Fig.\ref{integration} shows setup of the integration measurement.
A 64 ch MAPMT on a PDM frame is connected to the ASIC test board which
is controlled by a PC.
The MAPMT is illuminated by a blue LED fired by an Agilent waveform
generator 33250A to give roughly 1.8 PEs on average in 200 $ns$.
The trigger of waveform generator is synchronized to the GTU clock
on the ASIC test board.
Fig.\ref{PMTresponse} shows the distribution of number of photoelectrons
(PEs) which is obtained by one of the 64 anodes.
The width of the broad peak in the figure is roughly consistent
with an error caused by the Poisson fluctuation.
\section{Conclusion}
The first version of Front-End ASIC for the focal surface detector of
JEM-EUSO mission has been examined.
It has been shown that the fundamental functions of the ASIC work quite
well without any critical problems.
We also succeeded to readout MAPMT signals by the ASIC and we obtained
the distribution of number of detected photons.
More examinations such as searching for the optimum configuration
parameters to achieve the largest dynamic range, best signal to noise
ratio and minimum power consumption are still ongoing.
Also, further work to improve the chip itself is ongoing and in the next
design, the power consumption will be improved to be about 0.8 mW/ch, the
noise level will be further suppressed.\\

\clearpage


\newpage
\normalsize
\setcounter{section}{0}
\setcounter{figure}{0}
\setcounter{table}{0}
\setcounter{equation}{0}



\title{High Voltage System for JEM-EUSO Photomultipliers}

\shorttitle{Philippe Gorodetzky \etal JEM-EUSO High Voltage}

\authors{Jacek KARCZMARCZYK$^{1}$, Philippe GORODETZKY$^{2}$,
 Yoshiya KAWASAKI$^{3}$,\\ Jacek SZABELSKI$^{1}$, Tadeusz WIBIG$^{1}$
 for JEM--EUSO Collaboration }
\afiliations{
 $^1$NCBJ, Cosmic Ray Laboratory, PL-90-950, Lodz, Poland\\
 $^2$APC, University Paris Diderot, 10 rue A. Domon et L. Duquet,
   75013 Paris, France\\
 $^3$RIKEN Advanced Science Institute, 2-1 Hirosawa, Wako, Saitama 351-0198, Japan }
\email{js@zpk.u.lodz.pl}

\abstract{
 JEM-EUSO, the UV telescope to be installed on the ISS, has a camera
 (focal surface) composed of 4932 Hamamatsu new M64 photomultipliers,
 making a total of 315648 pixels. One pixel (2.88 x 2.88 mm) represents
 on the Earth surface a square of 500 m side.\\
 Two major specifications of JEM-EUSO are:
 \begin{enumerate}
     \setlength{\itemsep}{-12pt}
     \vspace{-8pt}
   \item[a)]
     the total power allocated for all the instrument should not be
     above 1000 W, so that the power allocated to polarize with high
     voltage should be less than 50 W (using normal resistive voltage
     dividers requires nearly 2~kW!).\\
   \item[b)]
    the light intensity reaching JEM-EUSO has a dynamic range larger
    than 10$^6$ going from the background (one photo--electron per pixel
    per 2.5 $\mu$s) to storm lightnings.
 \end{enumerate}
 \vspace{-8pt}
 Solution for a) is to use separate power supplies for each dynodes,
 regrouping identical dynodes at the same power supply.
 The groups can be at the level of the Elementary Cell
 (4 PMTs) for a total power of 50~W. 
 The solution chosen for b) is not to shut off the telescope
 at the approach of the storm,
 but to study the events producing a lot of light
 (allowing for instance to measure meteors or TLEs).
 The voltage applied to the cathode is reduced in a fast
 ($<$~3~$\mu$s) switch driven by the integrating parts 
 of the front--end ASICs.
 The focusing properties of the tubes are modified in such a way
 that the \lq \lq gain'' is reduced in a range of 10$^6$, 
 by steps of 10$^2$.
  }

\maketitle

 \section{JEM-EUSO telescope -- TPC.}
 JEM-EUSO telescope is going to measure the highest energy cosmic
 rays by monitoring the atmosphere (at dark side of the Earth)
 not from the ground, as has been observed for many years,
 but from the top: from the altitude of the International 
 Space Station (ISS)~\cite{tebisID1628, gmtID0956, fkajinoID1216}. 
 The advantage is the huge geometrical factor,
 as the area of radius of about 200~km would be monitored.
 The UV telescope (lenses and focal surface) will have the
 nearly circular area of about 2.3 meters in diameter.
 The light detector must be very fast, to monitor extensive
 air shower (EAS) development, which typically lasts about 30 $\mu$s,
 and the angular resolution (pixelisation) should be fine
 enough to see space development of EAS from 350--400 km
 altitude of ISS. To meet these requirements the light detector
 consists of 4932 multianode photomultipliers (MAPTM), each
 with 64 anodes (315648 pixels,
 each corresponding to about 500m x 500m at the Earth's ground level).
 In the basic mode, single photo--electrons would be counted
 in each pixel, and integrated every 2.5~$\mu$s (GTU - Gate Time Unit) 
 (400 thousand times per second). 
 The JEM--EUSO telescope would work as a TPC (time projection chamber)
 allowing 3D reconstruction of the EAS. 

 \section{Focal Surface detector structure. \label{sec:structure}}

 Hamamatsu -- the manufacturer of M64 Multi-Anode Photomultiplier Tubes
 (MAPMTs) for JEM--EUSO 
 -- developed 12 stage photomultipliers with additional grid
 near to anodes for better focusing of internal photo--electron (pe) cascades.
 As the telescope will be open viewing the night atmosphere,
 most of measured light would come from the UV background
 in the telescope field of view. We expect to measure
 on average about 1~pe in each pixel
 during GTU (for UV background analysis see~\cite{pbobikID0886}). 
 The gain will be about 10$^6$. So we expect
 the anode current of 4.1~$\mu$A from each MAPMT.
 4 MAPMTs form the basic unit -- Elementary Cell (EC),
 and 9 ECs form Photo Detection Module (PDM), where 
 there are 6 x 6 photomultipliers. There would be 137 PDMs
 on the Focal Surface (FS).\\
 This kind of MAPMTs requires a high voltage of about -900 -- -1000 V
 at the cathode with grounded anodes, to achieve the gain
 of 10$^6$.   
 One High Voltage Power Supplier (HVPS) is planned for
 every EC (4 MAPMTs). 
 The anode current (background measurements)
 for one  PDM is equal to 0.147 mA, and for the whole 
 FS it is equal to 20.2 mA.

 \section{High voltage supply for JEM-EUSO photomultipliers -- standard approach.
        \label{sec:standard}}

 Standard approach requires the resistive voltage divider
 in parallel to the MAPMT,
 and, to provide stability and linearity of MAPMTs,
 the current in the divider should be larger than 100 times
 the anode current. The superposition principle acts here,
 so the required divider current (or sum over all dividers)
 shall be 100 x 20.2 mA = 2.02 A. As the required voltage
 is -900~V, the power which would go to the divider
 would be more than 1.8 kW. This value is nearly twice the limit
 for JEM--EUSO power consumption for all instrument devices.
 Therefore the standard approach to powering photomultipliers
 is excluded.

 \begin{figure}[!t]
  \vspace{5mm}
  \centering
  \includegraphics[width=3.in]{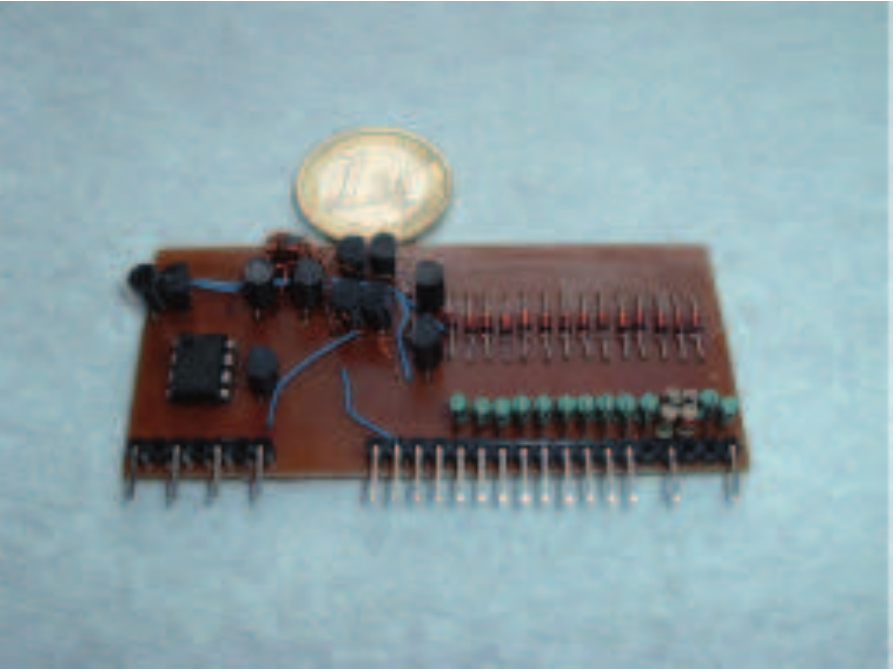}
  \caption{The Cockcroft--Walton high voltage
    photomultiplier supplier.
    }
  \label{fig:CWhvpsPhoto}
 \end{figure}

 \section{A photomultiplier model. \label{sec:model}}

 In photomultiplier the single pe emitted from the cathode
 is accelerated in electric potential U$_1$ between the cathode and the first
 dynode D$_1$. On average k$_1$ electrons are emitted from D$_1$ for each 
 electron arriving from the cathode. In the next step each electron emitted
 from D$_1$ is accelerated in electric potential U$_2$ between the first and the
 second dynode (D$_2$), and k$_2$ electrons are emitted on average from D$_2$
 per every electron from D$_1$, and so on. The last multiplication
 takes place at D$_{12}$, and electrons emitted from D$_{12}$
 are collected by anode. Assuming that U$_n$ = constant (which is very common
 for all dynodes but 1 or 2), and k$_n$ is proportional to U$_n$,
 we might express the phototube gain as equal to k$^{12}$, and
 for k = 3.16 we get the gain 10$^6$. Constant k implies constant U$_n$,
 and for 13 steps we would have U$_n$ = 900 V / 13 = 69~V.\\
 It is important to notice that on the last step D$_{12}$ -- anode 
 we have the largest current {\em i}$_A$, then between D$_{11}$ -- D$_{12}$
 the current {\em i}$_{12}$ is k$_{12}$ times smaller, then between 
 D$_{10}$ -- D$_{11}$ {\em i}$_{11}$ is still k$_{11}$ times smaller etc. 
 Similarly the power released in the photomultiplier itself
 is the largest at anode: {\em i}$_A$ $\times$ U$_A$,
 then at D$_{12}$ is {\em i}$_{12}$ $\times$ U$_{12}$
 then at D$_{11}$ is {\em i}$_{11}$ $\times$ U$_{11}$ etc.
 60\% of power is deposited at anode, 20\% at D$_{12}$, 
 7\% at D$_{11}$, 2\% at D$_{10}$ and 1\% at other dynodes.
 
 \begin{figure}[!t]
  \vspace{5mm}
  \centering
  \includegraphics[width=3.in]{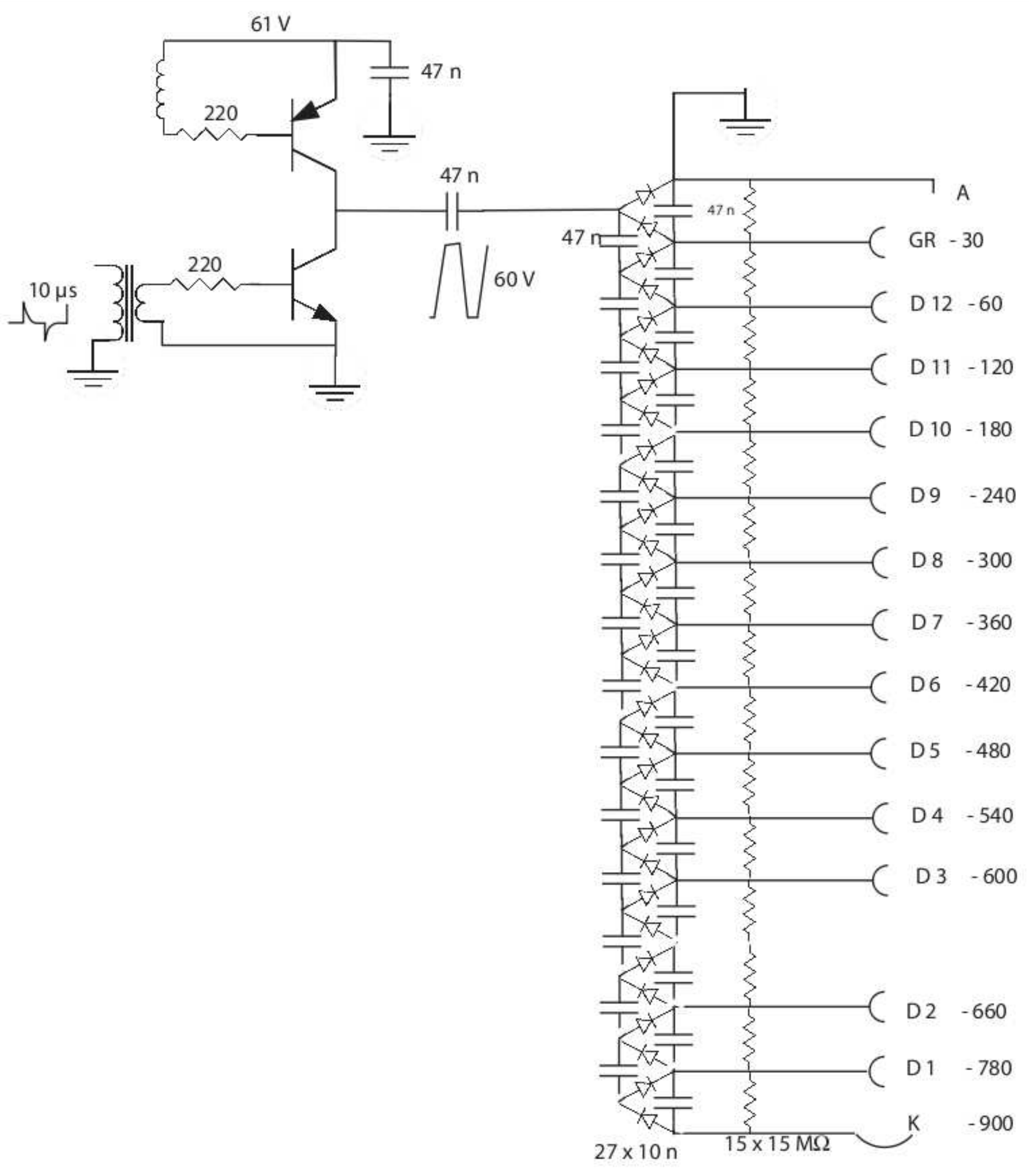}
  \caption{The Cockcroft--Walton solution for high voltage
    photomultiplier supplier without resistor divider.
    }
  \label{fig:CWhvpsScheme}
 \end{figure}

 \section{The Cockcroft--Walton voltage multiplier -- solution to the power problem.}

 The recomended photomultiplier voltage ladder is very near to the 
 simplified description presented
 in the Section~\ref{sec:model}. 
 The Cockcroft--Walton scheme shown in the Figure~\ref{fig:CWhvpsScheme}
 is close to the real voltage ladder.
 For dynodes with large indexes 
 the $\Delta$U are constant, and on the first 2-3 dynodes
 $\Delta$U are larger but the currents (and power) there are very small.
 We have made a high voltage power supply (HVPS) using the Cockcroft--Walton 
 voltage multiplier circuit with constant $\Delta$U approximately equal to 60~V.
 The idea is presented in the Figure~\ref{fig:CWhvpsScheme}.
 First steps near to anode (grounded) have constant $\Delta$U and are directly connected
 to corresponding dynodes and near to cathode we use resistive voltage divider
 (however required currents are so small, that power losses there are negligible).\\
 Large resistors presented in the Figure~\ref{fig:CWhvpsScheme}
 are for diode polarization and to discharge the HVPS when turned of.
 For the background load as described in the Section~\ref{sec:structure}
 the HVPS power consumption per PDM is about 120~mW
 (which corresponds to about 16~W per all FS, 
 i.e. 100 times less than a standard solution presented
 in the Section~\ref{sec:standard}).\\
 We have found that similar solutions for high voltage power suppliers
 for photomultipliers were used in the past,
 e.g. in CERN~\cite{neumaierHVPS}.
  
 \section{Higher photon fluxes.}

\begin{figure}[t]
  \vspace{5mm}
  \centering
  \includegraphics[width=3.in]{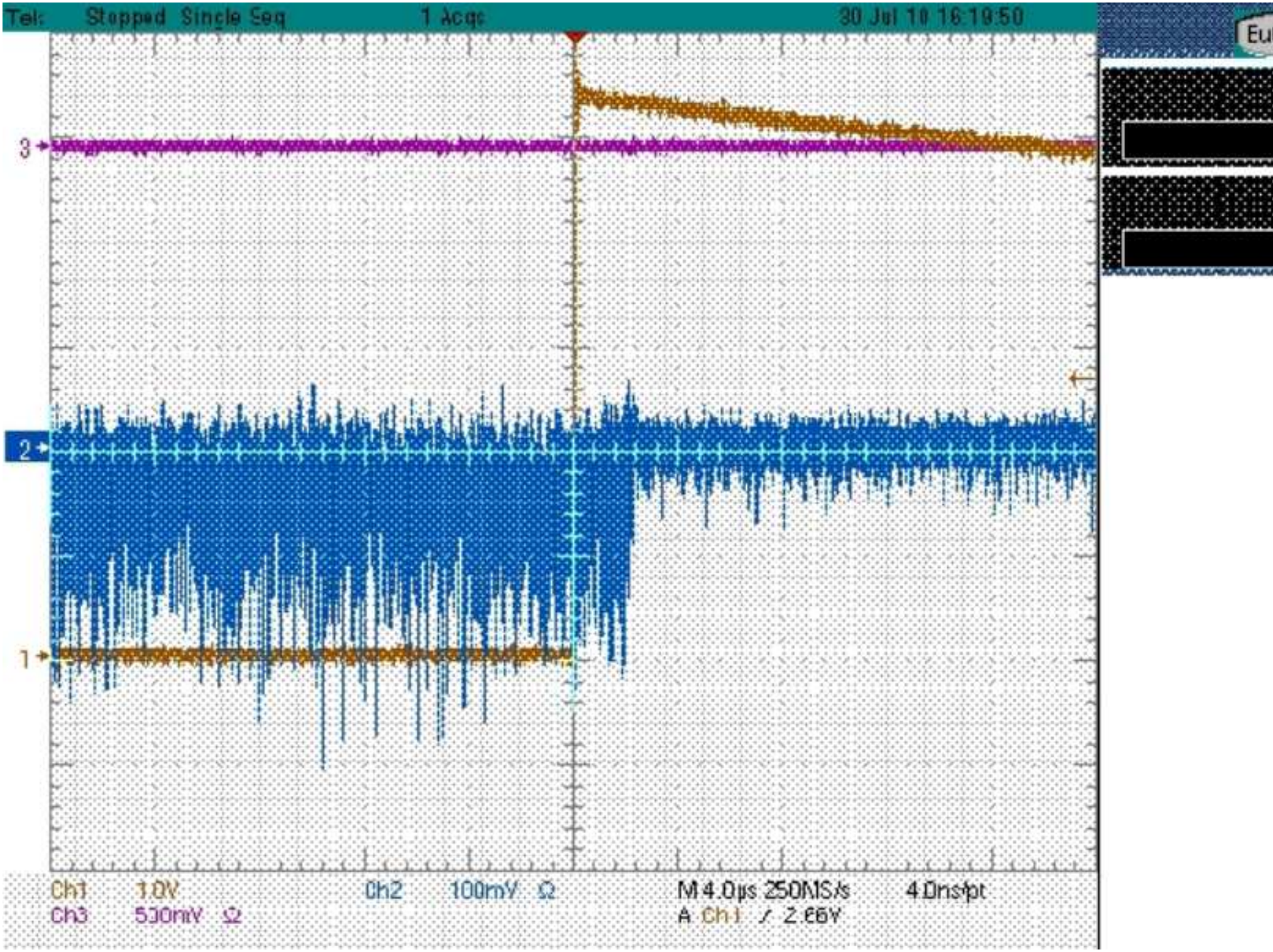}
  \caption{Oscilloscope showing the square pulses (yellow)
  to control the switch, and in blue the PMT anode pulses.
  One horizontal division is 4 $\mu$s.
    }
  \label{fig:switchAction}
 \end{figure}

 Lightnings, meteors, transient luminous events (TLE), or man made light (cities)
 could be very bright. 
 During the JEM--EUSO mission we might expect to meet photon fluxes 
 even 10$^6$ times higher than the background level (see Section~\ref{sec:structure}).
 Some of them can last tens of milliseconds (i.e. long compared with GTU = 2.5 $\mu$s).
 As we still like to measure them, following method is applied.\\
 When the light intensity is growing in 10 millisecond scale the HVPS
 should provide enough power to keep linearity of measurements
 up to about 200 times the background level.
 Laboratory measurements with calibrated light sources showed that
 our HVPS is capable to fulfil these conditions.\\
 However, when the light intensity rises above 100 times the background
 value, then we would reduce
 collection efficiency
 in the whole PDM by about 100 times,
 still going on with anode current measurements. 
 When the intensity still rises by the next 100 times, 
 we would reduce the
 efficiency
 in the PDM by another
 100 times, and the next such step can be performed
 (see Figure~\ref{fig:switchAction}).
 This reduction shall be fast (within 1-3 GTU) to secure
 the tubes against potential damages or wear by large 
 currents or charges at last dynodes and anodes.
 However, we would still measure the anode current and the higher
 gain would be restored when light flux falls below the harmful level.

 \newpage
 \section{Fast reduction of MAPMT efficiency.}

 In the case of increase of photon flux about 100 times above
 the background level we would reduce the efficiency
 of pe collection on the first dynode D$_1$.    
 We would apply new voltage on the cathode below
 the voltage on the first dynode D$_1$
 keeping all other voltages unchanged, e.g. changing the
 cathode voltage from -900~V to -750~V, and keeping the D$_1$
 voltage on the level -800~V. These would provide the opposite
 electric field and reduce the efficiency by about 100 times.
 The next 100 times efficiency reduction for M64 tubes requires
 cathode potential equal to 500~V, and with cathode grounded
 the efficiency would be still at the level about 3$\cdot$10$^{-5}$.\\
 In the M64 MAPMT the cathode and metal housing box are connected.
 Therefore the cathode capacitance of 36 photomultipliers (PDM)
 is large, about 1~nF. To change the electric potential of cathode by 500~V
 within one GTU (2.5 $\mu$s) requires a current of 0.3 A
 for that time, which is a large value.\\
 We made a two way switch controlled by low voltage circuit
 which has galvanic isolation from the high voltage part.
 It works as a current source (this or other way) providing
 large nearly constant current during the GTU. It requires
 large capacitors (0.2 $\mu$F on high voltage) in the required
 step of HVPS. The switch takes about 4 mW while not switching.
 
 \section{Conclusions.}

 We made several prototype models of HVPS and high voltage gain
 switches which have been successfully tested in laboratory conditions
 with M64 photomultipliers.
 Still further tests will be performed in a process 
 of preparing engineering models and then space qualified models.\\
  
 {\bf Acknowledgement}\\
 This work was supported in part by the Polish-French collaboration 
 COPIN-IN2P3 (09-135).

\clearpage


 \hspace{4cm}
 \newpage

\newpage
\normalsize
\setcounter{section}{0}
\setcounter{figure}{0}
\setcounter{table}{0}
\setcounter{equation}{0}



\title{The Cluster Control Board of the JEM-EUSO mission}

\shorttitle{author \etal paper short title}

\authors{
	J\"org Bayer$^{1}$,
	Mario Bertaina$^{2}$,
	Giuseppe Distratis$^{1}$,
	Francesco Fenu$^{1}$,
	Andrea Santangelo$^{1}$,
	Thomas Schanz$^{1}$,
	Christoph Tenzer$^{1}$
	on behalf of the JEM-EUSO collaboration
	}
	
\afiliations{
	$^1$IAAT, Kepler Center f\"ur Astro- und Teilchenphysik, Universit\"at T\"ubingen , Sand 1, 72076 T\"ubingen, Germany\\
	$^2$Dipartimento di Fisica Generale, Universit\`a di Torino, Via P. Giuria 1, 10125 Torino, Italy
	}
\email{bayer@astro.uni-tuebingen.de}

\abstract{
The Cluster Control Board (CCB) is one of the key elements of the JEM-EUSO read-out
electronics, which manages the data received from eight units of the Photo Detector
Module (PDM) of JEM-EUSO, performing data selection and transmission. To reduce the
large amount of data produced at the detector level and to discriminate the good
Extensive Air Shower (EAS) events from the spurious ones, a hierarchical trigger
scheme over two levels has been developed. The first trigger level consists of three
sub-levels which are implemented within the front-end Application Specific Integrated
Circuit (ASIC) and the PDM electronics.
The second trigger level is implemented in the CCB electronics. After the
onboard processing of the data received from 8 PDMs at a rate of around 57 Hz,
potentially good events are transmitted to the onboard CPU at the level of around 5
mHz. In this paper, we will firstly present the algorithm developed for the second
trigger level, focusing on its implementation in hardware. The algorithm aims at
distinguishing the unique patterns produced on the focal surface by the EAS from the
ones produced by background events. It is based on the scan of a predefined set of
directions, which covers the complete parameter space, to find good patterns associated with
the EAS. The final set of directions has been carefully optimized to adapt the
algorithm to the limited on-board computing power. To fulfill the requirement on the
processing time, the algorithm was implemented in a Field Programmable Gate Array in
order to make use of its parallel processing capabilities. After presenting the current
architecture of the CCB and discussing the complex interfaces with the other elements
of the read-out electronics, we will report on the performance of the laboratory model.}


\maketitle

\section{Introduction}
\label{sec:Introduction}
The planned Extreme Universe Space Observatory (EUSO) - attached to the Japanese
Experiment Module (JEM) of the International Space Station (ISS) -
is a large Ultra Violet (UV) telescope to investigate the nature and origin of
the Ultra High Energy Cosmic Rays (UHECR) by observing the fluorescence light
produced in Extensive Air Showers (EAS).

The main instrument of JEM-EUSO is a super-wide $\pm 30^{\circ}$ Field of View (FoV)
telescope, which will be able to trace the fluorescence tracks generated by the
primary particles with a timing resolution of 2.5 $\mu$s and a spatial resolution
of $0.07^{\circ}$ (corresponding to about 550 m on ground), allowing to reconstruct the
incoming direction of the UHECR with an accuracy better than a few degrees \cite{stakahashi}.

The Multi Anode PhotoMultiplier Tube (MAPMT) used as the basic detector element was
developed by RIKEN in collaboration with Hamamatsu Photonics K.K. and has 64
channels in an 8x8 array.

As the electronics has to handle over $3.15\cdot 10^5$ pixels, the Focal Surface (FS) has
been partitioned into subsections - the Photo-Detector Modules (PDMs) - and a
multi-level trigger scheme has been developed.
In the current baseline design, there are two trigger levels: the 'first level'
trigger (L1), which is implemented within the PDM electronics and the 'second level' trigger (L2)
which is implemented in the CCB electronics.

\section{L2 Trigger Algorithm}
\label{sec:L2TriggerAlgorithm}

\subsection{Overview}
\label{subsec:L2Overview}
The fluorescence light from EASs (induced by UHECRs) looks like a thin
luminous disk, which travels ultra relativistically on a straight
path through the atmosphere. As the EAS produces more particles, the luminosity
of the disc will also increase until the shower reaches its maximum and then fades
out. A typical proton induced shower with an energy of $10^{20}$eV, will be
detected as several photons per pixel and $\mu$s during a typical duration
of tens to hundreds $\mu$s.
As the detector has some exposure time for each image - the Gate Time Unit (GTU) -
the fluorescence light will appear as a small spot, which moves on a straight line
from image to image.
The speed and the direction of this moving spot is obviously depending on the
incoming direction of the UHECR that is from the shower axis.

The principle of the L2 trigger algorithm is therefore trying to follow the
movement of this spot over some predefined time, to distinguish this unique
pattern from the background. In the case of an L1 trigger, the PDM electronics
(see \cite{park} for details)
will send (together with the frame data) a starting point, which contains the pixel
coordinates and the GTU which generated the trigger - also called 'trigger seed'.
The L2 algorithm will then define a small box around this trigger seed, move
the box from GTU to GTU and integrate the photon counting values.
This integrated value is then compared to a threshold above the background
and an L2 trigger will be issued if the threshold is exceeded.

It should be stressed, that an effective implementation of the L2 algorithm is
constrained by the limited available computing power on board, due to power-,
weight- and size-requirements and the space-qualification of the hardware.

Currently it is foreseen to have a total of 375 starting points for the integration,
which are distributed equally over time and position around the trigger seed.
Each integration will be performed over $\pm$7 GTUs for a predefined set of directions
(see Section \ref{subsec:L2NumberOfDirections}).

In order to follow the movement of the spot on the detector, the speed and the
direction in terms of detector pixels has to be calculated.

We define in the following $\theta$ as the zenith angle ($\theta = 0^{\circ}$ means
nadir direction of JEM-EUSO) and $\phi$ as the azimuth angle of the EAS
($\phi = 0^{\circ}$ means direction of the ISS movement).
\begin{figure}[!t]
  \vspace{5mm}
  \centering
  \includegraphics[width=1.2in]{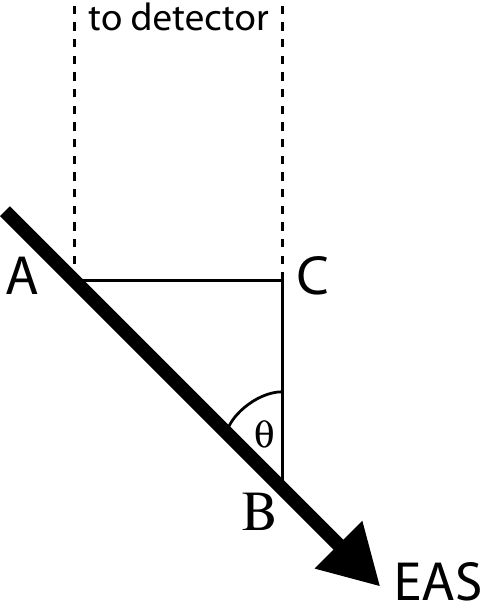}
  \caption{Kinematics of an EAS}
  \label{pic:EAS_kinematic}
\end{figure}
Since the EAS travels at the speed of light, the photons reaching JEM-EUSO
from any point on the EAS are lagging behind the passing EAS-front by
the time:
\begin{eqnarray*}
	\Delta t	= \frac{\overline{AB}+\overline{BC}}{c}
\end{eqnarray*}
with $c$ being the speed of light and $\overline{AB}+\overline{BC}$ as defined in
Figure \ref{pic:EAS_kinematic}. Together with
	\[
		\sin\theta = \frac{\overline{AC}}{\overline{AB}},\quad
		\tan\theta = \frac{\overline{AC}}{\overline{BC}},
	\]
	\[
		\tan\theta = \frac{\sin\theta}{\cos\theta} \quad\mbox{and}\quad
		\tan\frac{\theta}{2} = \frac{\sin\theta}{1+\cos\theta}
	\]	
it follows
\begin{eqnarray*}
	\Delta t	= \frac{\overline{AC}}{c\cdot \tan\frac{\theta}{2}}
\end{eqnarray*}
as can be derived from Figure \ref{pic:EAS_kinematic}. If we define
now $\Delta x$ and $\Delta y$ as the projections on the focal surface along $x$ and $y$
expressed in number of pixels and $\Delta L$ as the FoV at ground of a pixel
in a time $\Delta t$ (chosen as the GTU) we find
\begin{eqnarray*}
	\overline{AC}	&= \Delta L\cdot\sqrt{\Delta x^2+\Delta y^2}
\end{eqnarray*}
and therefore
\begin{eqnarray}
	\theta &=& 2\arctan\left(\frac{\Delta L}{c\cdot\Delta t}\cdot\sqrt{\Delta x^2+\Delta y^2}\right)\\
	\phi &=& \arctan\left(\frac{\Delta y}{\Delta x}\right)
\end{eqnarray}

\subsection{Number of Directions}
\label{subsec:L2NumberOfDirections}
Due to the fact, that the incoming direction of the EAS is unknown, the approach of the L2
trigger algorithm is simply to try out a set of directions which covers the complete
space ($\theta = 0^{\circ}\ldots 90^{\circ}$ and $\phi = 0^{\circ}\ldots 360^{\circ}$).
The integrated count value will have a maximum when we 'hit' the (nearly) correct
direction, because in this case the integrating box will follow the spot.

As the integration over the directions is a time consuming task, a trade-off between the
number of directions and the available computing power has to be found. In addition,
as the hardware only works on whole pixels, some directions will produce the same offsets
$\Delta x$ and $\Delta y$ - for example, in the extreme case $\theta = 0^{\circ}$
(which means the shower axis is parallel to the optical axis of the telescope), the spot
is not moving at all there is no need to integrate over different $\phi$ angles.

Therefore, a minimum set of directions (containing 67 $\theta$-$\phi$ combinations in total)
was selected and evaluated with simulations aiming to optimize the design according to
the constraints mentioned above.

\section{Cluster Control Board}
\label{sec:ClusterControlBoard}

\subsection{Overview}
\label{subsec:CCBOverview}
In case an L1 Trigger is issued, the data of the ring buffers from 8 PDMs are transferred
to the CCB and the L2 trigger algorithm is executed. The data from the PDMs are
processed independently, as it is not intended to have a fixed geometrical relationship
between the PDMs connected to one CCB.
In case an L2 trigger is found within any of the 8 PDMs, the complete data are transferred
to the mass memory module of the Mission Data Processor (MDP).

Due to the necessity for parallel processing and the need for a high number of
input/output pins (for the eight data interfaces to the PDMs and the interface
to the MDP, besides various other interfaces for control and housekeeping)
it is foreseen to implement the L2 trigger algorithm inside an Field Programmable
Gate Array (FPGA) chip on the CCB.
As a baseline, it is planned to use a radiation tolerant Xilinx Virtex-4QV FX-140
and as an advanced option the use of a radiation hard Virtex-5QV is currently under
investigation (see \cite{xilinx} for the specification).

\begin{figure*}
	\centering
	\includegraphics[width=0.9\textwidth, keepaspectratio=true]{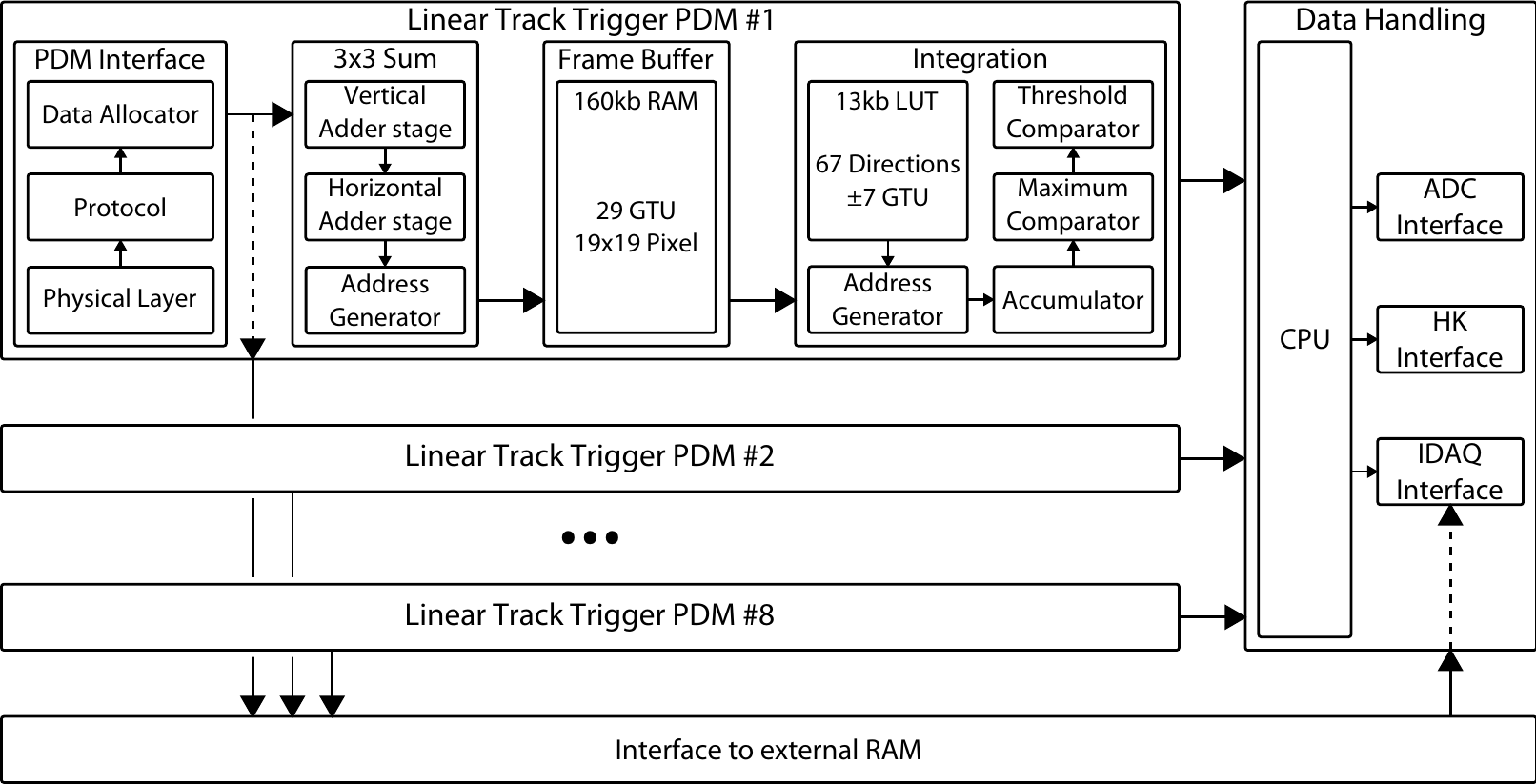}
	\caption{The current block diagram of the CCB FPGA.
		Most of the logic resources are needed for the eight 'Linear Track Trigger'
		modules which perform the L2 trigger algorithm and handle the data transfer
		from the PDM. As it is not possible to hold the data from all eight PDMs
		inside the FPGA RAM an interface to an external memory will be needed.
		Additionally, it is currently foreseen to use the integrated microcontrollers
		of the Virtex-4 FX140 to control and monitor the trigger modules and to
		handle the communication with the MDP and the housekeeping submodule.}
	\label{pic:FPGA_block}
\end{figure*}

The current block diagram for the FPGA is given in Figure \ref{pic:FPGA_block},
which is basically the implementation of the L2 trigger algorithm.
Besides the eight Linear Track Trigger (LTT) modules it is planned to use the
microcontroller cores for 'low speed' purposes, such as the configuration of the
LTT modules, housekeeping and the communication with the MDP.

In order to perform the necessary calculations as fast as possible, the hardware
architecture of the trigger is highly pipelined and parallelized.
As it is not possible to store the raw data completely in the internal RAM of the
FPGA, the data will be written to an external memory upon arrival. Only the data necessary
for the trigger calculation will be stored inside the FPGA to reduce the
probably slow read/write access for the external RAM. A first data 'selection'
will be done by the 'Data Allocator'-module. The data will then be passed to the
'3x3 Sum'-module which performs the summation of the 9-pixels blocks in
two stages (horizontal and vertical) for the whole frame which reduces redundant
3x3 summation to a minimum. The new frame generated this way is then trimmed to
a 19x19 pixel frame around the seed and stored in the 'Frame Buffer' for
$\pm 14$ GTUs.

After the 'Frame Buffer' is filled completely (it contains the necessary
information for the integration), the 'Address Generator' allocates the
different 3x3 sums - depending on the starting point for the integration and the
offsets for the various directions which are stored in a Look Up Table (LUT) - to the
'Accumulator'. After each direction-integration, the accumulated value is compared
to the current maximum to select the overall maximum value and is then stored along
with the information to which starting-point this value belongs ('Maximum Comparator').
When the integration process is finished for all starting points,
the maximum value is compared to the trigger threshold ('Threshold Comparator')
and an L2 trigger is issued if the threshold is exceeded.
In this case the raw data buffer (external RAM) is sent to the mass memory module
of the MDP, whereas the information at which pixel the maximum value was found,
could be used by the MDP to calculate the pixel coordinates on the focal surface.
This information then could be used as a target for the LIDAR to monitor the
atmospheric conditions around the EAS (see \cite{neronov} for more details).

\subsection{Interfaces}
\label{subsec:CCBInterfaces}
The interface between PDM and CCB for the scientific data from the detector
and for controlling/monitoring the PDMs is a critical part in the processing
chain as many other parts rely on the performance of this interface (e.g. the
ring buffer on the PDMs, the implementation of the L2 algorithm and the dead-time
of the instrument). In order to reduce the dead-time of the instrument, the event
data (around 2.7 Mbit per PDM and event) has to be transferred as fast as possible
within the hardware constraints. The current baseline is to implement a parallel
bidirectional interface with 20 data channels running at 40 MHz.
Due to the fact that the distance between the PDM and the CCB is about
1 m - as it is foreseen that the CCB is mounted on the back of the focal surface
- it was decided to use the Low Voltage Differential Signaling (LVDS) standard
which results in a more reliable interface, but doubles the number of lines needed.
In order to reduce the weight of the harness, an advanced option is currently 
under investigation which uses a double buffer approach whereby a total of 4 data
channels will be sufficient.

As a baseline for the interface between CCB and MDP (which will be controlled by an
intermediate board - the Interface Data AcQuisition (IDAQ) board) the SpaceWire
standard was chosen due to its approved reliability. To reduce the overhead of the
standard, a modified protocol will be studied as advanced option.

\subsection{Results from the performance tests}
\label{subsec:CCBResults}
To verify the correct operation of the 'Linear Track Trigger' module, a first
simple test was developed where the 'PDM Interface' was replaced by a 'PDM Simulator'
module.
This module consists of a small
Finite State Machine (FSM) and an 'Event Buffer' which is basically a large RAM,
filled with the simulated data of an event generated by the EUSO Simulation and
Analysis Framework (ESAF).

The data from the 'Event Buffer' are sent to the input of the 'Linear Track Trigger'
module where it will be processed as described in Section \ref{subsec:CCBOverview}.
To have a more detailed look into the processing, the 'Accumulator', 'Maximum Comparator'
and 'Threshold Comparator' are connected either to a ChipScope Virtual Input/Output (VIO)
or an Integrated Logic Analyzer (ILA) module.

The output of the 'Accumulator' is sampled by the ILA module every time the integration
of one direction is finished. This allows to compare every single integrated value
with the ones coming from simulations and to verify the correct operation of the
'Linear Track Trigger'.
Unfortunately, the sample buffer of the ILA module is limited (as it uses the
internal RAM of the FPGA) and it was not possible (with this simple test) to store
all 25125 integrator values. Therefore, only around 16000 integrator values
could be recorded. But as all sampled values (including the overall maximum) are
equal to the ones coming from simulation, it is assumed that this test was passed
successfully - also due to the fact, that the correct behavior of the the 'Linear
Track Trigger' module was verified in advance with hardware simulations.

According to the current implementation, the decision of an L2 trigger could be
performed within approximately 5 ms. The time needed for the data transfer
between PDM, FPGA and external RAM is not included in this number.
It should be stressed, that the actual calculation time depends on the system
clock of the FPGA, therefore, on the overall implementation.
However, even if the system clock has to be reduced by a factor of 2 the calculation
is still fast enough to meet the requirements, as 18 ms were allocated for
the L2 trigger calculation.

\section{Conclusion \& Outlook}
\label{sec:ConclusionOutlook}
The developed hardware is still in phase A and a long way is necessary to reach
the final space-qualified Cluster Control Board.
The hardware implementation of the 'Linear Track Trigger' algorithm as L2 trigger
is performing accordingly to the requirements. Based on the parallelization of the
calculation process, the requirement on the processing time of the L2 trigger could
be exceeded by a factor of 20. It should be stressed, that this value is highly
dependent on the overall implementation of the CCB FPGA but at least the requirement was met.

In our current plan of the CCB development are the implementations of the
interfaces to the Photo-Detector Module, to the Mission Data Processor
and to the Housekeeping. We will then start the design of a first laboratory model of the
CCB Printed Circuit Board. More information can be obtained from \cite{thesis_bayer}.


\clearpage


\newpage
\normalsize
\setcounter{section}{0}
\setcounter{figure}{0}
\setcounter{table}{0}
\setcounter{equation}{0}



\title{The Housekeeping subsystem of the JEM-EUSO instrument}

\shorttitle{G. Medina-Tanco \etal Housekeeping subsystem of JEM-EUSO instrument}

\authors{G. Medina-Tanco$^{1}$, J. C. D'Olivo$^{1}$, A. Zamora$^{1}$, H. Silva$^{1}$, L. Santiago Cruz$^{2}$, F. Trillaud$^{2}$, M. Casolino$^{3}$, K. Tsuno$^{4}$  for the JEM-EUSO Collaboration}
\afiliations{
$^1$Instituto de Ciencias Nucleares, UNAM, Circuito Exteriror S/N, Ciudad Universitaria, M\'exico D. F. 04510, M\'exico.\\
$^2$Instituto de Ingenieria, UNAM, Circuito Exteriror S/N, Ciudad Universitaria, M\'exico D. F. 04510, M\'exico.\\
$^{3}$  \\
$^{4}$RIKEN Advanced Science Institute, Japan.
}

\email{gmtanco@nucleares.unam.mx}

\abstract{
The JEM-EUSO instrument is a refractive telescope being proposed for attachment 
to the Japanese Experiment Module, Kibo, onboard ISS. The instrument is substantially complex, including large Fresnel lenses, an focal surface covered by 4932 MAPMTs of 64 pixels, atmospheric monitoring subsystems (IR camer and LIDAR), low and high voltage power supply subsystems, tilting mechanism and a 
lid. All these subsystems must be turned on and off and monitored, and telemetry has to be conveyed between them and the principal CPU. The housekeeping subsystem (HKSS) is in charge of those tasks. In this contribution we describe the requirements and design of the JEM-EUSO HKSS.  
}

\maketitle

\section{Introduction}

The JEM-EUSO instrument is a large refractive telescope to be installed at the Kibo module of the International Space Station (ISS) for the observation of extreme energy extensive air showers, using the fluorescence technique \cite{mission,science}. The instrument as a whole is described in \cite{instrument} and references there in.

\begin{figure*}[!t]
\centering
\includegraphics[width=17cm]{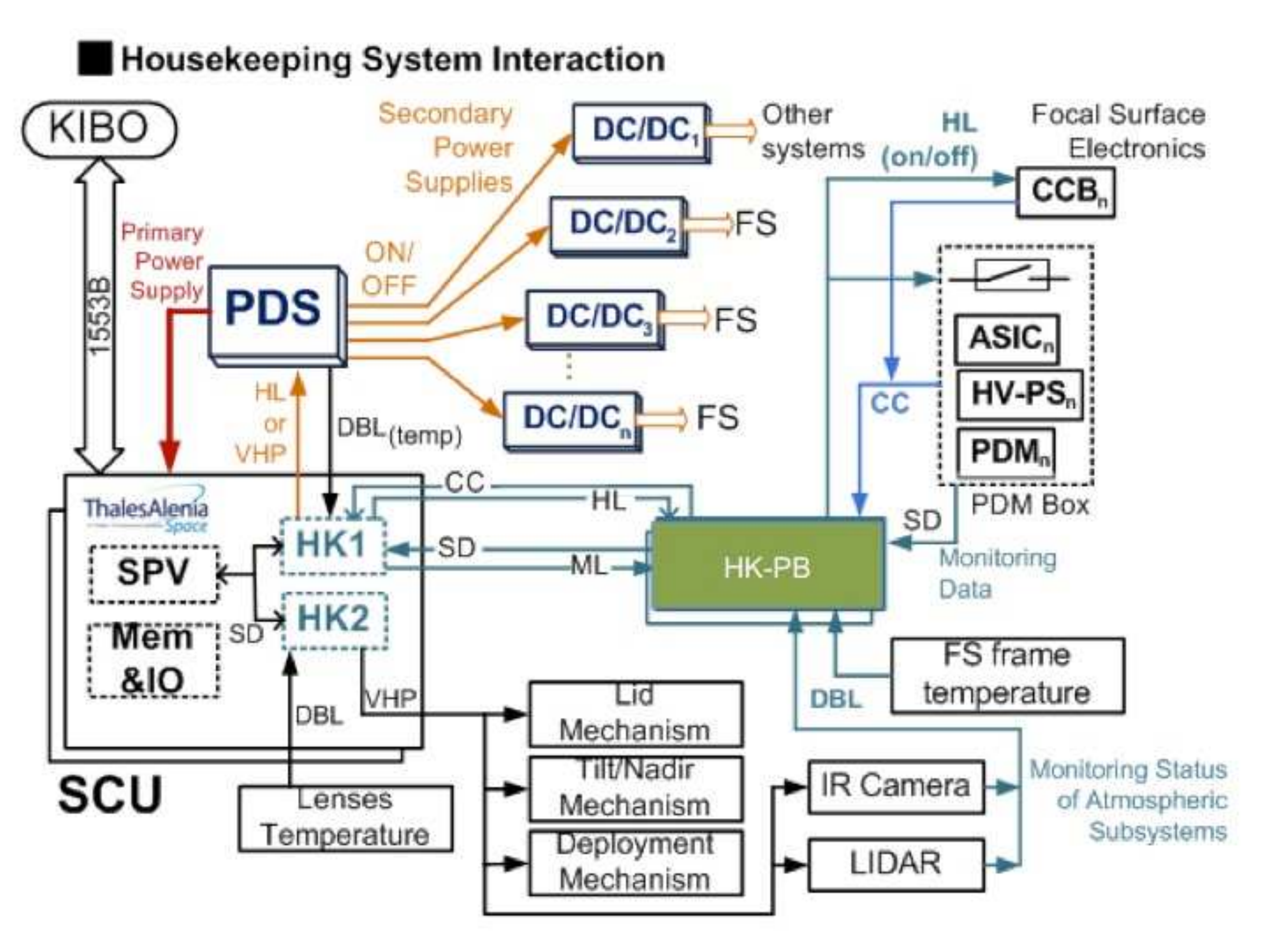}
\caption{General architecture of the JEM-EUSO housekeeping system}
\label{fig01}
\end{figure*}

The overall purpose of the Housekeeping Subsystem (HK) is to monitor and to relay control commands to the several subsystems that constitute the JEM-EUSO instrument. The HK sub-system is subservient to the CPU and all its activities are defined as slow control, i.e., with reaction time scales typically larger than a second. The HK subsystem architecture is conditioned by the wide variety of subsystems that constitute the JEM-EUSO instrument and with which it has to interact. The HK performs several tasks: (a) sensor monitoring of different subsystems in order to detect faults, (b) generation of alarms for the CPU, (c) distribution of telecommands to several subsystems, (d) telemetry acquisition from all subsystems, (e) monitoring of the status of the various electronic systems of the Focal Surface (FS), (f) switching between main and spares boards when appropriate, and (g) interaction with the power distribution system of the telescope, in order to turn ON and OFF the secondary power supplies, and therefore the FS, and verify adequate levels of power consumption. Fig. \ref{fig01} shows schematically the architecture of the HK and its interaction with various elements of the telescope. The HK prefix denotes the main boards that constitute the HK subsystem. The core of the HK is the HK principal board (HK-PB), which centralizes most tele-command distribution and telemetry gathering. The HK-PB is the direct responsible for monitoring the FS and other subsystems, as well as providing on/off and status verification for every single component of the instrument.

\begin{figure}[!h]
\centering
\includegraphics[width=8.5cm]{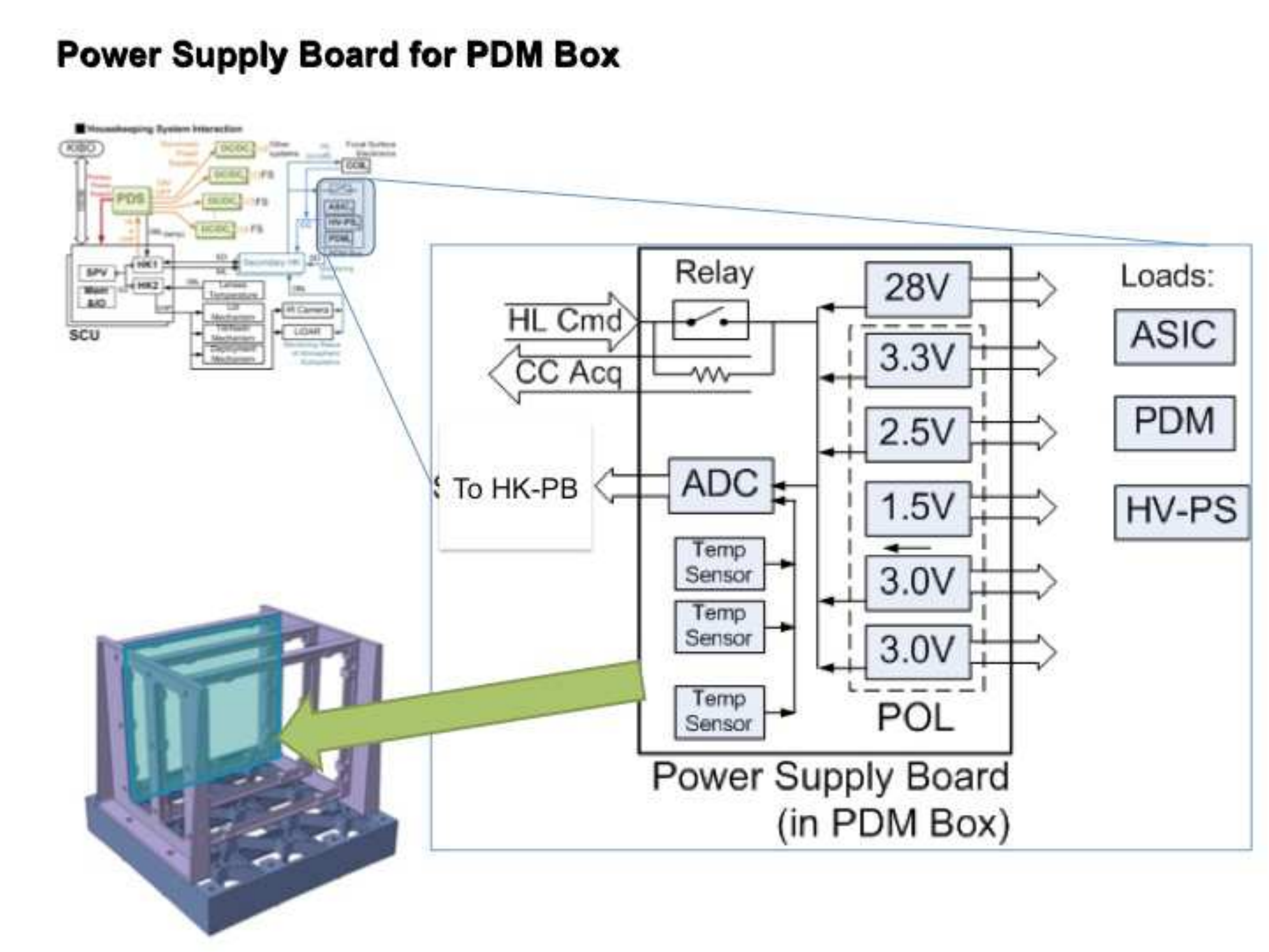}
\caption{Focal surface monitoring scheme.}
\label{fig02}
\end{figure}

\begin{figure}[!ht]
\centering
\includegraphics[width=8.5cm]{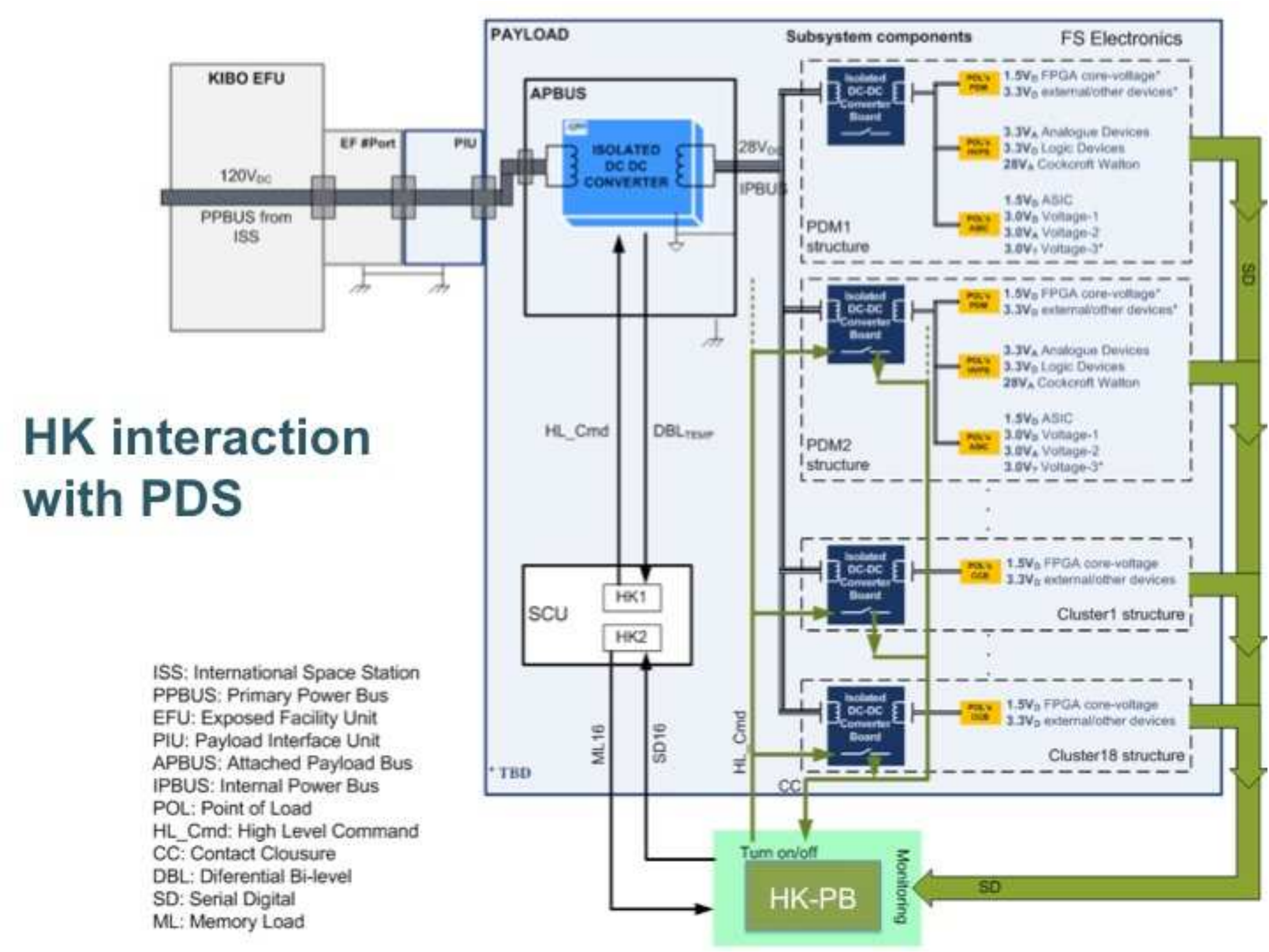}
\caption{Interaction between the JEM-EUSO housekeeping system and the power distribution system (PDS).}
\label{fig03}
\end{figure}

The anchor points of the HK in the focal surface are the Power Supply Boards (PSB), as shown in Figure \ref{fig02} and \ref{fig03}.  
The later board contains the relays that turn on and off the Points of Load (POLs) that generate the lowest voltages  (1.5V, 2.5V, 3.0V and 3.3V) required by the different components of the FS (e.g., ASIC, Photo-detector Module (PDM) and High-Voltage Power Supply (HV-PS)). The PSB also contains a footprint of the HK for telemetry: the ADC responsible for the digitalization of, mostly, the temperature sensors of the FS.

Due to its pervasive interaction with different subsystems, the HK is spatially distributed throughout the telescope.
Thus, the HK sub-system also possesses a secondary stage directly installed inside the main computer, the System Control Unit (SCU). The latter is divided in two boards, one of which (HK1) centralizes communication to and from the HK-PB and Power Distribution System (PDS), turning on and off and monitoring status of all the DC/DC converters of the Secondary Power Supply. The second module (HK2) receives lenses temperature data and relays tele-commands to the lid-mechanism, the tilting mechanism, the deployment mechanism, and the IR camera and LIDAR of the Atmosferic Monitoring System (AMS). 

The CPU and HK are turned on at the same time. HK has a direct interface with the CPU to distribute tele-commands sent from the ground by CPU users, it collects telemetry from different subsystems responsible for tasks such as atmosphere monitoring, telescope tilting, temperature measurement of lenses, and so on. Subsystem anomalies detected by the HK are reported as alarms to the CPU. In principle, HK does not take any action on these alarms, waiting for further commands from the CPU. At most, on critical instances, it can switch off a subsystem that triggered an alarm, in order to avoid possible further damage, while waiting for commands from the CPU for either restarting the reported system or its spare. The commands received and executed by the HK are either generated by the CPU itself or sent from the ground through the CPU. The HK is an auxiliary subsystem for tele-command distribution, monitoring status of the telescope's subsystems and alarms reporter.

The core of the HK subsystem is the master board HK-PB, to which all other portions of the HK report. Only slow data transmission rate is allowed between the HK-PB and any of its other components. Thus, for example, if an element of the FS is monitored at a particular time by a HK-CB, that same element will not be subsequently monitored for at least one or more seconds after the HK has established its status. FSÕ faults, or indeed any other subsystemÕs critical faults which require immediate shutdown in order to avoid damage, are the responsibility of the corresponding subsystem, which shall not wait for a reaction from the HK. The philosophy of monitoring and troubleshooting established for monitoring the FS will be by polling, in order to minimize power consumption by turning on sensors only when required.

An exception to the previous rule is the interaction with the power distribution subsystem. In this case, due to the critical nature of the subsystem, fault detection will be implemented by logical interruptions, which will allow fast detection and addressing of DC/DC converter failures and the consequent off-lining of the power system component in trouble. Concomitantly, HK will generate an alarm to the CPU, who is responsible for reporting operating conditions to the ground and/or deciding on an appropriate action to deal with the condition that triggered the alarm.In order to improve the offline reconstruction of the acquired scientific data, the HK will monitor the temperature of the lenses during observation. The HK will periodically activate and deactivate the conditioning and digitization board that interfaces the HK with the output signals from 15 analog temperature sensors, which will be located at five different points on each of the three lenses of the optical system. It is estimated that cycles of measurements will take place around a dozen times per observation run.HK also includes a General Subsystems I/F Board (HK-GSB) to interact with atmospheric monitoring subsystems, and mechanical subsystems of the telescope. The HK interaction will be limited to turn on and off subsystems, to monitor any physical parameters necessary to determine whether the subsystem is within an acceptable working range and to relay commands received from the CPU.

\clearpage


  \hspace{4cm}
 \newpage

\newpage
\normalsize
\setcounter{section}{0}
\setcounter{figure}{0}
\setcounter{table}{0}
\setcounter{equation}{0}



\title{Data Acquisition System of the JEM-EUSO project}

\shorttitle{casolino \etal DAQ of JEM-EUSO}

\authors{M. Casolino$^{1,2}$, T. Ebisuzaki$^{1}$, on behalf of the JEM-EUSO collaboration}
\afiliations{$^1${RIKEN Advanced Science Institute, Wako 351-0198, Japan} \\
$^2$ {INFN, Structure of Rome Tor Vergata, Via Della Ricerca Scientifica 1, 00133, Rome, Italy}\\
}
\email{marco.casolino@riken.jp}

\abstract{
The Extreme Universe Space Observatory on the Japanese Experiment Module (JEM-EUSO) of the International Space Station (ISS) is the first mission that will study  from space  Ultra High-Energy Cosmic Rays (UHECR). JEM-EUSO will observe  Extensive Air Showers (EAS) produced by UHECRs traversing the Earth's atmosphere from above. For each event, the detector will make accurate measurements of the energy, arrival direction and nature of the primary particle using a target volume far greater than what is achievable from ground. The corresponding increase in statistics will help to clarify the origin and sources of UHECRs as well as the environment traversed during production and propagation. Possibly this  will bring new light onto particle physics mechanisms operating at energies well beyond those achievable by man-made accelerators.
The spectrum of scientific goals of the JEM-EUSO mission includes as exploratory objectives the detection of high-energy gamma rays and neutrinos, the study of cosmic magnetic fields, and tests of relativity and quantum gravity effects at extreme energies. In parallel JEM-EUSO will systematically perform  observation of the  surface of the Earth in  the infra-red and ultra-violet ranges, studying also atmospheric phenomena (Transient Luminous Effects). 
The apparatus is a 2 ton detector using Fresnel-based optics to focus the UV-light from EAS on a focal surface  composed of about 6,000 multianode photomultipliers for a total of
  $\simeq 3\cdot 10^5$ channels. A multi-layer parallel  architecture has been devised to
   handle the data flow and select valid triggers, reducing it to a rate compatible with downlink constraints.
 Each processing level filters the event with increasingly complex algorithms using ASICs, FPGAs and  DSPs in this order
 to  reject spurious triggers and reduce the data rate.

}

\maketitle

\section{Introduction}

JEM-EUSO  is a   Fresnel-optics refractive telescope devoted to the observation of  Ultra High Energy Cosmic Rays  showers in the Earth's atmosphere (\cite{Kajino2010422}).
 This remote-sensing instrument will orbit the
Earth every $\simeq $  90 minutes on board of the International
Space Station (ISS) at an altitude of $~$ 330 - 400 km.
Its goal is the study of the sources of UHECR  and the determination of the origin and nature of these particles with high precision, thanks to the increase in statistics due to the larger exposure. The observation principle is the detection of  fluorescence light emitted by particles showering in the atmosphere. 

The  scientific goals of JEM-EUSO are described in detail by \cite{2009arXiv0909.3766M,2009NJPh...11f5009T,Ebisuzaki2008237} and elsewhere at this conference.

\begin{figure}
\begin{center}
\includegraphics[width=8.3cm]{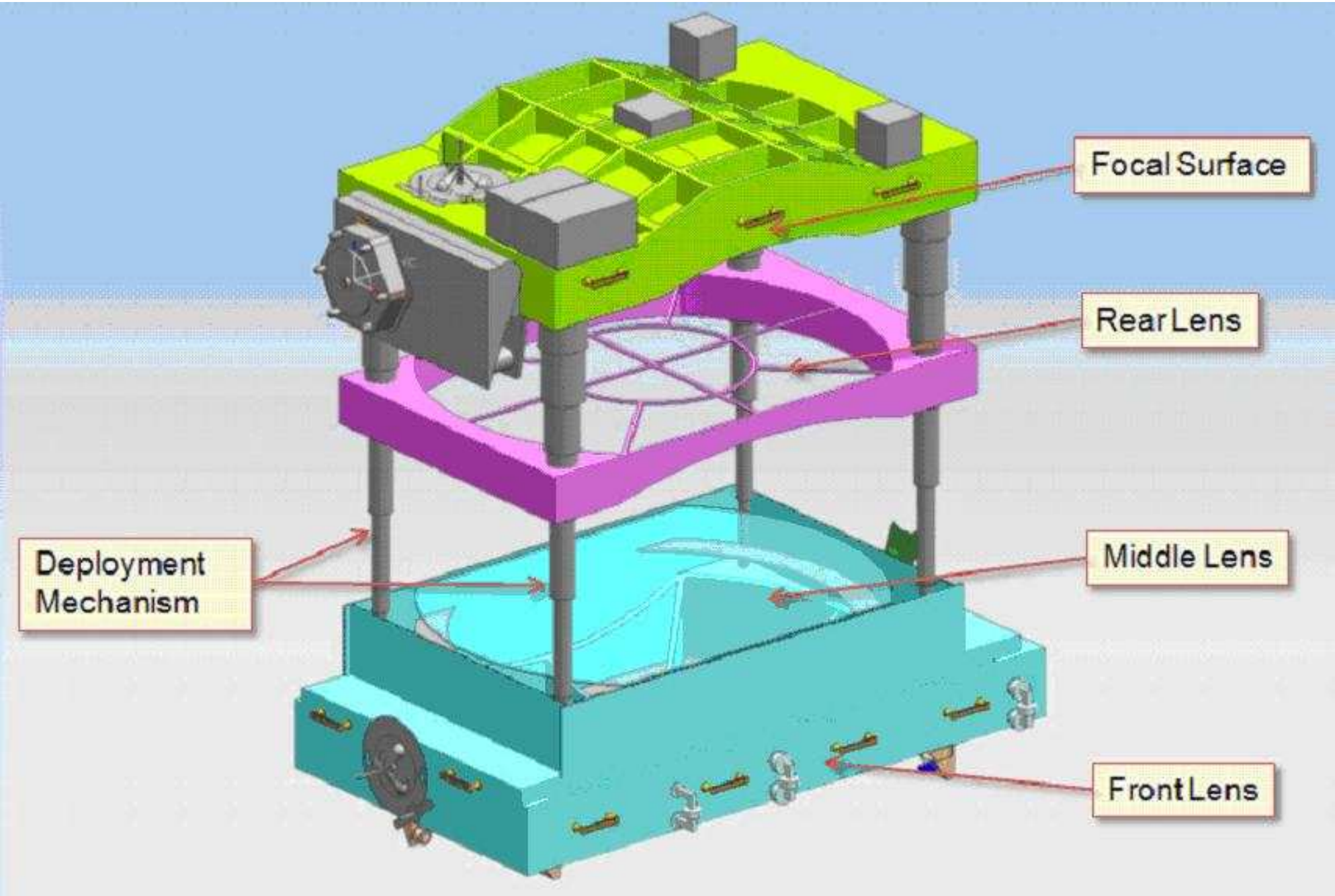}
 \caption{CAD model of the JEM-EUSO  structure. The detector is attached to the ISS through the focal surface. The front lens (on bottom of the picture) looks toward Earth.(Courtesy of IHI)} \label{zzfig4}
\end{center}
\end{figure}

\section{Detector characteristics and principle of observation}

JEM-EUSO (see Fig. \ref{zzfig4}) will observe  from space  the Earth's night atmosphere. It will measure  the UV (300-400 nm) fluorescence tracks and the Cherenkov reflected signal of the Extensive Air Shower induced by UHECR  interaction in the atmosphere.
 JEM-EUSO captures and reconstructs the temporal and
spatial evolution  of  the track through the fluorescent UV photon
component of the EAS in the atmosphere.
The light is focused through a Fresnel lens diffractive optics with  a  wide field-of-view ($\pm 30^{o} $). The light is detected by the focal plane electronics  which records the track of the EAS
with a time resolution of 2.5 $\mu $s and a spatial resolution of
about 0.75 km (corresponding to $0.1^o$). These
time-segmented images allow to determine the energies and
directions of the primary particles. 

From the UV profile the instrument can reconstruct
the incoming direction of the extreme energy particles with
accuracy better than several degrees.  The instantaneous geometrical area is (in nadir pointing mode) a circle of 500 km diameter, which converts to an instantaneous aperture of $6\cdot 10^5\: km^2\: sr$. The atmospheric mass monitored, assuming the 60-degree field-of-view, is about $1.7 \cdot 10^{12}$ ton. The target mass for upward neutrino detection is $~ 5 \cdot 10^{12}$ ton.
	 A  particle of  $E\simeq 10^{20}$ eV particle penetrating the Earth's atmosphere has an interaction length  of  $~ 40$ g/cm$^{2}$  and   generates a shower of secondary particles. The number of these secondary particles ($N\simeq 10^{11}$) is proportional to the shower maximum and is largely dominated by electrons/positrons. The total energy carried by the charged secondary particles is converted into fluorescence photons through the excitation of the air nitrogen molecules. The fluorescence light is isotropic and proportional to the number of charged particles in the EAS.

The instrument is designed to reconstruct the incoming direction of the ultra high-energy particles with an accuracy better than a few degrees.

 The size of the instantaneous geometrical area  depends on the tilt of the telescope, the angle between the telescope axis and nadir. The increase of  geometrical area from the nadir mode to the tilted mode is a factor of 2 - 5 and depends on the energy of the events.

The depth of maximum development of a shower ($X_{max}$, expressed in g/cm$^2$) increases with energy. For a given energy, the value of $X_{max}$ provides information on the nature of the primary particle. The JEM-EUSO objective is to reach a $X_{max}$ resolution of $\simeq 120$ g/cm$^2$, which is comparable to the differences in $X_{max}$ between showers initiated by protons and by Fe nuclei, making a distinction between protons and Fe nuclei with this kind of experiments possible.  
\section{Electronics}

\begin{figure}
\begin{center}
\includegraphics[width=8.3cm]{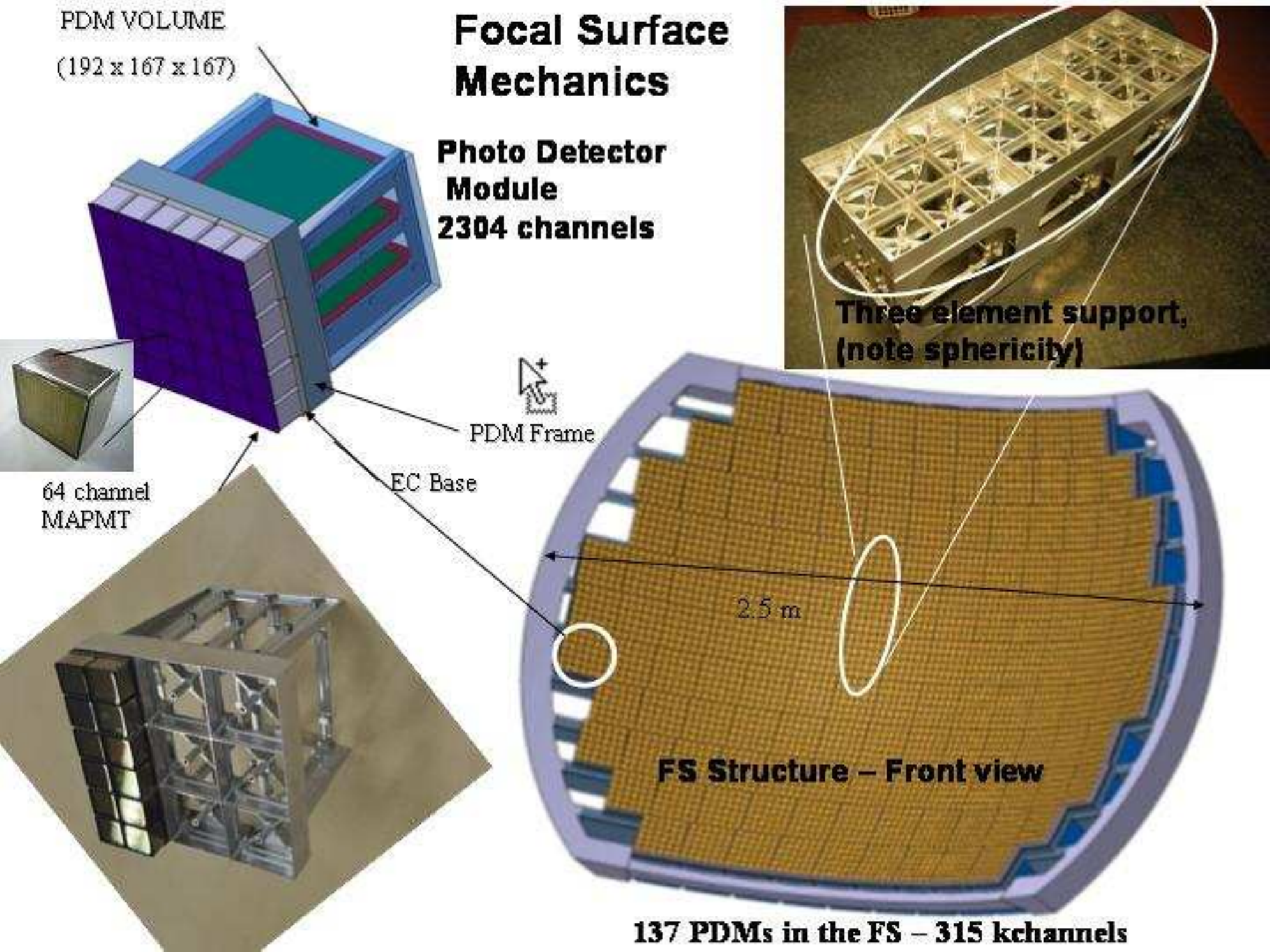}
 \caption{Bottom Right: Mechanical structure of the focal surface. The 2.5 m plane is divided in 137 PDM modules. Each PDM (Top Left) contains 36 Multi-Anode Photomultipliers (Hamamatsu Ultra-Bialkali R11265-64), each with 64 independent channels. The bottom left corner shows the prototype of the mechanical structure with two rows of 12 PMT installed. In the Top Right corner a sub-element of support beams containing three PDM is shown. } \label{zzfig5}
\end{center}
\end{figure}

\begin{figure}
\begin{center}
\includegraphics[width=8.3cm]{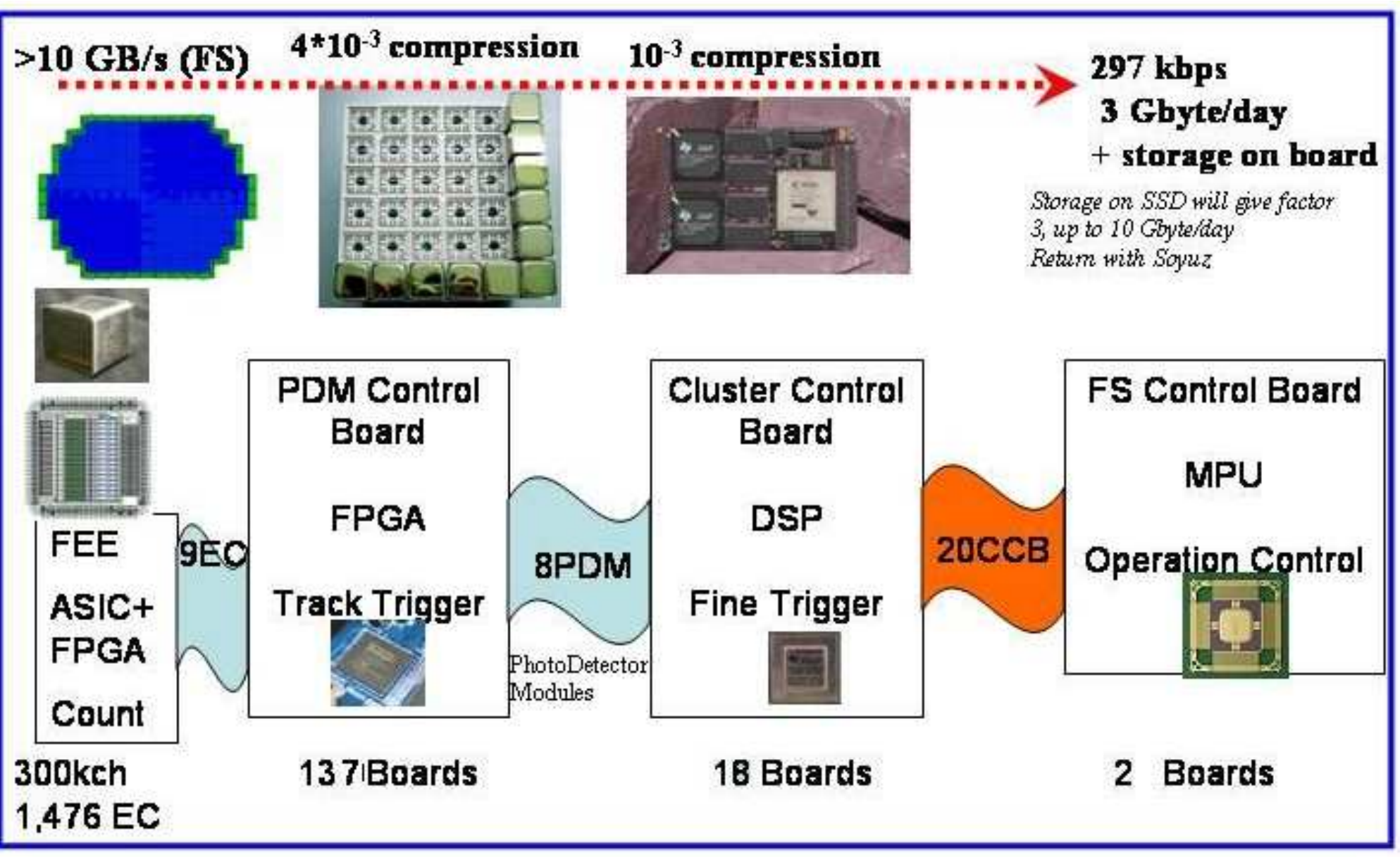}
 \caption{Data Reduction scheme. Each of the $\simeq $ 6000 Multi-Anode Photo Multiplier (MAPMT) of the focal surface is read by an ASIC digitizing the photoelectron signal.  A 6*6 array of MAPMT is present and read by each of the 137 PDM modules, where an FPGA performs first level triggering and rejects noise by three orders of magnitude.  8 PDMs are read by a  Cluster Control Board, each with an high performance DSP which rejects noise by other three orders of magnitude. The general acquisition and data storage is performed by   the main CPU (right). } \label{zzfig6}
\end{center}
\end{figure}

The data acquisition and handling system is designed to
maximize detector observation capabilities to meet the
various scientific goals, monitor system status, autonomously
taking all actions to maintain optimal acquisition capabilities
and handle off-nominal situations. CPU and electronics are based
on hardware successfully employed in space experiments such as
PAMELA, Altea, Sileye-3, etc., taking into account recent
  developments in microprocessors and FPGA technology.
Acquisition techniques and algorithms are also derived from the
technological development performed in these missions. Rad-hard technology will
be employed, with ground beam tests at accelerator facilities such as GSI, Dubna, HIMAC 
 to qualify and test resistance of new devices. Space qualified
devices will be employed for mission-critical items.  

The general approach is   to   use   off-the-shelf technologies in
the development of the laboratory models and breadboard systems
  to refine and test the various trigger and data reduction
algorithms in parallel to hardware development and construction.
The same approach will be followed in the use of communication
protocols and interfaces (e.g. VHDL, spacewire, 1553, 1355
protocols) and in the realization of the ground support equipment.
This will allow for a fast development of the software in parallel
to the engineering and production of flight boards, reducing costs and
integration time.

Hot/Cold redundancy will be implemented in all systems and in all
stages of data processing with the exception of intrinsically
redundant devices such as the focal surface detectors.

\subsection{Data Acquisition and Reduction}

The Data acquisition system  (Fig. \ref{zzfig6}) is based on an
architecture capable of reducing  at each level the amount of data
through a series of triggers controlling an increasingly growing
area of the focal surface (\cite{Casolino2010516}). It is necessary to reduce the 10 Gbyte/s output
on the focal surface (FS) to the   3 Gbyte/day  which can be
downlinked on the ground. Each board and data exchange protocol is
compliant to handle the data and send them to the higher level of processing if
they satisfy the trigger conditions. 

An ASIC chip performs  photo-electron signal readout and conversion for the 64 channels of the MAPMT. An FPGA
handles   first level trigger   data on a PDM level (reading 36 MAPMTs). The data are stored in a 100 GTU buffer (each GTU corrsponds to a 2.5 $\mu $s frame, for a total sampling of 250 $\mu $s) upon which the triggering and noise reduction algorithms are implemented. Background  events are rejected by a factor $10^3$. Second level triggering algorithms are implemented by the 18 CCB (Cluster
Control Boards), DSPs with about 1Gflop computing capability which
further process triggers coming from   8 PDMs.  At this level background is rejected by another factor $10^3$.
The CPU has a
relatively low processing power (100 MHz) since it  is charged of
the general handling of the experiment. The CPU is part of the Storage and Control Unit System (SCU), the evolution of a similar system used for PAMELA (\cite{cpu}) and   composed of a number of
boards devoted to different tasks: 1.  CPU mainboard 2.  Mass
Memory (8 Gbyte) 3.  Internal and external kousekeeping interfaces (CAN bus)
4.  Interfaces to ISS (1553 and Ethernet) 5. Fast bus interface for
event acquisition.
 The CPU  is devoted to 
the control of the apparatus and the general optimization of the
performance of the instrument in terms of data budget and detector
status. It is expected to function autonomously and to reconfigure
the working parameters with little or no intervention from the
ground. It is capable of  handling alarms and contingencies in real time
minimizing possible damage to the instrument. Long term mission
operation and observation planning will be
implemented from the ground with specific telecommands used to
overrule the specific operation parameters of the instrument. By
sending immediate or time-delayed telecommands it will be possible
to define the various operation parameters of the instrument in
terms of specific physics objectives or specific situations.

The main CPU tasks are: 1)   Power on/off of
all subsystems. 2)  Perform periodic calibrations. 3) Start
acquisition / Run.  4)  Define Trigger mode acquisition.
5) Read Housekeeping. 6)  Take care of real time contingency
planning. 7) Perform periodic Download / Downlink. 8)
Handle (slow control) 1553 commands.

\subsection{Housekeeping module}
  The housekeeping module is connected to the CPU
with the task to distribute commands to the various detectors  and to collect
telemetry for them in order to monitor in real-time the status of the experiment and optimize its 
observational parameters.

There are two modules, one internal to the CPU (I-HK), devoted to monitor critical
systems, power on/off of secondary power supply etc. I-HK is
turned on together with the CPU and enables power on to all
subsystems. The external housekeeping board (E-HK)  is devoted to the general slow
control and monitoring of the status of the apparatus.

I-HK functional module capable of handling both single (upon
request) or cyclic (periodic) acquisition/commanding operating
both is possible according to the   acquisition program and
status. Different acquisitions and controls are foreseen. For
instance all relays to switch on / off secondary power supplies and
subsystems are controlled by High Level signals. This approach has
the advantage of a great degree of flexibility keeping at the same
time a strong robustness and reliability.

Some of the main   electrical interfaces monitored by the module
are: 1.  Voltage monitor (Primary - 120V 28V; Secondary: +-5V +12V,
 +3.3V -700 V 2. Current monitor 3. Temperature monitor 4. Contact
closure (Lid status, relays) 5. Digital Communication Protocol.

\subsection{Communication Protocol}

Communication between different layers of the data acquisition chain operate with LVDS (differential signal) to
minimize interference and reduce power consumption. All lines are
redundant, with each line employing double connectors at each
end to increase reliability of the system and resistance to
vibrations and thermal stresses. High level communication protocol
between CCB and CPU is based on a simplified version of the SpaceWire.

\subsection{Commands from the ground}

Slow control communication from/to ground is based on
the MIL-STD-1553B standard. 1553 is a slow speed (1Mbit/s) bus used
in space and aeronautics for transmission / reception of critical
information. In JEM-EUSO the 1553 bus is employed to:
 \begin{enumerate}
 \item Switch on/off the instrument or part of its sections.
\item
Issue telecommands from the ground.
\item  Set general acquisition parameters based on
detector status. Furthermore they can be used to patch (reprogram)
part of the software at CPU, DSP or FPGA levels and dump the
memory of each level in case of debugging.
\item  Reception of
keep-alive information from the detector, of nominal events,
alarms.
\item Switch from main to spare channel (acquisition,
power supply, etc.).
\end{enumerate}

\subsection{Storage, downlink, download}

Data stored in the mass memory of JEM-EUSO are periodically sent to ISS via a
  high speed link based on Ethernet protocol. Data are subsequently downlinked  to the
ground via TDRSS satellite  link or stored on hard disks. Data
transmitted to the ground consists of: 1. Cosmic ray data from the
focal surface 2. Atmospheric Luminous Phenomena, lightning etc... 3. Housekeeping
information 4. Alarm 5. Calibration data 6. Ancillary information.

 Data are sent to the ground with
highest priority given to housekeeping and alarm information.
Experimental data are sent to ground with main priority to high
energy particle data and special trigger (e.g. Transient Luminous Phenomena,
meteoroids, lightning, etc... ).  The amount of data downlinked to the Earth is $\simeq 3 Gbyte/day$, amounting to about $20\%$ of the data budget.  The rest of the data is stored on board ISS on a dedicated
disk server. Disks are then periodically returned to the ground with
Soyuz capsules. In the current configuration, it is expected to have $\simeq 5 TByte/6\: months$ sent on the ground. Even though the  UHECR event rate is very low, the
background occupies a large part of the data. This is especially
true at low energies, where shower development is shorter and more
difficult to sort with on board algorithms. A higher memory
capability allows to increase the trigger efficiency at low energies (around) $
3-4\cdot 10^{19}$ eV and improve the data bandwidth devoted to atmospheric physics (IR and UV channels).

\section{Acknowledgements}
This paper is dedicated to  the memory of Y. Takahashi. 

This work has been partially supported by the Italian Ministry of Foreign
Affairs, General Direction for the Cultural Promotion and Cooperation.

\clearpage


  \includepdf[pages=-]{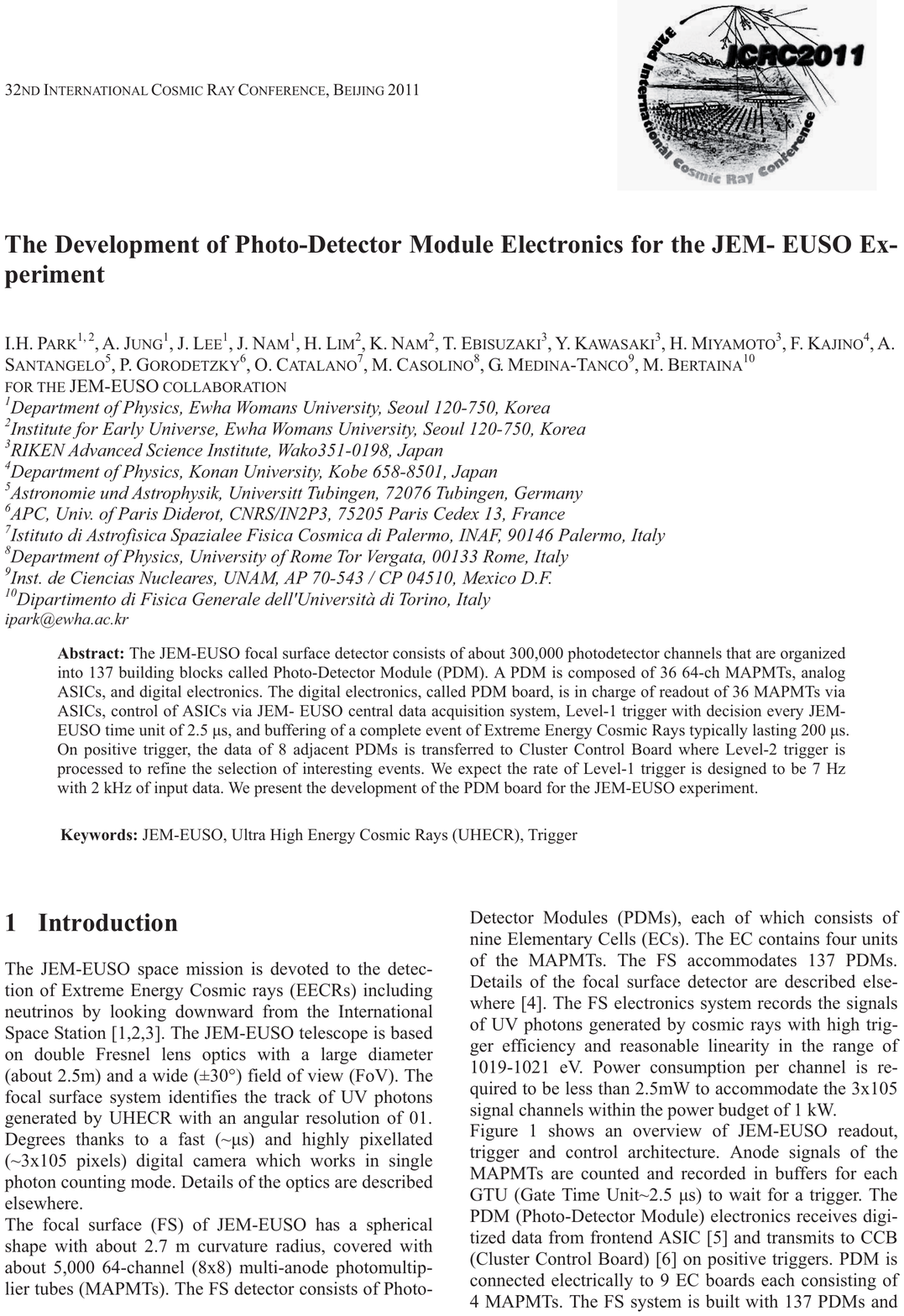}

  \includepdf[pages=-]{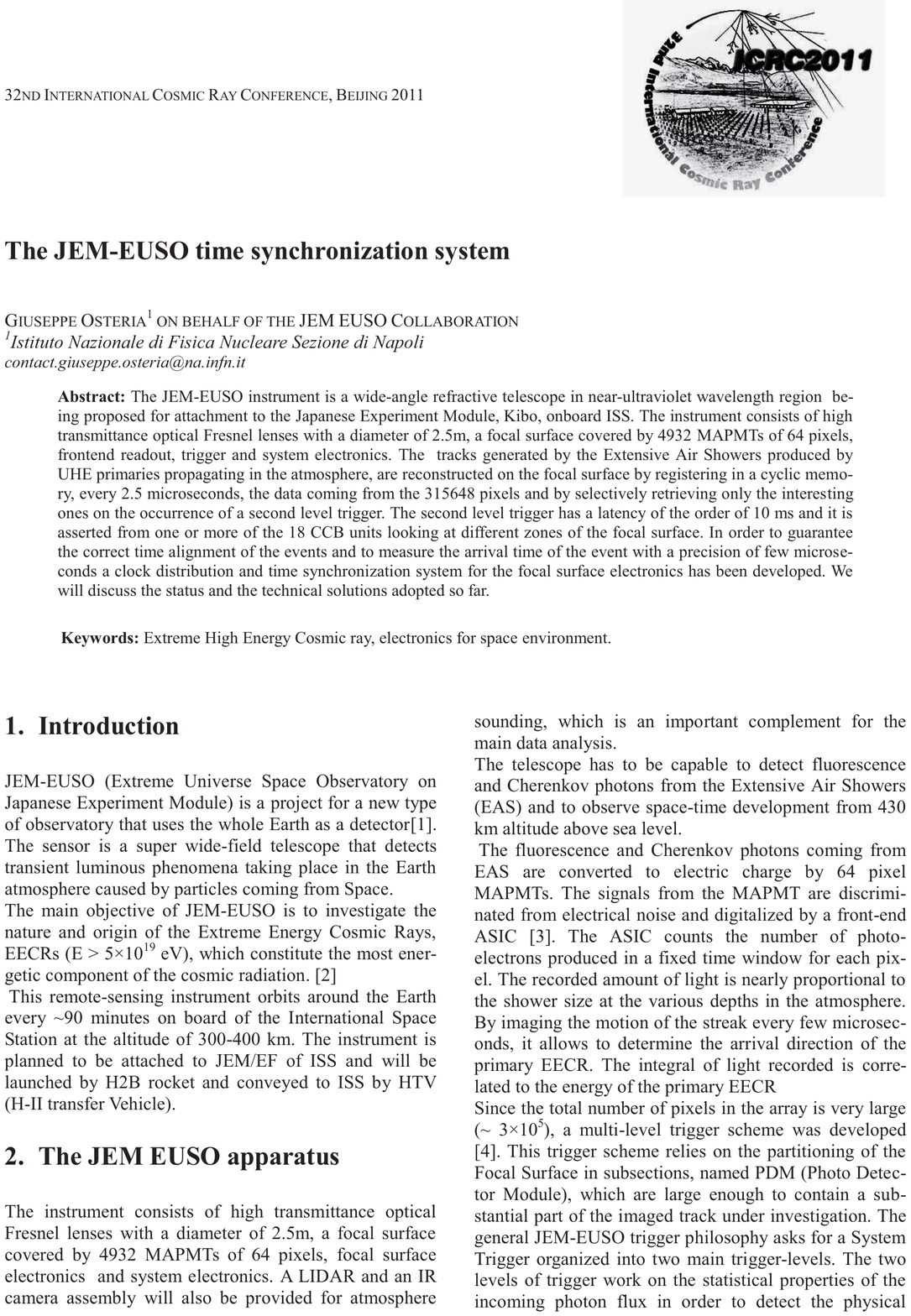}

  \includepdf[pages=-]{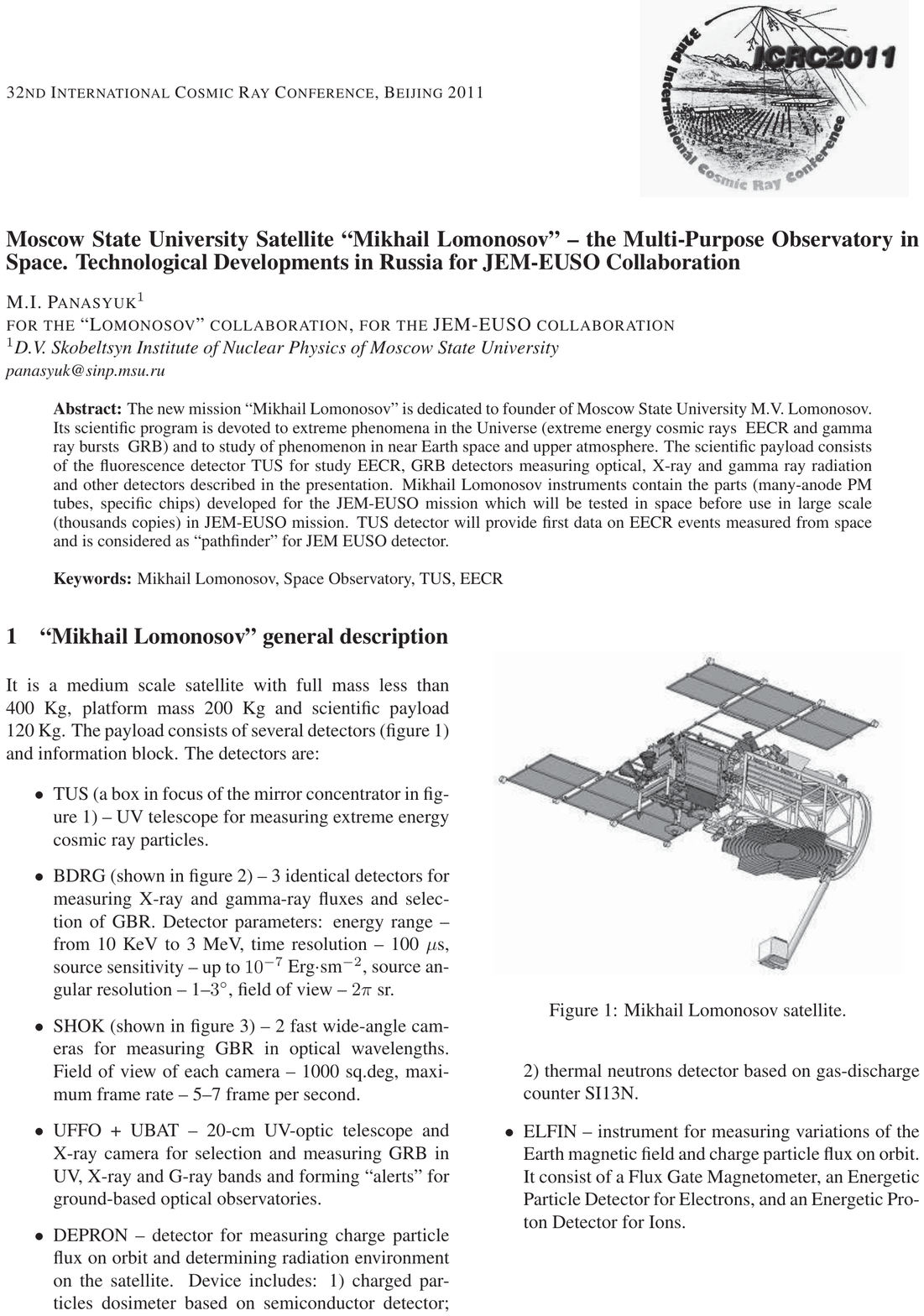}

\newpage
\normalsize
\setcounter{section}{0}
\setcounter{figure}{0}
\setcounter{table}{0}
\setcounter{equation}{0}



\title{Calibration of JEM-EUSO photodetectors}

\shorttitle{P.Gorodetzky \etal Calibration of JEM-EUSO photodetectors}

\authors{P. Gorodetzky$^{1}$, N. Sakaki$^{2}$, M. Christl$^{3}$ for the JEM-EUSO collaboration}
\afiliations{%
 $^1$Laboratoire AstroParticule et Cosmologie, APC, Paris, France\\
 $^2$Department of Physics and Mathematics, Aoyama Gakuin University, Sagamihara, Kanagawa 252-5258, Japan\\
 $^3$NASA Marshall Space Flight Center, Huntsville, Alabama 35812 USA}
\email{Philippe.Gorodetzky@cern.ch}

\abstract{The calibration of the JEM-EUSO focal surface (FS) about 5000
photomultipliers (PMTs) we will made in two steps: a) on earth, PMTs 
will be sorted out according to their gain and efficiency. 
Then identical PMTs will be used to make the Photo Detector Modules (PDM)
 consisting of 36 PMTs (2304 pixels). Immediately after the PDM assembly, 
the gain and absolute efficiency of each pixel will be measured
 with the PDMs own front end electronics working in single pho-electron mode.
 A X-Y-Z-$\theta$-$\phi$ (PDM has a spherical shape) movement will support
 the illuminating device consisting of UV LEDs with wavelength of 330-430~nm
 inside an integrating sphere
 whose exit port will feed a collimator with 0.3 mm holes.
 The light will be monitored with a NIST photodiode mounted on a third port
 of the sphere. This set-up will be calibrated by replacing the PMT
 with another NIST.
b) in space, during the day, when the JEM-EUSO lid is closed, 
the focal surface  will be illuminated in a uniform way by a set of
 1 inch spheres set on the periphery of the last lens 
(at one meter from the FS). The spheres will be equipped with LEDs and
 monitored by NIST photodiodes. Another set of identical spheres
 will be put at the FS periphery and will illuminate the lenses. The light will
 bounce back on the lid covered with diffusive reflector, to reach the FS which
 has been previously calibrated.
Other means of in flight calibration will use ground sources (Xenon flashers)
 and finally the moon light reflected by earth albedo.
}

\maketitle

\section{Introduction}

JEM-EUSO is a mission to observe ultra high energy cosmic rays (UHECRs)
above
$10^{20}$~eV\cite{sjemeuso1,sjemeuso2}. The JEM-EUSO telescope will be
attached to the International Space Station (ISS) and will detect
fluorescence photons from extensive air showers (EASs) induced by 
UHECRs.
It is necessary to understand the instrument very well
to discuss their origins. The JEM-EUSO instrument consists of Fresnel
lens optics with a diffractive lens and the focal surface detector with
photon counting
capability\cite{sjemeuso-instr1,sjemeuso-instr2}.

The number of observed photo-electrons ($\Delta S$) from a luminous
phenomena (with emitting number of photons of $\Delta Q$)
at distance $r$ is expressed as:

\begin{equation}
\Delta S = \frac{\epsilon \kappa \eta T_l T_f T_e T_\alpha A}{4\pi
 r^2}\Delta Q,
\end{equation}
where,

\begin{tabular}{ll}
 $\epsilon$ & quantum efficiency of the detector\\
 $\eta$ & collection efficiency of the detector\\
 $\kappa$ & the probability to be contained in a pixel\\
 $T_l$ & throughput of the Fresnel lens system\\
 $T_f$ & transmission of the optical filter\\
 $T_e$ & trigger efficiency of the electronics\\
\end{tabular}

\begin{tabular}{ll}
 $T_\alpha$ & atmospheric transmission\\
 $A$ & aperture of the telescope.\\
\end{tabular}
\\

The terms related to the instrument among above are
$\epsilon, \eta, \kappa, T_f, T_e$ and $ T_l$.
Here, let's consider the calibration of the instrument with the
following four cases.

\begin{itemize}
 \item Pre-flight calibration
 \item On-board calibration
 \item In-flight Calibration with ground light sources
 \item Atmospheric monitor
\end{itemize}

For the atmospheric monitor, a dedicated subsystem is organized\cite{sjemeuso-ams}.
A ultra violet(UV) LASER and an infrared camera are under preparation
to measure the cloud coverage in the field of view and the cloud heights.
The other three calibrations are included in the ``calibration system''
and are being prepared by the collaboration of Japan, France, United States,
 Italy and Mexico.

The energy of $10^{20}$~eV has not been reached until now artificially,
so that it is difficult to determine the absolute energy scale directly
by any calibration. However, JEM-EUSO is expected to detect more than 1000
UHECRs in the mission period. If there are sources in our vicinity,
the energy spectrum for each source will be obtained. Since ultra high energy
cosmic rays interact with cosmic microwave background photons and lose
energy, the spectrum will be suppressed above $\sim 4\times10^{19}$~eV
(GZK suppression). If the suppression threshold energy is obtained as a
function of distance, the absolute scale might be determined.

\section{Pre-flight calibration}
\subsection{Outline}
The JEM-EUSO optics consists of two Fresnel lenses, a diffractive lens
and a focal surface\cite{sjemeuso-instr1,sjemeuso-instr2,sjemeuso-optics}. 
The focal surface is covered with about 5000 pieces
of 1'' multi-anode photomultiplier tubes (MAPMTs) developed for
JEM-EUSO\cite{sjemeuso-fs1,sjemeuso-fs2}.
The focal surface
detectors are grouped, to make it easy for manufacturing and the data
handling. A module of four PMTs is called ``Elementary Cell'' (EC)
and a photo-detector module (PDM) consists of 9 ECs. Each PDM has a
capability to detect EASs induced by cosmic rays
by itself. The stand-alone performance and the functions of optics, 
electronics, etc. will be checked in each dedicated subsystem in
principle. The calibration
subsystem is in charge to measure the efficiency of the
focal surface, especially efficiency of the MAPMTs.

\begin{figure*}[tbh!]
\begin{center}
 \includegraphics[width=0.9\textwidth]{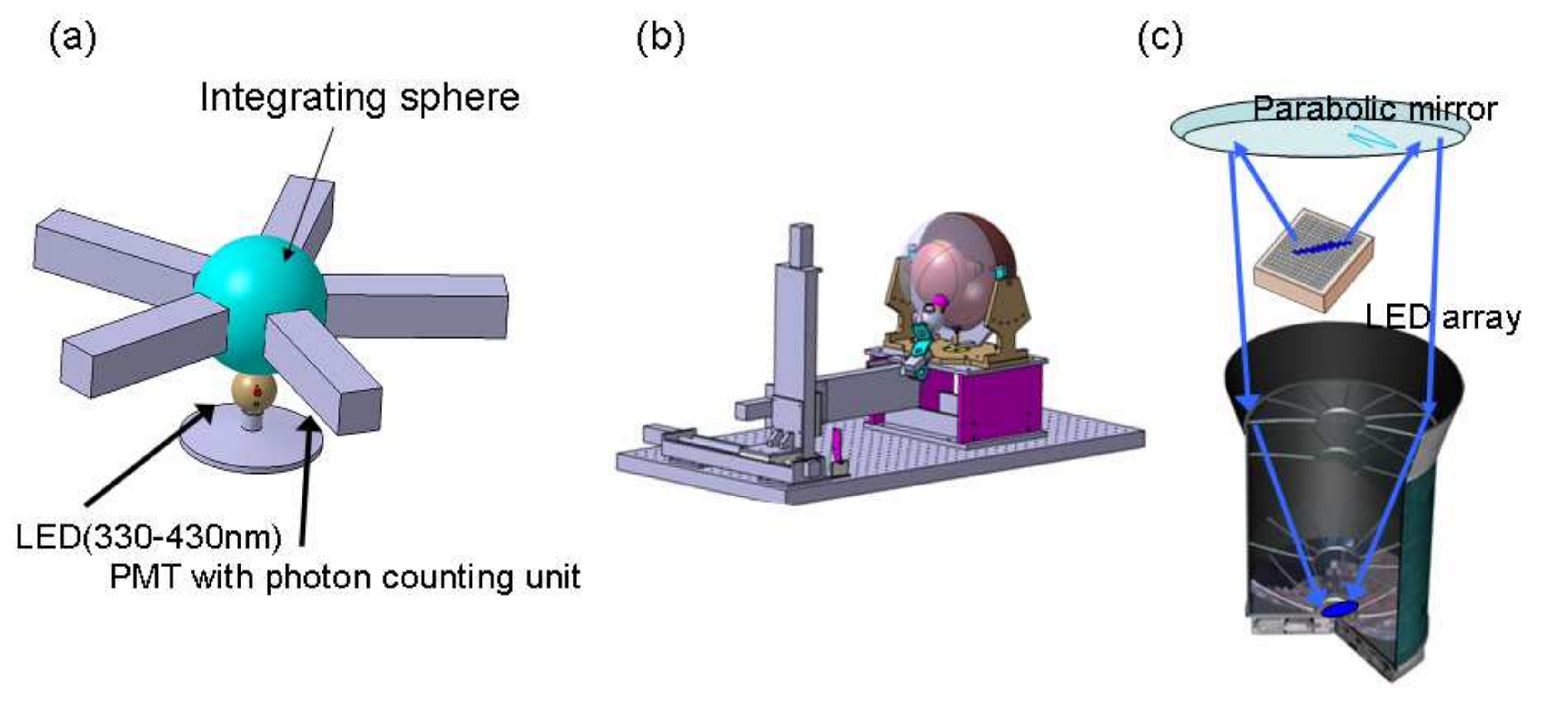}
\end{center}
\caption{Apparatus for the pre-flight calibration. (a) Uniform light made
 with an integrating sphere illuminate ten MAPMTs. Single
 photo-electron spectrum will be taken with photon counting and MAPMTs
 will be sorted according to the gain and the efficiency. (b) A light
 source made of an integrating sphere is mounted on a stage to measure
 the PDM efficiency at various positions and with various incident
 angles. (c) An emulate air shower image with a LED array will be seen
 with the JEM-EUSO instrument to check total performance.}
\label{fig:preflightcalib}
\end{figure*}

\subsection{Calibration of MAPMTs}
The gain of a MAPMT will be measured at various voltage to determine
an appropriate voltage for the input to the front-end electronics. As shown in
 Figure~\ref{fig:preflightcalib}(a), LED light will be diffused with an
 integrating sphere to illuminate the photo-cathodes of 5-10 MAPMTs.
Three kinds of wavelength will be used in the range of 330~nm and
430~nm. Pulse height distribution will be taken with photon-counting
method with the electronics developed for night glow measurement
by the EUSO Italy group.
All the PMTs will be sorted out by their gain and efficiency,
and every four PMTs with similar characteristics will be packed as an EC.

\subsection{Calibration of PDMs}
In order to measure the position and angular dependences of
the photon detection
efficiency of PDMs, an integrating sphere with UV LED at 375~nm and a
monitor photo-diode is mounted on a XYZ$\theta\phi$ stage and the PDM
surface will be scanned with 1~mm step with various incident
angles (Figure~\ref{fig:preflightcalib}(b)). The light is collimated to 1~mm in
diameter. The photo-diode is calibrated precisely by the manufacturer.
The efficiency of not only MAPMTs but the whole system 
(=quantum efficiency$\times$collection efficiency$\times$electronics
efficiency) can be obtained. In this measurement,
the variation of the intensity is monitored by a well-calibrated
photo-diode, and that the attenuation of light to single
photo-electron level is determined by the geometrical factor of the
collimator. Therefore, the efficiency can be obtained better than a few
percent.

Next, to check the trigger efficiency of PDM,
fluorescence image from a cosmic ray air shower will be emulated with
an array of UV LEDs. UV beam reflected by a rotating mirror is
another candidate to emulate EAS image.
The rotating speed will be adjusted to reproduce the light spot speed of
EAS on the focal surface.

\subsection{End-to-End calibration}

The total performance of the instrument will be checked at this stage.
One of the possible methods is that EAS image generated with a LED array
for the PDM calibration is projected to the entrance pupil 
by a large parabolic mirror(Figure~\ref{fig:preflightcalib}(c)).
UV LASER light reflected by a rotating
mirror is another candidate light source at this stage, too.

\section{In-flight calibration}
\subsection{Outline}
Absolute values of efficiency, gain, etc. will be measured
on ground before launch, and only the relative changes will be monitored
in flight in principle. Several light sources will be put in the JEM-EUSO
instrument to monitor the efficiency and the detector gain.
Ground light sources and the
reflection of the moonlight by the Earth will also be utilized
to make the calibration more reliable.

\begin{figure*}[bth!]
\begin{center}
\includegraphics[width=.8\textwidth]{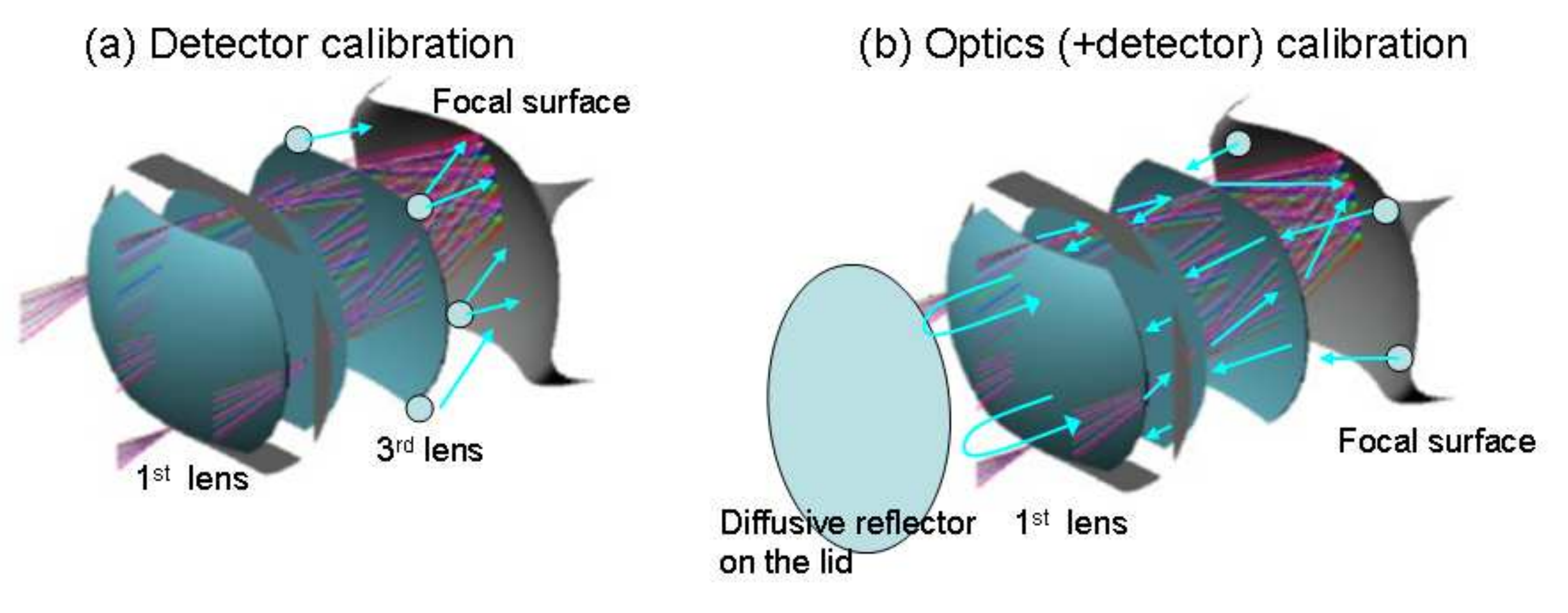} 
\end{center}
\caption{On-board calibration system. Diffused UV light sources made of
 integrating spheres will be set at the position shown in the panels (a)
 and (b), and the time variation of the efficiency of the optics and the
 detector will be monitored. (a) Several light sources will be set along
 the edge of the rear lens to illuminate the focal surface directly. The
 relative change of the detector efficiency will be taken. (b) The same
 light sources are placed along the edge of the focal surface to
 illuminate the rear lens. The light is reflected back at the diffuse
 surface on the lid and is detected by the focal surface detector. Here,
 convolution of the efficiency of the optics and that of the detector
 will be obtained.}
\label{fig:onboardcalib}
\end{figure*}

\subsection{On-board calibration}
The light source consists of a small integrating sphere equipped with UV
LED in 330-430~nm and a NIST photo-diode to monitor the variation of the
light intensity. Several identical light sources will be settled behind
 the rear lens  and illuminate the whole focal surface
(Figure~\ref{fig:onboardcalib} (a)). The intensity will be
set at single photo-electron level and the photon detection efficiency
of the system and the gain of MAPMT will be measured. If large change of
gain is found, the threshold level for the counting will be adjusted.

Other several light sources will be set along the edge of the focal
surface to illuminate the rear lens. The light passes through the lenses
and is reflected back at the diffuse surface on the lid. A certain
amount of the emitted light will be detected by the focal surface detector.
The time variation of the efficiency of the optics and the detector will
be obtained in this measurement. Therefore,
after subtracting the degradation of the detector itself,
the decrease in the optics throughput will be obtained.

\subsection{From-ground calibration}

\begin{figure}[h]
\includegraphics[width=.48\textwidth]{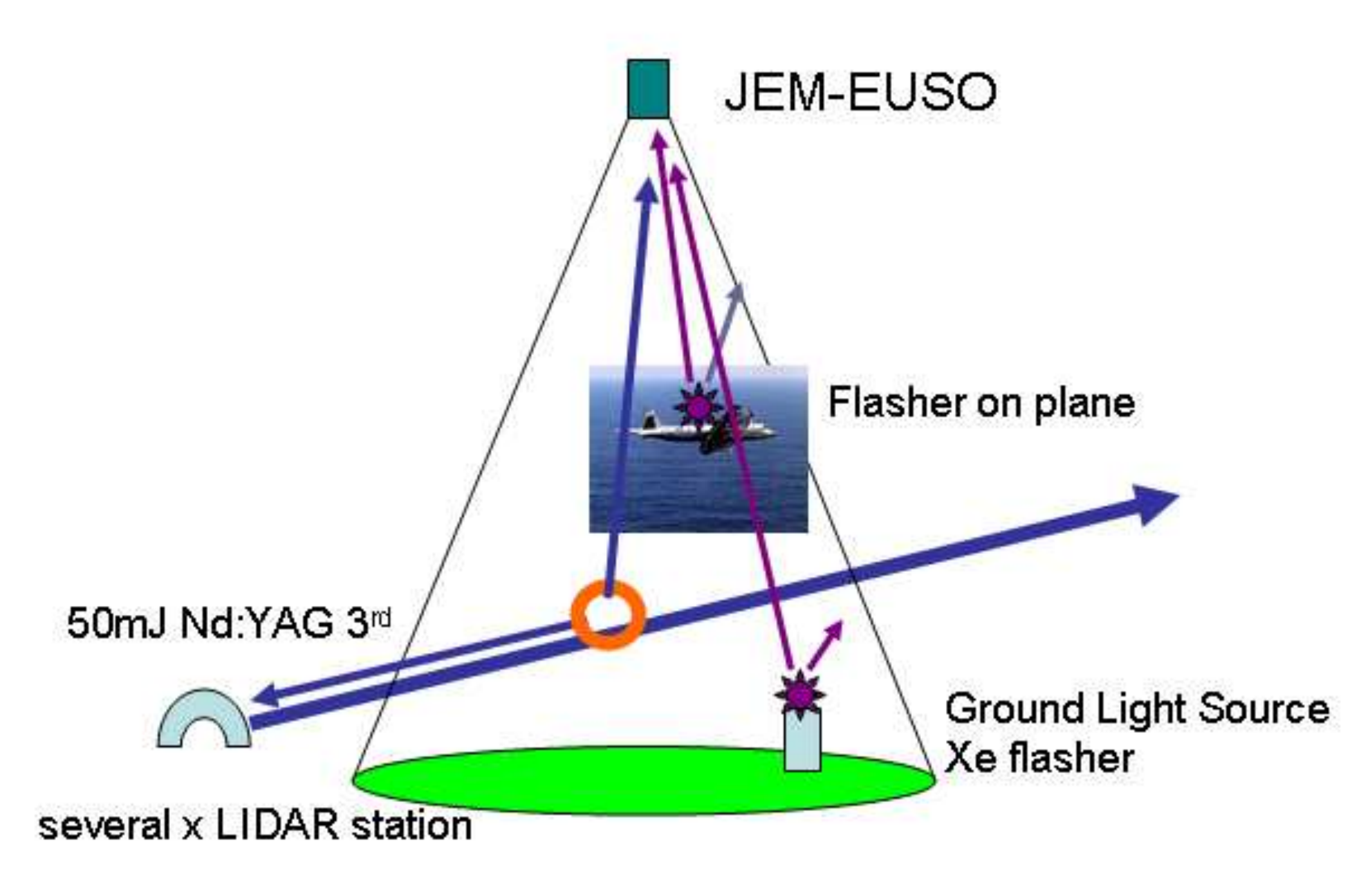}
\caption{In-flight calibration with ground light sources. JEM-EUSO will
 fly over one of the 10-20 Xe flasher stations in average every night,
 and atmospheric transmittance
 and the JEM-EUSO efficiency will be checked. Artificial extensive air
 showers generated by LIDAR will be used to study the reconstruction
 accuracy experimentally.}
\label{fig:fromgroundcalib}
\end{figure}

\subsubsection{Ground light source (Flasher)}
There will be a dozen ground-based units deployed at host stations in
different geographical locations to cover various
atmospheric conditions, and one airborne unit.
ISS will fly over one flasher in average every night, 
the lamp will be lit by remote control for the
cross-check of the JEM-EUSO photon detection efficiency, atmospheric
transmittance, focusing quality of the JEM-EUSO optics. 
The airborne unit is to be installed on an upward directed portal of a
P3B research aircraft stationed at NASA Wallops Flight Facility. It
flies under the orbit of ISS at the altitude of 1-6 km above both land
and sea every month during the JEM-EUSO mission.
The Hamamatsu flash lamp L6404 has an light intensity of 2J per
flash. The expected signal detected by JEM-EUSO is about 500
photo-electrons for clear nights. The maximum
flash-to-flash variation for this lamp is ~3\% and the spatially
non-uniformity is less than 5\% over a 60$^\circ$ field of view.
 The duration of over-flights range from 5 to 70
seconds, so that typically 100 flashes per over-flight will be observed by
JEM-EUSO. Atmospheric transmittance will be determined with a few percent
accuracy by repeating measurements.
Each ground flasher consists of four lamps with band pass filters at
337~nm, 357~nm and 391~nm, which corresponds to the wavelength of main N$_2$
fluorescence lines, and one broad band filter similar to that on
JEM-EUSO. The intensity of the lamps is monitored by a photo-diode
precisely calibrated at NIST. These lamps will be controlled by way of
the Internet and be flashed at $\sim$30 Hz when JEM-EUSO flies over the
station. The image of a ground flasher moves at $\sim$0.03$^\circ$/0.03 s. If a
light source passes through at the center of the FoV, about 2000 flashes
will be observed by JEM-EUSO.

\subsubsection{Ground LIDAR}
Since we can emulate EASs with the third harmonic of NdYAG Laser
(355~nm), ground LIDARs may be an effective tool for calibration. In
order to emulate the EAS of $3\times10^{20}$eV
 with an elevation angle of $\sim$20$^\circ$,
we need the output of 50 mJ at least. Once the power and the elevation
angle of the Laser are fixed, we can determine the size of the receiver
(~1 m in diameter). The signals back-scattered at 30 km and 60 km are about 800
photo-electrons and 20 photo-electrons in GTU (2.5 $\mu$s).
If we shoot 100 times,
for example, more than 1000 photo-electrons will be observed and the
atmospheric properties are determined well. We can measure
the transparency with an accuracy of 5-10\% after 100 shots.
As a ground light source the shot in the elevation angle of 20-30$^\circ$ is
optimum and probably the fixed directional Laser may be robust and
minimize the maintenance of the mechanical parts of the system. The
Laser can be tunable up to 10-30~Hz. In these horizontal shots, the
Laser beam reaches the top of the atmosphere after traveling ~30 km,
where the Rayleigh scattering is dominant. The beam travels in pure
molecular region for another 30 km and we can get the boundary condition
for the LIDAR equation, because we can know the ratio of the
back-scattered intensity to the beam intensity in the pure molecular
region. Then we can solve the LIDAR equation to obtain the transmittance
of the atmosphere as a function of height.
The Laser beam with an elevation angle of 20$^\circ$ can be seen as a track of
30-50 km long from JEM-EUSO. If the scattering is dominated by the
Rayleigh process, the number of photons at the entrance of JEM-EUSO can
be calculated. The scattering angles of photons that JEM-EUSO will
receive are always larger than 40$^\circ$. In such large scattering angles,
Rayleigh process usually dominates under good weather condition. We will
use photons scattered above 3 km where the scattering is better
described only by the Rayleigh process. Simultaneous operation of the
on-board LIDAR system and the ground LIDAR system gives us more detailed
information about the atmosphere and more redundant measurements. It
will also reduce the systematic error in the measurement
significantly. The systematic errors and the resolutions of arrival
direction and energy determinations by JEM-EUSO can be
evaluated experimentally by reconstructing LIDAR events.

\section{Summary}
Calibration of the instrument plays a key role to open a new era of
``Particle Astronomy'' by JEM-EUSO. In order to achieve
better than a few \% accuracy, several methods for pre-flight and
in-flight calibrations are proposed and the preparation is in progress.


\section*{Acknowledgments}
This project was partially supported by JSPS and CNRS
under the Japan-France Research Cooperative Program.


\clearpage


  \includepdf[pages=-]{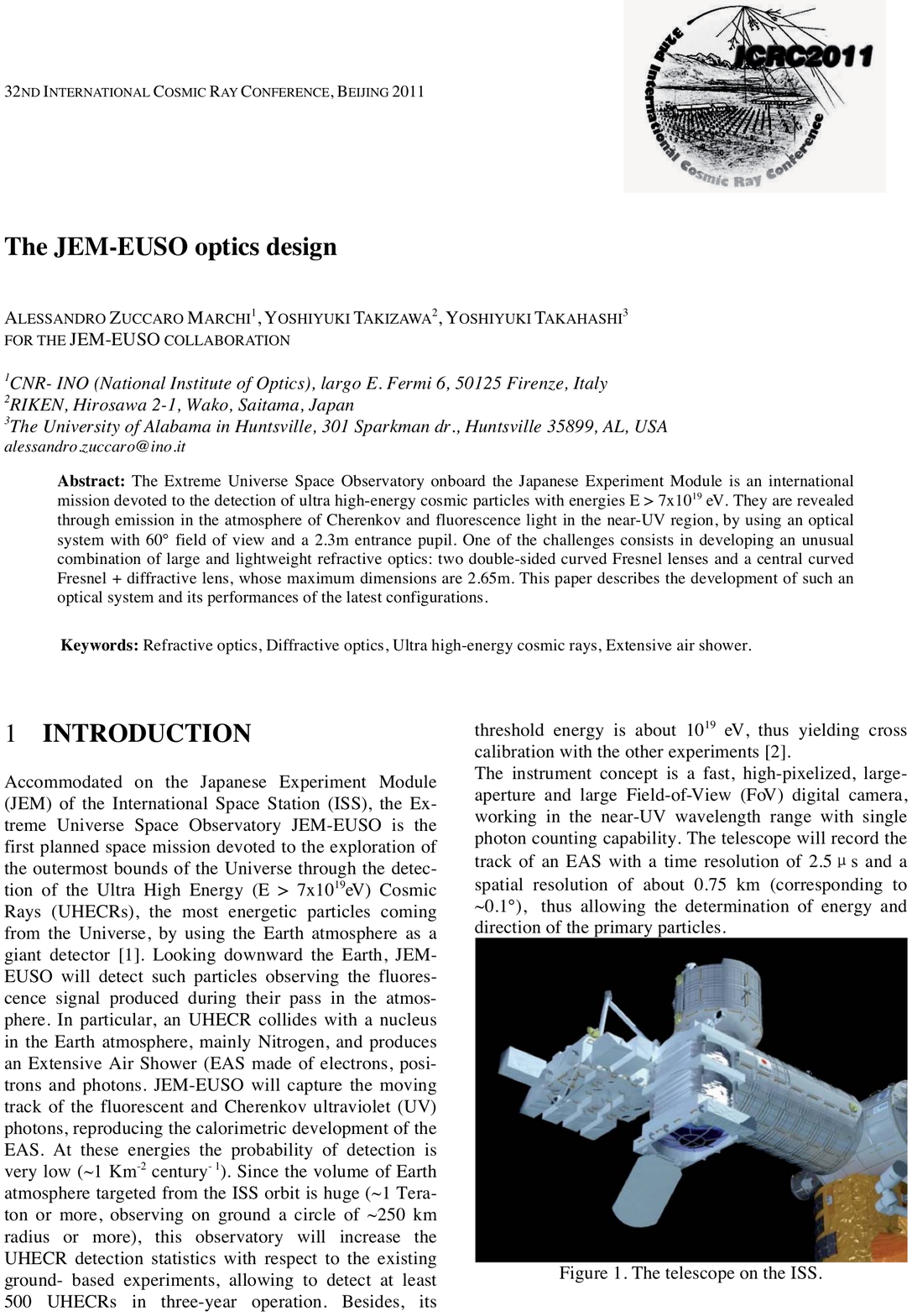}

  \includepdf[pages=-]{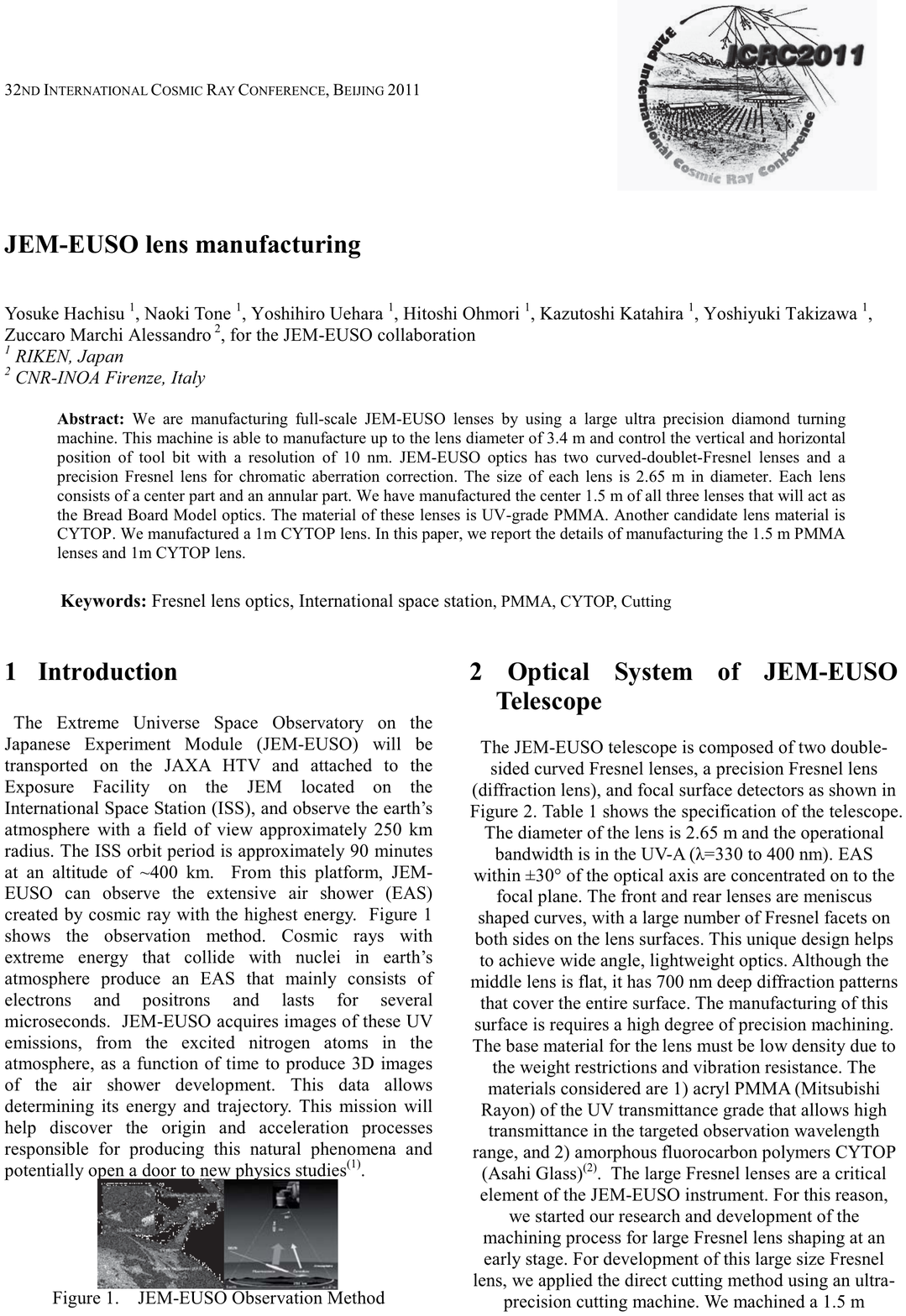}

  \includepdf[pages=-]{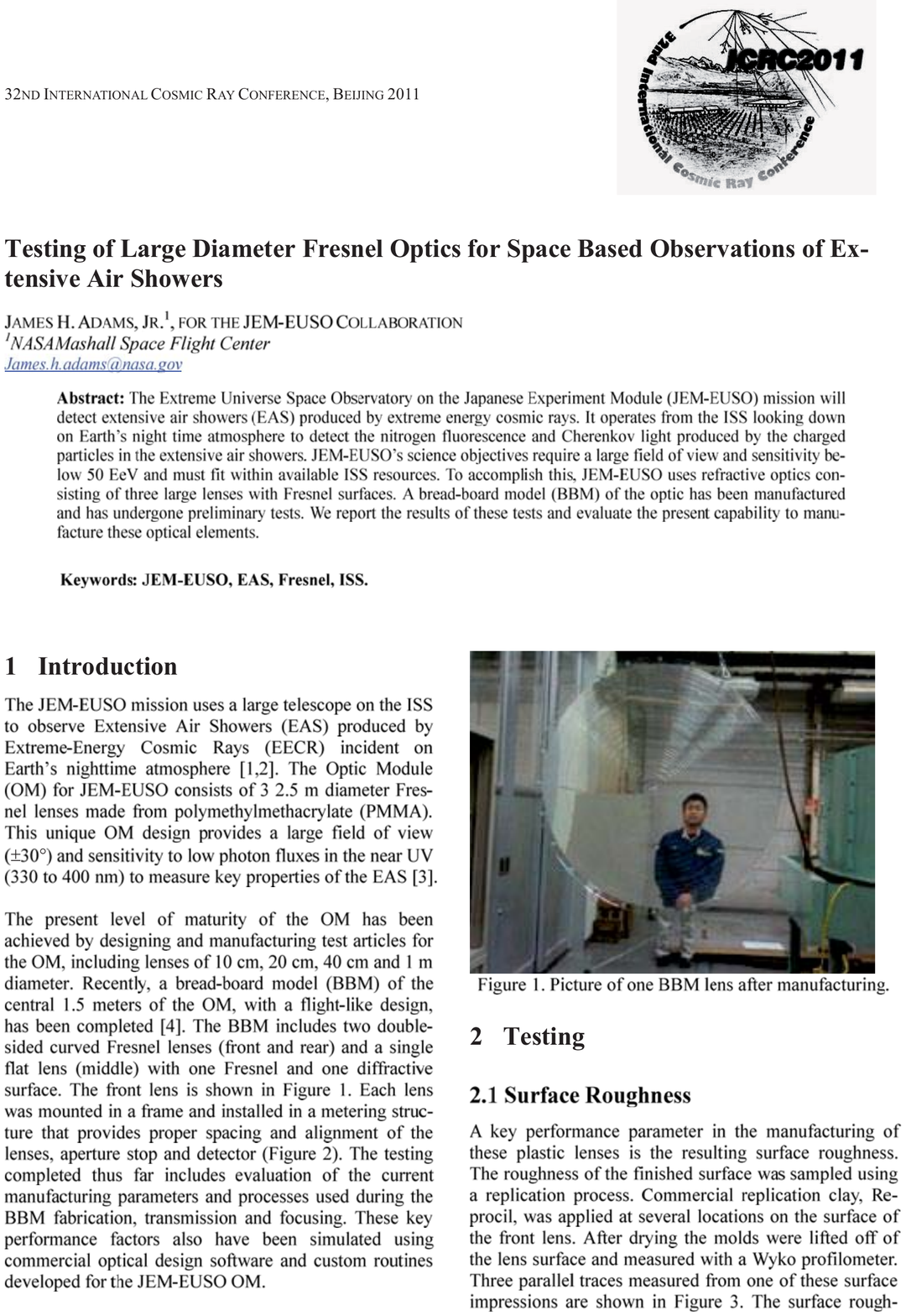}

\newpage
\normalsize
\setcounter{section}{0}
\setcounter{figure}{0}
\setcounter{table}{0}
\setcounter{equation}{0}



\title{Atmospheric Monitoring System of JEM-EUSO}

\shorttitle{Neronov \etal Atmospheric Monitoring of JEM-EUSO}

\authors{A.Neronov$^{1}$, S.Wada$^{2}$, M. D. Rodr\'iguez Fr\'ias$^{3}$, A. Morales de los R\'ios$^{3}$, G. S\'aez Cano$^{3}$, H. Prieto$^{3}$, J. Pi\~neiro$^{3}$, L. del Peral$^{3}$, J. Hern\'andez$^{3}$, N. Pacheco$^{3,4}$, M. D. Sabau$^{5}$, T. Belenguer$^{5}$, C. Gonz\'alez$^{5}$, M. Reina$^{5}$, S. Briz$^{6}$, A. de Castro$^{6}$, F. Cort\'es$^{6}$, F. L\'opez$^{6}$, G. Herrera$^{7} $, J. Licandro$^{7}$, E. Joven$^{7}$, M. Serra$^{7}$, O. Vaduvescu$^{7}$ A.Anzalone$^8$, F.Isgr\`o$^9$, R.Cremonini$^{10}$, C.Cassardo$^{10}$, A.Maurissen$^{11}$, C.Urban$^{11}$, T.Ogawa$^2$ K.Shinozaki$^2$ for JEM-EUSO collaboration}
\afiliations{$^1$ISDC Data Centre for Astrophysics, Versoix, Switzerland\\ 
$^2$RIKEN Advanced Science Institute, Japan\\
$^3$ SPace \& AStroparticle (SPAS) Group, UAH, Madrid, Spain\\
 $^4$Instituto de F\'isica Te\'orica (IFT), Universidad Aut\'onoma de Madrid (UAM), Spain\\
$^5$ LINES laboratory, Instituto Nacional de Tecnica Aeroespacial (INTA), Madrid, Spain\\
$^6$ LIR laboratory, University Carlos III of Madrid (UC3M), Spain.\\
$^7$ Instituto Astrofisico de Canarias, Tenerife, Spain\\
$^8$ INAF-IASF, Istituto di Astrofisica Spaziale e Fisica Cosmica,Palermo, Italy\\
$^9$ Dipart. di Scienze Fisiche, Universit\`a degli Studi di Napoli Federico
II, Napoli, Italy\\
$^{10}$ Dip.di Fisica, Universit\`a degli Studi di Torino, Italy\\
$^{11}$ Centre Suisse d'Electronique et Microtechnique, Switzerland
}
\email{Andrii.Neronov@unige.ch}

\abstract{
JEM-EUSO telescope on International Space Station will detect UV fluorescence emission from Ultra High Energy Cosmic Rays (UHECR) induced Extensive Air Showers (EAS) penetrating in the atmosphere. The accuracy of reconstruction of the properties of the primary UHECR particles from the measurements of UV light depends on the extinction and scattering properties of the atmosphere at the location of the EAS and between the EAS and JEM-EUSO. The Atmospheric Monitoring system of JEM-EUSO will use the LIDAR, operating in the UV band, and an infrared camera to detect cloud and aerosol layer features across the entire 60$^\circ$ field of view of JEM-EUSO telescope, to measure the cloud top altitudes with the accuracy of 500 m and the optical depth profile of the atmosphere in the direction of each EAS with the accuracy $\Delta \tau\le 0.15$ and resolution of 500 m. This should ensure that the energy of the primary UHECR particles  and the depth of EAS maxima are measured with the accuracy better than 30\% and  120 g/cm$^2$, respectively. }

\maketitle

\section{Introduction}

JEM-EUSO is a next-generation fluorescence telescope for detection of Extreme Energy Cosmic Rays (EECR, cosmic rays with energies $\sim 10^{20}$~eV and higher) which will be installed at International Space Station (ISS) in 2016 \cite{jem-euso,takahashi}. It is a refractive telescope with the aperture $\simeq 2.5$~m which will detect fluorescence UV emission from Extensive Air Showers (EAS) produced by EECR penetrating in the atmosphere within the $60^\circ$ Field of View (FoV). At the altitude of the ISS $H\simeq 400$~km the area over which the EAS events will be detected is $\simeq \left(400\mbox{ km}\right)^2$. The ISS orbits the Earth with the period $P\simeq 90$~min along an inclined orbit extending between $\pm 52^\circ$ from Equator.

\begin{figure*}[th]
  \centering
  \includegraphics[width=4in]{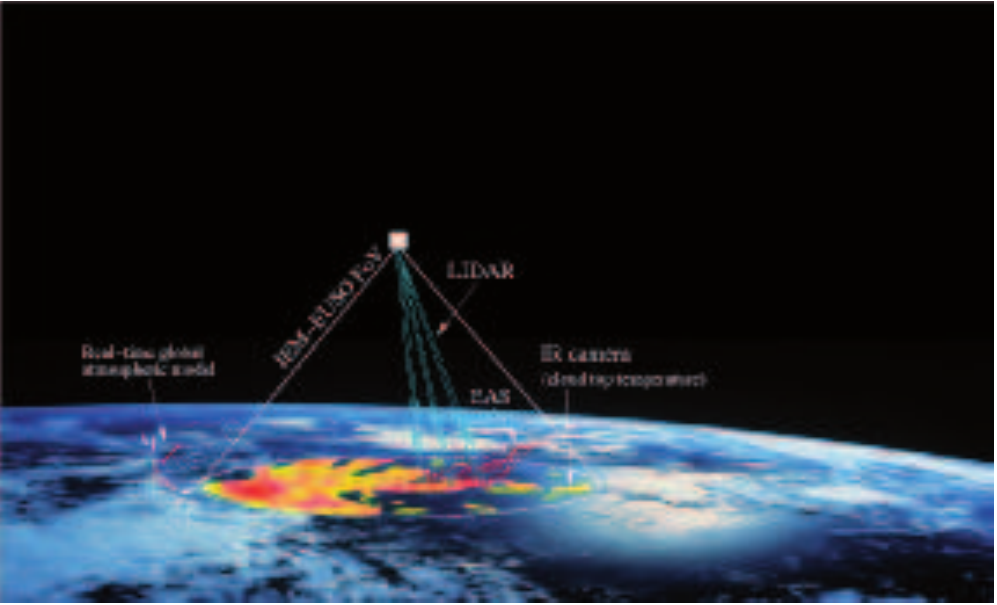}
  \caption{The principle of Atmospheric Monitoring in JEM-EUSO
    }
  \label{fig:AMS}
 \end{figure*}
 
The properties of the primary EECR particles (energy, type, arrival direction) will be derived from the imaging and timing properties of the UV emission from the EAS trace in the atmosphere. The Earth atmosphere absorbs and scatters UV light. The amount of absorption and scattering depends on the air column density between the emission and detection point and also on the type of absorbing and scattering centers. Scattering and absorption properties of the atmosphere are strongly affected by the presence of clouds and aerosol layers \cite{Spain1}. Cloud- and aerosol-induced variations of the scattering and absorption properties at the locations of EAS events distort the UV signal from EAS detected by JEM-EUSO. In the absence of detailed information on the presence and physical properties of the cloud and aerosol layers in JEM-EUSO FoV, distortions of the UV signal from EAS lead to systematic errors in determination of the properties of EECR from the UV light profiles. 

In particular, presence of optically thin cloud layers between the EAS and JEM-EUSO telescope reduces the overall intensity of UV light leading to an under-estimate of the EECR energy. EAS penetration into an optically thick cloud produces strong enhancement of the scattered Cherenkov light emission from EAS, which can be misinterpreted as Cherenkov light reflection from the ground/sea. This again leads to a wrong estimate of the depth of EAS maximum in the atmosphere. Statistics of the Earth cloud coverage known from the satellite measurement \cite{chepfer10,garino} indicates that as much as 70\% of EAS profiles might be affected by the presence of the clouds.
 
Cloud and aerosol layer induced distortions of the EAS profiles could be corrected if the detailed information on distribution and optical properties of the cloud/aerosol layers in JEM-EUSO FoV is known. This information will be provided by the Atmospheric Monitoring (AM) system of JEM-EUSO. The ISS orbital speed is $\simeq 7$~km/s so that the atmospheric volumed monitored by JEM-EUSO changes every $400\mbox{ km}/7\mbox{ km/s}\simeq 60$~s. This means that the distribution of clouds and aerosol layers in JEM-EUSO FoV is continuously changing. 
The AM system will continuously monitor the variable atmospheric conditions in JEM-EUSO FoV during all EECR data taking periods. 

In this contribution we describe the set up of the AM system of JEM-EUSO and  its expected performance.  

\section{Atmospheric Monitoring system}

 The goal of the AM system of JEM-EUSO is to provide information on the distribution and optical properties of the cloud and aerosol layers within the telescope FoV. The basic requirements on the precision of measurements of the cloud and aerosol layer characteristics are determined by the requirements on the precision of measurement of EAS parameters \cite{santangelo}: {\it (A1)} measurement of EECR energy with precision 30\%; {\it(A2)} measurement of the depth of the shower maximum with precision 120~g/cm$^2$.

Precision of the measurement of the energy of EECR is affected by the absorption of UV light cloud and aerosol layers. Precision of the measurement of the depth of shower maximum is additionally affected by the uncertainties of location of clouds and aerosols in the atmosphere. Imposing the requirements on the performance of the AM system:
{\it (B1)} measurement of the optical depth of atmospheric features with precision down to $\Delta\tau\le 0.15$;
{\it (B2)} measurement of the altitude of the boundaries of atmospheric features with precision $\Delta H\le 500$~m.
assures that the systematic error of the measurement of the energy and the depth of the EAS maximum introduced by the uncertainty of atmospheric conditions is significantly below that of requirements {\it A1, A2}.

The required precision of measurement of the altitude and optical depth of the cloud and aerosol layers will be achieved with the following  dedicated AM system which will consist of (Fig. \ref{fig:AMS})
\begin{enumerate}
\item LIght Detection And Ranging (LIDAR) device,
\item Infrared (IR) camera and 
\item global atmospheric models from the post-analysis of all available meteorological data by global weather prediction services like ECMWF \cite{ecwmf} and GMAO \cite{gmao}. 
\end{enumerate}

JEM-EUSO will take the cosmic ray data during the ISS nighttime. To reveal the overall picture of cloud distribution in the FoV an IR camera will be used. The IR camera is an infrared imaging system used to detect the presence of clouds and to obtain the cloud coverage and cloud top altitude during the observation period of the JEM-EUSO main instrument. Measurement of the temperature of the clouds will be used to estimate the altitude of the cloud top layers. Such an estimate is possible in the troposphere in the altitude range $0 - 10$~km where the atmosphere is characterized by a steady temperature gradient of $dT/dH\simeq 6^\circ/$km. To achieve the precision of measurement of the cloud top altitude $\Delta H\simeq 0.5$~km the precision of the temperature measurements by the IR camera will be $\Delta T=(dT/dH)\Delta H=3$~K. 

The AM system will additionally use a LIDAR device. The LIDAR will measure the optical depth profiles of the atmosphere in selected directions, with the ranging accuracy of $375/\cos(\theta_z)$~m, where $\theta_z$ is the angle between the direction of the laser beam and nadir direction. The power of the laser will be adjusted in such a way that cloud/aerosol layers with optical depth $\tau\ge 0.15$ at 355 nm wavelength will be detectable. 

The IR camera and LIDAR measurements will provide complementary information with the amount of details sufficient to 
\begin{itemize}
\item select the EAS events appearing in the clear sky conditions;
\item provide information on the optical properties of the clouds needed for correction of the cloud affected EAS profiles which could be retained for further analysis;
\item reject EAS events occurring in the complicated atmospheric conditions (multi-layer cloud/aerosol structures).
\end{itemize}

\section{IR camera}

The IR camera on board of JEM-EUSO will consist of a refractive optics made of germanium and zinc selenide and an uncooled microbolometer array detector \cite{Spain2}. Interferometer filters will limit the wavelength band to 10-12 $\mu$m. In the current configuration, two $\delta \lambda=1\ \mu$m wide filters will be used centered at the wavelengths 10.8~$\mu$m and 12~$\mu$m two increase the precision of the radiative temperature measurements.The FoV of the IR camera is $60^\circ$, totally matching the FoV of the main JEM-EUSO telescope. The angular resolution, which corresponds to one pixel, is about $0.1^\circ$. A temperature-controlled shutter in the camera and mirrors are used to calibrate background noise and gains of the detector to achieve an absolute temperature accuracy of 3~K. Though the IR camera takes images continuously at a video frame rate (equal to 1/30 s), the transfer of the images takes place every 30 s, in which the ISS moves half of the FoV of the JEM-EUSO telescope.

 \begin{figure}[!t]
  \includegraphics[width=0.8\columnwidth]{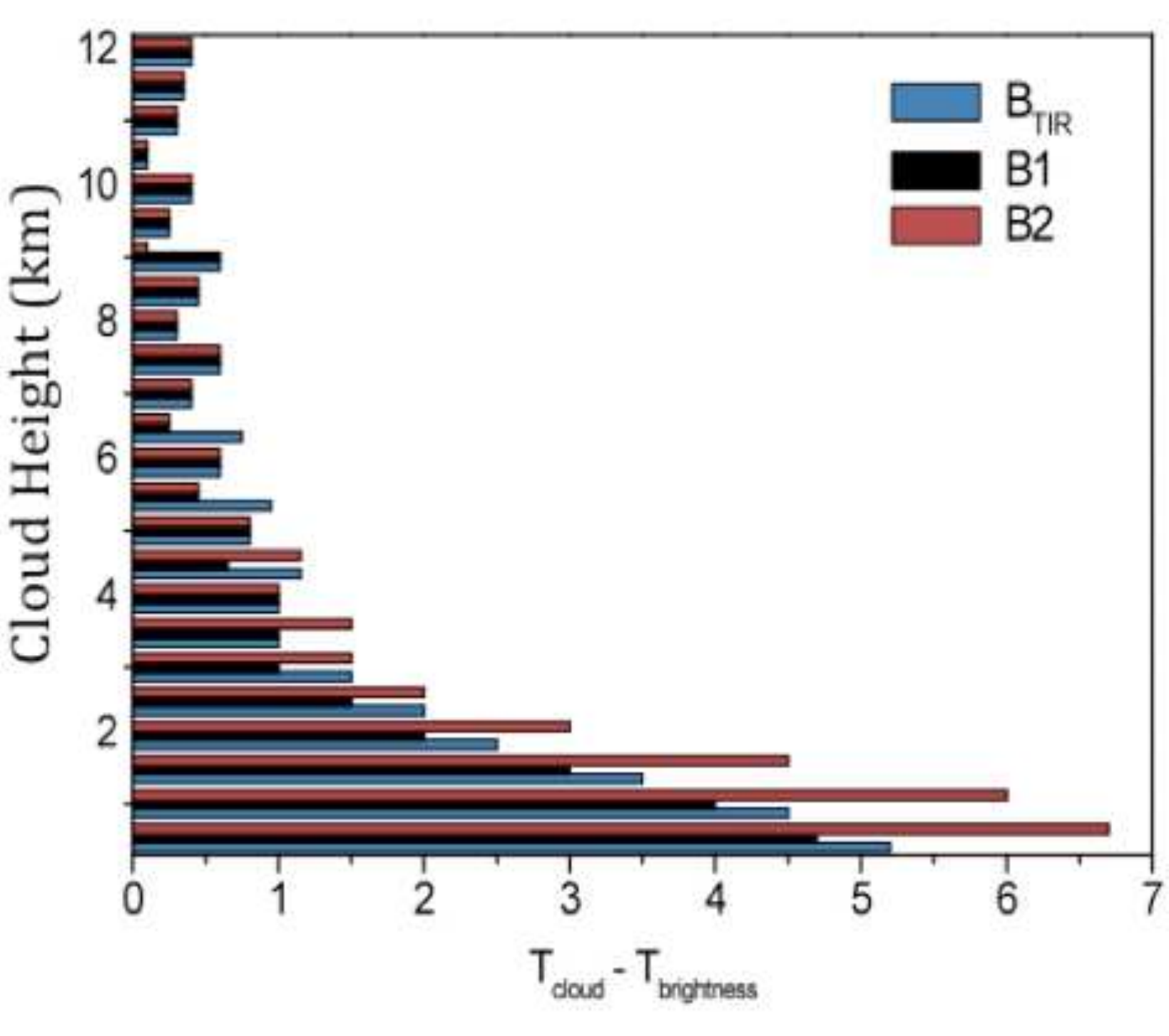}
  \caption{Error in the determination of temperature of clouds with the IR camera as a function of the cloud altitude.}
  \label{fig:temperature}
 \end{figure}

Fig. \ref{fig:temperature} shows the precision of the measurement of the cloud temperature for different cloud altitudes, reachable with the current IR camera design. The precision is within the required $3$~K limit almost everywhere down to the altitude of $\sim 1$~km. Three different columns show the error derived from the measurements in the shorter  (B1) and longer (B2) wavelength bands as well as in the combined ($B_{\rm TIR}$) measurement (see \cite{anzalone} for more details). 

\section{LIDAR}

The most relevant information about the absorption and scattering properties of clouds and aerosols is at the locations around the EAS events. To get this information, the LIDAR will have a re-pointing capability. The laser beam will be repointed in the direction of EAS candidate events following each EAS trigger of JEM-EUSO telescope. The average trigger rate of JEM-EUSO will be $\sim 0.1$~Hz. During the time interval between subsequent triggers, the LIDAR will 
\begin{itemize}
\item re-point to the direction in which the EAS trigger occurred and
\item take the measurements of laser backscattering signal in several directions around the supposed EAS maximum.
\end{itemize}
 \begin{figure}[!t]
  \includegraphics[width=0.9\columnwidth]{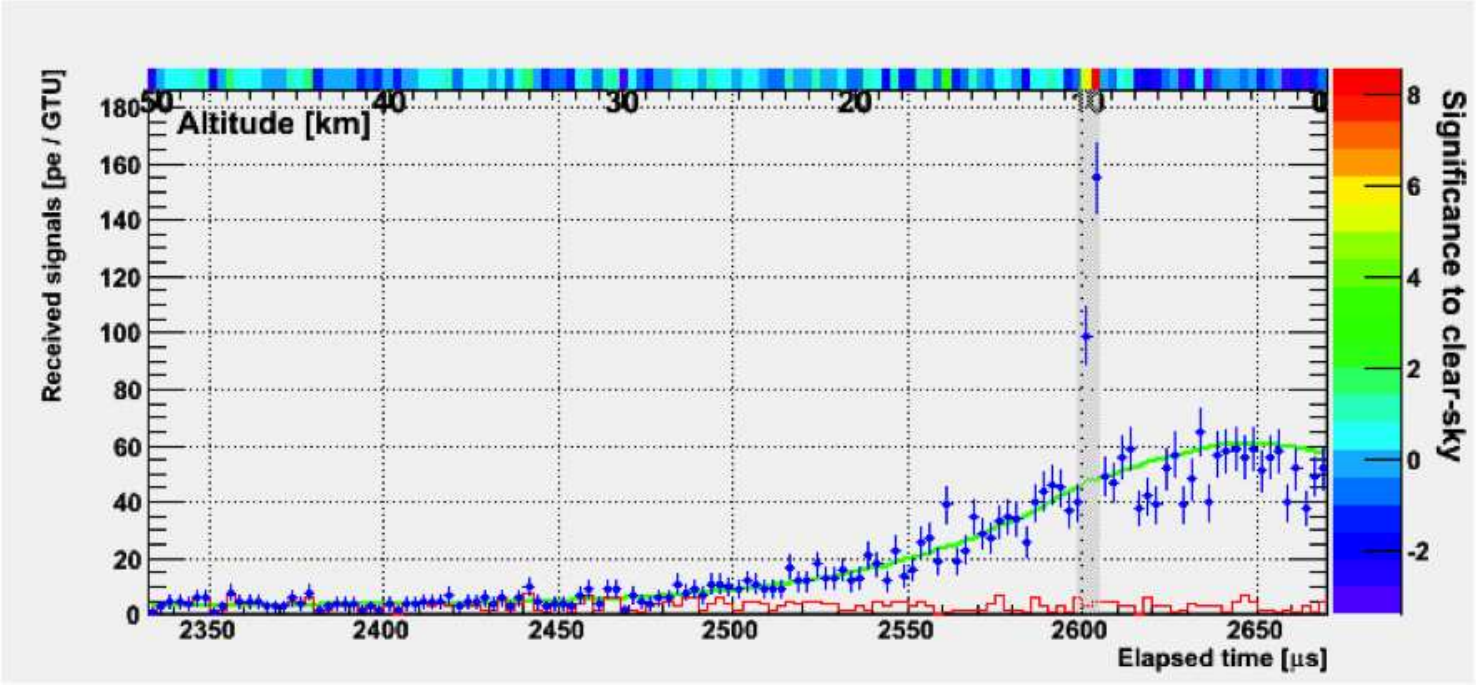}
  \includegraphics[width=0.9\columnwidth]{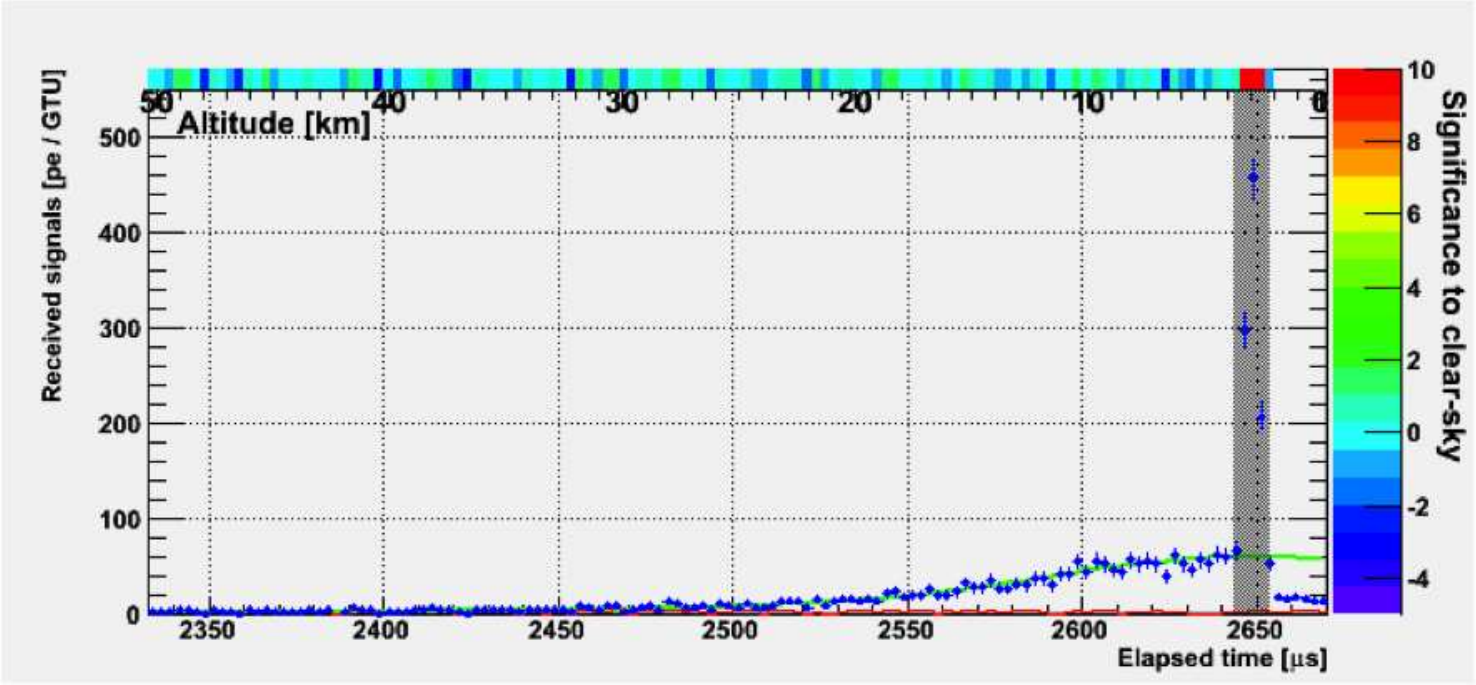}
  \caption{Simulated LIDAR backscatter signal in the presence of an optically thin (top) and optically thick (bottom) clouds.}
  \label{fig:simul}
 \end{figure}

Re-pointing of the laser beam will be done with the help of a steering mirror with two angular degrees of freedom and maximal tilting angle $\pm 15^\circ$.  The laser backscatter signal will be received by the main JEM-EUSO telescope which is well suited for detection of the 355 nm wavelength. Any Multi-Anode Photo-Multiplier Tube (MAPMT) in the focal surface of JEM-EUSO telescope could temporarily serve as the LIDAR signal detector, a special LIDAR trigger is foreseen in the Focal surface electronics of JEM-EUSO detector \cite{zfs}. Measurements of the laser backscatter signal with time resolution of 2.5 $\mu s$ (time unit of the focal surface detector) will provide ranging resolution of 375 m in nadir direction. The energy of the laser pulse will be adjusted in such a way that the backscatter signal will have enough statistics for the detection and measurement of the optical depth of optically thin clouds with $\tau \le 0.15$ at large off-axis angles. Examples of the simulated laser backscatter signal as it would appear in the JEM-EUSO detector are shown in Fig. \ref{fig:simul}. The upper panel shows the signal in the presence of an optically thin cloud with $\tau =0.06\pm 0.04$ ($\tau=0.05$ was assumed in the simulation) at the altitude of 10~km. Laser pulse energy $E=20$~mJ is assumed. Bottom panel shows an example of optically thick low altitude cloud. Assuming the same laser pulse energy $\tau=0.8\pm 0.2$ is derived from the simulated data, while $\tau=0.9$ was assumed in the simulation.

Parameters of the LIDAR system of JEM-EUSO are summarized in Table \ref{tab:lidar}.

\begin{table}[t]
\begin{center}
\begin{tabular}{|l|l|}
\hline
{\bf Parameter} & {\bf Value} \\
\hline
Wavelength & 355 nm\\
\hline
Pulse repetition rate & $>1$~Hz\\
\hline
Pulse width & 15 ns\\
\hline
Pulse energy & 20 mJ/pulse\\
\hline
Steering of output beam& $\pm 30^\circ$\\
\hline
Receiver & JEM-EUSO telescope\\
\hline
Detector & MAPMT (JEM-EUSO)\\
\hline 
Range resolution & 375 m\\
\hline
Mass & 17 kg\\
\hline
Power & $<70$~W\\
\hline
\end{tabular}
\caption{Characteristics of JEM-EUSO LIDAR}
\label{tab:lidar}
\end{center}
\end{table}

\section{Global Atmospheric Model data.}

Precision of the analysis of both IR camera and LIDAR data is largely improved when the basic atmospheric parameters of the atmosphere (temperature and pressure profiles, humidity etc) in the monitored region are known. Such parameters will be systematically retrieved from the global atmospheric model resulting from the post-analysis of weather models, calculated on regular basis by global meteorological service organizations (GMAO, ECMWF). These models also provide information on the presence and altitude distribution of cloud and aerosol layers, information which is directly relevant for JEM-EUSO data analysis. This justifies the incorporation of real time global atmospheric models in the AM data  of JEM-EUSO. 

\section{Conclusions.}

The AM system of JEM-EUSO, which includes the IR camera, the LIDAR and global atmospheric model data will provide sufficient information on the state of the atmosphere around the location of EAS events. This information will be used to correct the profiles of cloud-affected EAS events for the effects of clouds and aerosol layers, so that most of the cloud-affected events could be retained for the in the EECR data analysis.


\clearpage


\newpage
\normalsize
\setcounter{section}{0}
\setcounter{figure}{0}
\setcounter{table}{0}
\setcounter{equation}{0}



\title{The IR-Camera of the JEM-EUSO Space Observatory}

\shorttitle{J.A. Morales de los Rios \etal The IR-Camera of the JEM-EUSO Space Observatory}

\authors{J.A. Morales de los R\'ios$^{1}$, G. S\'aez-Cano$^{1}$, H. Prieto$^{1}$, L. del Peral$^{1}$, J. Pi\~neiro$^{1}$, K. Shinozaki$^{1,2}$, J. Hern\'andez$^{1}$, N. Pacheco$^{3}$, M.D. Sabau$^{4}$, T. Belenguer$^{4}$, C. Gonzalez$^{4}$, M. Reina$^{4}$,
S. Briz$^{5}$, A.J. de Castro$^{5}$, F. Cort\'es$^{5}$, F. Lopez$^{5}$, J. Licandro$^{6}$, E. Joven$^{6}$, M. Serra$^{6}$, O. Vaduvescu$^{6}$, G. Herrera$^{6}$, S. Wada$^{2}$, K. Tsuno$^{2}$, T. Ogawa$^{2}$, O. Catalano$^{7}$, A. Anzalone$^{7}$, M. Casolino$^{8, 2}$ and M.D. Rodr\'iguez-Fr\'ias$^{1}$, for the JEM-EUSO collaboration.}
\afiliations{$^1$SPace $\&$ AStroparticle (SPAS) Group, UAH, Madrid, Spain.\\ 
$^2$RIKEN, 2-1 Hirosawa, Wako, Saitama 351-0198, Japan.\\
$^3$Instituto de F\'isica Te\'orica (IFT). Universidad Aut\'onoma de Madrid (UAM), Spain.\\
$^4$LINES laboratory, Instituto Nacional de T\'ecnica Aeroespacial (INTA), Madrid, Spain.\\
$^5$LIR laboratory, University Carlos III of Madrid (UC3M), Spain.\\
$^6$Instituto de Astrof\'isica de Canarias (IAC), Tenerife, Spain.\\
$^7$INAF/IASF Istituto di Astrofisica Sapziale e Fisica Cosmica di Palermo, Italy.\\
$^8$University of Rome Tor Vergata, Rome, Italy.}
\email{josealberto.morales@uah.es}

\abstract{The JEM-EUSO space observatory will be launched and attached to the Japanese
module of the International Space Station (ISS) in 2016. Its aims is to
observe UV photon tracks produced by Ultra High Energy Cosmic Rays (UHECR) and Extremely High Energy Cosmic Rays (EHECR) 
developing in the atmosphere and producing Extensive Air Showers (EAS).
JEM-EUSO will use our atmosphere as a huge calorimeter, to detect the
electromagnetic and hadronic components of the EAS. The Atmospheric
Monitoring System plays a fundamental role in our understanding of the
atmospheric conditions in the Field of View (FoV) of the telescope and it
will include an IR-Camera for cloud coverage and cloud top height detection.}

\maketitle

\section{Introduction}

JEM-EUSO (Extreme Universe Space Observatory on Japanese Experiment Module) \cite{jemeuso_colaboration} is an advanced observatory onboard the International Space Station (ISS) that uses the Earth's atmosphere as a calorimeter detector. The instrument is a super wide-Field Of View (FOV) telescope that detects UHECRs and Extremely High Energy Cosmic Rays (EHECRs) with energy above 10$^{19}$ eV. This 
instrument orbits around the earth every $\approx$ 90 minutes on board of the International Space Station (ISS) at an altitude of $\approx$ 430 km. \\
An extreme energy cosmic ray particle collides with a nucleus in the Earth's atmosphere and produces 
an Extensive Air Shower (EAS) that consists of a huge amount of secondary particles generating flourescence light of atmospheric N$_2$.
JEM-EUSO captures the moving track of the fluorescence UV photons and reproduces the calorimetric development of the 
EAS \cite{jemeuso_collaboration2}, \cite{jemeuso_collaboration3}. At the energies observed by JEM-EUSO, above 10$^{19}$ eV, the existence of clouds will blur the 
observation of UHECRs. Therefore, the 
monitoring of the cloud coverage by JEM-EUSO Atmospheric Monitor System (AMS), is crucial to estimate the effective exposure 
with high accuracy and to increase the confidence level in the UHECRs and EHECRs events just above the threshold energy of the 
telescope. Therefore, the JEM-EUSO mission have implemented the AMS as far as the impact onto mass and power 
budget is insignificant. It consists of 1) Infrared (IR) camera, 2) LIDAR, 3) slow data of the JEM-EUSO telescope.\\ 
The Atmospheric Monitoring System (AMS) IR Camera is an infrared imaging system used to detect the presence of clouds and to 
obtain the cloud coverage and cloud top altitude during the observation period of the JEM-EUSO main instrument. Cloud top height retrieval can 
be performed using either stereo vision algorithms (therefore, two different views of the same scene are needed) or accurate 
radiometric information, since the measured radiance is basically related to the target temperature and therefore, according 
to standard atmospheric models, to its altitude \cite{swada}.

\section{Requirements for the infrared camera measurements}
The Atmospheric Monitoring System (AMS) IR Camera is an infrared imaging system used to detect the presence of 
clouds and to obtain the cloud coverage and cloud top altitude during the observation period of the JEM-EUSO main instrument. 
Moreover, since measurements shall be performed at night, it shall be based on cloud IR emission. The observed radiation is 
basically related to the target temperature and emissivity and, in this particular case, it can 
be used to get an estimate of how high clouds are, since their temperatures decrease linearly with height at 
6 K/km in the Troposphere. Table \ref{table_req} summarizes the current scientific and mission requirements for the JEM-EUSO AMS IR camera.

\begin{table}[t!]
\begin{center}
\begin{tabular}{|l|cc|}
\hline
Parameter & Target value &  Comments  \\
\hline
\hline
Measurement &  & Annual variation \\ range  & 200 K - 320 K &  of cloud  \\ &  & temperature plus \\ & & 20 K margin    \\
\hline
  &  & Two atmospheric \\ Wavelength & 10-12 $\mu$m  & windows available: \\ & & 10.3-11.3 $\mu$m \\ & & and 11.5-12.5 $\mu$m    \\
\hline
FoV     & 60$^o$ &  Same as \\  & & main instrument \\
\hline
Spatial & 0.25$^o$@FOV center & Threshold values. \\ resolution &  0.22$^o$@FOV edge  &  \\
\hline
Absolute &  & 500 m in cloud \\ temperature & 3 K & top altitude \\ accuracy &  &  \\
\hline
Mass  & $\leq7$ kg & Inc 30$\%$ margin.    \\
\hline
Dimensions     & $200\times280\times320$ & $300\times300\times500$ \\ & mm. & mm. Max  \\
\hline
Power    & $\leq11$ W  &  Inc 30$\%$ margin.       \\
\hline
\end{tabular}
\caption{Requirements for the IR camera.}\label{table_req}
\end{center}
\end{table}

Although there are no formal requirements for data retrieval, it has been assumed that the IR camera retrieval 
of the cloud top altitude could be performed on-ground by using stereo vision techniques or radiometric algorithms 
based on the radiance measured in one or several spectral channels (i.e. split-window techniques). 
Therefore, the IR camera preliminary design should be complaint with both types of data processing.
Moreover, in this work, we have considered two methods for the data retrieval based on the use of one or two IR bands.

\section{The infrared camera preliminary design}
The IR-Camera can be divided into two main subsystems:
a) Opto-mechanical unit, including the components necessary for the manipulation of the optical signal (magnification, spectral filtering…), the detection of light (sensor) and the calibration of the sensor.
b) Electronic subsystem, providing the instrument control and HouseKeeping (HK) functions, scientific data processing, redistribution of the power supply to IR camera components and electronic interfaces with the JEM-EUSO Instrument.

\begin{figure}[!t]
  \vspace{0mm}
  \centering
  \includegraphics[width=2.5 in]{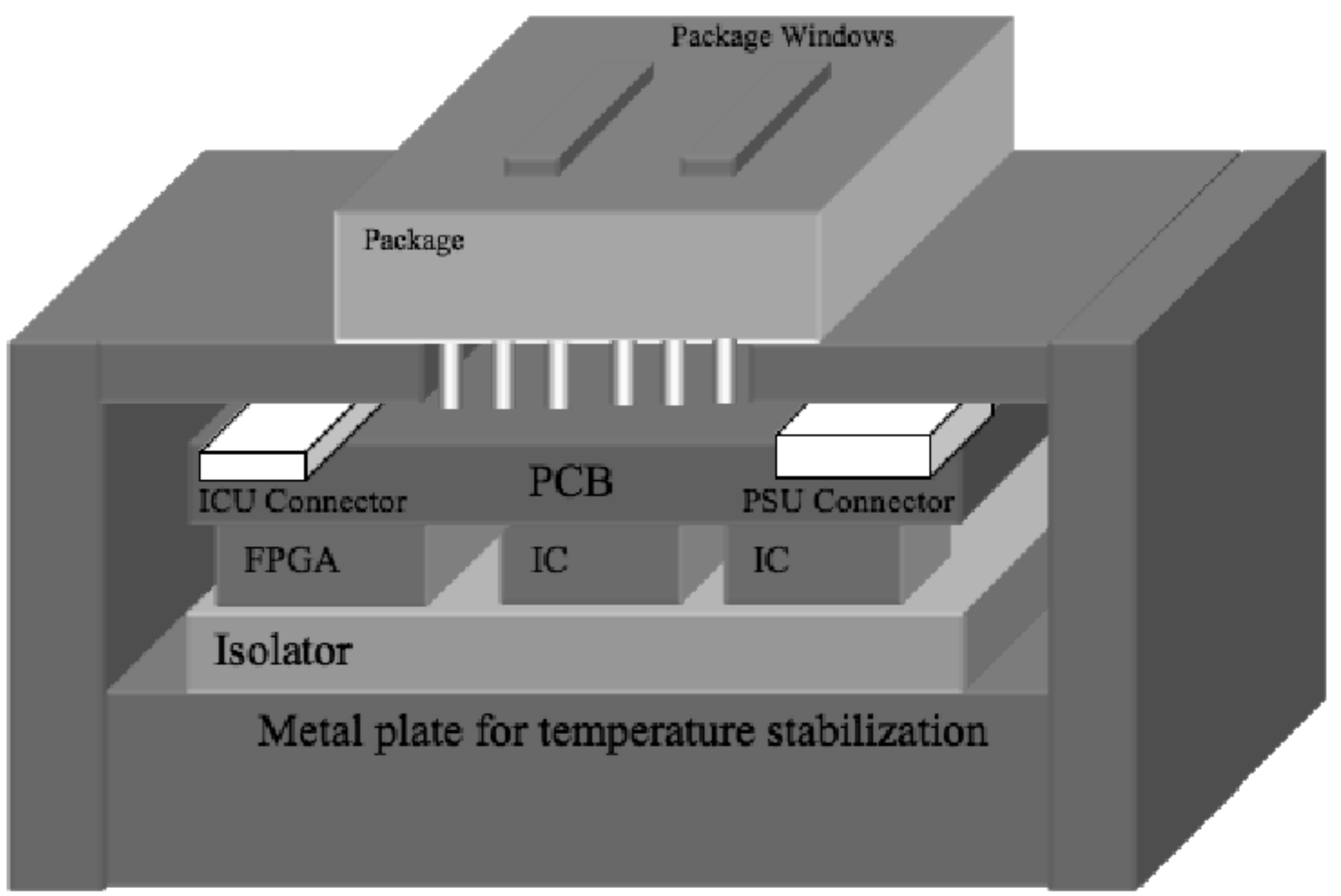}
  \caption{Detail view of the temperature stabilizer and the arrangement of the diferent components.}
  \label{cameradesign1}
 \end{figure}

The optical system will be a refractive objective based on a modified IR Cooke triplet, with an aperture of 15mm, made of composed 
materials to save weight. This type of lens is especially interesting because has enough available degrees of freedom to allow the 
designer the correction of primary aberrations. For the detector, the baseline is an uncooled microbolometer pixel array, with a 
size of 640 x 480 pixels, and a Read Out Integrated Circuit (ROIC) as well, that is in charge of reading the values of the photo-detector 
array. The ROIC has on-chip programmable gain for optimization of the performance over a wide range of operating conditions. The system is uncooled, 
although a temperature stabilizer is required with a thermistor close to the detector array 
for accurate temperature measurements, and a Peltier cooling system, as is show in figure \ref{cameradesign1}.
 
In the Electronic subsystem, we can find the image channel wich is formed by an optic element, used to focus the image, and the IR detection unit (detector plus video 
electronics). The data generated by the image channel is processed by the Data Processing Unit (DPU) within the Instrument Control 
Unit (ICU), which is in charge of controlling several aspects of the system management such as the electrical system, the thermal 
control, the calibration subsystem and the communication with the platform computer. The actuators of the instrument are managed by 
the ICU through an interface with the Centralized Control Unit (CCU). The Power Supply Unit (PSU) provides the required power 
regulation to the system. The management of the PSU is controlled by the ICU as well.

Most of the digital circuit implementation is based on FPGAs. The design contains two FPGAs which are located inside the ICU and inside 
the CCU respectively. Both FPGAs offer a control interface to the microprocessor to deal with basic functions such as actuator manipulations 
and data processing functions. All the commands are transmitted through a common data interface.

Therefore, the system architecture can be split into four different blocks: the optic with the detector and video electronics, 
the CCU with the power drivers and the mechanisms controller FPGA, the PSU which is in charge of provide the power supply to the 
instrument, and the ICU that controls and manages the overall system behavior. A cold redundancy scheme has been selected for the design.

\section{Infrared data retrieval}

The radiation emitted by the Earth's surface and the atmospheric clouds is measured by the IR sensor and the system 
retrieves the temperature of the emitter from this measurement. However the radiation collected by the 
IR sensor is not emitted by one single source. On the contrary, the atmosphere between the emitter 
and the sensor absorbs and emits energy. Therefore, the temperature obtained directly is not exactly the 
temperature of the emitter. These effects involve some uncertainties in the emitter temperature obtained 
from the direct radiation measurements.
The objective of this part is the estimation of the errors associated to several factors: temperature 
and water vapor profiles deviations and cirrus effect. For this purpose some retrievals simulations have been 
carried out. 

\begin{figure}[!t]
  \vspace{0mm}
  \centering
  \includegraphics[width=3.5 in]{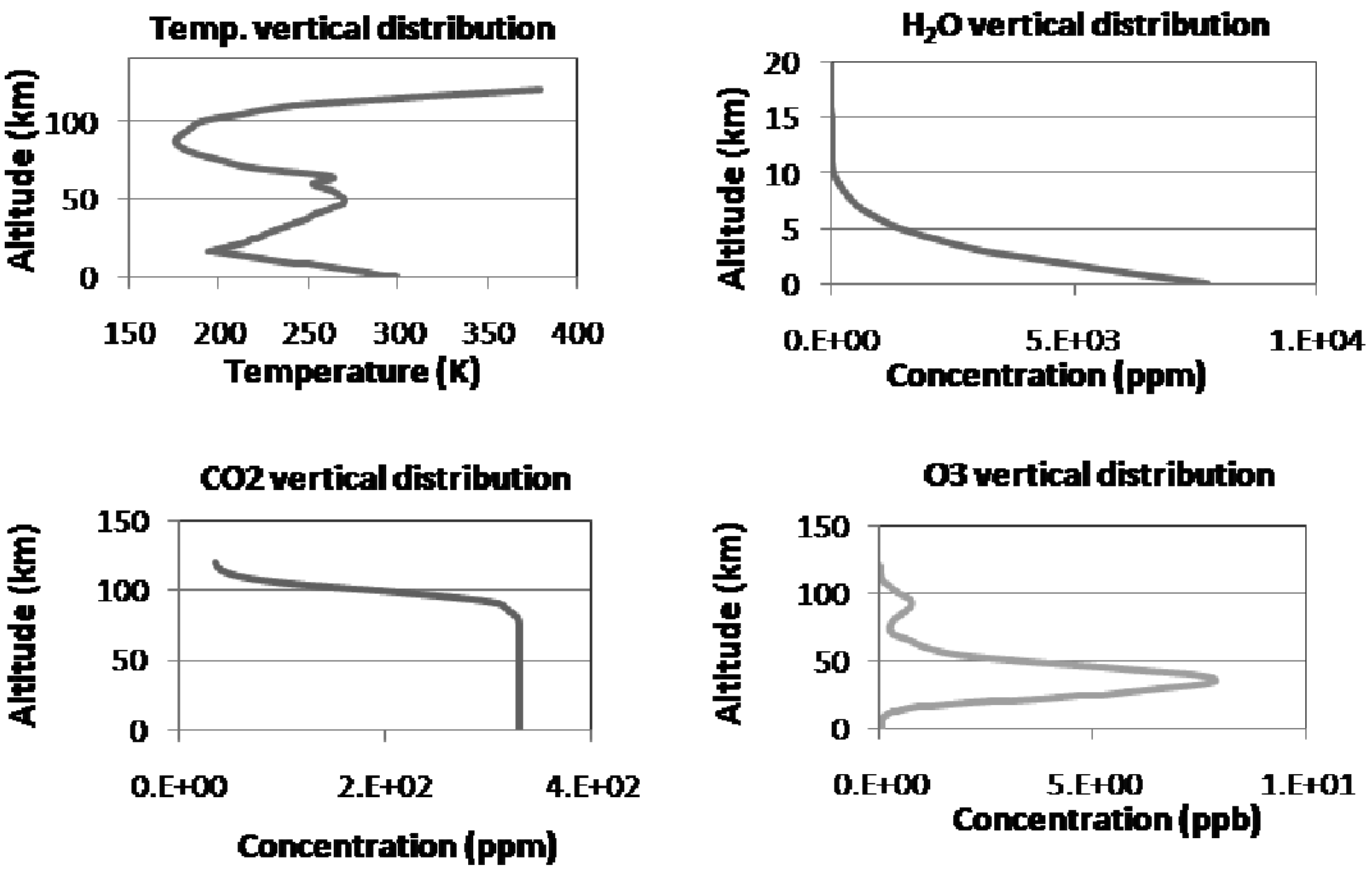}
  \caption{Examples of vertical profiles that describe the atmosphere model used in radiance calculations.}
  \label{susana1}
 \end{figure}

The simulations are based in a radiative model that consists of an atmospheric model, with the Earth surface emitting at 300K 
and a cloud at a certain height.
In order to define the atmospheric model, the atmosphere is split into layers and values of temperature, 
pressure and gases concentrations have to be assigned. In this study, the atmosphere has been divided in 
0.5 km thick layers from the bottom to the top of the atmosphere assumed at 150 km. Far away from this 
altitude, it is assumed that there is not physical effects on the IR radiation transport through the medium. 
In this way, the atmospheric model is 
described by vertical profiles of temperature, pressure and density. As a good approximation, clouds 
can be considered as blackbody emitters. For this reason, the clouds would absorb the energy emitted by the 
Earth surface and by the atmosphere beneath the clouds. For the same reason, the cloud can be modeled by a 
thin layer located at the top of the cloud that behaves as a blackbody at the temperature of 
the atmospheric layer at the same level. Figure \ref{susana1} shows the vertical profiles describing the atmosphere 
model used.

\subsection{Results of the one-band analysis}
The main conclusions can be summarized as follow: a)The effect of the temperature vertical profiles is not significant (errors $<$3 K).
b)The effect of water vapor vertical profile is significant for low-level clouds and atmospheres with high water vapor concentrations.
c)The effect of thin clouds (cirrus) cannot be neglected since errors in retrieved temperatures are higher than 3 K for low and medium-level clouds.
d)The temperatures retrieved by only one band are not accurate enough due to the effect of water vapor profiles and thin clouds.

\subsection{Results of the two-band analysis}
In order to take advantage of the two-bands, a Split-Window Algorithm (SWA) has to be applied to the brightness temperatures 
retrieved from B1 and B2 bands. These algorithms have been used since latest 70s to measure the Earth's surface temperature from 
satellites to minimize the effect of the atmosphere. 
There are plenty of SWA that have been developed to retrieve the surface temperature from satellite measurements 
\cite{king} \cite{wan}. All these algorithms are based on linearization of Planck's law and on the Radiative Transfer 
Equation (RTE). They have been applied to radiances obtained in two spectral bands and all of them consist of linear or quadratic functions 
of the temperatures retrieved in two bands. A comparison between retrieved temperatures by the one-band option and the SWA is shown
in table \ref{susanatable1}. For blackbody clouds, the coeficients only depend on the atmospheric transmittance and they 
can be calculated because atmospheric profiles are known in simulations studies.

\begin{table}[t!]
\begin{center}
\begin{tabular}{|l|cccc|}
\hline
Cloud  & T$_{cloud} (K)$ &  T$_{band-1} (K)$ & T$_{band-2} (K)$ & T$_{SWA} (K)$ \\
Height & & & & \\
\hline
0.5 km & 296.7 & 293.6 & 292.0 & 297.4 \\
1 km & 293.7 & 291.5 & 290.3 & 294.0 \\
2 km & 287.7 & 286.7 & 286.1 & 287.7 \\
\hline
\end{tabular}
\caption{Comparison between retrieved temperatures by the one-band algorithm and the SWA. Although clouds have been studied from top 
heights of 0.5 km to 12 km in 0.5 km steps, in this table only the worst cases are shown}\label{susanatable1}
\end{center}
\end{table}

For real cases, the transmittance is not always known and, for this reason, different algorithms have been developed by different 
authors. The differences between algorithms lies in the SWA parameters and different authors propose different parameters to retrieve 
the surface temperatures in different conditions \cite{galve}. The 
algorithms have to be validated for different examples and environmental conditions but there is not an universal algorithm that can 
be applied to any problem with enough accuracy. The same methodology can be applied to measure clouds temperatures, especially to 
low-level and thick clouds. In fact, there are also some SWAs devoted to retrieve clouds temperatures from satellites such 
as AVHRR, MODIS, etc. \cite{heidiger}. These algorithms are able to retrieve top-cloud temperature (Nieman, 1993), 
cloud emissivity and type of clouds (Pavolonis, 1985 and Inoue, 1987) and cloud microphysics (Inoue 1985). However the results attained 
when semitransparent cirrus are found in the FOV are not so accurate \cite{heidiger}. Therefore still open points remain to be adressed in order to retrieve top-cloud temperatures accurately.

Summarizing the results of this SWA preliminary study we can state: a) SWA is as accurate as the transmitances calculated with specific known 
atmospheric profiles. b) In order to study the effect of a cirrus (semitransparent cloud) in the temperature retrieval of a blackbody cloud, some simulations 
have been performed considering a cirrus between the cloud top and the IR camera. The examples show that the one-band option temperature 
retrievals have stronger uncertainties than SWA option, although SWA error is still above 3K. c) Not all SWAs from the bibliography always give good results. 
d) For higher clouds the coefficients of SWA have to be checked because the distance between the cloud top and the sensor decreases.
These are only preliminar results, other factors like partially-covered pixels, semitransparent clouds, and some other issues will be estudied in the future.

\section{End to end simulation of the IR-camera}

End to end simulation of the infrared camera will give us simulated images of those we expect to obtain with the instrument. Therefore 
this simulation together with the data analysis will be included in the AMS detector
simulation module of the JEM-EUSO analysis software. First the simulated radiation produced by 
the Earth's surface and atmosphere have been considered, taking in account the effect of the optics and the detector, and finalizing in
the electronics and image compresion algorithm. A data analysis module is foreseen to 
take the data from simulator, and real data from the IR-Camera to perform the analysis tasks with the algorithms
for data retrieval. The output from this analysis module will be used as an input in the official codes for the event reconstruction of the main telescope.

\begin{figure}[!t]
  \vspace{0mm}
  \centering
  \includegraphics[width=3.5 in]{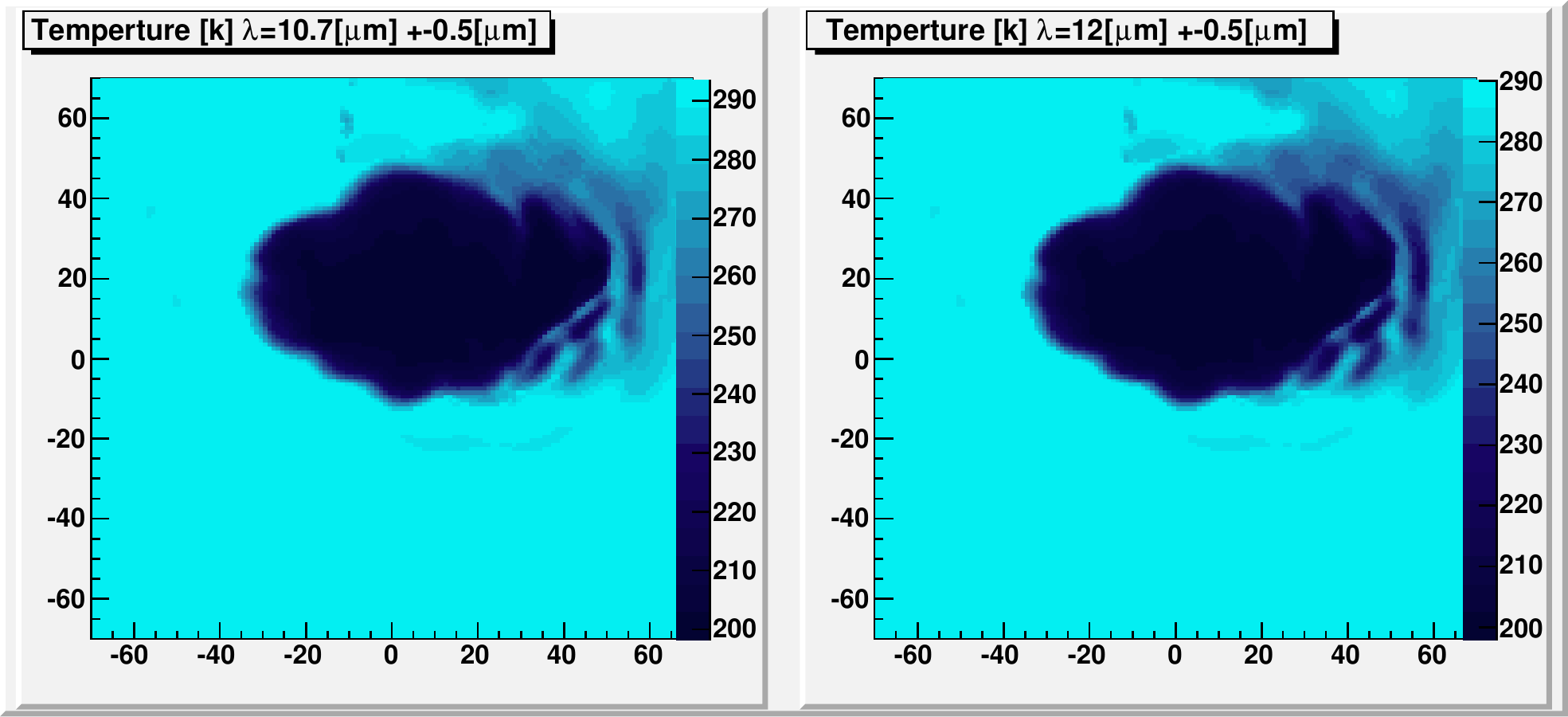}
  \caption{Examples of simulated cloud with SDSU modified for the IR-Camera bands.}
  \label{jose1}
 \end{figure}

So far we have just started with the IR simulations, emitted by the ground and the atmosphere, using a modified 
version of the SDSU \cite{sdsu} software developved in the Hydrospheric Atmospheric Research Center, Nagoya University,  
to simulate the wavelenght of our detector, and thanks to the capabilities of this code we are simulating the UV(Ultra-violet) 
range of our main instrument (250-500 $\mu$m) to get an aproach for the slow data of JEM-EUSO. At the end, the output from this work will be images similar to what we 
expect from our camera, that would allow us to test the data retrieval algorithms and calculate correction factors for the 
IR-Camera. A preliminary example of the IR simulations done by the SDSU software is shown in the figure \ref{jose1}.

\section{Acknowledgments}
The Spanish Consortium involved in the JEM-EUSO Space Mission are funded by MICINN under projects AYA2009-06037-E/ESP, AYA-ESP
2010-19082, CSD2009-00064 (Consolider MULTIDARK) and by Comunidad de Madrid (CAM) under project S2009/ESP-1496.
J.A. Morales de los R\'ios wants to acknowledge the University of Alcal\'a for his Phd fellowship.


\clearpage


\newpage
\normalsize
\setcounter{section}{0}
\setcounter{figure}{0}
\setcounter{table}{0}
\setcounter{equation}{0}



\title{Cloud Coverage and its Implications for Cosmic Ray Observation from Space}

\shorttitle{F.Garino, A. Guzman \etal Cloud Coverage and its implications for CR observation from Space}

\authors{
F.~Garino$^{1}$,
A.~Guzman$^{2}$,
M.~Bertaina$^{1}$,
C.~Cassardo$^{1}$,
R.~Cremonini$^{1}$,
C.~De Donato$^{2}$,
G.~Medina-Tanco$^{2}$,
K.~Shinozaki$^{3}$,
on behalf of the JEM-EUSO Collaboration
}
\afiliations{
$^1$ Dipartimento di Fisica Generale dell' Universit\`a di Torino, Torino, Italy\\
$^2$ ICN-UNAM, Mexico City, Mexico\\
$^3$ RIKEN, Wako, Japan\\
}
\email{tapozio@gmail.com,bertaina@to.infn.it}

\abstract{JEM-EUSO is an International mission planning to observe Extreme High Energy Particles from Space.
Flying on the ISS, during its operation, JEM-EUSO will experience all possible weather conditions inside 
its field of view. In order to estimate the effective aperture of the detector, one key point is the 
evaluation of the role of clouds, in particular their frequency as a function of altitude and optical depth 
for the different geographical areas of the planet. The probability of occurrence of a defined atmospheric 
condition has been assessed in this study by means of different meteorological databases: TOVS, 
ISCCP and CACOLO data. Because of the specific peculiarities of each dataset, the comparison of the 
different results is used to assess the systematic uncertainty on the derived conclusions.  
}

\maketitle

\section{Introduction}
JEM-EUSO \cite{yYoshi,yToshi} is a new type of observatory under development with the aim of detecting 
Extreme Energy
Cosmic Rays (EECR) from the International Space Station (ISS), by using the whole Earth as a detector. 
JEM-EUSO telescope will orbit around the Earth every $\sim$90 minutes at the
altitude of 350-400 km to capture the moving track of the Ultra Violet (UV) photons produced
during the development of Extensive Air Showers (EAS) in the atmosphere.
The telescope has a super-wide ($\pm$30$^\circ$) Field-of-View with optics composed by Fresnel lenses 
\cite{yTakky}.
The telescope records the track of an EAS with a time resolution of 2.5$\mu$s and a spatial resolution of
about 0.5 km (corresponding to 0.07$^\circ$) in nadir mode by using a highly pixelized focal surface
(3$\times$10$^5$ pixels) \cite{yKajino}. These time-segmented images allow determining energy and direction 
of the primary particles \cite{yAndrea}.\\
During its operation JEM-EUSO will experience all possible weather conditions. 
The amount of both fluorescence and Cherenkov signals reaching JEM-EUSO depends on the extinction and
scatteringt of UV light in atmosphere. Correct reconstruction of EECR energy and of the type of the
primary cosmic ray particle requires, therefore, information about absorption and scattering properties
of the atmosphere. 
For this reasons the JEM-EUSO observatory will include an Atmospheric Monitoring System (AM) 
\cite{yAndrii} which 
will consist of an infrared camera \cite{ySpain}, and a LIDAR device. Moreover, it will benefit from 
the real time global atmospheric models like those generated by the Global Modeling and Assimilation 
Office (GMAO) \cite{yGMAO}, the European Center for Medium range Weather Forecasts (ECMWF) \cite{yECMWF}
and other similar services.\\
The cloud coverage in the FoV of JEM-EUSO will be continuously monitored by the infrared camera, which
will also measure the altitude of the top of optically thick clouds. LIDAR will determine the detailed
scattering and extinction properties of the atmosphere at the location of each triggered EAS event.
Real-time models of the atmosphere are used to deduce the parameters relevant for the modeling of the
transmission and extinction properties of the air, neeeded for the analysis of the LIDAR data, for the
calibration of the infrared camera, and for the modeling of development of EAS in the atmosphere.\\
The peculiarity of the observation from space is the possibility of observing CR also in some cloudy
conditions, which is tipically not the case for ground-based telescopes.
In a simplified way, we can assume that if the maximum of the shower is above the cloud layer the 
reconstruction of the shower parameters will be possible. It is clear that the same top cloud layer will 
affect in a different way showers of various inclination or originating from the different type of 
primary particles (i.e. neutrino will develop much deeper in the atmosphere compared to EECR).
Thin clouds ($\tau <$ 1, typical of cirrus) will affect the energy estimation but the measurement of the 
arrival direction will still be possible with acceptable
uncertainty. Thick clouds ($\tau >$ 1) will compromise, or 
prevent, the measurement only if located at high altitudes. 
As an example, 60$^\circ$ zenith-angle inclined showers will have the shower maximum at 6-7 km altitude, 
much higher than the typical range of stratus \cite{yLupe}.\\
As the location of the clouds will affect either the duty cycle or the effective aperture
of the instrument, and, consequently, the exposure, a detailed analysis of the probability of occurrence
of the differect atmospheric conditions has been evaluated by means of different meteorological databases.
The main reason is that each database has its own peculiarity, therefore, their comparison will allow to 
study the variability of the results and assess an uncertainty on the cloud distribution.\\

\section{The meteorological databases}
TOVS, ISCCP and CACOLO meteorological databases have been used in the following analysis. 
The NASA project TOVS (TIROS Operational Vertical Sounder) \cite{yTOVS} on board
NOAA's TIROS series of polar orbiting satellites consists of three instruments:
a high-resolution infrared radiation sounder modification 2 (HIRS/2), a 
stratospheric sounding unit (SSU) and a microwave ounding unit (MSU). The three
instruments have been designed to determine the radiance needed to calculate
temperature and humidity profiles of the atmosphere from the surface to the
stratosphere. These data have a good spectral distribution and provide optical
depth and altitude of clouds. They are distributed irregularly and to obtain 
a complete data-set, the application of the transport radiative model has been
necessary. In this study, data from January 1988 to December 1994 have been 
used, divided between land and ocean data.\\
The International Satellite Cloud Climatology Project (ISCCP) \cite{yISCCP} was established in 1982 as part of 
the World Climate Research Program (WCRP) to collect and analyze satellite 
radiance measurements to infer the global distribution of clouds, their properties, and their diurnal, seasonal and 
interannual variations. Data collection is still on. The 
resulting data-sets and analysis products are being used to 
improve the understanding and the modeling of the role of clouds in climate,
the primary focus being the elucidation of the effects of clouds on the
radiation balance. These data can be also used to support a number of other 
cloud studies, including the understanding of the hydrological cycle. The data
are collected from the suite of weather satellites operated by several nations
and processed by groups in government agencies, laboratories, and universities. 
ISCCP has developed cloud detection schemes using visible and infrared window
radiance (infrared during nighttime and daytime, while visible during daytime).
The data from July 1983 to June 2008 have been used in this analysis. The data have 
the following characteristics:
a) possibility to obtain monthly, seasonal and annual means; b) the cloud types 
are defined by the VIS/IR 
(visible/infrared) top pressure and optical depth and they are divided in 3 levels (low clouds with a top 
pressure greater than 680 mb, about 3.2 Km, high clouds with a top pressure minor than 440 mb, 
about 6.5 Km, and middle clouds with a pressure between the other types); c) no division between ocean data and land data; d)  frequency of occurrence of cloudy conditions in 
individual satellite image pixels, each of which covers an area of about 4 to 49 square kilometers; 
e) data are given on a 2.5 degree square latitude-longitude grid, so we obtained a 
map divided in 10368 boxes (144 X 72 - longitude X latitude). As the data of this dataset can be extracted 
also on a monthly basis, they allow to reconstruct the interannual variability of cloud coverage for low, 
middle and high clouds.\\ 
The CACOLO (Climatc Atlas of Clouds Over Land and Ocean data) database \cite{yCACOLO} 
presents maps introduced in the atlases of cloud climatological data obtained 
from visual observations from Earth. The cloud averages presented on these maps
have been extracted from a digital archive of gridded land and ocean cloud 
climatological data. Maps are given for total cloud cover, clear-sky frequency,
and the average of nine cloud types within the low, middle, and high levels of
the troposphere. The amount of cloud is defined as the fraction of the 
sky-hemisphere covered by the cloud. Maps of precipitation frequency are also 
included. Monthly, seasonal, and annual averages are given for both daytime
and night-time. Land and ocean data have been analyzed separately, and are
mapped separately for most quantities. Two grid sizes are used to display the
cloud averages. Most data are given at 5-degree latitude-longitude resolution.
A 10-degree grid is used to map some ocean data. The land data are based on
analysis of 185 million visual cloud observations made at 5388 weather stations
on continents and islands over a 26-year period (1971-1996). The ocean maps are
based on analysis of 50 million cloud observations made from ships over a 
44-year period (1954-1997).\\ 
In this sense CACOLO is a truly complementary database compared to the other 
two as the information is coming from ground observations instead from space.

\section{Data analysis}
A first study has been conducted in order to evaluate the differences between
night-time and daytime, oceans and lands, using the TOVS data-set. Data have
been used only in the range of latitudes 50N-50S since this is the range of latitudes
spanned by the ISS. Clouds have been classified into 16 categories, according to
their top altitude ($h$) ($h<$3 km, 3$<h<$7 km, 7$<h<$10 km, $h>10$ km) and
optical depth (OD) ($OD<0.1$, $0.1<OD<1$, $1<OD<2$, $OD>2$).  
Table~\ref{qtab1} reports the results of the occurence of each cloud tipology
for oceans during daytime. 
 \begin{table}[th!]
  \caption{Relative occurrence (\%) of clouds between 50$^\circ$N and 50$^\circ$S 
latitudes on TOVS database in the matrix of cloud-top altitude vs optical 
depth. Daytime and ocean data are used for the better accuracy of the 
measurement.}
   {\scriptsize
  \centering
  \begin{tabular}{|l|c|c|c|c|c|}
  \hline
Optical Depth&\multicolumn{4}{c|}{Cloud-top altitude}\\ \hline
 &$<$3km&3-7km&7-10km&$>$10km\\ \hline 
$>$2 & 17.2 & 5.2 & 6.4 & 6.1\\ \hline
1-2 & 5.9 & 2.9 & 3.5 & 3.1\\ \hline
0.1-1 & 6.4 & 2.4 & 3.7 & 6.8\\ \hline
$<$0.1 & 29.2 & $<$0.1 & $<$0.1 & 1.2\\ \hline
  \end{tabular}
  \label{qtab1}
   }
 \end{table}
This configuration has been chosen as data taken during daytime are in general
more reliable and the same applies to the ocean data compared to the land ones. The 
comparison between day and night has been performed then on lands as higher
variations are expected on land surface compared to oceans. Slight differences 
among the tables exist, however, the general trend seems to be independent from 
the geographical and temporal conditions. Table ~\ref{qtab2} shows the highest
deviations from tab.~\ref{qtab1} obtained in all possible combinations spanned
with TOVS data (oceans, land, day, night).
 \begin{table}[th!]
  \caption{Highest deviations from tab.~\ref{qtab1} obtained in all possible 
combinations spanned with TOVS data (oceans, land, day, night)}
   {\scriptsize
  \centering
  \begin{tabular}{|l|c|c|c|c|c|}
  \hline
Optical Depth&\multicolumn{4}{c|}{Cloud-top altitude}\\ \hline
 &$<$3km&3-7km&7-10km&$>$10km\\ \hline 
$>$2 & -5.1 & +1.6 & -0.6 & +2.7\\ \hline
1-2 & -2.9 & -0.2 & +0.4 & +0.3\\ \hline
0.1-1 & -2.0 & -0.8 & -0.4 & +0.3\\ \hline
$<$0.1 & +7.6 & +0.1 & $<$0.1 & +1.4\\ \hline
  \end{tabular}
  \label{qtab2}
   }
 \end{table}
The results of tab.~\ref{qtab1} can be classified in the following way.
OD$<0.1$ corresponds to clear sky and it accounts for $\sim$30\%. Clouds below
3 km height do not hamper the measurements as the shower maximum will develop
at higher altitudes, regardless of their OD and they account for another 
$\sim$30\%,
which gives a total of $\sim$60\% of the time when the measurement is clearly
possible. Thick (OD$>$1) and high (h$>$7km) will prevent the possibility of
measurement, and they account for $\sim$19\%. The remaining $\sim$21\% will
limit the measurement to very inclined showers (zenith angle $>$60$^\circ$, 
which by the way correspond to the best category of data in terms of light
intensity, angular accuracy and energy resolution - see \cite{yAndrea}), or to 
the study of the arrival direction analysis, as the energy estimation will be
worsened by the shower attenuation in atmosphere.\\ 
The study performed with TOVS data is important to have a first estimation of 
the uncertainty of the cloud distribution and its effects on 
shower-reconstruction capabilities, however, possible systematic effects of the 
technique employed in the TOVS measurement can not be inferred. For this reason,
the same type of study has been applied to ISCCP and CACOLO data and results 
have been compared. As previously explained, the ISCCP and CACOLO data divide the 
clouds only in low, middle and high type, without distinguish 
according to their OD.
In order to compare these data with the TOVS ones, the latter data were grouped 
only on the basis of their top altitude: clear sky, low clouds (h$<$3 km),
middle clouds (3-7km), high clouds (h$>$7km). 
Tab.~\ref{qtab3} shows the comparison between the 3 data sets in the case of lands
and oceans during day-time.
 \begin{table}[th!]
  \caption{Comparison among TOVS, ISCCP and CACOLO databases for the relative cloud occurence (\%) in the
different meteorological situations. Data refer to day-time, with a weighted average between oceans and
lands.}
   {\scriptsize
  \centering
  \begin{tabular}{|l|c|c|c|c|}
  \hline
Sky condition &\multicolumn{3}{c|}{Database}\\ \hline
 & TOVS & ISCCP & CACOLO \\ \hline 
high clouds & 32.7 & 23.3 & 17.9 \\ \hline
middle clouds & 8.4 & 16.0 & 25.0 \\ \hline
low clouds & 28.4 & 26.0 & 40.4 \\ \hline
clear sky & 30.5 & 34.7 & 16.7 \\ \hline
  \end{tabular}
  \label{qtab3}
   }
 \end{table}
Results look quite different at a first glance. However, if clear sky and low clouds are averaged
together, they give almost similar results, with a minimum of 57.1\% for CACOLO to a maximum of 60.7\%
in case of ISCCP. As a consequence also the sum of middle and high clouds gives similar results. 
More in detail, TOVS data seem to overestimate high clouds meanwhile CACOLO data tend to overestimate
the low ones. This overestimation might be due to the fact that CACOLO data are taken by ship and weather 
stations in the visual band only (so they tend to underestimate high clouds, especially in presence of low
and middle clouds), while TOVS data are taken by satellites (for a similar reason, the low and middle clouds
tend to be underestimated, because 'masked' as high clouds). ISCCP data are a sort of average of the other 
two data sets, since they are taken from satellite in the visual and infrared bands 
and this fact facilitates to distinguish the various levels. In this sense, as the TOVS data provide the
highest value for high clouds, the results presented before can be considered as a conservative estimation
of the fraction of events that could be measured by JEM-EUSO.\\
Finally, ISCCP data have been used to check the dependence of the above mentioned results according to their
geographical area. Data have been divided into 5 latitude layers, separating among equatorial area, 
tropics and middle latitudes. Results are provided in tab.~\ref{tab4}.
 \begin{table*}[th!]
  \caption{Distribution of the cloud properties (\%) for 5 different geographical areas using ISCCP data.
Data refer to daytime.}
   {\scriptsize
  \centering
  \begin{tabular}{|l|c|c|c|c|c|c|c|c|c|c|c|}
  \hline
Sky condition &\multicolumn{10}{c|}{Geographical area}\\ \hline
 &\multicolumn{5}{c|}{ocean}&\multicolumn{5}{c|}{land}\\ \hline
 & 51.6-35N & 35-15N & 15N-15S & 15-35S & 35-51.6S & 51.6-35N & 35-15N & 15N-15S & 15-35S & 35-51.6S \\ \hline 
high clouds & 23.5 & 18.9 & 27.8 & 17.2 & 20.1 & 26.1 & 19.4 & 33.5 & 24.2 & 28.1\\ \hline
middle clouds & 24.2 & 12.4 & 12.1 & 13.6 & 24.2 & 22.0 & 12.4 & 14.8 & 11.8 & 18.9\\ \hline
low clouds & 35.7 & 28.7 & 21.0 & 33.1 & 38.9 & 18.4 & 18.6 & 15.2 & 16.0 & 22.8\\ \hline
clear sky & 16.6 & 40.0 & 39.1 & 36.1 & 16.8 & 33.5 & 49.6 & 36.5 & 48.0 & 30.2\\ \hline
  \end{tabular}
  \label{tab4}
   }
 \end{table*}
In general the combination of low clouds and clear sky is slight higher onto oceans, which by the way 
account for the higher fraction of time. High clouds are particularly frequent in the equatorial region.
This is normal, as it is correlated also with the big storms occurring in that area.

\section{Conclusions}
Flying on the ISS, during its operation, JEM-EUSO will experience all possible weather conditions inside
its field of view. In order to estimate the effective aperture of the detector, one key point is the
evaluation of the role of clouds, in particular their frequency as a function of altitude and optical depth
for the different geographical areas of the planet.\\ 
The probability of occurrence of a defined atmospheric
condition has been assessed in this study by means of different meteorological databases: TOVS, ISCCP and CACOLO data. 
The peculiarity of the observation of cosmic rays from space is
the fact that the presence of low clouds (h $<$ 3 km) is de facto equivalent to clear-sky conditions as the shower maximum
will be located at altitudes higher than the cloud top altitude. 
The results of the present analysis, which is based on visible
and infrared data, indicate that showers will develop in the
atmosphere in clear-sky conditions
 for at least $\sim$60\% of the time. The results are marginally
dependent on the database adopted in the analysis ($\sim$5\%).
A precise evaluation of the effective fraction of time in
which shower observation will be possible as a function of
the arrival direction of the primary cosmic rays and how this
will impact on the exposure of the experiment is reported 
in \cite{yLupe}.\\ 
In the future we plan
to extend the analysis by using other databases such as MERIS 
and CALIPSO. Furthermore,
CALIPSO data will be used to assess the effects of the ISS 
orbital displacement on the inferred cloud
structure along the effectively probed line of sight.

\vspace*{0.5cm} \footnotesize{
{\bf Acknowledgement:} 
This work was partially supported by the Italian Ministry of Foreign Affairs, General Direction for the
Cultural Promotion and Cooperation.}

\

\clearpage


\newpage
\normalsize
\setcounter{section}{0}
\setcounter{figure}{0}
\setcounter{table}{0}
\setcounter{equation}{0}



\title{A study of different cloud detection methods for the JEM-EUSO
atmospheric monitoring system}

\shorttitle{A. Anzalone \etal cloud detection for JEM-EUSO}

\authors{A. Anzalone$^{1}$, M. Bertaina$^{2}$, R. Cremonini$^{2}$, M.D. Fr\'{\i}as Rodr\'{\i}guez$^{3}$, F. Isgr\'o$^{4}$ for the JEM-EUSO collaboration}
\afiliations{$^1$INAF-IASF, Palermo, Italy\\
$^2$Dipartimento di Fisica, Universit\'a degli Studi di Torino, Italy\\
$^3$SPACE \& AStroparticle (SPAS) Group, UAH, Madrid, Spain \\
$^4$Dipartimento di Scienze Fisiche, Universit\'a degli Studi di Napoli Federico II, Italy}
\email{anna.anzalone@ifc.inaf.it}

\abstract{
The observation of the atmosphere is a crucial task for the JEM-EUSO
mission, and a module for the atmospheric monitoring is included
in the design of the whole system. In this paper the retrieval of
cloud coverage in the field of view of the telescope is addressed
considering both radiative methods commonly used in the meteorological
field and methods of image analysis, with the aim of studying the
feasibility of these approaches to the data that the JEM-EUSO infra
red camera will provide. The complementarity of the two approaches
will be further investigated, together with  a different set of
techniques, to contribute to achieve the best cloud estimation in
JEM-EUSO. 
}

\maketitle

\section{Introduction}
The strength of the fluorescent light and the Cherenkov signal
received from EAS, as well as the reconstruction efficiency and
errors, depend on the transparency of the atmosphere, the cloud
coverage and the height of the cloud top. A crucial task for the
success of the JEM-EUSO mission 
\cite{Tascio11} is to observe the
conditions of the atmosphere in the field of view of the telescope. To
this end a dedicated atmospheric monitoring (AM) system
\cite{Neronov11} is being designed. The system includes an infrared
camera, that will be used to estimate cloudiness and height maps in
the field of view of the telescope.

This paper reports on current work to identify optimal cloud detection
algorithms from infrared data, that will be implemented into the
JEM-EUSO observing system 
for accurate estimations of cosmic-ray energy.  To this end here we
revise the performance of different methods for cloud
detection: threshold algorithms, radiative, and methods exploiting
image analysis techniques. The experiments are run on scenes under
different conditions, retrieved by operational atmospheric sensors
similar to the JEM-EUSO atmospheric monitoring system. 

\section{Radiative methods}
Geostationary (i.e. GOES, MSG) and LEO satellites (i..e. Terra/Aqua,
HIRS) provide multi-spectral observations with good spatial and
temporal resolution. CALIPSO mission combines an active lidar
instrument with passive infrared and visible images to probe the
vertical structure and properties of thin clouds and aerosols over the
globe. The cloud mask (CMa) allows the identification of cloud free areas
where other products (total or layer precipitable water, stability
analysis imagery, snow/ice cover delineation) may be computed. The
main aim of the CMa is therefore to delineate all cloud-free pixels in
a satellite scene with a high confidence. In addition, the typical CMa
product provides information on the presence of snow/sea ice, dust
clouds and volcanic plumes.
SEVIRI is a multi-band sensors operating
on MSG satellite series by EUMETSAT \cite{alab3}: starting from SERIVI
radiance observations, it has been developed an algorithm for
identifying cloud presence and cloud contamination. The algorithm is
based on several and differential band threshold tests, using only
infra red bands as JEM-EUSO work during nighttime: some difference band
tests are specific for thin cirrus detection. Thresholds depend on
pixel background (land, water and coast) and on Numerical Weather
Prediction (NWP) model temperature at surface and at standard levels. 
Starting from four categories of tests a probability of clear sky is
defined as follow:\linebreak  

\begin{displaymath}
cloud\,sky\,probability=^{3}\sqrt{P_{ir} \cdot P_{thin} \cdot P_{diff-ir}}
\end{displaymath}

Where $P_{ir}$ test is the probability that for a given band/threshold
the pixel is cloudy or cloud contaminated; $P_{thin}$ and
$P_{diff-ir}$ have the same meaning but for band differences and thin
cirrus specific tests. As JEM-EUSO will work with at least with a two
band infrared camera, the tests performed on SEVIRI have been limited to
10.8 and 12.0 $\mu$m bands. Single infrared algorithm can detect only
thick and extended clouds, with better performance during summer and
over warm sea due to high thermal contrast \cite{alab1}. 

 \begin{figure}[!t]
  \vspace{5mm}
  \centering
  \includegraphics[width=.5\textwidth]{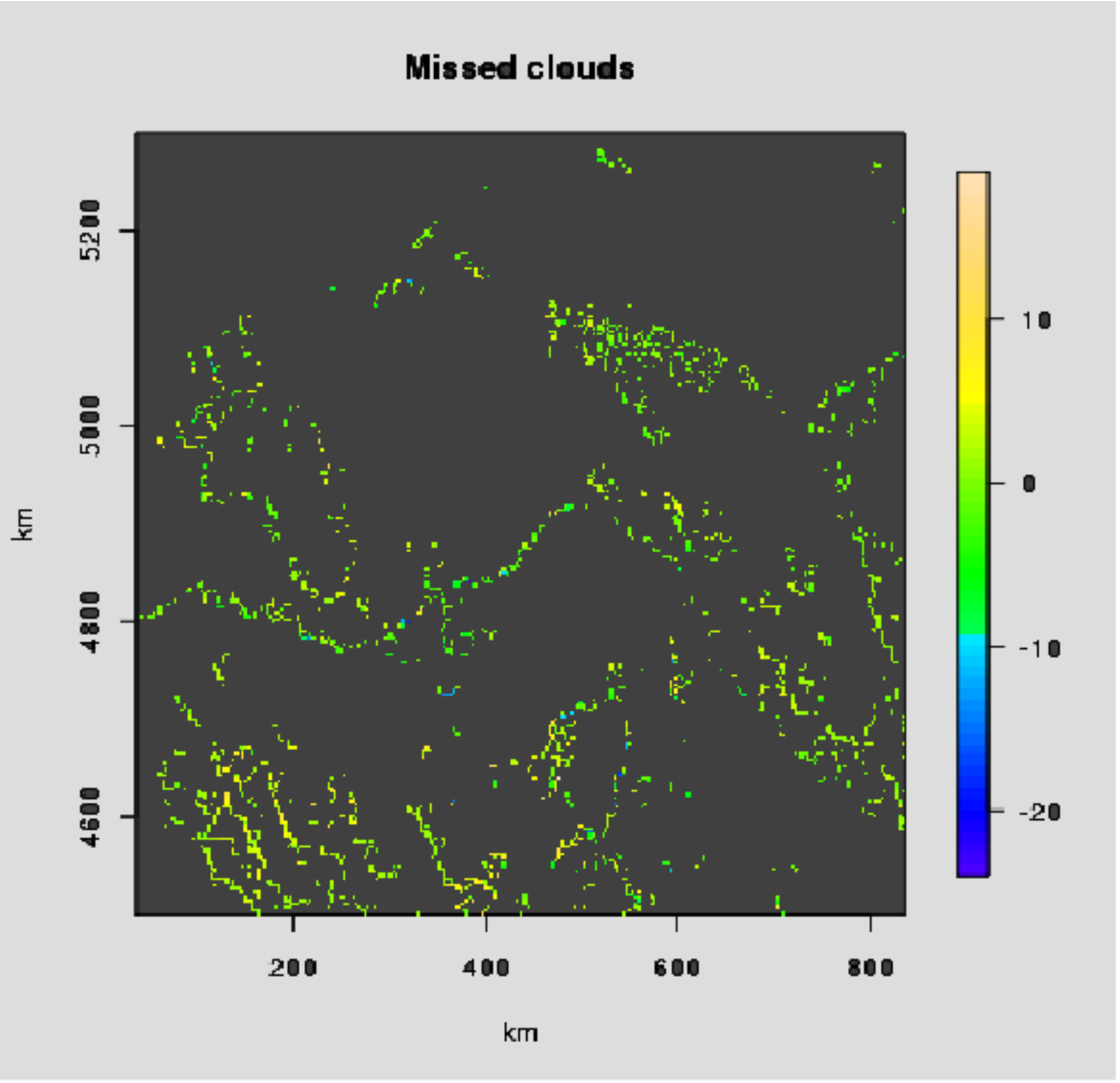}
  \caption{Single IR band cloud detection respect all IR channel CMa}
  \label{afig01}
 \end{figure}

Figure \ref{afig01} shows undetected clouds when only 10.8 $\mu$m IR band is
used: major problems occur near coastal borderline, with thin cirrus
and at border of cloud desk, where pixels and not
fulfilled. Well-known split window channels are essential to detect
thin cirrus or broken clouds and to estimate Earth's surface and cloud
top temperatures. The 12.0 $\mu$m band is more sensitive for high thin
cirrus but it is not easy to recognise through visual inspection,
while it is highlighted by band difference \cite{alab2}. 

 \begin{figure}[!t]
  \vspace{5mm}
  \centering
  \includegraphics[width=.5\textwidth]{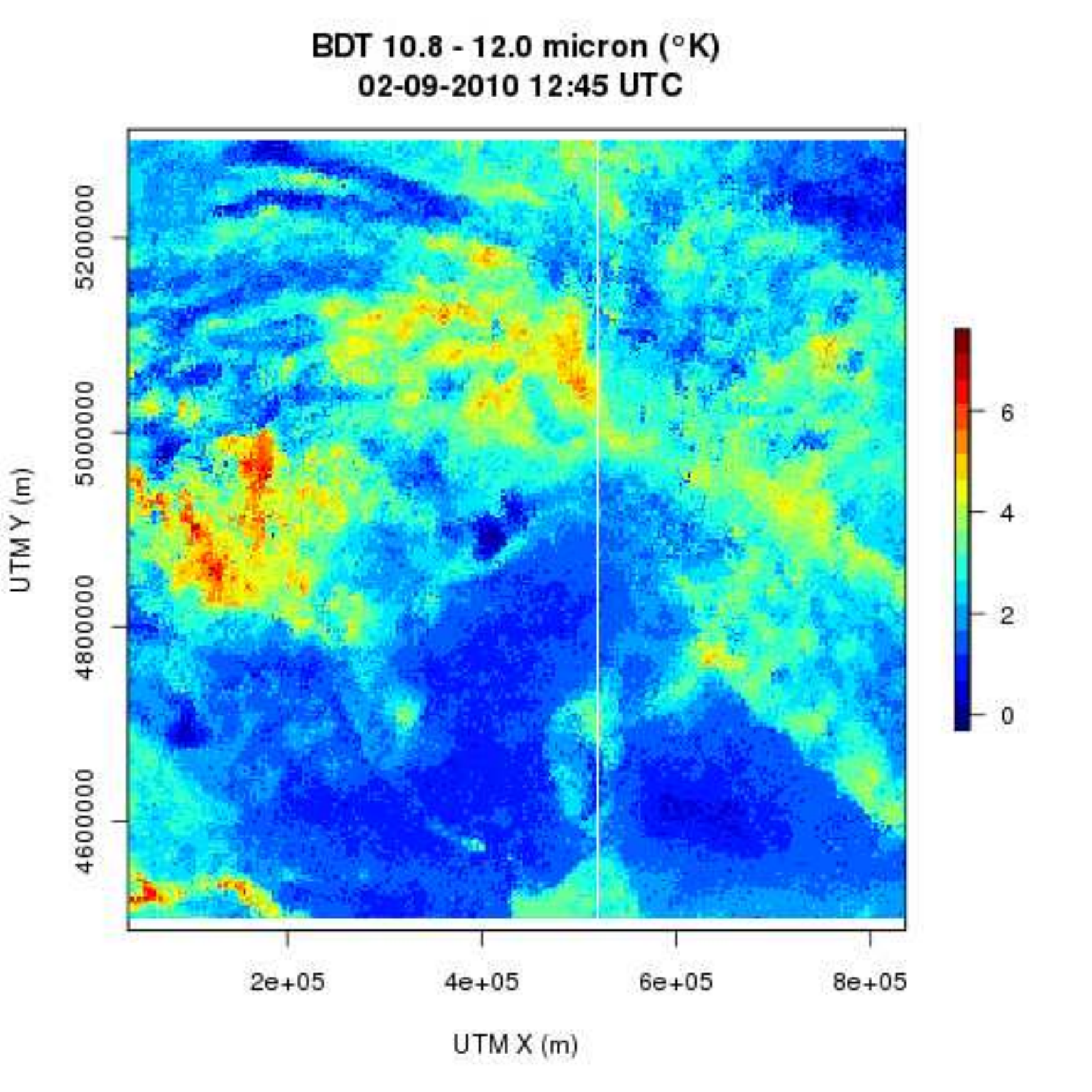}
  \caption{Band difference from SEVIRI for 10.8 $\mu$m and 12.0 $\mu$m at 12:45 on 2th sept 2010}
  \label{afig02}
 \end{figure}

Figure \ref{afig02} shows the difference between Brightness Temperature (BT)
at 10.8 $\mu$m and 12.0 $\mu$m at 12:45 on 2th sept 2010: BT
differences greater than 3$^{\circ }K$ distinguish between
semi-transparent thin clouds and thick ones.

\section{Image analysis methods}
The methods discussed in this Section exploit the image content of the
infrared data, not considering its physical meaning. That is that the
content of a pixel is regarded as a colour information only, and not
as related to the temperature. 

In this Section we consider two different methods: a supervised image
segmentation algorithm, and a second method that follows a geometric
approach.

\subsection{Feature based method}
In this  method the classification of the cloudy pixels is
performed using a state of the art machine learning tool: a Support
Vector Machine (SVM) \cite{Cristianini2000}, for which we chose a
Gaussian kernel. In particular we use a public
available implementation of the SVM \cite{Chung2001}.

The classification using SVM is a supervised method, and it needs a
training set of data. This means that a relatively large number of
pixels must be manually labelled. A simple graphical interface has been
created to facilitate this task.

The classification is not performed in the gray-level space, but each
image pixel is mapped into a higher dimensional space, that is
usually called feature space. In our case the feature vector
associated to each pixel (i,j) is
\begin{displaymath}
\mathbf{f_{i,j}} = (v, \mu, \sigma, D_x, D_y, h_1, h_2, h_3, h_4, h_5, h_6,
h_7, h_8)
\end{displaymath}
where $v$ is the gray-level of pixel (i,j), $\mu$ and $\sigma$ are
mean and standard deviation respectively in a $5 \times 5$
neighbourhood centred in the pixel, $D_x$ and $D_y$ are the gradient
components, and $h_1$
to $h_8$ are the entities of the eight-bin histogram of the $5 \times 5$
neighbourhood in the image.\\

\begin{figure}[htb]
\begin{center}
\includegraphics[width=0.5\textwidth]{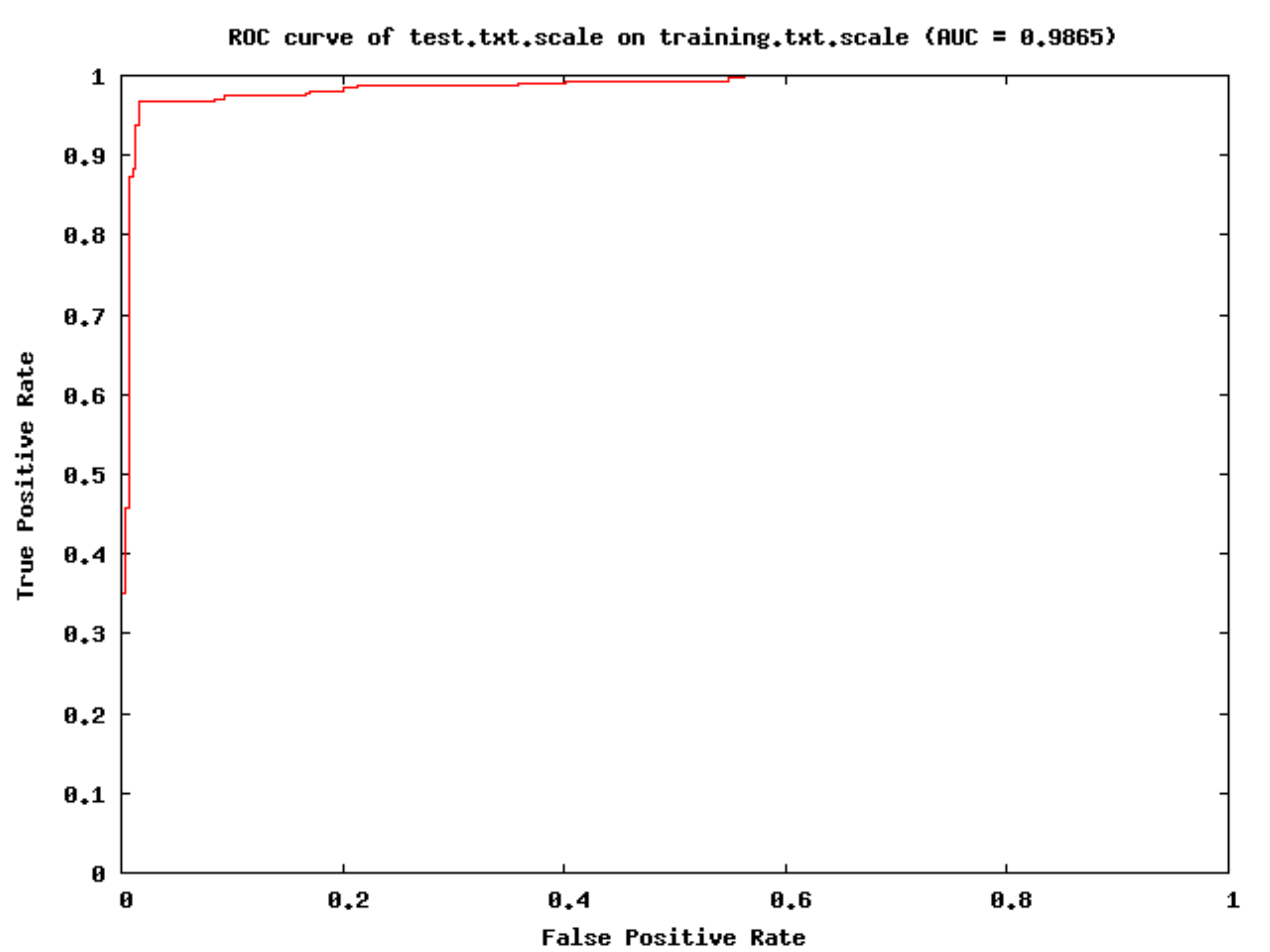}
\end{center}
\caption{ROC curve for the SVM classifier.}
\label{fig:med2.roccurve}
\end{figure}

The classifier has been tested on more than 2000 images from different
sensors. The training
set includes $610$ points belonging to both \emph{cloud} and \emph{not-cloud} classes,
manually selected from the first  $1000$ frames of the image set.
From the remainder of the sequence we selected further $609$ points
that have been used as test set for evaluating the performance of the
classifier. As it can be seen from the ROC curve shown in Figure
\ref{fig:med2.roccurve} the performances are particularly good, since
the test set is temporally close to the training set. We are planning
to use more data for a better evaluation.

\subsection{Stereo based method}
In this section the possibility of retrieving maps of cloudiness from maps of heights is presented supposing that stereo acquisition was enabled by the use of the infra red camera.
Using stereo methods, the depth of the imaged points can be recovered
from two, or more, images, in this way, in presence of clouds, the cloud-top height (CTH) can be recovered \cite{Seiz06,Muller07,AnzIsgrTeg04}. Stereo
could be achieved in JEM-EUSO exploiting the ISS
movement. While the ISS flies along its orbit, the IR camera acquires
an image of the FoV at every fixed time interval.

Exploiting this information, and the fact that the images can be
geo-located, we can mark as clouds all those pixels for which the recovered height
is higher than the altitude of the corresponding ground. 

We analyse the feasibility of this method studying the
theoretical reconstruction error for the depth, as a large error in
reconstruction may lead, especially for lower clouds, to
mis-classification.

The depth $Z$ of a point can be obtained by triangulation as
\begin{displaymath}
\hat{Z} = \frac{b}{\hat{d}_x}
\end{displaymath}
where $b$ is the baseline of the stereo system (the distance covered
by the ISS between the two views), and $\hat{d}_x$ is the estimated
disparity \cite{Trucco98}. If $d_x$ and $Z$ are the true disparity and
the true depth respectively we have that 
\begin{displaymath}
\hat{d}_x = d_x + \delta x
\end{displaymath}
and
\begin{displaymath}
Z = \frac{b}{d_x}
\end{displaymath}

We can
write the depth error $\delta Z$ as a function of $\delta x$ by Taylor expansion as
\begin{displaymath}
\delta Z = \frac{b}{d_x^2} \delta x = \frac{Z^2}{b} \delta x
\end{displaymath}
which shows that for a fixed baseline $b$ and $\delta x$ then the
error in the depth measurement rises as the square of distance from
the camera. Therefore we need a large baseline $b$ to get a good depth
resolution, but also we can expect a poor depth resolution for distant objects.

\begin{figure}
\begin{center}
\includegraphics[width=0.5\textwidth]{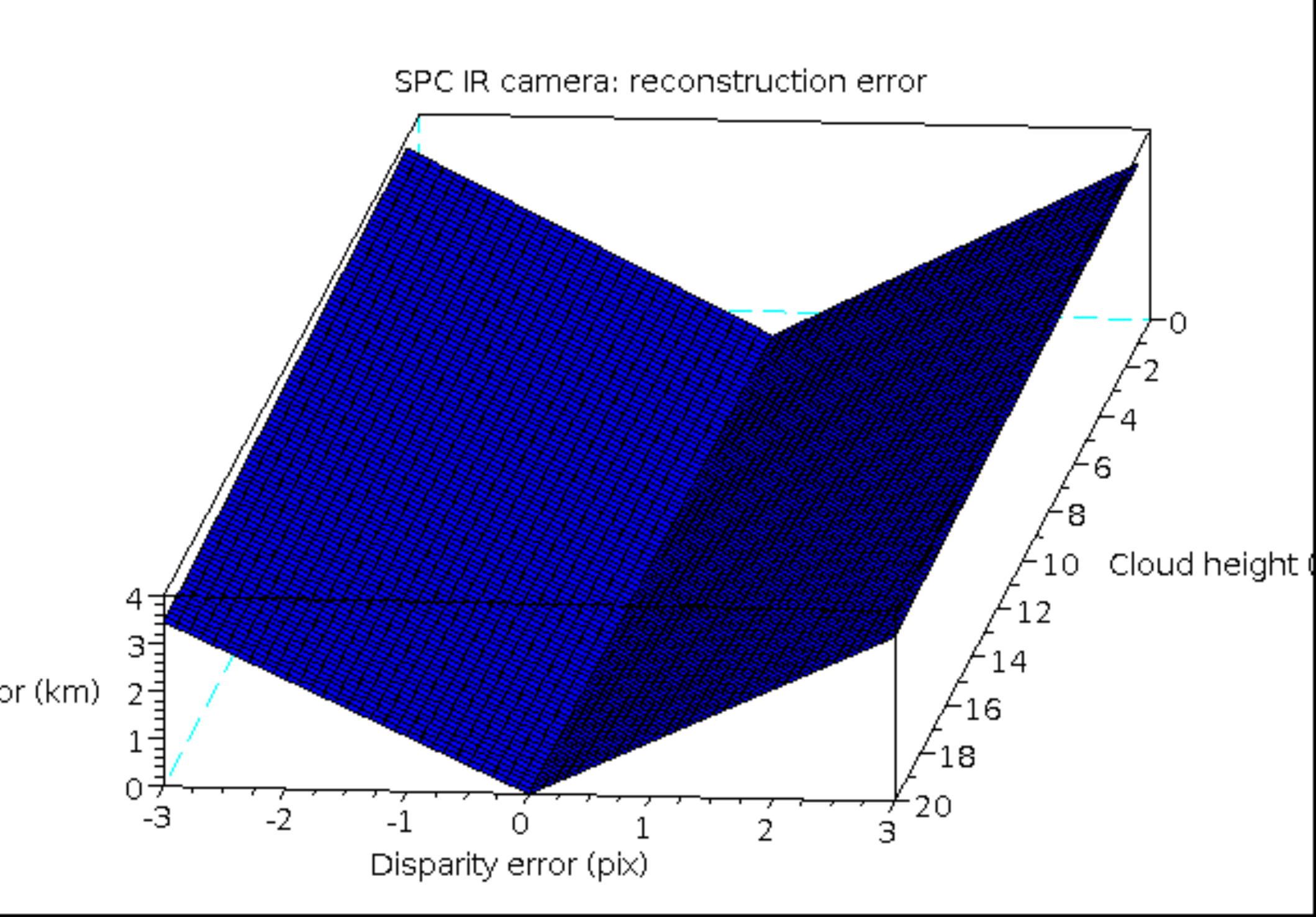}
\end{center}
\caption{Plot of the CTH estimate error against the disparity error and the
  true CTH. ISS station altitude at 430 Km.}
\label{fig:error.plot}
\end{figure}

The error function is plotted in Figure \ref{fig:error.plot}. For this
simulation we used the specs for the infrared camera given in Table
\ref{tab:SPIR}, the altitude of the sensor fixed at 430 Km, and a time
interval between the two images of 32 sec, which ensures a 50\%
overlap.

\begin{table*}
\begin{center}
\begin{tabular}{|c|c|c|c|c|c|c|}
\hline
 & FoV & IFoV & Pixel & Number of & Focal & Pixel \\
 &     &      & resolution & of pixels & length &  pitch \\
\cline{2-7}
 & $60^\circ$ & $0.1^\circ$ &  & $640 \times 480$ & 15 mm& 20$\mu$m \\
\hline
ISS$_h$=350 Km & & & $\simeq 0.58$ Km& & & \\
\hline
ISS$_h$=430 Km & & & $\simeq 0.72$ Km& & & \\
\hline
\end{tabular}
\end{center}
\caption{Specification for the IR camera used for this experiment}
\label{tab:SPIR}
\end{table*}

From the analysis of the results of the simulations we can conclude
that we have an accuracy within 500 meters with a disparity error of
0.5 pixels, that can be achieved for most pixels with a good matching
strategy. With higher disparity error, say 1-2 pixels, we have an
error in the depth estimation that is within 2 Km.

\section{Conclusions}
Radiative and image methods for cloud detection and cloud height
estimation have been preliminary considered as candidates for JEM-EUSO
atmospheric monitoring system. While performance of radiative methods
depend on IR camera thermal resolution and available bands - especially when the scene is
thin cirrus contaminated - the image methods described in this contribution depend on spatial
angular resolution of the sensors, and on the quality of the features or on the quality of the matching. 
Both radiative and image methods need to be deeper
investigated and moreover other well known techniques and features in image
analysis will be investigated. Radiative and image approaches can be
considered as complementary, an their integration to best achieve cloud coverage estimation for JEM-EUSO will be considered.    




  \includepdf[pages=-]{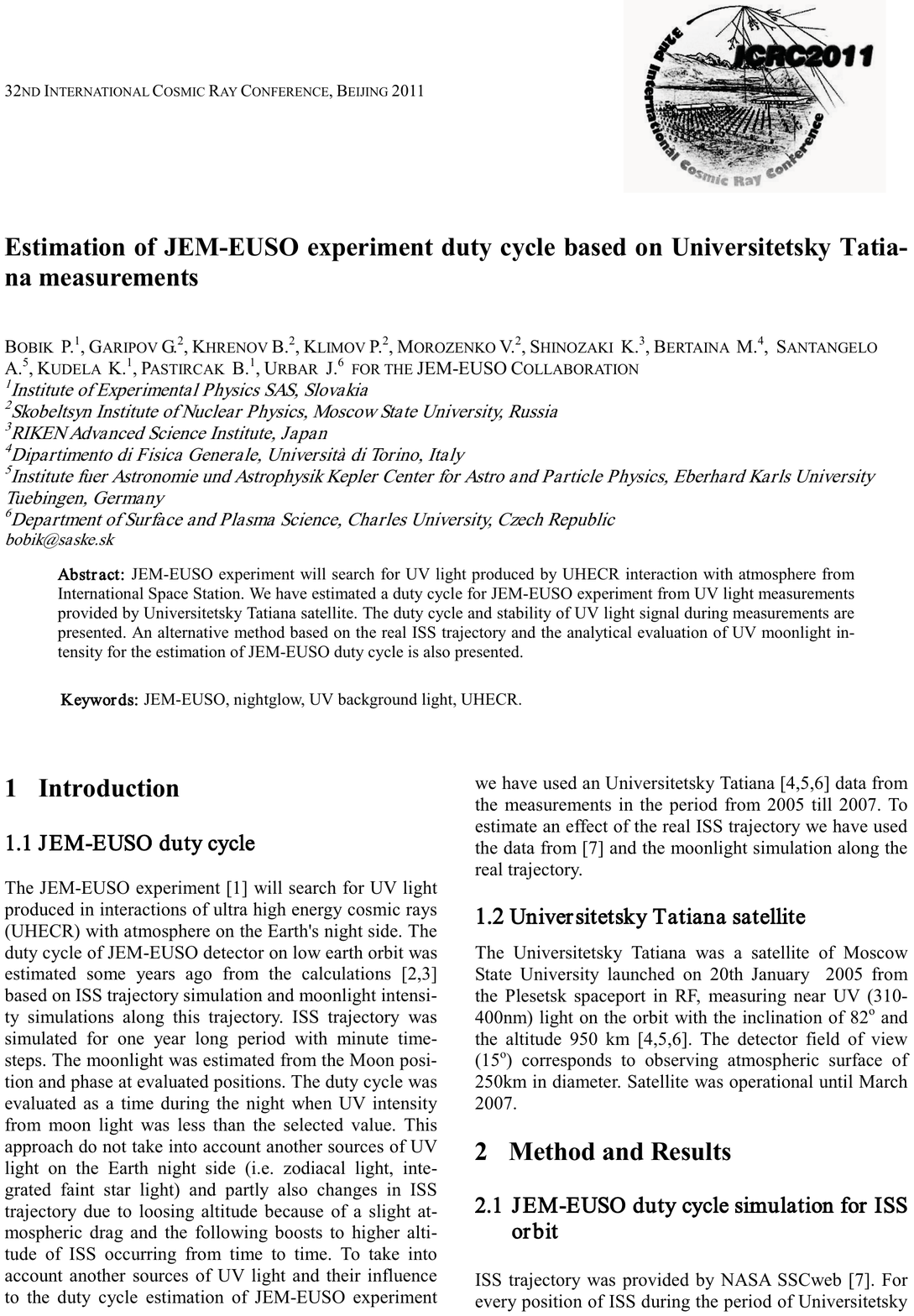}

\newpage
\normalsize
\setcounter{section}{0}
\setcounter{figure}{0}
\setcounter{table}{0}
\setcounter{equation}{0}



\title{Observation of Ultra-High Energy Cosmic Rays in cloudy conditions by the JEM-EUSO Space
 Observatory}

\shorttitle{S\'aez Cano \etal, Observation of UHECR in cloudy conditions by JEM-EUSO}

\authors{G. S\'aez Cano$^{1}$, J. A. Morales de los R\'ios$^{1}$, 
K. Shinozaki$^{2,1}$, F. Fenu$^{3}$, H. Prieto$^{1}$, N. Pacheco G\'omez$^{4}$, 
J. Hern\'andez$^{1}$, L. del Peral$^{1}$, 
A. Santangelo$^{3}$ $\&$ M. D. Rodr\'iguez Fr\'ias$^{1}$ for the JEM-EUSO Collaboration}
\afiliations{$^1$SPace \& AStroparticle (SPAS) Group, University of Alcal\'a, Madrid. Spain.\\ 
  $^2$RIKEN, 2-1 Hirosawa, Wako 351-0198, Japan.\\
  $^3$Institut f\"ur Astronomie und Astrophysik, Eberhard-Karls Universit\"at T\"ubingen, T\"ubingen, Germany.\\
  $^4$Instituto de F\'isica Te\'orica (IFT), Universidad Aut\'onoma de Madrid, Spain. }

\email{lupe.saez@uah.es}


\abstract{Source of Ultra-high Energy Cosmic Rays (several times $10^{19}$ 
eV) are still unidentified.  Overcoming their extremely small fluxes, a 
detector with huge observation areas is needed to investigate the energy and 
arrival direction distribution of EECRs. JEM-EUSO is a unique experiment 
that will be located in the International Space Station to observe extensive 
air showers (EAS) by monitoring night part of Earth atmosphere. In addition 
to clear sky condition, the extensive air showers in cloudy 
condition are also observable by taking advantage of the certain fraction
of EAS develop above the cloud. In the preset work, using Monte Carlo 
simultions for test clouds, the cloud impact to the trigger efficiency was 
estimated taking into account the statistics of cloud property.}


\maketitle

\section{Introduction}

Cosmic rays origin is not identified, specially for Ultra-High Energy Cosmic 
Rays (UHECRs), despite of the limited numbers of astrophysical objects that 
can accelerate particles to such energies \cite{uHillas}.  Properties of the 
primary UHECR can be measured by the observation of Extensive Air Showers 
(EAS). These EAS are developed when cosmic rays come through the atmosphere. 
The primary energy is shared among secondary particles. Most of them are 
electrons which carry about 90$\%$ of the primary energy. These electrons 
excite nitrogen molecules in the atmosphere that results in fluorescence light 
through the de-excitation of the molecules. Also Cherenkov component is 
produced, due to the relativistic velocity of the particles. These components 
have been measured by ground-based ultra-violet (UV) telescopes.
However, with a steep power-law energy spectrum and possible 
Greisen-Zatsepin-Kuzmin effect, UHECR flux at highest 
energy (above ${\sim 5\times 10}^{19}$ eV) \cite{ugzk1,ugzk2}
is such small, that their origin cannot be investigated by these 
ground-based experiments.

JEM-EUSO (Extreme Universe Space Observatory on Japanese Experiment 
Module) is a new type, space-based experiment that will be launched 
in 2017, aiming to identify origin sources by detecting UHECRs at 
large statistics \cite{ueuso_ICRC,ueuso1}. JEM-EUSO telescope will cover 
a much larger area than ground-based experiments with a wide field of 
view (FoV) of 60$^{\circ}$. From the orbit on the International Space 
Station (ISS) an altitude of $\sim 400$ km, it will search the 
Earth's atmosphere as a detector for light produced by EAS. In order 
to determine the energy and arrival direction of the primary particle 
as well as its composition, the light profile is needed to be measured.  
This profile depends on atmospheric conditions, such as absorption and 
scattering. The atmosphere is also the source of UV background such as 
subsistence airglow and  transient luminous events, artificial sources, 
etc. Therefore, JEM-EUSO will need a dynamical trigger system capable 
of continuously adapting the triggering requirements 
\cite{uCatalano,uBayer}. Furthermore, the FoV of JEM-EUSO varies as ISS 
orbits with sub-satellite speed of $\sim 7$ km/s and therefore presence 
and properties of clouds change significantly. To acquire such 
information, LIDAR (LIght Detection And Ranging) device and an infrared 
(IR) camera will be installed on JEM-EUSO \cite{uAMS_ICRC,uIR}.  The 
former will measure transmittance as a function of the altitude and the 
latter will be accommodated to obtain cloud coverage overview and 
cloud-top altitude ($H_C$) in JEM-EUSO's FoV that provide data for 
evaluate exposure of the observation.

In this paper, the impact to the trigger aperture was investigated by 
Monte Carlo 
simulation taking into account cloud conditions.  The presence of 
the clouds may affect varying by their altitude and optical depth $\tau.$ 
The effect due to clouds may also depend on the fraction of EAS develops 
above the typical altitude of cloud over the orbit. Effectively observing 
such EAS events helps increase the statistics of UHECR events.  
As conclusion, the cloud impact to the exposure will be presented. 

\section{Simulations}


In this work, ESAF (Euso Simulation and Analysis Framework) 
\cite{uESAFpaper} was used. It is a software framework to simulate 
space-based cosmic observations, including showers generation, 
emission and transport of photons, ray trace of optics, photodetector response and telemetry, as well as reconstruction. 
Key parts of ESAF were developed in EUSO project \cite{ueus} and nowadays, 
it is adapted and optimized for JEM-EUSO instrument \cite{uESAF2}. 

In the ESAF, EAS event is generated along with fluorescence and 
Cherenkov photons emission and their propagation in the atmosphere.
In present work, fluorescence yield, one of uncertainty in energy scale, 
is assumed by the measurement of Reference \cite{ungn} from the 
available options. Even in case of clear sky condition, UV photon propagation 
through atmosphere severely involves Rayleigh scattering 
and absorption by ozone in shorter wavelengths ($\sim 320$ nm).
The transmittance of these processes are modeled by 
LOWTRAN package \cite{ulwtrn}.

In consideration of the effect of photon scattering in clouds which
consists of droplet below $\sim 8$ km altitude, Mie scattering is more 
dominant since the scattering particle size well larger than wavelength. 
Scattering can be considered as independent of wavelength range of our 
interest (300-450 nm). The same behavior is observed for cirrus, made 
of ice crystals \cite{ubun}. In the software, analytical formulation of 
scattering process including phase function is modeled and implemented 
in ESAF \cite{uESAFpaper}.

To include clouds in ESAF, there are two different options in its 
atmospheric model: with TOVS (TIROS Operational Vertical Sounder) 
database \cite{utovs} including $\tau$ and $H_{C}$ or as a uniform and 
homogeneous layer. The database was analysed to understand the global 
distribution of clouds within the range of the JEM-EUSO orbit and was 
analyzed in \cite{ugar}. For the last option, physical parameters 
considered for the 
cloud layer are the optical depth $\tau$, that yields  transparency
by $\exp(-\tau)$, the top altitude of the cloud and its physical thickness. 
For our study, the latter option was chosen for the discrete test values 
in $\tau$ and $H_{C}$. 

\section{Results}

\subsection{Shower simulation in cloudy conditions}

\begin{figure}[t]
  \vspace{5mm}
  \centering
  \includegraphics[width=3.25in]{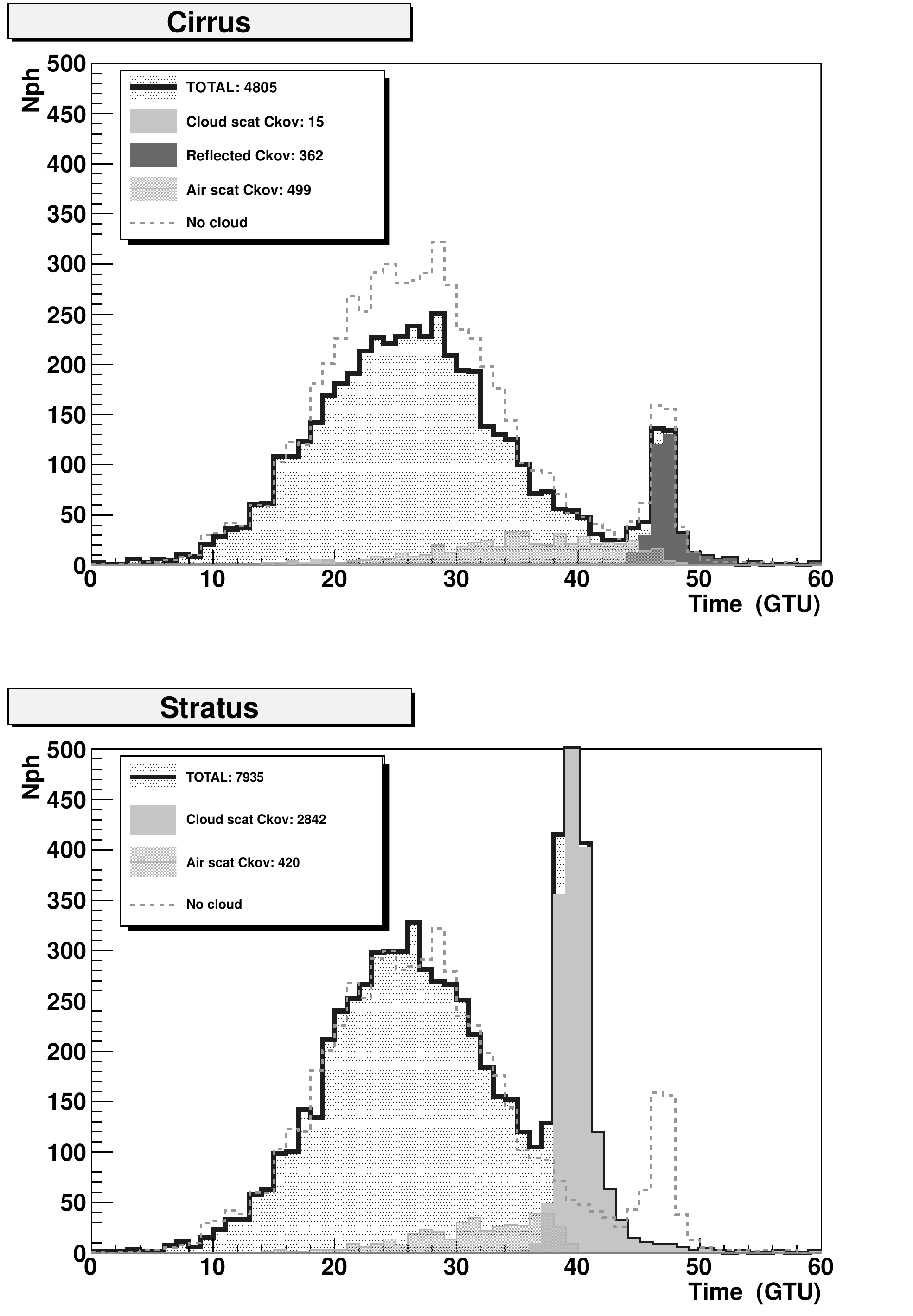}
  \caption{Light curve (arrival time distribution of photons to the 
    telescope pupil) of typical EAS events. Note that the horizontal 
    axis is in unit of GTU (gate time unit) corresponding 2.5 
    $\mu$s. Top and bottom panels correspond to the cases of cirrus- 
    and stratus- like test clouds, respectively. In each panel, dark 
    most shaded histogram denotes the total number of photons.  Other 
    histograms indicates the different components of Cherenkov photons.
    For comparison, light curve for clear sky for EAS
  at similar energy is drawn (dashed histogram).}
 \end{figure}

In Figure 1, light curves (arrival time distribution of photons 
to the telescope pupil) of typical EAS events with zenith angle 
of $60^\circ$ are shown for cirrus- (top panel) and stratus- like test 
clouds (bottom). Note that the horizontal axis is in unit of GTU (gate time unit = 2.5 $\mu$s.
In each panel dark most shaded  histogram denotes the total number of 
photons.  Other histograms indicate the different components of 
scattered Cherenkov photons in atmosphere of from cloud or Earth's
surface . For comparison, light curve for clear sky for EAS
at similar energy is drawn (dashed histogram).

For cirrus-like cloud at lower altitudes, signals from EAS are attenuated 
according to the optical depth, while the shower image and its time 
evolution will allow the arrival direction analysis. The scattered 
signals of Cherenkov from the ground is also observed.
  
For stratus-like clouds with large $\tau$ at lower altitudes, most of 
signals from EAS are observed without attenuation when the altitude of 
the cloud is well below the altitude of EAS development. Such clouds 
also produce a very intense reflected Cherenkov signals and the 
detected signal is even larger, due to higher albedo of clouds, than 
that for the clear sky case. This may enhance the better capability of 
triggering for particular case such as low zenith angle event. It is more 
pronounced in reconstruction 
of the EAS geometry since the location of the impact on the cloud
is more accurately determined.

\subsection {Trigger efficiency in cloudy conditions}

In order to evaluate the impact of clouds in FoV into trigger efficiency, 
shower simulations for different cloudy cases were made. To characterize 
the test cloud property, four altitudes have been considered ($H_C = 2.5
$ km, 5 km, 7.5 km and 10 km), as well as four optical depths 
($\tau$= 0.05, 0.5, 1.5 and 5). 
For each of the sixteen cases, incident angles from 0$^{\circ}$ to 
90$^{\circ}$ and energies of the primary particles (protons) with
energies of $\log E = 19.5$ to 21 have been considered. For comparison, 
simulation for clear sky case was also made. 

In this paper, `trigger efficiency' $\epsilon(E)$ is referred to the 
ratio to the trigger aperture at energy $E$ in comparison to the nominal 
semi-saturated aperture. The semi-saturated aperture is meant to be the 
product of solid angle ($\pi$ for $\theta = 0^\circ-90^\circ$) and 
observation area determined by the result of the optical ray trace 
simulations \cite{uzuc,uks}. Note that the efficiency can be 
slightly higher than 1 since some EAS that cross a part of FOV may 
trigger. To quantitatively estimate the effect of clouds, we first calculate the ratio
$\epsilon$ of given cloudy condition to that of clear sky case. 
The ratio, to be called `cloud impact' hereafter that represents 
the ratio of the number of events in comparison to the one expected for 
clear sky condition. For a given cloud condition $(H_0,\tau)$, the average 
cloud impact $\epsilon(E; H_0,\tau)/\epsilon(E;\rm clear)$ is defined
taking into account the assumed UHECR flux.

\begin{table}
\caption{Average cloud impact for different types of clouds 
for energy ranges above $5\times 10^{19}$ eV (top) and $7\times 10^{19}$ eV
eV (bottom). In each case, a differential spectrum of $dN/dE \propto 
E^{-3}$ was assumed.}
\label{table_single}
\begin{center}
\begin{tabular}{l|cccc}
\hline
\hline
& \multicolumn{4}{c}{$H_C$}\\
\cline{2-5}
\parbox{2.5cm}{$E>5\times 10^{19}$ eV} & 2.5 km &  5 km   &  7.5 km & 10 km\\
 \hline
$\tau=5$    & \multicolumn{1}{c|}{88\%} & \multicolumn{1}{c|}{66\%} & 37\% & 18\%     \\
$\tau=1.5$  & \multicolumn{1}{c|}{89\%} & \multicolumn{1}{c|}{69\%} & 43\% & 26\% \\
\cline{4-5}
$\tau=0.5$  & \multicolumn{1}{c|}{88\%} & 82\% & 74\% & 70\%  \\
$\tau=0.05$ & \multicolumn{1}{c|}{90\%} & 89\% & 89\% & 90\%   \\
\hline\hline
\multicolumn{5}{c}{}\\
\hline\hline
& \multicolumn{4}{c}{$H_C$}\\
\cline{2-5}
\parbox{2.5cm}{$E>7\times 10^{19}$ eV} & 2.5 km & 5 km & 7.5 km & 10 km \\
\hline
$\tau=5$    & \multicolumn{1}{c|}{98\%}& \multicolumn{1}{c|}{77\%} & 44\% & 21\%     \\
$\tau=1.5$  & \multicolumn{1}{c|}{99\%} & \multicolumn{1}{c|}{83\%} & 54\% & 39\% \\
\cline{4-5}
$\tau=0.5$  & \multicolumn{1}{|c|}{100\%} & 95\% & 88\% & 84\%  \\
$\tau=0.05$ & \multicolumn{1}{|c|}{99\%} & 100\% & 100\% & 99\%    \\
\hline\hline
\end{tabular}

\end{center}
\end{table}

In Table~\ref{table_single}, the average cloud impacts are 
summarized for the different tested clouds with different energy 
thresholds of $5\times 10^{19}$ (top) and $7\times 10^{19}$ eV
(bottom) with an assumed differential spectrum of $dN/dE \propto E^{-3}$.

In case of optically thick clouds with $\tau \geq 1$,
the presence of clouds affect the trigger efficiency depending on
$H_C$. Especially high-altitude clouds absorb EAS signals emitted beneath
the cloud that significantly result in lowering the trigger efficiency. 
In the middle altitudes such as $\sim 5$ km, the influence
of the clouds are limited to EAS from lower zenith angles, which
develop even lower altitudes. 

In the presence of similarly high clouds but with $\tau < 1$, signal 
from EAS below such clouds is only attenuated by a factor of 
$\exp(-\tau)$ and the effect to the trigger efficiency is limited.

If $H_C$ is well below the altitudes where EAS develops, the clouds 
do not attenuate the EAS signals.

Comparing different energy thresholds, the difference of the 
cloud impact slowly increases with energy, while it stays marginal 
in the energy of interest. 

\section{Discussion}

In order to estimate the overall impact due to clouds, one needs
to take into account how often the different types of clouds appear 
in FoV. From the TOVS data analysis for JEM-EUSO orbit, statistical
distribution of clouds is summarized in Table~\ref{tab:fig} (see 
\cite{ugar} for further details). The undesired clouds such as 
one with $\tau >1$ and $H_C>7$ km accounts for 20\%, while $\sim$ 60\%
cases are only low altitude clouds with $H_C <3$ km whose influence to 
EAS is limited.

\begin{table}
  \caption{Statistical distribution of clouds for $\tau$ and $H_C$ 
    from TOVS database \cite{utovs}  analyzed taking in
account JEM-EUSO orbit \cite{ugar}.} 
  \label{tab:fig}
  \begin{center}
    \begin{tabular}{l|cccc}
      \hline
       & \multicolumn{4}{c}{$H_C$}\\
      \cline{2-5}
      $\tau$& $<3$ km & 3--7 km 
      & 7--10 km & $>10$ km \\
      \hline
      $ >2$  & 17.2\%  & \multicolumn{1}{|c|}{5.2\%}  & 6.4\%  & 6.1\%     \\
      $ 1-2$  & 5.9\%  & \multicolumn{1}{|c|}{2.9\%}  & 3.5\%  & 3.1\% \\
      \cline{4-5}
      $0.1-1$  & \multicolumn{1}{c|}{6.4\%}  & 2.4\%   & 3.7\%  & 6.8\%  \\
      $< 0.1$ & \multicolumn{1}{c|}{29.8\%} &  0.03\% & 0.01\% & 1.2\%    \\
      \hline
    \end{tabular}
  \end{center}
\end{table}

By the convolution of $\epsilon(E; H_C,\tau)$ with such information 
of cloud property distribution, the expected cloud impact on 
the trigger efficiency is obtained.  In presence of cloud, however,
the triggered events are needed to be selected with proper criteria
of quality cut. In this work, we assumed observed EAS as `quality event' 
if its maximum of development lies above the cloud-top altitude.
In the case of clouds with $\tau <1$, all triggered events are
also accepted since the the maximum of development is measurable even
with attenuated EAS signals. In such a case, angular reconstruction 
is little affected since it is based on the angular speed of moving 
spot corresponding to EAS track. 

\begin{figure}[t]
  \vspace{5mm}
  \centering
  \includegraphics[width=3.25in]{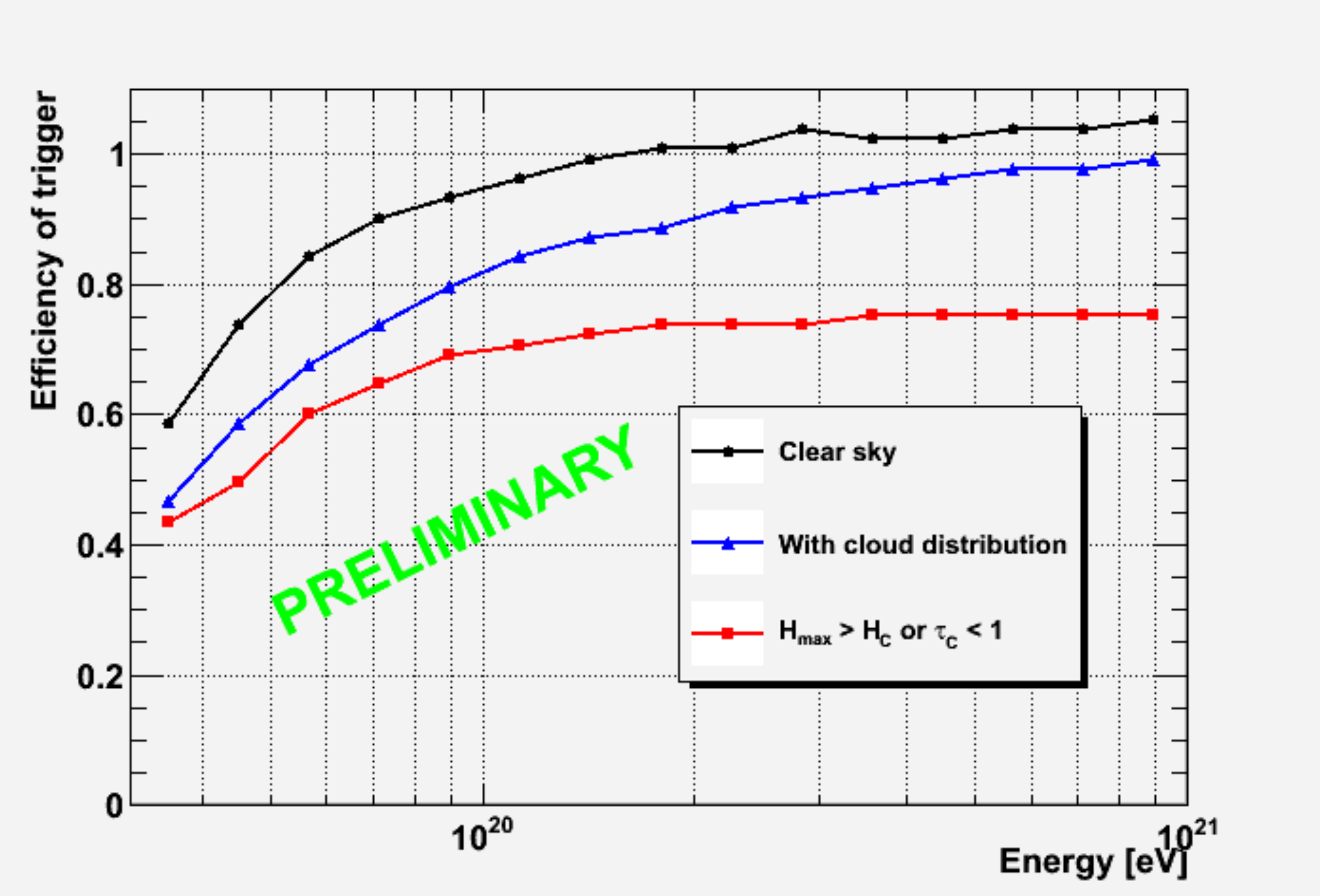}
  \caption{Preliminary trigger efficiency vs energy. The clear sky 
    case is denoted by circles. Cloud-statistics average case
    is shown by triangles. The case for quality event is indicated
    by squares.}
  \label{simp_fig1}
\end{figure}

In Figure 2, the trigger efficiency is shown as a function of 
energy. For cloudy cases, all triggered events as cloud-statistics 
average is shown by triangles. The case of selection of 
quality event into those triggered events is also indicated by squares.
For comparison, the clear sky case is shown by the circles.

For all triggered events, the efficiency increases with energy
and approaches that of clear sky case at highest energies. 
For quality events, it increases up to $\sim 10^{20}$ eV and becomes
almost constant at higher energies. This is because a certain 
fraction of the clouds with $\tau>1$ exists at higher altitudes. 
From Table 2, for example,  such clouds with $H_C>7$ km accounts 
for 20\%. Therefore a part of EAS  develops below such type of 
clouds. The cloud impact for overall cloud-statistics is estimated
to $\sim 70\%$ above $\sim 3 \times 10^{19}$ eV. This value is an
important factor when one estimates the effective exposure over the
mission (see \cite{uks,usan}). Similar estimate was carried out in
EUSO mission and this result is in fair agreement apart from 
detailed difference in selection criterion.  Currently, detailed 
study on reconstruction is in progress to take into account 
configuration of the JEM-EUSO mission. It should be mentioned that 
the main telescope of JEM-EUSO will be operated along with AM system. 
Utilization of these subsystem is investigated in parallel 
\cite{uAMS_ICRC,uIR}.

\section{Summary}

In this work, the impact of the clouds in observation of UHECRs
by the JEM-EUSO mission is investigated using ESAF simulation
package with a test cloud assumption. The light curves for
typical EAS with a stratus-test cloud shows the intense scattered
signals of Cherenkov photons from the cloud-top. In the case of
cirrus-like test cloud, there is  attenuation of EAS signals 
that are emitted or scattered below the test cloud corresponding to 
the transmittance determined by the cloud's optical depth. For 
various cases, the trigger efficiency was estimated and compared 
with that of clear sky case. In the case of optically thick
and high cloud, EAS signals are generally attenuated that results
in smaller trigger efficiency. For optically thin $(\tau < 1)$ cloud, 
a part of EAS does not trigger, while it keeps good visibility of EAS maximum.
For the low altitude cloud, the influence is limited especially at higher 
energies. Taking into account the statistics of 
cloud property and the observability of the EAS maximum, the cloud impact 
to trigger aperture is $\sim 70\%$ above $3\times 10^{19}$ eV. The results herein are preliminary and further detailed studies are 
in progress along with utilization of the atmospheric monitoring system.

\section*{Acknowledgements}

S\'aez Cano thanks to University of T\"ubingen and RIKEN 
for their kind hospitality during her research stays.  This
work is supported by  MICINN under projects AYA2009-06037-E/AYA, 
AYA-ESP 2010-19082, CSD2009-00064 (Consolider MULTIDARK) $\&$ 
AYA2011-29489-C03-01 and by Comunidad de Madrid under project 
S2009/ESP-1496. For computation, SPAS-UAH cluster was used.

\clearpage


\newpage
\normalsize
\setcounter{section}{0}
\setcounter{figure}{0}
\setcounter{table}{0}
\setcounter{equation}{0}



\title{THE ESAF SIMULATION FRAMEWORK FOR THE JEM-EUSO MISSION}

\shorttitle{author \etal paper short title}

\authors{F. Fenu$^{1,2}$, T. Mernik$^{1,2}$, A. Santangelo$^{1,2}$,
  K. Shinozaki$^{2}$, M. Bertaina$^{3}$, L. Valore$^{4}$, S. Biktemerova$^{5}$,
  D.Naumov$^{5}$, G. Medina Tanco$^{6}$ on behalf of the JEM-EUSO collaboration }
\afiliations{$^1$Eberhard Karls Universit\"at T\"ubingen, Sand 1, 72076
  T\"ubingen, Germany \\ $^2$RIKEN 2 - 1 Hirosawa, Wako 351 - 0198, Wako,
  Japan\\$^3$Universit\`a degli studi di Torino, Via Pietro Giuria 1 - 10125
  Torino, Italy\\$^{4}$Universit\`a degli studi di Napoli "Federico II"-Istituto
  Nazionale di Fisica Nucleare - Complesso Universitario di Monte Sant'Angelo
  - via Cintia 80126 Napoli, Italy \\$^4$JINR Joliot Curie 6, 141980 Dubna, Russia\\$^5$UNAM
  Ciudad Universitaria, Circuito de investigacion cientifica, Mexico D.F., Mexico  }
\email{fenu@astro.uni-tuebingen.de}

\abstract{ESAF, the EUSO Simulation \& Analysis Framework, was originally developed as the
simulation and analysis software for the  Extreme Universe Space Observatory - EUSO
mission of ESA. More recently, ESAF has been extended and modified to simulate the
JEM-EUSO mission.
ESAF consists of several independent modules, which perform the shower simulation,
the light transport in the atmosphere, the instrument and telemetry simulation, and
eventually the analysis of the observed track in order to reconstruct the energy,
arrival direction and Xmax of the event.
In this paper,  we present the ESAF event simulation structure.
In particular we describe the shower generators, the atmospheric modeling, the
simulation of the JEM-EUSO optics, sensors and electronics including the trigger
algorithms developed to discriminate the good event signals from the background,
allowing a fake trigger rate compliant with the JEM-EUSO telemetry constrains.
We will also show some event describing it step by step through the entire detector. }

\maketitle


\section{Introduction. ESAF the Euso Simulation and Analysis Framework }
\begin{figure}[!h]
  \centering
  \includegraphics[width=7cm]{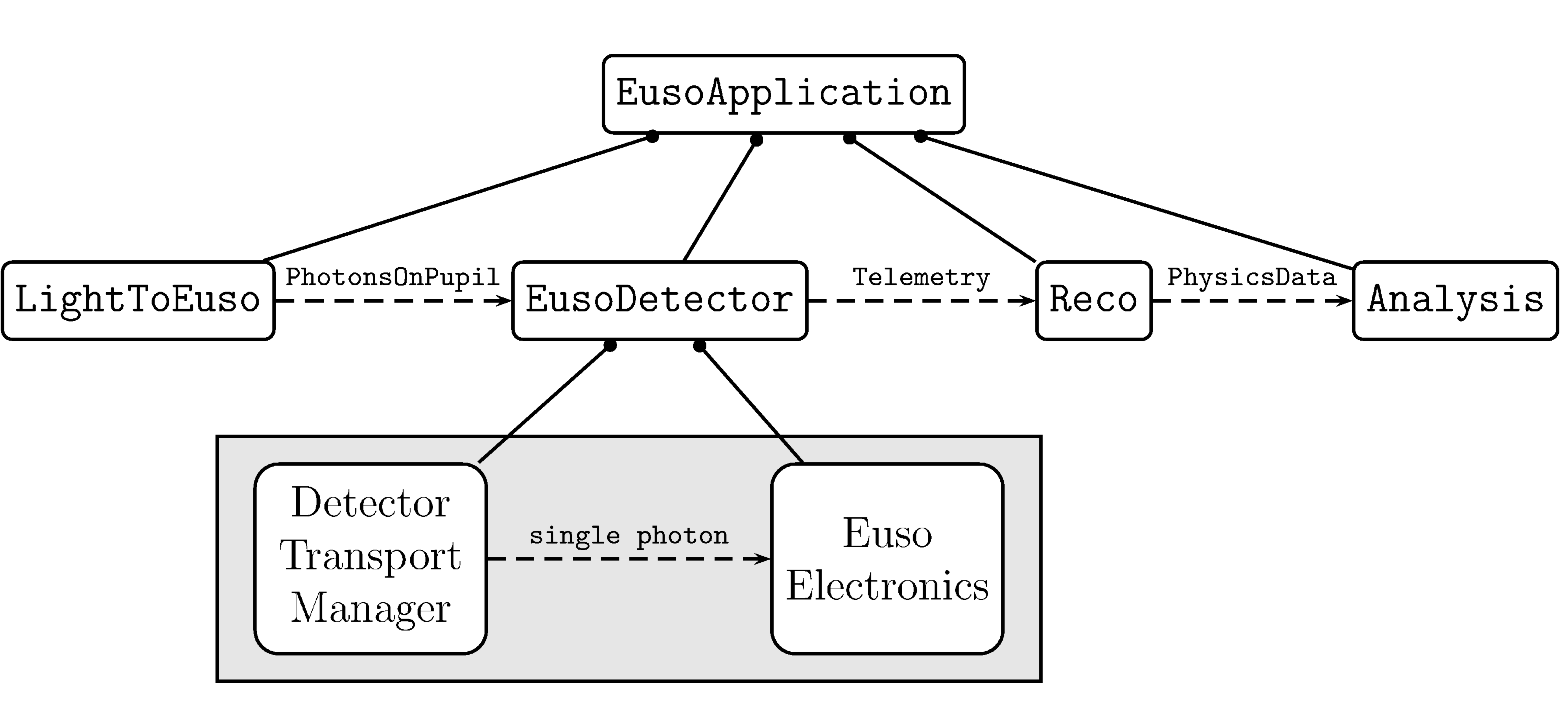}
  \caption{Basic ESAF scheme. Taken from \cite{lab2}}
  \label{ESAF_STRUC}
 \end{figure}

The Euso Simulation and Analysis Framework (ESAF) is currently used as the simulation and
analysis software for the JEM-EUSO mission \cite{lab0}\cite{lab1_1}. It has been developed in the
framework of the ESA-EUSO mission \cite{lab1}\cite{lab1_3}. ESAF performs the simulation of the
shower development,
 atmospheric transport, detector optics and electronics simulation.
Furthermore, algorithms and tools for the reconstruction of the shower properties are
included in the ESAF package. In the framework of the JEM-EUSO Phase-A
study, we took all the necessary steps to implement the JEM-EUSO mission
configuration.
This is carried out in a coordinated effort between several groups which are actively collaborating in the software development
and on the mission performances assessment.
A general sketch of the ESAF structure is given in fig \ref{ESAF_STRUC}.
ESAF is a C++, Object Oriented, root\footnote{Package developed at CERN for
  particle physics data analysis. \cite{lab1_4}} based software. It has been
written in a modular way in order to cope with the high complexity of the
mission and with the rapidly changing instrumental design and science requirements.
It consists of several independent modules: the LightToEuso, the EusoDetector,
the Reco and the Analysis module.
The LightToEuso module allows the simulation of the shower development
and of the light transport through the atmosphere to the detector.
The EusoDetector includes the simulation of all the detector components from
Optics to the  Electronics of the
JEM-EUSO telescope.
Once the trigger algorithms issued a trigger signal, the event is sent through
 telemetry to Earth for the event reconstruction.
At this stage (the Reco framework) the reconstruction of the arrival
direction, 
energy and type of primary particle is performed.
The Analysis module is being developed.
The executables of the software have been divided in two parts: the Simu and
the Reco file. The first performs the simulation of the real event from
shower generation to telemetry while the second takes care of the reconstruction.
This has been done in view of a future utilization of the reconstruction
module which could be used for the analysis of real data.

\section{The Simulation framework}
In this section we will describe the simulation framework of the ESAF
software. This part is meant to simulate all the physical processes which
are  related to the shower development, the light production
and propagation, the detector and eventually the telemetry. 
\subsection{Event generators}
Several shower simulators are implemented in ESAF, following parametrical
and Monte Carlo approach. As parametrical generator, the
Gaisser-Ilina-Linsley (GIL) function \cite{lab3}, is used to reproduce the profile as function of
Energy and slant depth. Other generators such as the Monte Carlo simulator Corsika \cite{lab4} and
the Monte Carlo Conex simulator \cite{lab5} are interfaced with ESAF. Consistency studies between all
the different approaches have been performed and the appearance of some small
inconsistency between the parameterization and the Monte Carlo simulators is
still under investigation. With the different shower generators we are now
able to generate showers of different primary. Neutrino showers can now
be generated with the Conex generator and then analyzed by ESAF.
Lidar events can be now generated in ESAF: specific methods have been
developed and implemented to simulate photons at 355 nm emitted by laser
sources, in parallel to methods in use for showers.
Other sources of light (lightnings, TLEs\footnote{TLE: Transient Luminous
  Event. Transient event in the high atmosphere responsible for the production
  of huge amounts of light in the UV range. In this category we can consider
  Sprites, Jets, Elves and many other phenomena.}, cities, meteors)  cannot be
simulated yet although test light sources can reproduce the effect of those
events up to first approximation.

\subsection{Atmospheric transport}
\begin{figure}[!h]
  \centering
  \includegraphics[width=8cm]{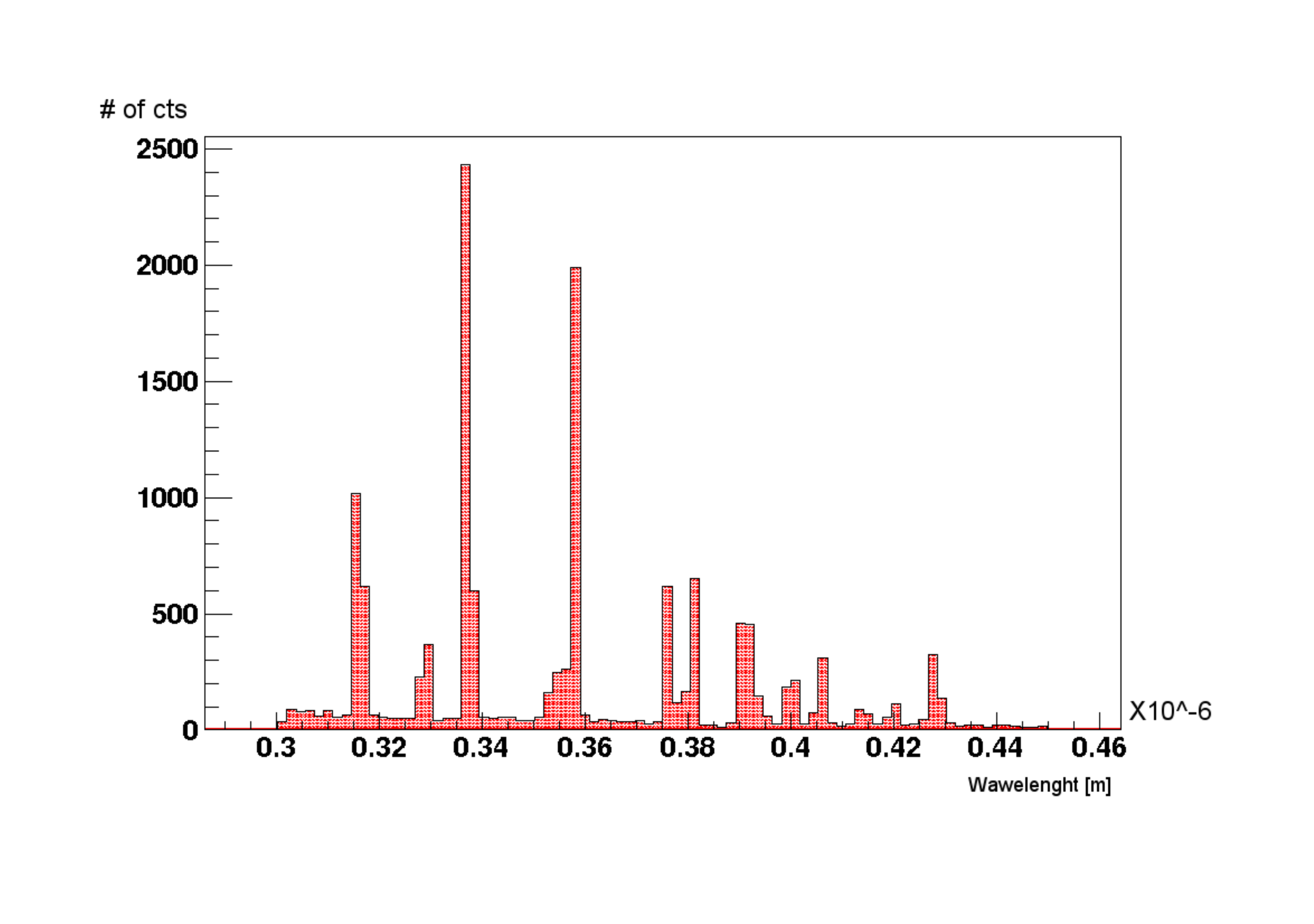}
  \caption{Photon spectrum simulated with ESAF. Both the typical
    Fluorescence emission lines and the Cherenkov (low level continuous spectrum)
    spectrum are visible. Fluorescence is calculated according to Nagano et
    al. \cite{lab6}}
  \label{nagano}
 \end{figure}

Both Fluorescence and Cherenkov production is taken into account in ESAF.
The simulated Fluorescence spectrum according to Nagano et al \cite{lab6}
is shown in Fig. \ref{nagano}. 
The Cherenkov production is taken into
account following the standard Cherenkov theory. Both  the ground
reflected and the backscattered component are considered. All the
photons are affected by Rayleigh scattering and ozone absorption. Furthermore,
photons can reach the detector in indirect way after scattering. Optionally
clouds can be simulated as constant layer of variable altitude thickness and optical
depth. Non uniform cloud coverage is also included in ESAF. The effect of
aerosols and dust has not been included in ESAF yet. 

\subsection{Detector}
\begin{figure}[h!]
	\begin{center}
		\includegraphics[height=3.cm]{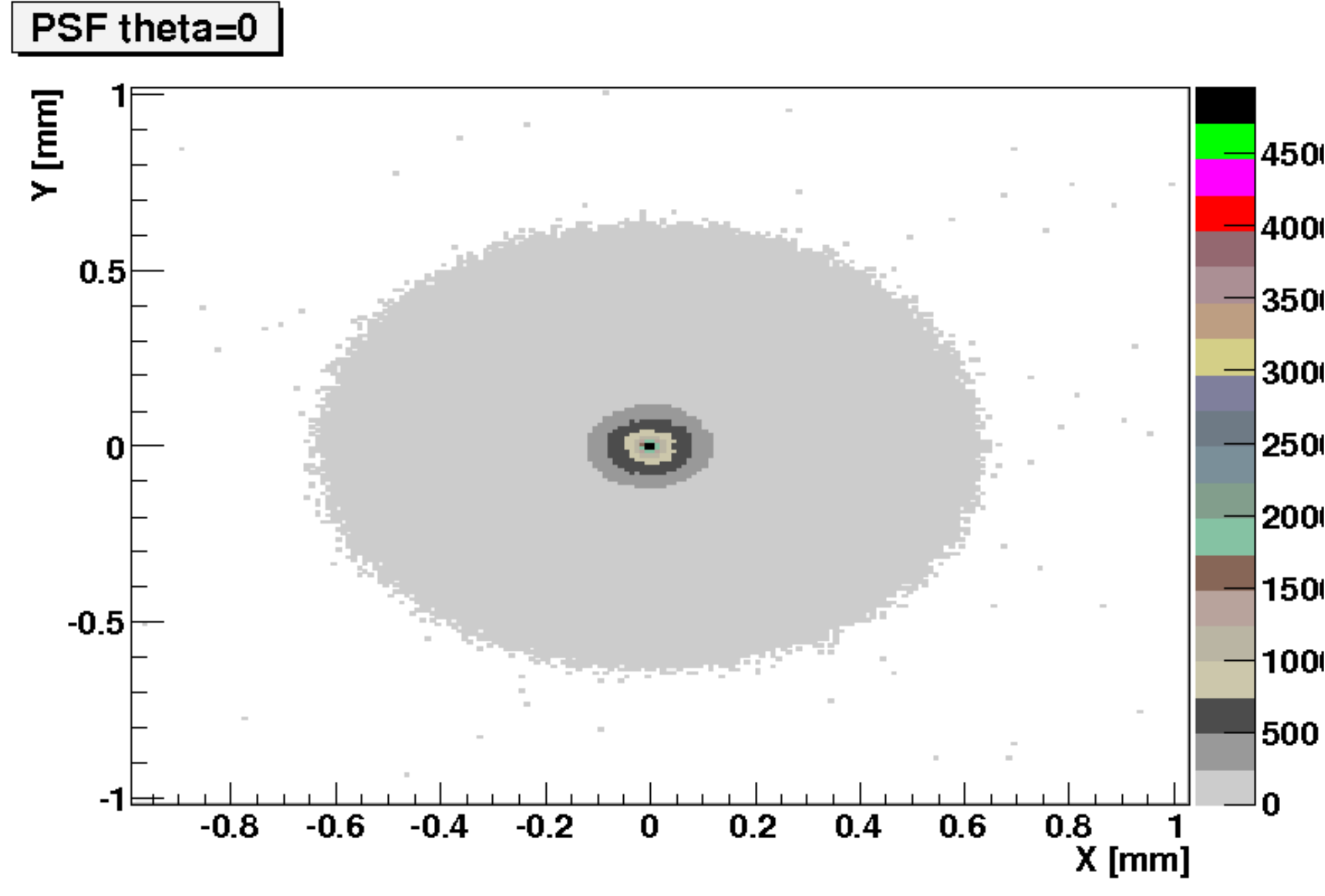}%
		\qquad
		\includegraphics[height=3.cm]{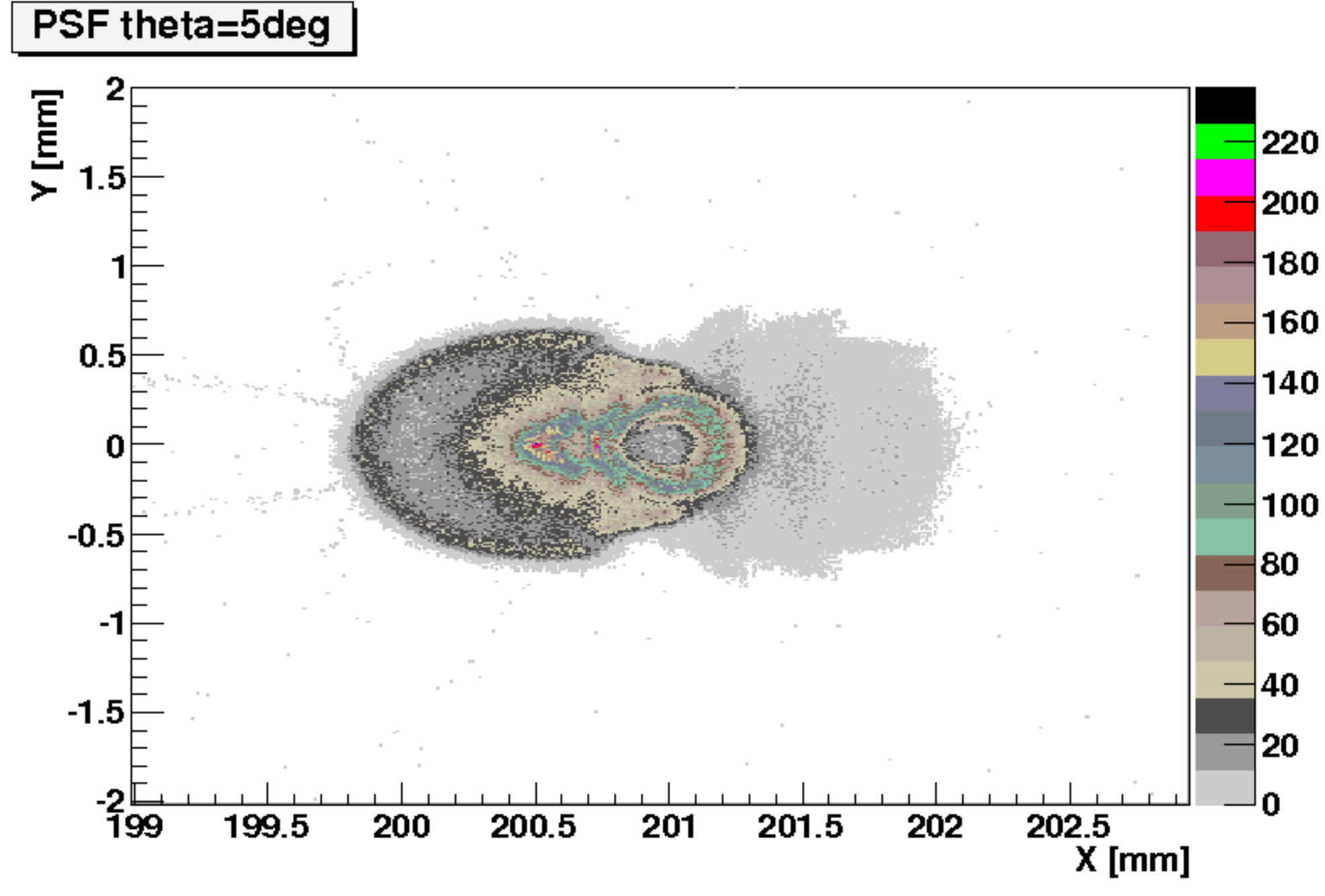}%
		\qquad
		\includegraphics[height=3.cm]{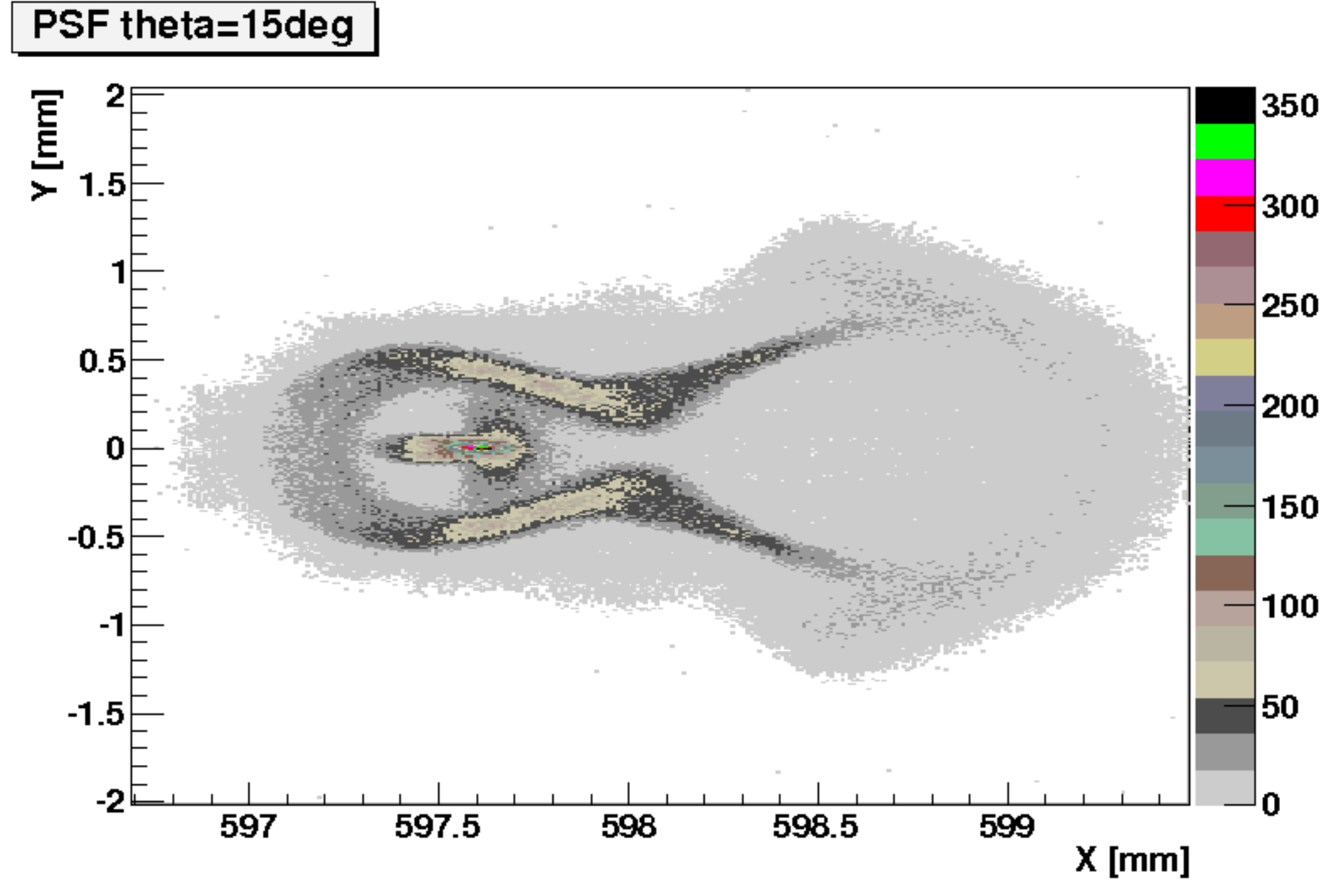}%
		\qquad
		\includegraphics[height=3.cm]{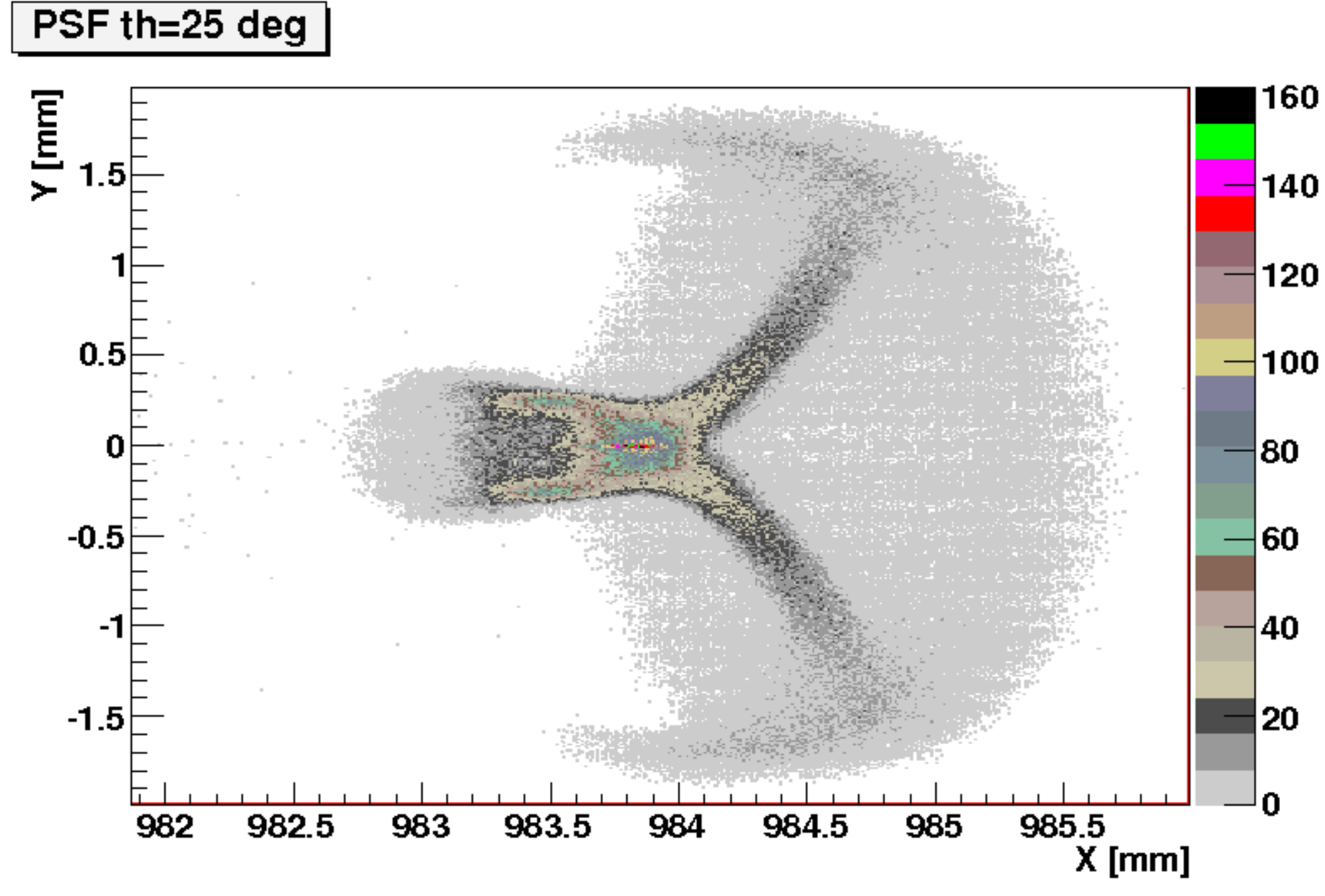}
		\caption{Point spread functions simulated with the RIKEN
                  ray-trace code interfaced with ESAF for several inclination
                  angles (0, 5, 15, 25
                   deg). On the axes the position in mm on the FS can be read.  }
		 \label{psf}
	\end{center}
\end{figure}

Once the photons reach the detector they are taken over by the optics
module. Several optics simulation approaches have been considered.
The parametrical simulation module calculates
analytically the position of the photon on the focal surface and adds to this
position a random spread. This is intended to be the first approximation of
the optics simulation and is basically the fast working tool to test the
features of the different optics designs. Furthermore, the optics simulation code developed in
RIKEN \cite{lab7} is included in the simulation code. This ray-trace code is
interfaced with the ESAF framework in order to transport every photon within
the optics through  a Monte Carlo simulation. In Fig. \ref{psf} an example of the generated RIKEN ray-trace Point Spread function
can be seen. Several optics
configurations have been included in the course of time to assess the performances.
Another optics module is the Geant 4 optics module \cite{lab8} which uses an interface
with the Geant simulator to transport the photons from pupil to the focal
surface.\\ 

\begin{figure}[!h]
  \centering
  \includegraphics[width=7cm]{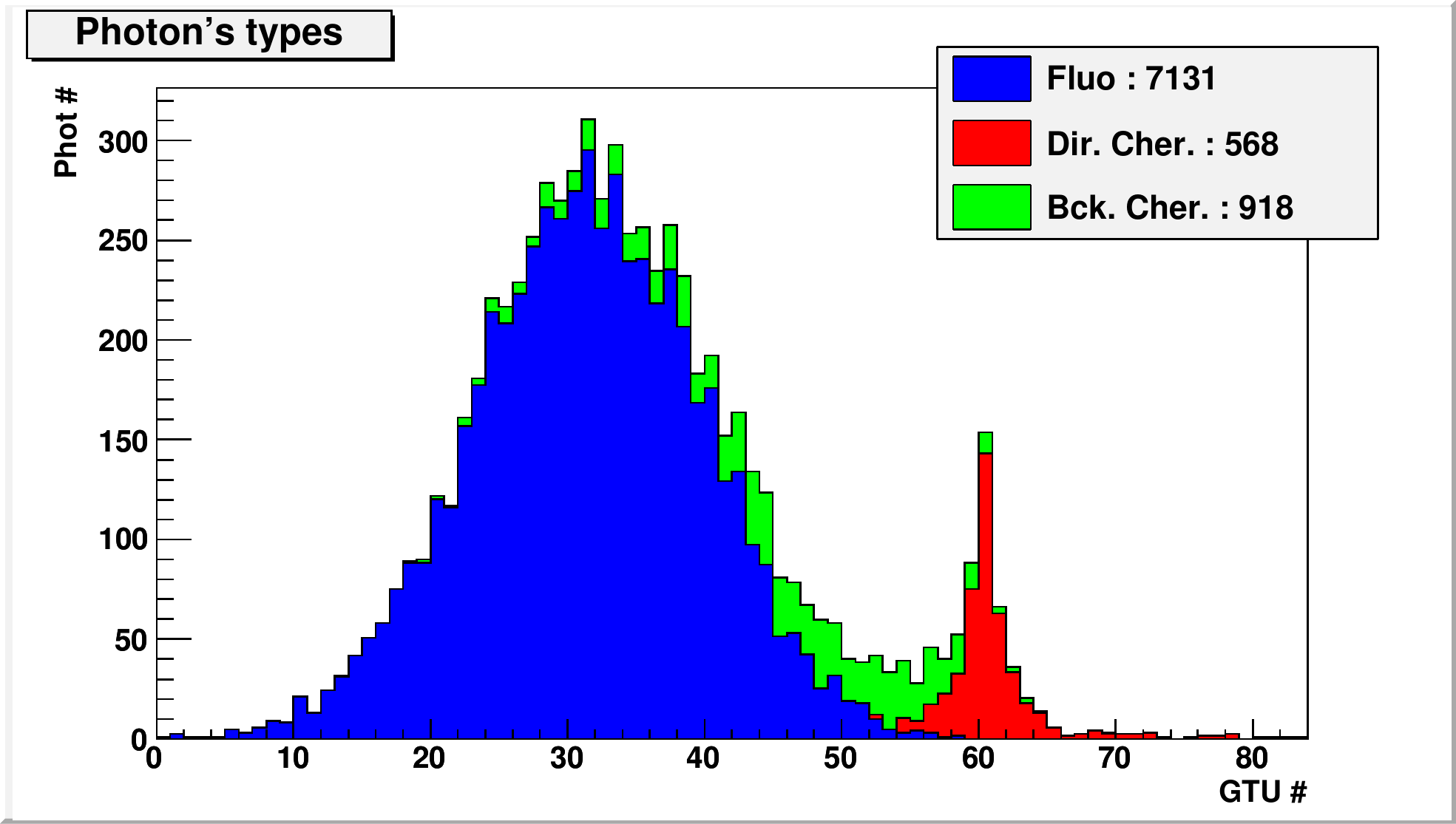}
  \caption{Composition of the photons at the detector pupil. Standard shower ($10^{20}$eV 60 $\deg$) as simulated by ESAF
  with the GIL parameterization.}
  \label{comp}
 \end{figure}

\begin{figure}[!h]
  \centering
  \includegraphics[width=7cm]{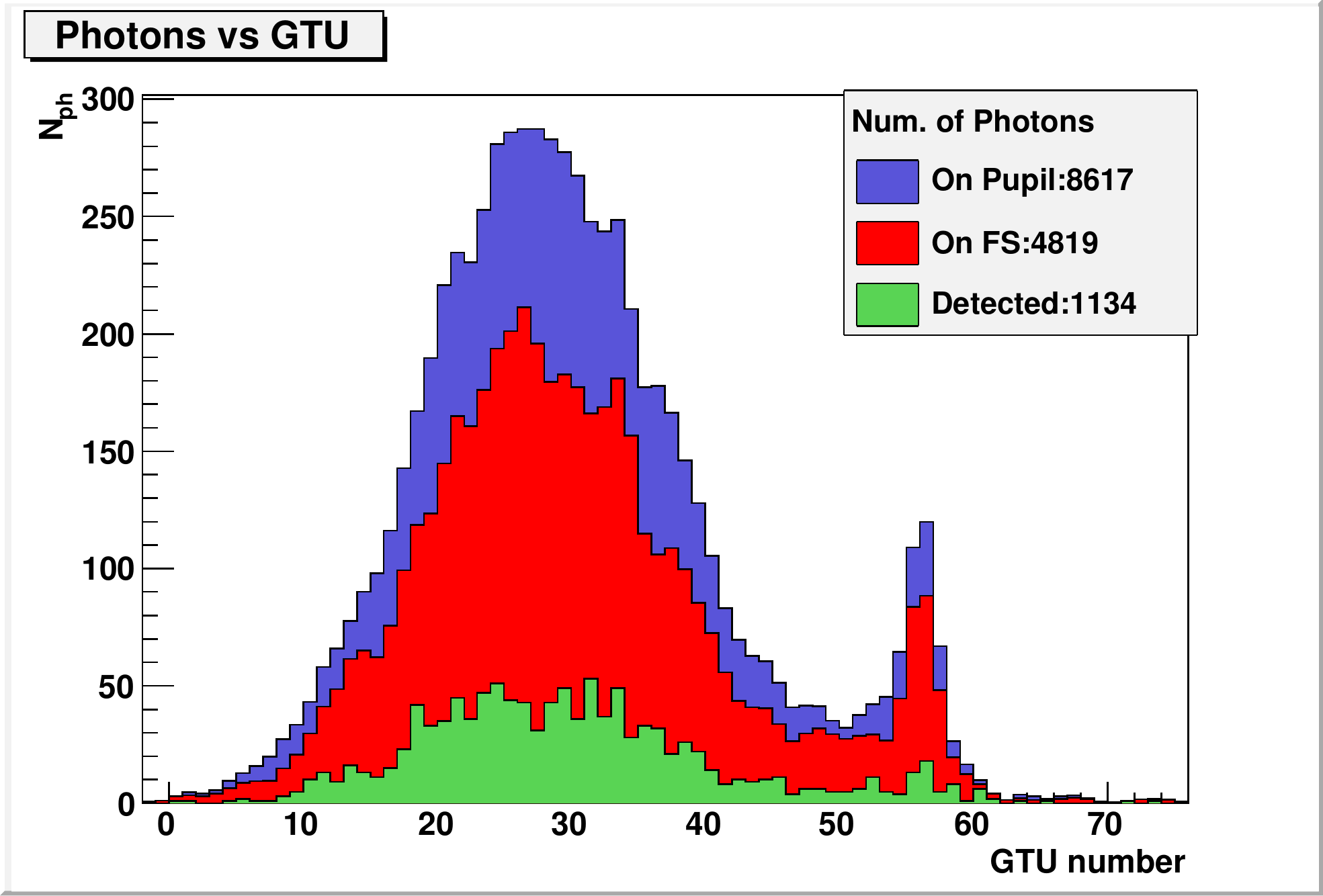}
  \caption{The event as seen through the detector. The Blue curve gives the
    number of photons as function of GTU \# at the pupil. The red one tells how
  many photons reach the FS. The green one represents the counts. Standard shower ($10^{20}$eV 60 $\deg$) as simulated by ESAF
  with the GIL parameterization.}
  \label{det}
 \end{figure}

 In Fig. \ref{comp} we analyze the composition of the photon
spectrum arriving at the pupil. As can be seen both direct fluorescence,
reflected and backscattered Cherenkov are visible.
Moreover we can observe in Fig. \ref{det} the event through the entire
detector from the pupil to detected counts regardless of the photon's kind.

Once the photons reach the focal surface they are transported through the
filter and the  optical adaptor before of reaching the photocathode. All the
relevant effect including geometrical losses, inefficiencies of the adaptor
(the BG3)
and of the filter are taken into account. A parameterization of the
photomultiplier is included in the electronics part. All the effects like
quantum efficiency (and its dependence from the photon inclination),
collection efficiency and cross talk are also taken into account pixel by
pixel within one Photomultiplier (PMT). The implemented Photomultiplier is the
M64 Photomultiplier of Hamamatsu. In table \ref{table_composition} we give a resume of the most relevant
parameters of the detector. More details can be found in \cite{lab8_1}. 
\begin{table}[ht]
\begin{center}
\begin{tabular}{c|c}
\hline
       Quantum Efficiency &    $\sim$ 39.6\%\\
       Collection Efficiency &  $\sim$ 80\% \\
       Cross talk  &     Negligible\\
       Pixel Area &     9 $mm^{2}$\\
       Number of Pixels &     64\\
\end{tabular}	
\end{center}
  \caption{The most relevant parameters for the implemented M64
    Photomultiplier}
	\label{table_composition}
\end{table}
The signal is then amplified by a
parameterized gain and the resulting output current is collected and treated
by the Front End Electronics. A threshold
is set on the PMT output current in order to accept or reject the signal count.

\subsection{Trigger}
The trigger algorithm's duty is to filter the background in order to increase
the signal to noise ratio. Being the telemetry limited, the instrument cannot
afford the transmission of the entire Focal Surface data to Earth. The entire
triggering scheme is therefore organized in a multiple step filtering. After
the Front End Electronics identified a photon count a first search for
persistency is done at the level of PDM \footnote{Part of the Focal
Surface consisting of 36 Photomultipliers.}. This is called the first level
trigger (L1).
After at this level a trigger signal is issued data are sent at the next level:
the so called Cluster Control Board \footnote{Electronics board which operates
on 324 Photomultipliers.} trigger. This is also called second level trigger (L2). Here the Fake
Trigger Rate must be further reduced to fit with the telemetry
constraints (from $\sim$ 1kHz to 0.1Hz on the entire Focal Surface).   
Several algorithms have been implemented and tested: the so called Linear Tracking
Trigger (LTT) scheme and the Progressive Tracking Trigger (PTT) as well as the
so called Cluster Control Board LTT trigger (CCB\_LTT) \cite{lab8_2}.
Several combinations of trigger schemes have been tested and
compared. Triggering efficiency for several detectors and different GTU length
have been produced. Once the trigger has been produced the triggered events
are sent through telemetry to the reconstruction framework.
A more comprehensive review on the trigger scheme is given in \cite{lab9}.

\subsection{Tilted mode}
In order to further increase the exposure it might be useful to tilt the
instrument.
In this way the surveyed area will be increased by a large
factor. Unfortunately the larger distance at which showers are observed under
these conditions will significantly increase the energy threshold.
Therefore tilting must be carefully studied in order to optimize the
inclination of the instrument. The tilting angle also deeply affects the scientific
output of the mission. Low tilting angles are more favorable to study Cosmic
Rays in the GZK region while higher inclinations give us a larger exposure
above $10^{20}$eV. 
We started some preliminary activity to assess the most proper mission
configuration and as a further step the tilted mode will be implemented in ESAF.

\section{The Reconstruction framework}
Aim of this framework is to analyze the detector response in order to identify
the direction of arrival, the energy and the type of the primary.
The first step consists in the identification of the signal inside the
transmitted data. For this purpose both a clustering and a Hough module have
been implemented. Then through fits procedures the direction of the primary is
calculated. Several different fits algorithms have been included in ESAF.
As last step the profile, $X_{max}$ and the energy are reconstructed. A more
comprehensive review of the reconstruction module is given in \cite{lab10}. 
In Fig. \ref{thomas} we see how the signal is treated after having been
identified.
A fit procedure is applied in order to find the arrival direction of the shower. 
\begin{figure}[!h]
  \centering
  \caption{[figure missing here: see ICRC proceedings]
    The standard event arrival direction is here reconstructed. The
    event is seen after the clustering procedure while a fit is performed in
    order to find the arrival direction. (T. Mernik) }
  \label{thomas}
 \end{figure}

\section{Conclusions}
In this paper we described the ESAF simulation framework. After the short historical
introduction we described the structure of the software and the physical
models implemented in it. We showed how an event is
treated by the ESAF simulation software by showing the key plots of the
simulated event through the various steps of the detector simulation and of
the reconstruction.
Moreover we wish to remember that the ESAF package is available under the svn
repository based in Lyon. We encourage interested people to contact the
accounts manager at the address naumov@numail.jinr.ru. 
\section*{Acknowledgments}
This work has been conducted by using the ESAF software which has been written by the ESAF developers 
team during the Phase-A study of the EUSO mission under the initiative by the European Space Agency ESA.
We wish to thank RIKEN, Japan, for an allocation of computing resources on the RIKEN Integrated Cluster of Clusters (RICC) system.
Moreover FF wishes to thank the Computational Astrophysics Laboratory for their kind hospitality.

\clearpage


\newpage
\normalsize
\setcounter{section}{0}
\setcounter{figure}{0}
\setcounter{table}{0}
\setcounter{equation}{0}



\title{The ESAF Reconstruction Framework of UHECR Events for the JEM-EUSO Mission}

\shorttitle{mernik \etal ESAF Reconstruction for JEM-EUSO}

\authors{Thomas Mernik$^{1,2}$, Francesco Fenu$^{1,2}$, Domenico D'Urso$^{3}$, Andrea Santangelo$^{1,2}$, Klaus Bittermann$^{1,2}$, Kenji Shinozaki$^{2}$, 
Mario Bertaina$^{4}$, Svetlana Biktemerova$^{5}$, Dmitry Naumov$^{5}$, Gustavo Medina Tanco$^{6}$ \\on behalf of the JEM-EUSO collaboration}
\afiliations{$^1$Institut f\"ur Astronomie und Astrophysik, Kepler Center, Universit\"at T\"ubingen, Sand 1, 72076 T\"ubingen, Germany\\ 
$^2$RIKEN, 2-1 Hirosawa, 351-0198 Wako, Japan\\ 
$^3$Universit\`a degli studi di Napoli and INFN Istituto Nazionale Fisica Nucleare, Complesso Universitario di Monte Sant'Angelo - Via Cintia 80126 Napoli, Italy\\
$^4$Universit\`a degli studi di Torino, Via Pietro Giuria 1 - 10125, Italy\\
$^5$JINR, Joliot-Curie 6, 141980 Dubna, Russia\\
$^6$UNAM, Ciudad Universitaria, Circuito de la Investigacion Cientifica, Mexico D.F., Mexico }
\email{mernik@astro.uni-tuebingen.de}

\abstract{JEM-EUSO is a space based UV detector that will be mounted on the International Space Station (ISS) to monitor the earth's 
atmosphere searching for UHECR induced extended air showers (EAS). 
By evaluating the fluorescence and Cherenkov signal on the focal surface of the instrument the arrival direction, energy and nature of the 
primary can be determined. Due to the instantaneous aperture of 10$^5$ km$^2$ sr JEM-EUSO will be able to measure several hundreds 
of events at energies  higher than 5*10$^{19}$ eV.   
ESAF is a software for the simulation of space based UHECR detectors. It is configured to 
cover the specific aspects of the JEM-EUSO mission and to estimate its expected performance.
ESAF can simulate every step of the generation and observation of an EAS - from the fluorescence track formation, the light transport 
in the atmosphere and through the instrument to the telemetry stage. The reconstruction chain covers the discrimination of the 
recorded track from background as well as the estimation of energy, arrival direction and Xmax for the determination of the UHECR species.
In this paper we present strategies and algorithms implemented to estimate the spatial and energy resolution of JEM-EUSO as well as a selection of 
examples demonstrating the expected performance.}


\maketitle

\section{Introduction}
The JEM-EUSO detector is a space based UHECR detector designed to be mounted on the Japanese Experiment Module ''Kibo'' on board the ISS \cite{wTaka}.
It will monitor the earth's atmosphere from above to search for extended air showers generated by of cosmic rays in the energy 
range of $10^{19}$ eV to $10^{21}$ eV and possibly beyond. JEM-EUSO will reach an instantaneous aperture of approximately $10^5$ 
km$^2$ sr \cite{wTosh} allowing a high statistics of events compared to ground based observations.  
Thus, JEM-EUSO is a key mission to explore the realms of extremely high energy cosmic rays
far beyond the capabilities of any ground based UHECR observatory. More details can be found in \cite{wPurplebook}.  

ESAF - the EUSO Simulation \& Analysis Framework, is a ROOT \cite{wroot} based, modular software designed to simulate space based UHECR detectors. 
It has been developed in the context of the former EUSO mission \cite{wRedbook}. 
Its modular structure allows to simulate any EUSO-like\footnote{We define EUSO-like a space borne detector for the 
measurement of UHECR by the measurement of the fluorescence and/or Cherenkov light of EAS.} instrument. 

The simulation comprises all physical processes relevant to UHECR measurement. Among these are the development of the resulting air shower, the production of 
fluorescence and Cherenkov light as well as propagation of photons towards the detector. Inside the instrument, simulations involve the 
propagation of photons through the optics, the response of the photomultiplier and electronics and the event reconstruction eventually. 
In this article we explain the reconstruction algorithms implemented in ESAF and present some results to demonstrate their achievement potentials.

\section{The Reconstruction Framework}
Cosmic ray induced EAS emit fluorescence light isotropically in all directions plus a beamed Cherenkov component. 
Parts of that light go directly to the telescope. Other components are reflected diffusely from ground or scattered towards JEM-EUSO.
The UV photons reaching the entrance pupil of the instrument propagate through the optics and 
activate the photomultiplier tubes arranged on the focal surface. When the readout electronics recognizes certain patterns a trigger is issued. Now the signal
is processed and transmitted to earth for analysis and reconstruction.
More details on the observation technique can be found in \cite{wTosh}. 

In ESAF different modules are dedicated to the single stages during the evaluation of the signal. First of all, the signal has to be disentangled from noise. 
Following that direction and energy reconstruction algorithms can be applied.   

\subsection{Pattern Recognition}
The fluorescence signal will appear as a faint moving spot of the instruments focal surface embedded in the background generated by night 
glow, city lights, weather phenomena and other sources. The extraction of the signal track and the 
determination of its spatio-temporal behavior remains crucial for any further analysis aiming at reconstructing the arrival direction or 
energy of the primary. There are two possible algorithms for the pattern recognition:  
\begin{itemize}

\item \emph{Clustering} of data points in space and time to disentangle causally related data points from those distributed randomly.
  
\item \emph{Hough Transform}, developed to identify prefixed shapes within noise by transforming the relevant parameters to the so called 
                              Hough space and back. 

\end{itemize}
Both are in principle capable to perform the required operation and have been implemented in ESAF.

\subsection{Clustering}
\begin{figure}[!h]
  \centering
  \includegraphics[width=3.in]{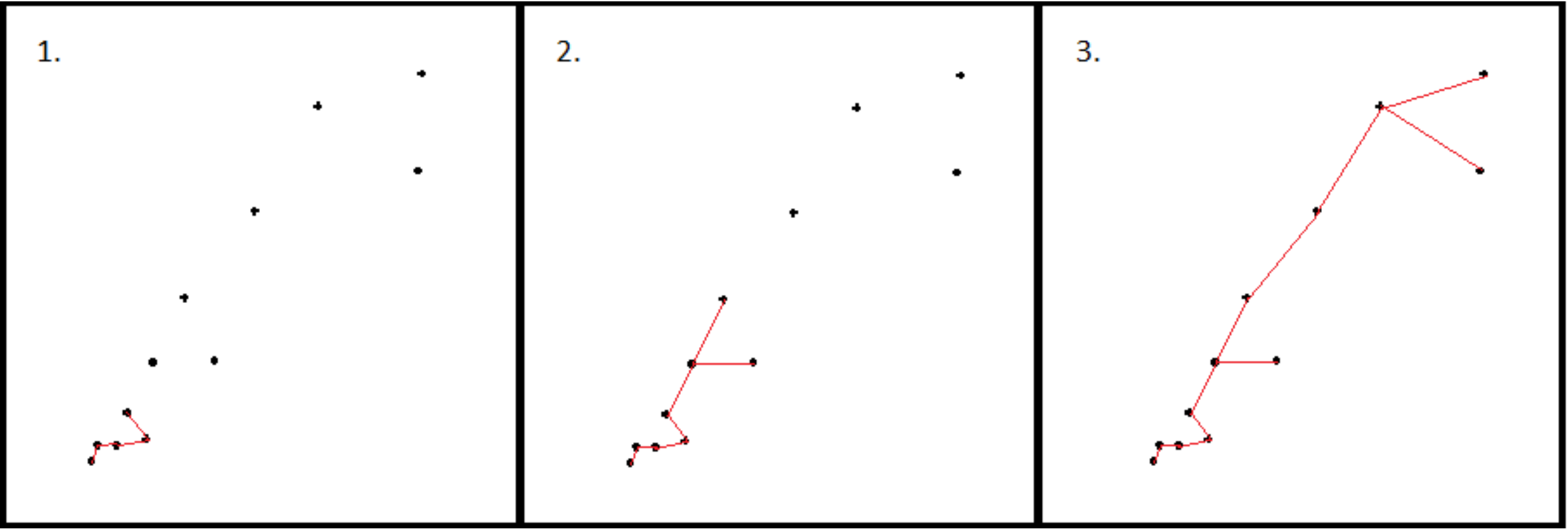}
  \caption{Clustering of data points: the size of the cluster depends on a threshold}
  \label{cluster}
 \end{figure}

The approach of the cluster technique is to arrange data points into causal patterns by analyzing their minimum spanning tree (MST) which is in 
this case made of the Euclidean distance between them (fig. \ref{cluster}).
A group of activated pixel is then identified as a cluster if their distance is less than a certain pre-adjusted threshold $\xi$. If this
cluster is regarded as significant (large with respect to others) a line fit is performed to estimate its geometrical parameters \cite{wMac1}. 

\subsection{Hough Transform}
\begin{figure}[!h]
  \centering
  \includegraphics[width=3.in]{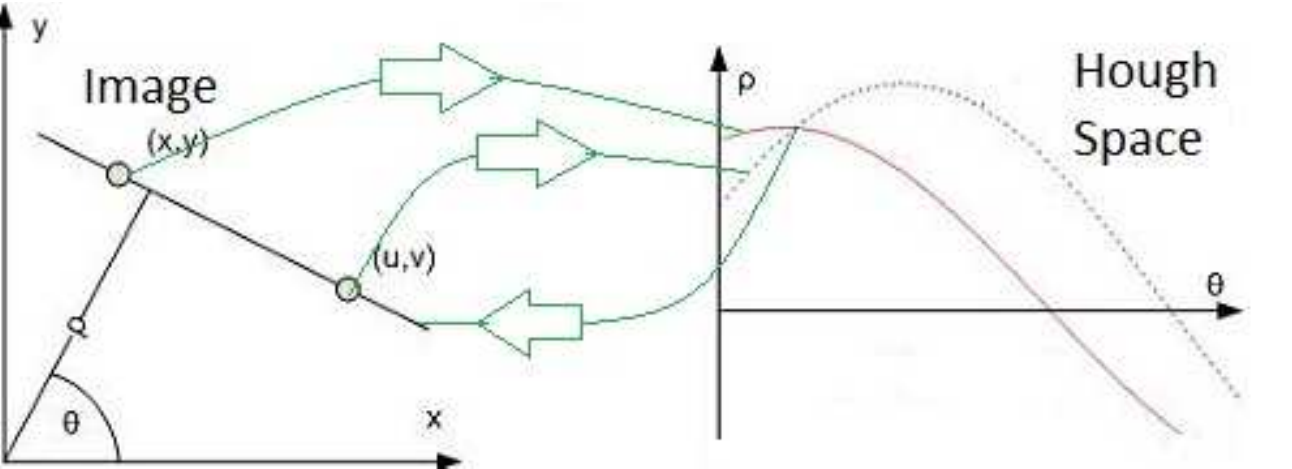}
  \caption{Simple example of a Hough transform for two data points.}
  \label{ht}
 \end{figure}

Initially designed for the detection of patterns in bubble chambers, the HT is an algorithm for the discrimination of certain shapes 
(even incomplete ones) from others, e.g. noise \cite{wHough}. Here the item sought-after is a longish pattern that can be abstracted as a straight 
line. For each data point the HT assumes a number of lines passing through it. These lines can be parametrized by their distance from the origin 
of the coordinate system $\rho$ and the angle $\theta$ between its normal and the x-axis (fig. \ref{ht}, left). Transformed into the Hough space, a 
two dimensional parameter space spanned by $\rho$ and $\xi$ each data point represents a sinusoidal curve (fig. \ref{ht}, right). The intersection points of the many
sinusoidals are summed up in an accumulator. The intersection point that drews in most of the counts is then transformed back into the image space, where it corresponds to a straight
line passing through as many data points as possible. Hence, when the signal track is identified a line fit estimates its parameters. 
This information is handed over to the direction reconstruction module.     

\subsection{Pulse Finder}
Performances of pattern recognition algorithms can be further improved
introducing a preliminary pulse finding step over each single camera pixel
defining the associated time window validity.
In each pixel, pedestal, variance and peaks of the recorded signal
as a function of time are evaluated.
Starting from a first guess time window defined around peaks,
start and stop positions of signal are identified maximizing
the signal to noise ratio.
The procedure allows to make a pre-rejection
of the noise and strongly reduces the probability to
misleadingly identify a noisy pattern as a track.

\section{Direction Reconstruction}
\begin{figure}[!h]
  \centering
  \includegraphics[width=3.in]{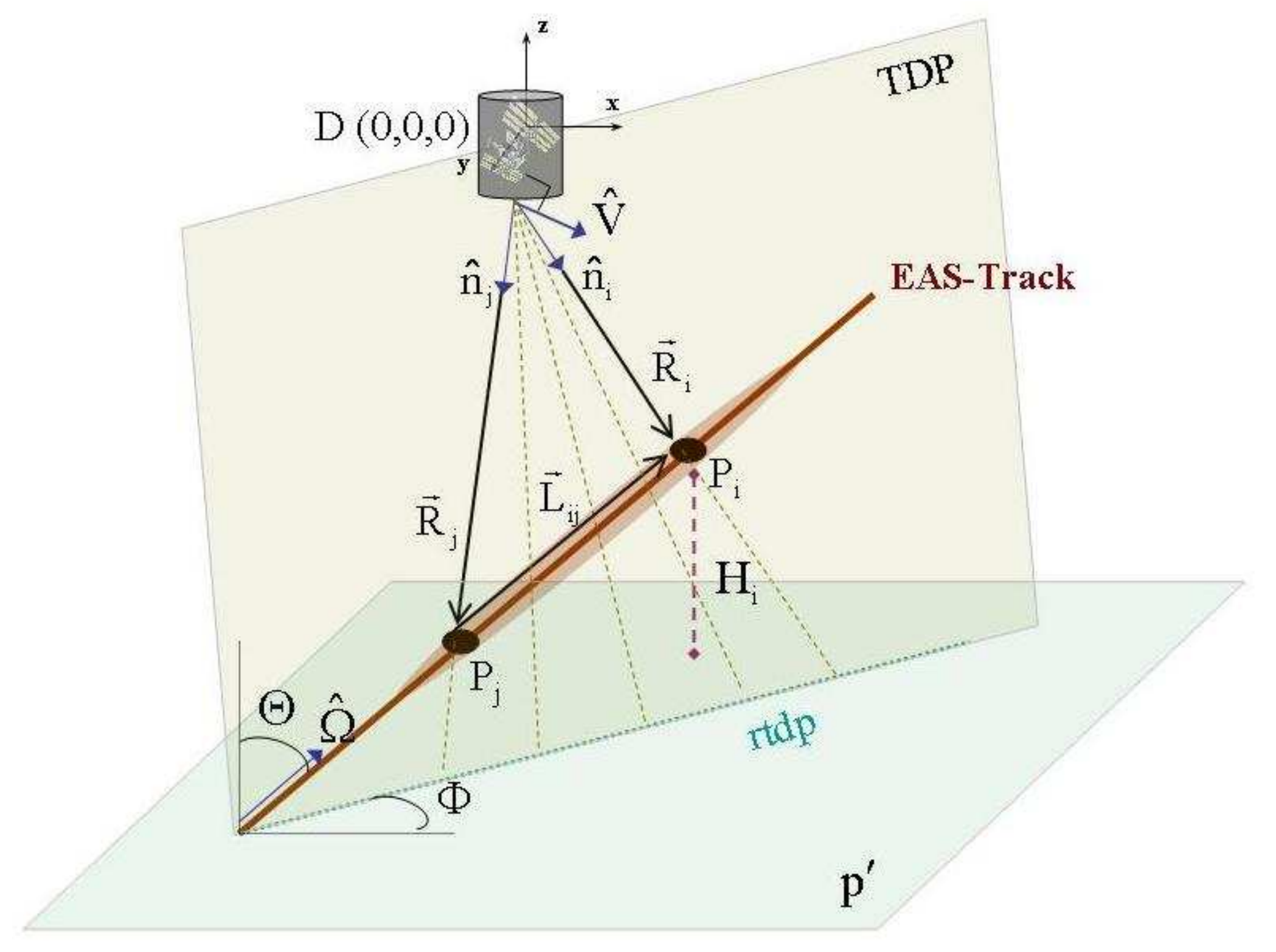}
  \caption{EAS observed with JEM-EUSO: Within the track-detector-plane (TDP), photons emitted at different times t$_j > t_i$ reach the 
detector from certain directions $\hat n_i$, $\hat n_j$ after traversing R$_i$, R$_j$ in atmosphere. From the timing information and arrival angle of the 
shower photons, the direction of the primary $\hat \Omega(\Theta,\Phi)$ can be determined.}
  \label{recosyst}
 \end{figure}

From the geometrical properties of the signal track on the focal surface the arrival direction of the primary can be computed by a variety 
of methods implemented in ESAF as described in more detail in \cite{wBer} and \cite{wBot}. Fig. \ref{recosyst} shows the system of the EAS and the detector.
In the current configuration there are 5 different algorithms implemented in ESAF. Two of them yield the most promising performance:
\begin{itemize}

\item \emph{Analytical Approximate 1}: angular velocities of the signal track int the x(t) and y(t) planes are linearly fitted. 
The arrival angle of the primary is derived by geometrical estimations. 

\item \emph{Numerical Exact 2}: a $\chi^2$ minimization is performed between the activation times of pixel induced by the actual 
signal to those induced by a signal track theoretically computed.  
\end{itemize}

\subsection{Energy Reconstruction}
The ESAF package has an extensive module dedicated to the reconstruction of the energy of the primary as further described in \cite{wCol}.

\begin{figure}[!h]
  \centering
  \includegraphics[width=3.in]{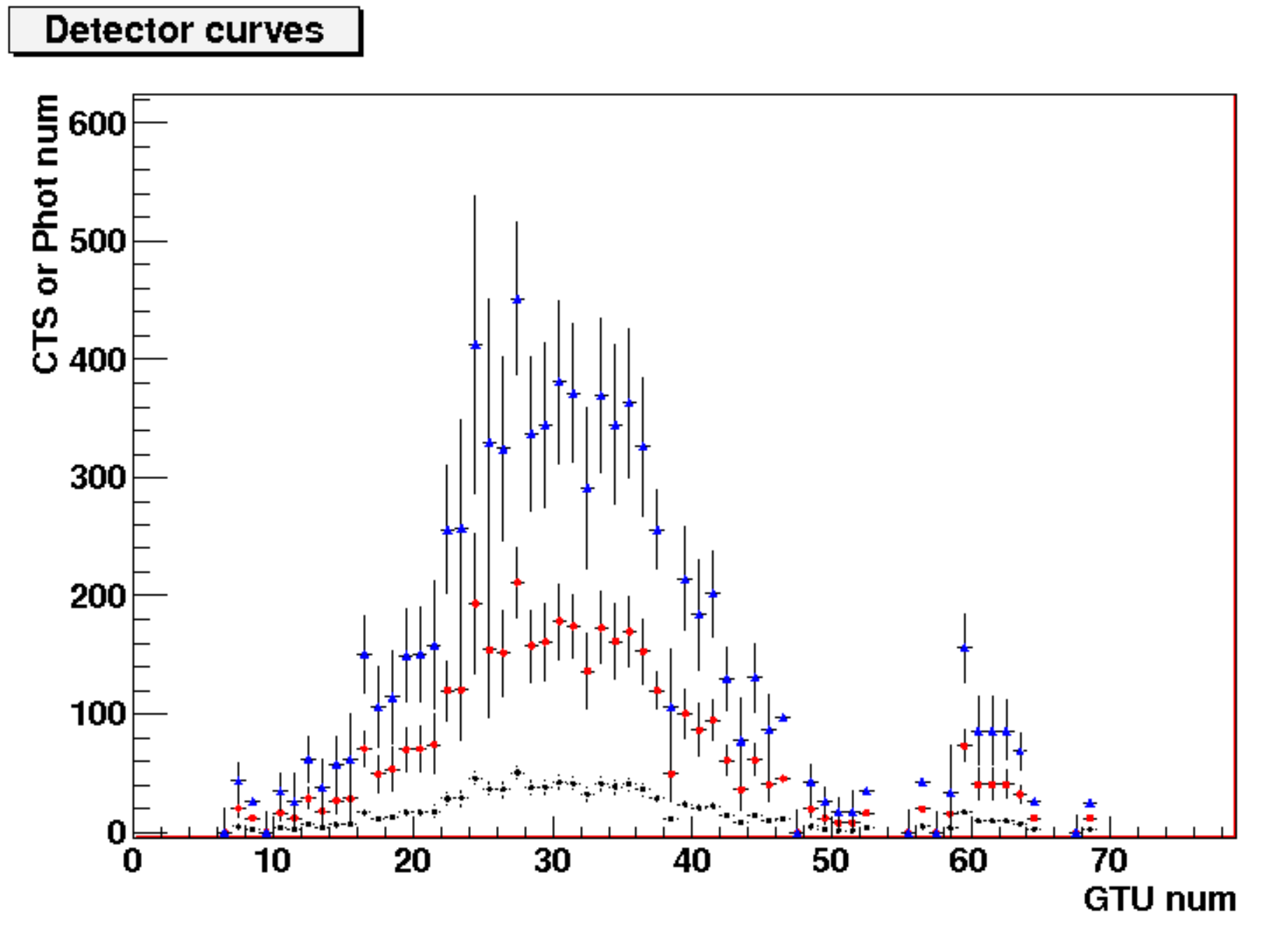}
  \caption{Energy reconstruction: Count distribution (black) is obtained by smoothing the clustering data. Focal surface light curve (red) 
and entrance pupil light curve (blue) are computed by taking into account the instruments efficiency.}
  \label{PhotCurve}
 \end{figure}

Moreover, an alternative energy reconstruction module have been implemented by the T\"ubingen group.
Starting from the reconstructed signal profile inherited from the pattern recognition we successively
correct for the instrumental losses such as optical absorption in the lens system or inefficiency of the photomultipliers 
in order to get the photon's curve at
 the level of the entrance pupil of the optics (fig. \ref{PhotCurve}). 
 A parameterization both for the photomultipliers and for the optics response is
 required at this step.
 Using the reconstructed primary arrival direction and, if present, the timing of the
 Cherenkov mark we then reconstruct the position of the shower at each
 time. Using this information we can now correct the atmospheric effects and
 the geometrical loss which dim the signal.
 Several methods to reconstruct the geometry of the shower \footnote{Geometry
 means the position of the shower in the FOV at each time.} have been
 implemented. Some of them use the timing of the Cherenkov mark, other make
 assumptions on the particle type or infer the position of the
 maximum from the shower width.
 Once the photon distribution within the shower has been reconstructed, assuming a certain fluorescence yield 
we can compute the number of charged particles in the shower.
 After that,  the backscattered Cherenkov component is calculated
 and then subtracted from the contaminated electron curve.
 Eventually, a fit with a parameterization for the shower profile is performed. 
Having already calculated the Xmax (at the
 level of the electron curve reconstruction) - energy is the only remaining
 free parameter. Performing the fit therefore allows to
 give an estimate of the energy of the primary.

\section{Some Examples}
Currently the ESAF code is being updated to the most recent JEM-EUSO configuration. 
This includes improvements in the optical system as well as latest trigger algorithms
(see \cite{wFenu}). Here we present a few examples of ESAF's reconstruction 
performance. For the angular resolution study only a certain subclass of events is considered as seen in tab. \ref{ttable_single}. 
For the energy resolution we limited outselves to standard events (E=$1*10^{20}$ eV, $\Theta$=60$^{\circ}$)
\begin{table}[t]
\begin{center}
\begin{tabular}{l|l}
\hline
              energy   & $7*10^{19}$ eV,  $1*10^{20}$ eV, $3*10^{20}$ eV              \\
             primary   & proton                       \\
inclination $\Theta$   & 30$^{\circ}$, 45$^{\circ}$, 60$^{\circ}$, 75$^{\circ}$                 \\
right ascension $\Phi$ & 0$^{\circ}$ -  360$^{\circ}$ \\
statistics             & 1000 events each E/ $\Theta$ config.                  \\
\hline
\end{tabular}
\caption{Configuration of study.}\label{ttable_single}
\end{center}
\end{table}
All simulations have been carried out assuming an ISS altitude of 430 km. 
The the recently introduced PulseFinder has not been used in this particular study. Thus the examples presented here 
are rather conservative and may be regarded as preliminary.

\subsection{Spatial Resolution}
$\gamma^{68}$ is a measure of the angular resolution. 68\% of all events have separation angle less than $\gamma^{68}$.
Here we show the separation angle as function of E and $\Theta$.
 \begin{figure}[!h]
  \centering
  \includegraphics[width=3.in]{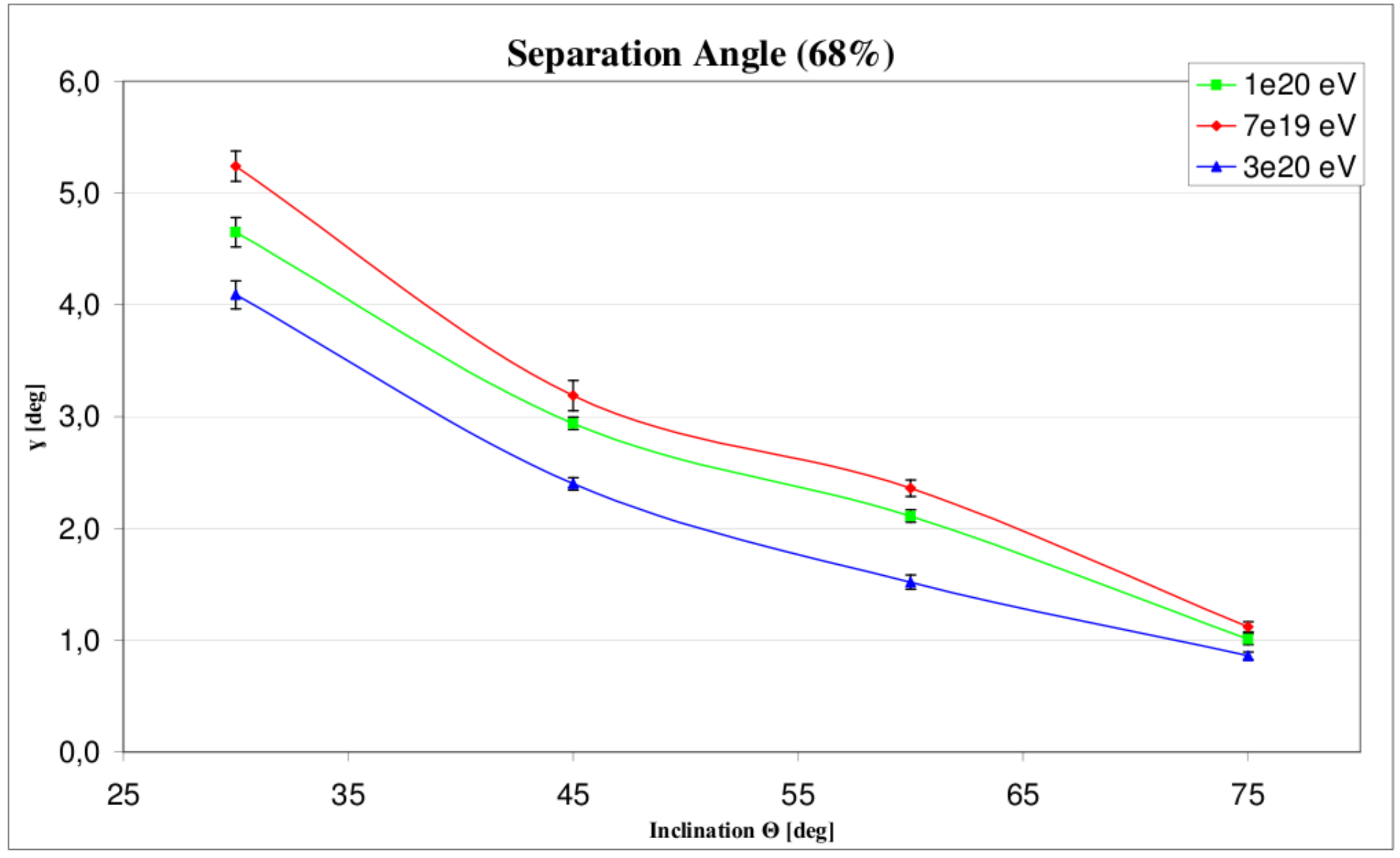}
  \caption{Angular resolution for different E ($7*10^{19}$,$1*10^{20}$, $3*10^{20}$ eV) and $\Theta$
(30$^{\circ}$, 45$^{\circ}$, 60$^{\circ}$, 75$^{\circ}$).}
  \label{theta}
 \end{figure}

\subsection{Energy Reconstruction}
\begin{figure}[!h]
  \centering
  \includegraphics[width=3.2in]{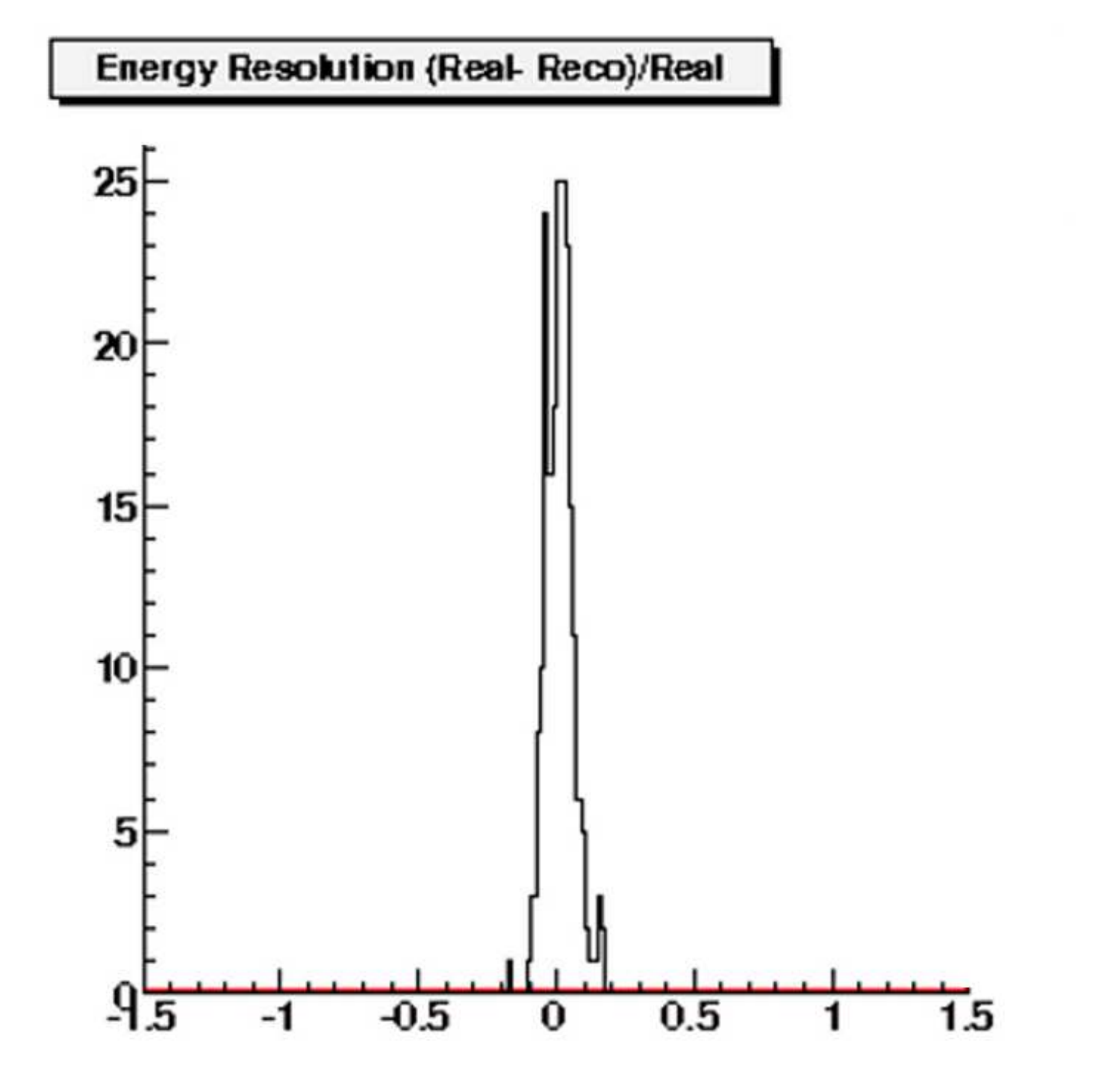}
  \caption{Energy resolution: $\frac{(E_{real}-E_{reco})}{E_{real}}$}
  \label{ERes}
 \end{figure}

In fig. \ref{ERes} we show the relative energy resolution for standard events (proton, E=$10^{20}$ eV, $\Theta$=60$^{\circ}$ ). 
For details see \cite{wSantangelo}.

\section{Conclusion}
ESAF is a powerful software and the tool of choice for the simulation and reconstruction of UHECR measeruments with space-based detectors. 
ESAF provides an independent and parallel assessment of the JEM-EUSO performance.
A complete End-To-End simulation and analysis of a larger number of events including studies about different primaries and changing atmospheric 
conditions such as cloud coverage is in progress (see \cite{wShinozaki}). Due to relatively recently introduced improvements such as the PulseFinder technique, we expect
significant improvements of the spatial and energy resolution in the near future.

\section*{Acknowledgements}
This work has been conducted by using the ESAF software which was written by the ESAF developers 
team during the Phase-A study of the EUSO mission under the initiative by the European Space Agency ESA.
We wish to thank RIKEN, Japan, for an allocation of computing resources on the Integrated Cluster of Clusters (RICC) system.
Moreover TM wishes to express his gratitude to the Computational Astrophysics Laboratory of RIKEN for their kind hospitality.


\clearpage


\newpage
\normalsize
\setcounter{section}{0}
\setcounter{figure}{0}
\setcounter{table}{0}
\setcounter{equation}{0}



\title{Simulation framework of STM code for development of JEM-EUSO instrument}

\shorttitle{author \etal paper short title}

\authors{Kazuhiro Higashide$^{1,2}$,Naoya Inoue$^{2}$,Takao
Shirahama$^{2}$,Keisuke Nagasawa$^{2}$,\\for JEM-EUSO Collaboration }
\afiliations{$^1$RIKEN, 2-1 Hirosawa, Wako, Saitama 351-0198, Japan\\
 $^2$The Graduate School of Science and Engineering, Saitama University, Saitama 338-8570, Japan}
\email{higashide@crsgm1.crinoue.phy.saitama-u.ac.jp}

\abstract{The Extreme Universe Space Observatory installed on the
Japanese Experimental Module at ISS(JEM-EUSO) is a science mission to
investigate the nature and origin of Ultra-High-Energy Cosmic-Rays
(UHECRs). The Saitama simulation is covering the End-to-End procedure
for the JEM-EUSO mission to study the telescope performance, therefore it has been used as a tool especially for the development of the optics and focal surface devices.
This implementation can contribute to the hardware development for
seeking the best performance of the telescope. The code offers the data
analysis routine for the simulated shower events, as well.
In this report, the framework of this simulation code will be introduced
and discussed.}

\maketitle

\section{Introduction}
The Extreme Universe Space Observatory installed on the
Japanese Experimental Module at ISS(JEM-EUSO) is a science
mission to investigate the nature and origin of Ultra-High-Energy
Cosmic-Rays(UHECRs)\cite{all}. 
An UHECR interacts with an atmospheric nucleus and
then produces an Extensive Air Shower (EAS). JEM-EUSO
on board of the ISS at the altitude of about 430$km$, captures the
moving track of the fluorescent UV photons along EAS development.

JEM-EUSO telescope
records EAS track with a time resolution of 2.5$\mu s$ and
a spatial resolution of 0.5$km$ at ground. These timesegmented
EAS images allow us determining EAS energies,
their arrival directions and longitudinal profiles.

The End-to-End simulation code has been developed considering hardware
characteristics of the JEM-EUSO optical system, the focal surface
detector and the output signal control circuit.

The accuracies of cosmic ray energy, arrival direction and longitudinal
development can be estimated by reconstructing EAS profiles with the
most suitable algorithm; the End-to-End simulation code is also used for
successive improvements of the hardware system to upgrade their
accuracies.


\section{STM code}
\begin{figure}[!h]
  \vspace{5mm}
  \centering
  \includegraphics[width=3.in]{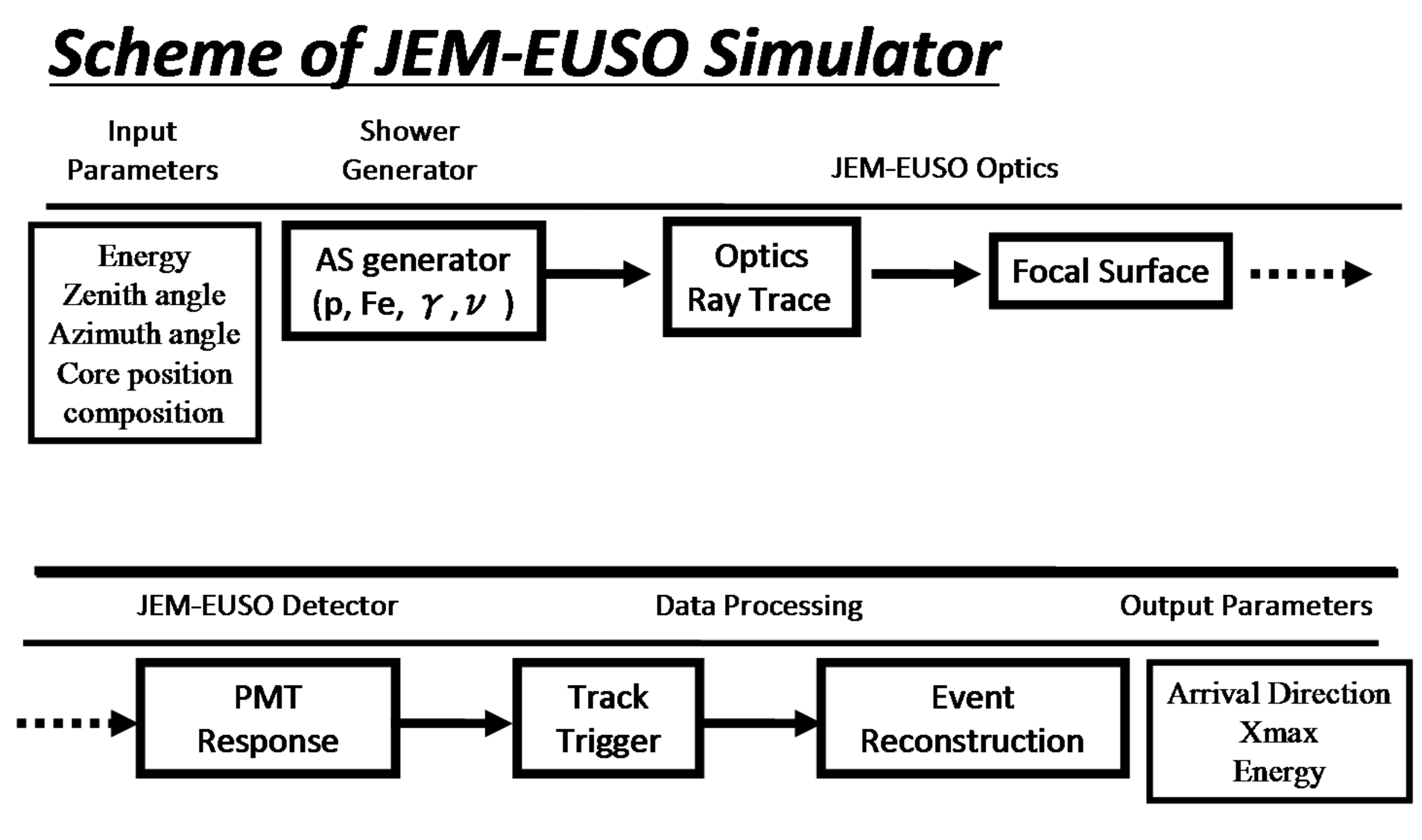}
  \caption{Scheme of STM code}
  \label{scheme}
 \end{figure}

The Shower and Telemetry Module(STM) code is End-to-End simulation code
for JEM-EUSO. The STM code have three
components(fig\ref{scheme}): EAS generation, detector simulation and
EAS event reconstruction part.This code was written by C language. Each
parts of STM code are independent and joined by Perl. 

\subsection{EAS generation part}
\begin{figure}[h]
  \vspace{5mm}
  \centering
  \includegraphics[width=3.in]{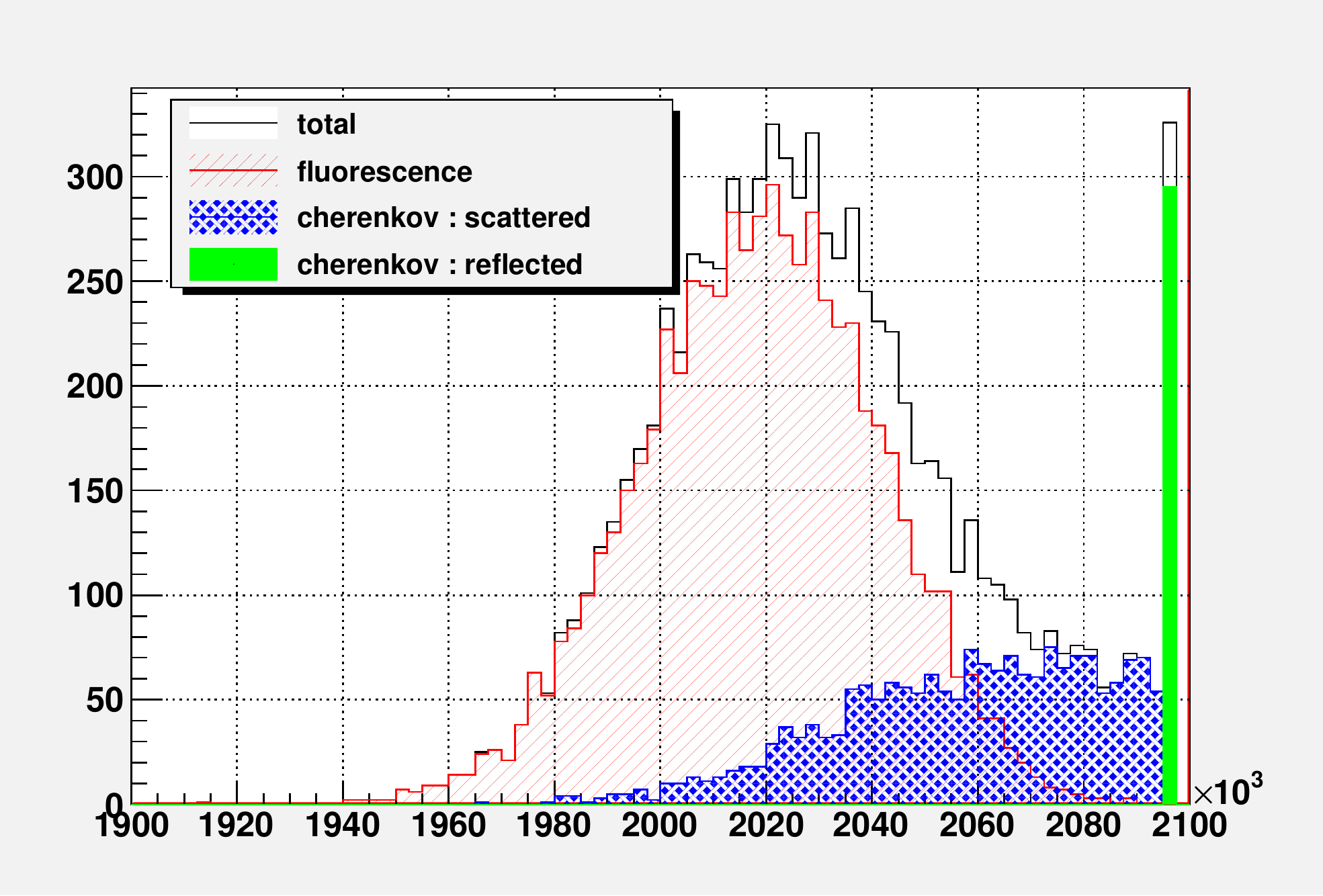}
  \caption{arivall time profile of photons on the optical lens of the telescope}
  \label{fig_time_profile}
 \end{figure}

EAS generation code generates EAS longitudinal profile in atmosphere
initiated by cosmic ray assumed chemical compositions, injected angle
and energy. EAS has been generated by executing the code for various EAS
longitudinal developments pooled in the EAS database made by AIRES or CONEX. 
Air fluorescence and Cherenkov light emissions are calculated
taking into account their yields \cite{fluorescence}.they have
wavelength 300-500$nm$.
Absorption and scattering in atmosphere(rayliegh scattering, mie
scattering and ozone absorption) are also calculated. And then the
characteristics of 
the photons (wavelength, arrival time and spatial position of emission)
on the optical lens of the telescope are evaluated.
Figure \ref{fig_time_profile} is sample of arrival time profile of
photons on the optical lens.
 
\subsection{Detector simulation part}
In the detector simulation code, characteristics of
hardware responses to incident photons, photoelectrons and
analogue/digital signals have been taken into account. 
The detector simulation have optical raytrece, PDM layout on focal surface,
PMT performance(QE, CE and etc.) and trigger algorithm for JEM-EUSO.

In optical raytrace simulation, main part of raytrace made by Y.Takizawa
and N,Sakaki. We converted for STM code. Optical raytrace simulation has
several lens design(old, baseline and advance design). Detail of Lens
design is 2.65m diameter side cut model(minimum diameter is 1.9m), and
PMMA or CYTOP material. 

\begin{figure}[h]
  \vspace{5mm}
  \centering
  \includegraphics[width=3.in]{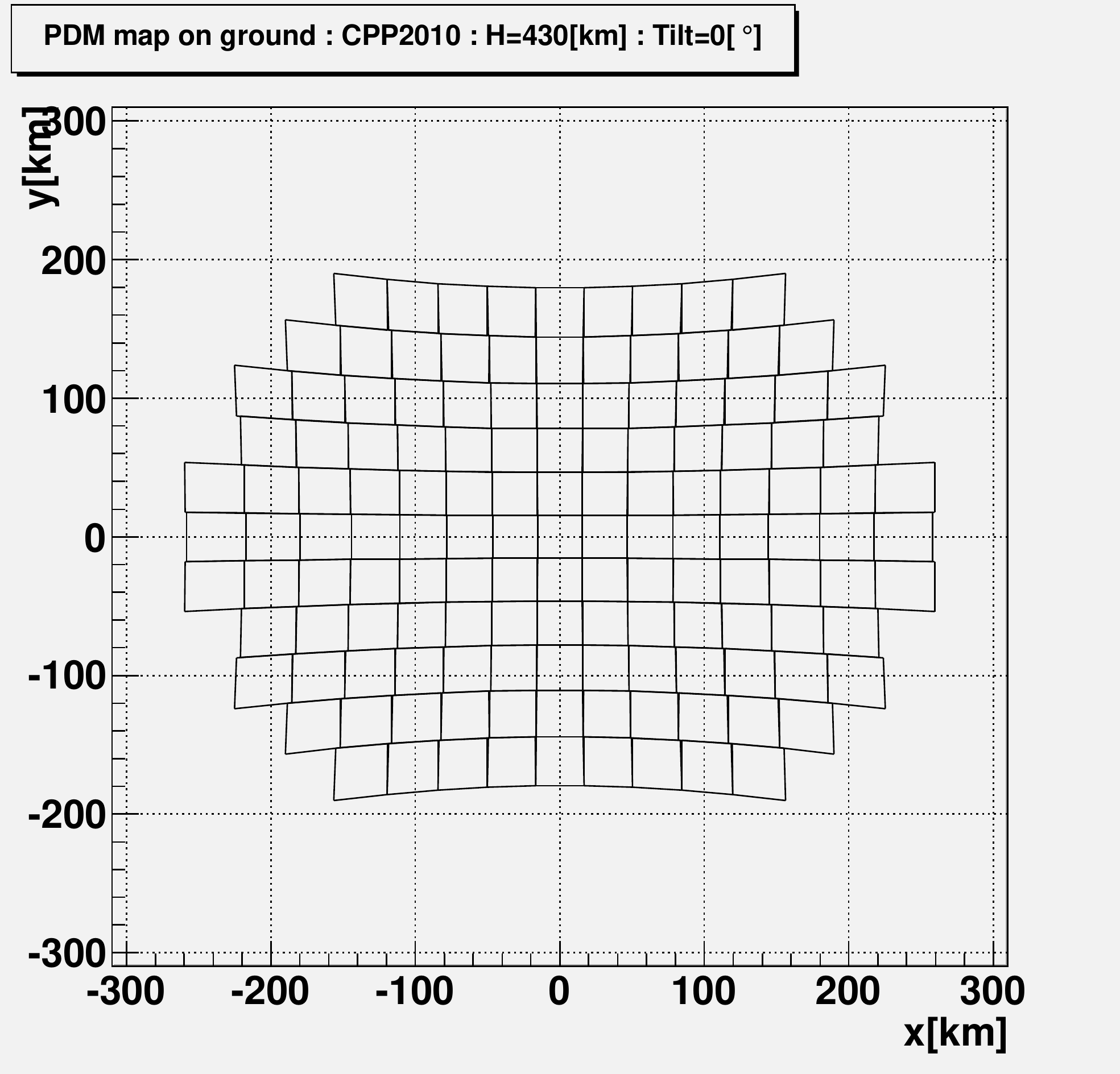}
  \caption{PDM layout thrown on earth}
  \label{fig_PDM_layout}
 \end{figure}

PDM layout on focal surface exists each lens design. It has gap of each
PDM and PMT. Number of PDM is 137(baseline) or 143(advance) on focal
surface and number of PMT is 36 on PDM. PDM length is 165mm and PMT
length is 26.04mm. Figure\ref{fig_PDM_layout} is PDM layout thrown on
earth. Area of a PDM on ground is about $40km \times 40km$ on center of
forcal surface. 

PMT simulation calculate number of photoelectron changed from photon with
quantum efficiency, collection efficiency of PMT. It has raytrace of
BG3 filter. 

Trigger scheme have two step. First step of trigger was called
progressive tracking trigger(PTT). Second step of trigger was called
linear tracking trigger(LTT). Detail of PTT and LTT can be found in
\cite{trigger}.    

\begin{figure}[h]
  \vspace{5mm}
  \centering
  \includegraphics[width=3.in]{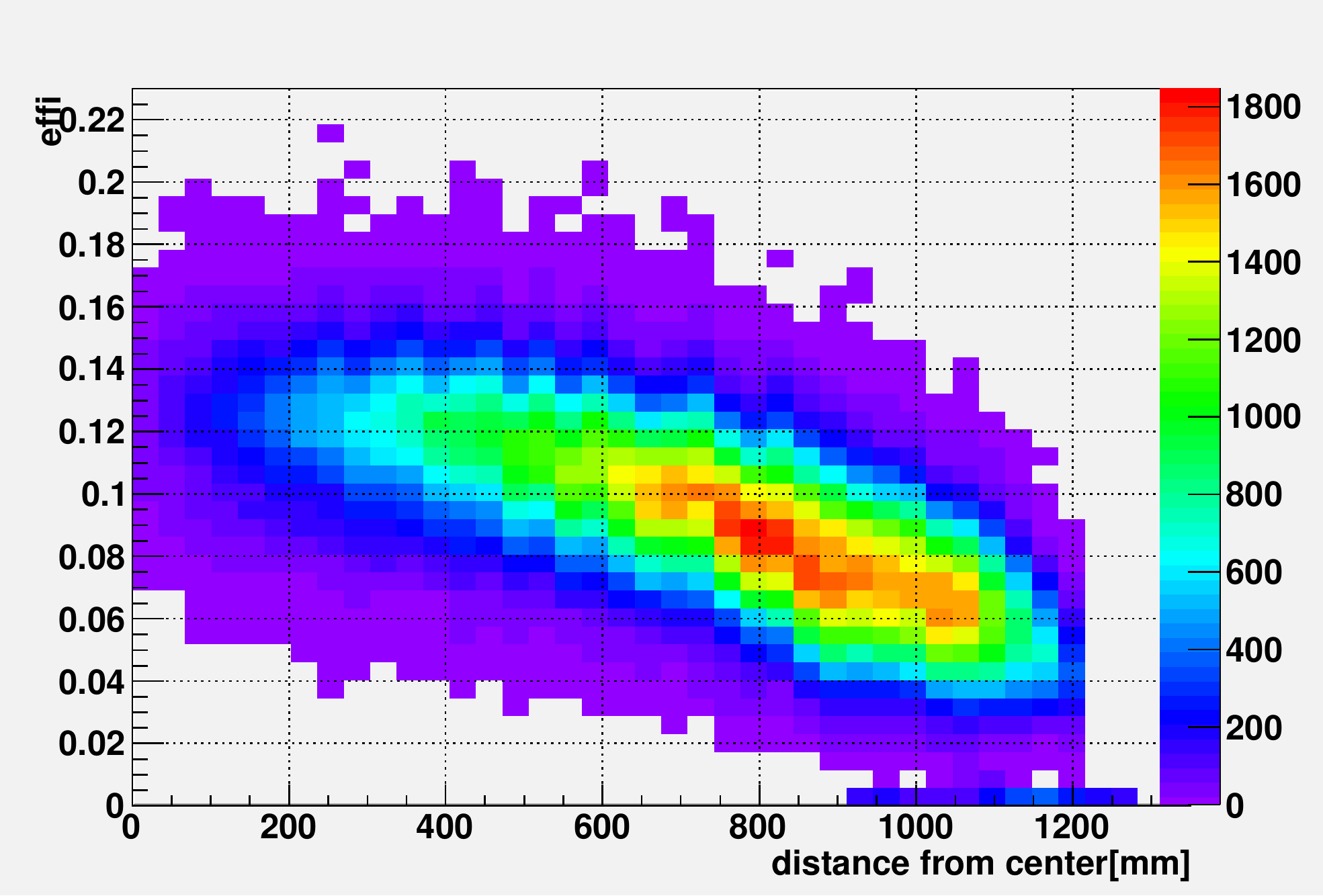}
  \caption{photon counting efficiency on pixel}
  \label{fig_effi_distance}
 \end{figure}

Finally pseudo-observational data including overall hardware
responses will be generated. Figure\ref{fig_effi_distance} is total photon
counting efficiency(from injected photon to detected photoelectron) on
pixel calculated by STM code. 

\subsection{EAS event reconstruction part}
\begin{figure}[!htbp]
  \vspace{5mm}
  \centering
  \includegraphics[width=3.in]{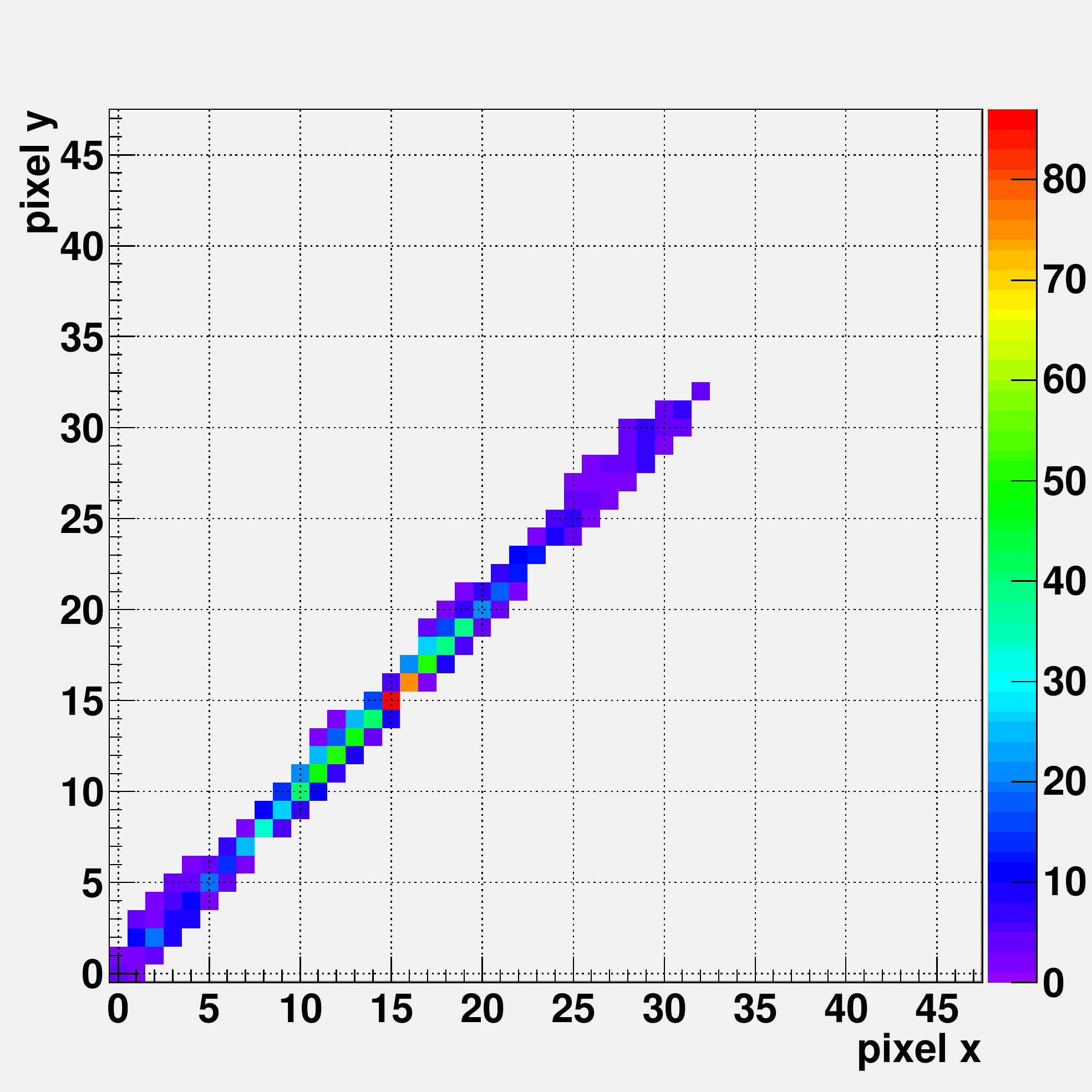}
  \caption{sample of reconstructed shower image calculated by STM
 code. energy=$10^{20}[eV]$. zenith angle=60[degree]]}
  \label{fig_shower_image_recon}
 \end{figure}

In addition, EAS event reconstruction code determines EAS energy,
arrival direction and longitudinal development from simulated
pseudo-observational data, and is used
for evaluating their accuracies. This part also contributes as a
feedback for the studies related to the development of analytical
algorithms and hardware improvements aiming at the excellent telescope‘s
capability with the best accuracies. Figure\ref{fig_shower_image_recon}
is a sample of reconstructed shower image calculated by STM code.

\section{Conclusions}
 The STM code is End-to-End simulation code for JEM-EUSO mission. In
 this paper we described the STM code framework.

\clearpage


  \hspace{4cm}
 \newpage

\newpage
\normalsize
\setcounter{section}{0}
\setcounter{figure}{0}
\setcounter{table}{0}
\setcounter{equation}{0}



\title{Estimation of effective aperture for extreme energy cosmic 
ray observation by JEM-EUSO Telescope}

\shorttitle{K. Shinozaki \etal, Estimation of aperture for EECR 
observation by the JEM-EUSO Telescope}

\authors{K. Shinozaki$^{1,2}$, M.E.~Bertaina $^{3}$, 
  S.~Biktemerova$^{4}$, P.~Bobik$^{5}$, F.~Fenu$^{6,1}$, 
  A.~Guzman$^{7}$, K. Higashide$^{8,1}$, G.~Medina Tanco$^{7}$, 
  T.~Mernik$^{6}$, J.A.~Morales~de~los~Rios~Pappa$^{2}$,  
  D.~Naumov$^{4}$, M.D.~Rodriguez~-~Frias$^{2}$, 
  G.~Sa\'ez C\'ano$^{2}$ and A. Santangelo$^{6}$ on behalf of 
  JEM-EUSO Collaboration}

\afiliations{
  $^1$Computational Astrophysics Laboratory, RIKEN Advanced 
  Science Institute, 2-1 Hirosawa, Wako 351-0198, Japan\\ 
  $^2$Space and Astroparticle Group (SPAS), University of 
  Alcala, E-28807 Alcal\'a de H\'enares, Spain\\
  $^3$Department of General Physics, University of Torino, 
  Via P. Giuria 1, I-10125 Turin, Italy\\
  $^4$Joint Institute for Nuclear Research, Joliot-Curie 6, 141980 
  Dubna, Moscow Region, Russia\\
  $^5$Institute of Experimental Physics, Slovak Academy of Science, 
  040 01 Kosice, Slovakia\\
  $^6$Institut f\"ur Astronomie und Astrophysik, Eberhard-Karls 
  Universit\"at T\"ubingen,  Sand 1, D-72076 T\"ubingen, Germany\\
  $^7$Institudo de Ciencias Nucleasres, Universitat Natcional 
  Autonoma Mexico, Mexico City 04510, Mexico\\
  $^8$Department of Physics, Saitama University, Saitama 338-8570, 
  Japan\\}
\email{kenjikry@riken.jp}

\abstract{JEM-EUSO (Extreme Universe Space Observatory on Japanese 
    Experimental Module) is a space-based new type observatory to 
    explore the extreme-energy-region Universe in particle channel.
    In the present work, we 
    estimated the effective aperture of the current baseline 
    configuration of the JEM-EUSO telescope in observing extreme 
    energy cosmic rays. We tested the effect of the qualty cut 
    among observed extensive air showers for cross-calibration with 
    other experiments. We also demonstrated several advanges for 
    the space-based JEM-EUSO observation.}


  
  \maketitle
  
  
  \section{Introduction}
  The origin and existence of extremely energetic cosmic rays 
  (EECRs; referred to as ones with energies $E_0$ several $\sim 
  10^{19}$ eV and higher) remains an open puzzle in the 
  contemporary astroparticle physics. Possible indications of 
  sources or excess of EECRs in Celestial Sphere have been 
  claimed by ground-based experiments \cite{tak,hir,cen}, 
  despite that capable sources are most powerful objects 
  within limited distances by the Greisen-Zatseptin-Kuzmin effect 
  \cite{gzk1,gzk2}. To investigate this puzzle, studies of energy 
  spectrum and arrival directions of EECRs against their 
  extremely low fluxes of 1 or fewer in km$^2$ per century, are 
  essential. The size of observation area is therefore critical 
  factor.

  JEM-EUSO (Extreme Universe Space Observatory on-board Japanese
  Experiment Module) is the observatory for EECRs 
  \cite{euso1,euso2}. The JEM-EUSO telescope will be accommodated 
  on JEM/Exposed Facility of the International Space Station 
  (ISS). The scientific objectives include astronomy and 
  astrophysics through EECR channel and other exploratory 
  objectives \cite{sci} such as detection of extreme energy gamma 
  rays and neutrinos. 
  
  By means of air fluorescence technique, the observation of EECRs 
  depends upon extensive air showers (EASs) phenomenon initiated 
  by primary EECRs.  This technique has been developed by several 
  ground-based fluorescence telescopes, however, never been 
  practiced in space.  From the orbit, EAS event is observed as a 
  luminous spot moving at the speed of light. For the event with
  an energy $E_0=10^{20}$ eV, for example, the EAS development 
  results in emission of an order of $10^{16}$ fluorescence photons 
  depending on the zenith angle  $\theta$ of EAS. The telescope 
  receives an order of thousands of photons per square meter 
  aperture.

  By monitoring night Earth with a wide field-of-view (FOV) 
  telescope, a series of advantages and scientific merits are 
  expected. When the JEM-EUSO telescope points to the nadir (nadir
  mode), unique geometry between EAS and telescope provides less 
  uncertainty in EAS reconstruction due to well-constrained 
  EAS-to-telescope distance. Observations over the orbit will 
  cover the entire Celestial Sphere that allows searching any 
  direction for EECR sources and for global arrival direction 
  distribution.  For scientific objectives, the most essential 
  merit is the observation area far larger than ground-based 
  telescope. We also plan to tilt the telescope off the nadir 
  toward the horizon (tilt mode) that enhances the  projected FOV 
  on the Earth's surface to allow more effective observation at 
  higher energies.

  In the present work, we focus on the aperture of the JEM-EUSO 
  trigger system for EECR observation.  We will discuss relevant 
  issues to estimate the exposure of the data. 

  \section{Apparatus and observation conditions}
  
  \paragraph{Apparatus}  
  The main part of the JEM-EUSO telescope consists of an 
  $\sim$ 4-m$^{2}$-aperture optics with three Fresnel lenses 
  \cite{opt} with aspherical curved focal surface (FS) covered by 
  about 137 photodetector modules (PDMs) \cite{fs}. Each PDM is 
  composed with 36 multi-anode photomultiplier tubes (MAPMTs) with 
  ultra-bialkali photocathode with 64 channels \cite{pmt}. PDMs 
  are aligned on FS to maximize the observation area. In the 
  baseline design, about 5000 MAPMTs are deployed and thus the 
  total number of pixels is $\sim 3\times 10^5$. symmetrically 
  cut with a $40^\circ$ segment. The spatial resolution for each 
  pixel corresponds to $\sim 0.07^\circ$  or $\sim 0.5$ km on the 
  Earth's surface for an orbit altitude $H_{\rm ISS} \sim 400$ km. 
  For each pixel, data is acquired with every 2.5 $\mu$s (gate 
  time unit) when the two consecutive levels of trigger schemes 
  are activated\cite{trg2}. These trigger schemes are referred to 
  persistent track trigger (PTT) and line track trigger (LTT). 
  Each scheme searches individual PDM for localized or aligned 
  excesses of signals. Threshold levels for PTT and LTT are 
  dynamically set to fit the rates within hardware requirement and 
  telemetry budget.

  \paragraph{Orbit and observation area}

  The orbit of the ISS has an inclination $i = 51.6^\circ$ with
  $H_{\rm ISS}$ ranging in 278--460 km by the operational limit. 
  The sub-satellite speed and period are $\sim 7$ km/s and $\sim 
  90$ minutes, respectively. Apart from effects by orbital decay 
  and operational boost-up, the ISS motion is approximated as a 
  circular motion with an eccentricity of practically 1. Among 
  these elements, $H_{\rm ISS}$ is widely variable  throughout 
  its operation and so far has range between $\sim 350$ and $\sim 
  400$ km. 

  The `observation area' of JEM-EUSO which depends upon tilting 
  angle $\xi$ off the nadir and $H_{\rm ISS}$ is estimated by ray 
  trace simulations  \cite{opt,stm} for isotropic light source 
  viewed by the FS detectors. In the following we defined it as
  the projected area on the Earth's surface from which the main 
  ray of photons are detected within outer most boundaries of the 
  FS detector.
  
\begin{figure}
    \vspace{5mm}
    \centering
    \includegraphics[width=2.9in]{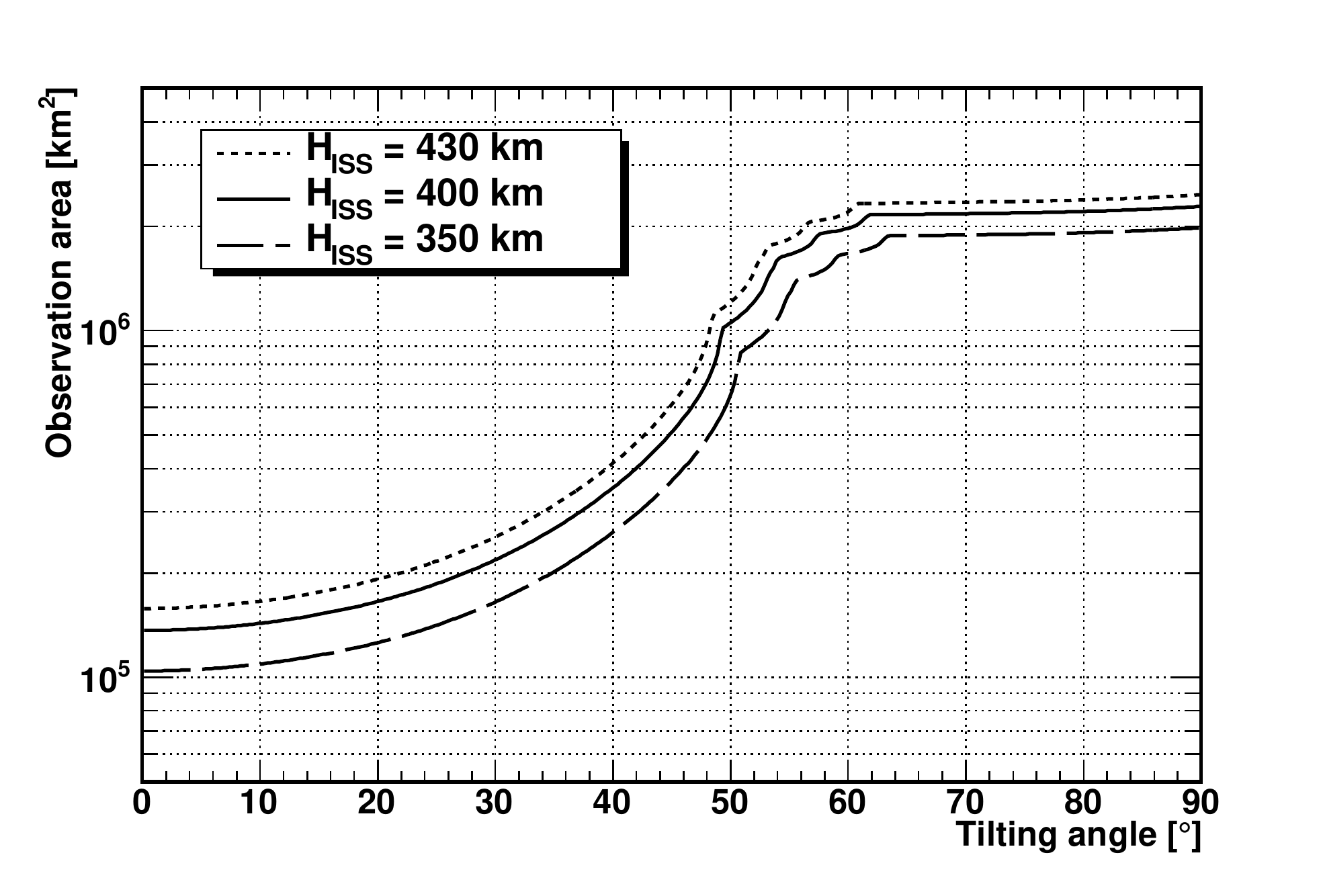}
    \caption{Observation area  as a function of tilting angle for 
      different altitudes of 400 km (solid line), 350 km (dashed 
      line) and 430 km (dotted line).}
    \label{fig1}
  \end{figure}

  Figure~\ref{fig1} shows the observation area as a function of 
  tilting angle for different $H_{\rm ISS} = 350, 400$ and 430 km.

  For the baseline layout of 137 PDMs, the observation area 
  $A_{\rm obs}^{\rm (nadir)}$ for nadir mode is a function of 
  $H_{\rm ISS}$ expressed by:

  \begin{equation}
    A_{\rm obs}^{\rm (nadir)}[{\rm km}^{2}] \approx 1.4 \times 
    10^{5} \cdot {\displaystyle \left(\frac{H_{\rm ISS}}{400[{\rm 
	    km}]}\right)^2}
  \end{equation}
  
  With tilting angles $\xi$ up to  $\sim 40^\circ$, the observation 
  area $A_{\rm obs}$ is approximated as follows:

  \begin{equation}
    A_{\rm obs}(\xi) \approx 
    A_{\rm obs}^{\rm (nadir)} (\cos\xi)^{-b}
  \end{equation}
  where $b$ ranges $3.2$--$3.4$ for the altitude of interest. In 
  this region $A_{\rm obs}$ increases with $\xi.$ Around $xi \sim 
  40-50$ degrees depending upon $H_{\rm ISS}$, a part of FOV views 
  the sky over the local horizon and $A_{\rm obs}$ saturates above 
  $\xi \sim 60^\circ$. 

  \paragraph{Background and cloud impact}

  The level of background (BG) noise is a key parameter to define 
  the observation and schemes that yields the observation
  duty cycle $\eta_0$ as well. The first order constraint for 
  $\eta_0$ is astronomically determined by the ISS transit over 
  terminator. For $H_{\rm ISS} \sim 400$ km, the average fraction 
  of nighttime is $\sim 33\%$  at the orbital altitude. By applying 
  the upper limit of the BG  flux in UV range of 300--400 nm less 
  than 1500 photon m$^{-2}$ sr$^{-1}$, $\eta_0$ corresponds to 
  $\sim 20\%$ (see \cite{bg} for details). In this criterion, the 
  average background flux is $\sim 500$ photons m$^{-2}$ sr$^{-1}$ 
  ns$^{-1}$ (referred to `average BG level').  Note that the 
  presence of the Moon with its phase close to New Moon is 
  included in operational time as JEM-EUSO telescope is only 
  affected by the illumination of Earth's surface. 
  
  The impact of clouds is estimated by the global secular 
  statistics of the optical depth and cloud-top altitude 
  \cite{trn} convolved with the trigger  probability for each 
  case. The trigger aperture for the time-average cloudy 
  condition is $\sim 80\%$ above $\sim 5\times 10^{20}$ eV in 
  comparison with that for the cloud-free case. Applying quality 
  cut for events with shower maximum above the optically thick 
  clouds, the overall impact factor is estimated to be 
  $\kappa_{\rm C} \sim 70\%$  above $3\times 10^{19}$ eV (see 
  \cite{saz} for details).
    
  \section{Simulation and results}

  \paragraph{Simulation}
  In the present work, we employed the ESAF (Euso Simulation and 
  Analysis Framework) \cite{esa1,esa2} adapted into the present 
  JEM-EUSO baseline configuration.  The software is written in C++ 
  using an object-oriented programming approach and runs on the 
  ROOT package \cite{root}. EAS generation is based on the GIL 
  (Greisen-Ilina-Linsley) formulation  \cite{slast} that 
  reproduced the longitudinal development of hadronic showers 
  simulated by CORSIKA \cite{cor} with QGSJET interaction model 
  \cite{qgs}. Fluorescence yield  is a well recognized uncertainty 
  for energy scale \cite{flu,flu2}. In the present work, we 
  assumed it by Nagano \etal \cite{ngn}. To estimate trigger 
  aperture, we simulated a large number of EAS uniformly injected 
  into an area far larger than $A_{\rm obs}^{\rm (nadir)}$. 
  $H_{\rm ISS}$ is set to be 350 and 400 km. Threshold levels for 
  PTT and LTT trigger judgements need to fit within 
  permissible fake trigger rates, while it is preferable to keep 
  as low as possible. For optimizations of those parameters, we 
  generated a large amount of noise simulations by STM code 
  \cite{stm}.

  \begin{figure}[!t]
    \vspace{5mm}
    \centering
    \includegraphics[width=2.9in]{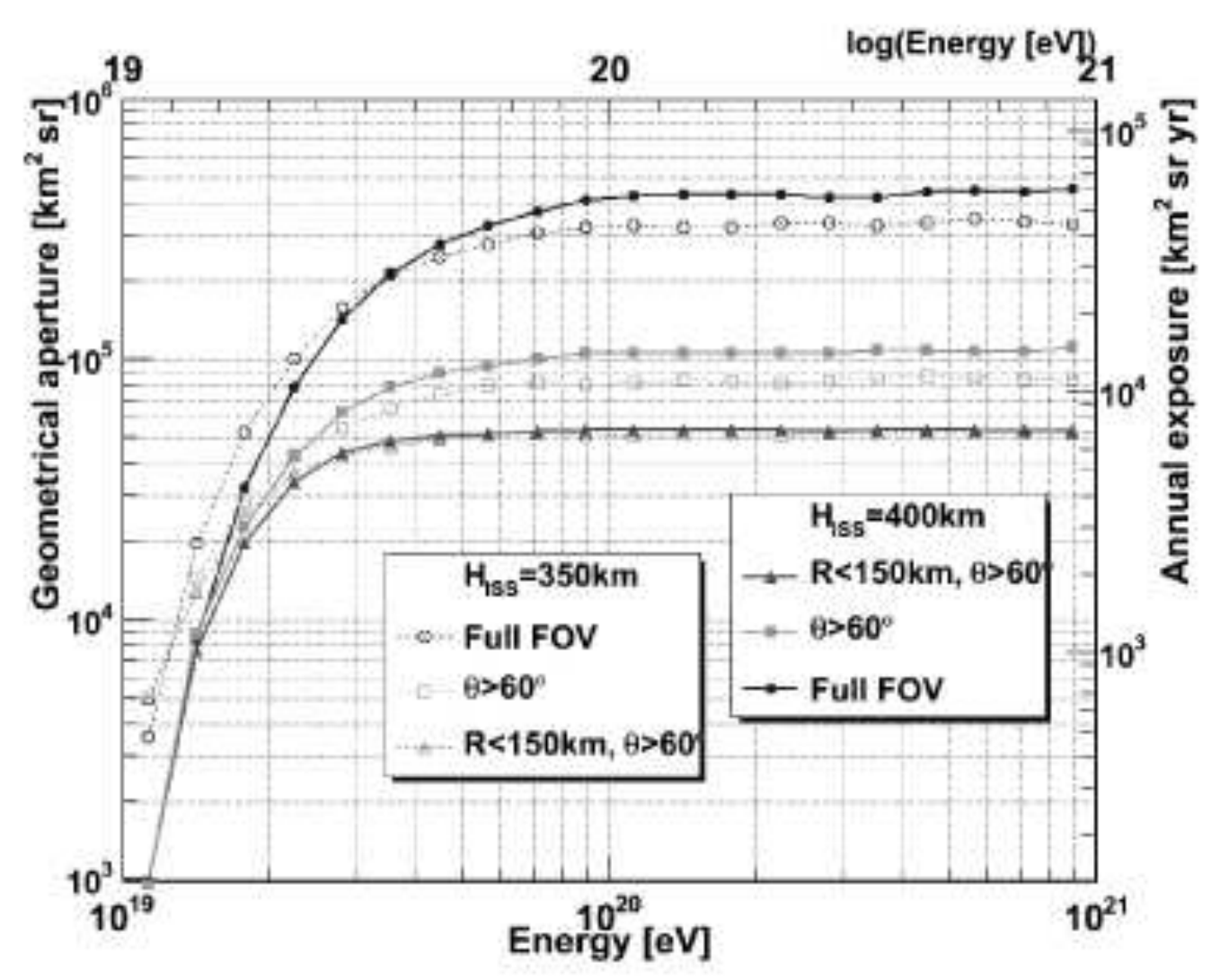}
    \caption{Geometrical aperture as a function of $E_0$. Open 
      and closed circles indicate geometrical apertures for the 
      ISS altitudes of 400 and 350 km, respectively. Squares and 
      triangles show the cases of different geometrical cuts of 
      $\theta>60^\circ$ and  $R<150$ km, respectively. The 
      vertical axis on the right represent annual exposure taking 
      into account observation duty cycle and cloud impact.}
    \label{fig2}
  \end{figure}
  
  \paragraph{Geometrical aperture}
  
  Unless otherwise noted, we define `geometrical aperture' based 
  on the probability satisfying second level LTT trigger condition
  by means of Monte  Carlo simulations. The time-variant 
  conditions such as cloud coverage or BG level are excluded in 
  definition. In the present work, we assume the clear sky 
  condition with average BG level. The exposure growth per given 
  time may be evaluated by a product of $\eta_0$ and 
  $\kappa_{\rm C}$ in the previous section.  The estimation herein 
  is a preliminary result for the current baseline detector 
  configuration for the nadir mode.

  For $N_{\rm trig}$ trigger events   among simulated 
  $N_{\rm inject}$ injected EECRs with an energy $E_0$, the 
  corresponding geometrical aperture $A(E_0)$ is defined as 
  follows:
  \begin{equation}
    A(E_0) = {\displaystyle  
      \frac{N_{\rm trig}}{N_{\rm inject}}  \cdot S_0 
      \cdot {\Omega_0}}
    \label{eqn:af}
  \end{equation}
  where $S_0$ and $\Omega_0 = \pi$ [sr] for $\theta = 
  0^\circ-90^\circ$ are 
  the area and the effective solid angle, respectively, in which 
  uniform EAS flux is assumed. To evaluate full geometrical 
  aperture, we applied $S_0 \gg A_{\rm obs}$ to take into account 
  EAS crossing FOV with a core location out of the observation 
  area.

  By applying the geometrical selection for good quality events 
  by core location distance $R$ from the center of FOV and lower 
  limit of zenith angle $\theta_{\rm cut},$ subset of geometrical 
  aperture for a given energy is expressed as follows:
  \begin{equation}
    A_{\rm sub}
    \propto {\displaystyle \int_{0}^{R_{\rm max}} 
      \int^{90^\circ}_{\theta_{\rm cut}}  \epsilon(\theta,\vec{r}) 
      \cdot\sin\theta\cos\theta d\theta \cdot r dr}
  \end{equation}
  where $\epsilon(\theta,\vec{r})$ is the probability of trigger at 
  the location of $\vec{r}$ with respect to the correspoding 
  position on Earth's surface to the center of FOV. The amount of 
  light produced in EAS increases with zenith angle since the 
  apparent EAS track becomes longer before being truncated at 
  Earth's surface. In the inner part of FOV, higher efficiency in 
  trigger is expected due to better focusing power of the optics 
  along with shorter EAS-to-telescope distance. 

  Figure~\ref{fig2} shows the geometrical aperture as a function 
  of $E_0$ for $H_{\rm ISS}= 400$ and 350 km.  Effects of different 
  geometrical cuts in $\theta$ and $R$ are also demonstrated. The 
  scale of annual exposure (growth in exposure by one-year 
  operation) is also shown on the right by taking into account 
  $\eta_0 =0.2$ and $\kappa_{\rm C}$ = 0.7 (see caption and legend 
  for details).
  
  At highest energies, the geometrical aperture for full FOV is 
  almost constant above $\sim (6-7)\times 10^{19}$ eV. The 
  saturated aperture is determined by $A_{\rm obs}$ for given 
  $H_{\rm ISS}$ and therefore the higher altitudes result in the 
  larger apertures.  Comparing annual exposure to the Auger (7000 
  km$^2$ sr yr) \cite{agbei}, it is expected to be $\sim$ 9 times 
  for $H_{\rm ISS} = 400$ km. 

  Applying $\theta_{\rm cut} = 60^\circ$ cut to full FOV, while 
  the effective solid angle reduces to $\pi/4$ [sr], almost 
  constant aperture is achieved above $\sim (4-5) \times 10^{19}$ 
  eV. In addition, more stringent $R_{\rm max}=150$ km cut extends  
  such range down to $\sim (2-3)\times 10^{19}$ eV. It is worthy to
  mention that for lower $H_{\rm ISS}$ shorter EAS-to-telescope 
  distances increases $\epsilon(\theta,\vec{r})$ for the same 
  energy.  This results in the larger apertures and enable better 
  comparison with other experiments in more extended energy range. 

  \begin{figure}[!t]
    \vspace{5mm}
    \centering
    \includegraphics[width=2.9in]{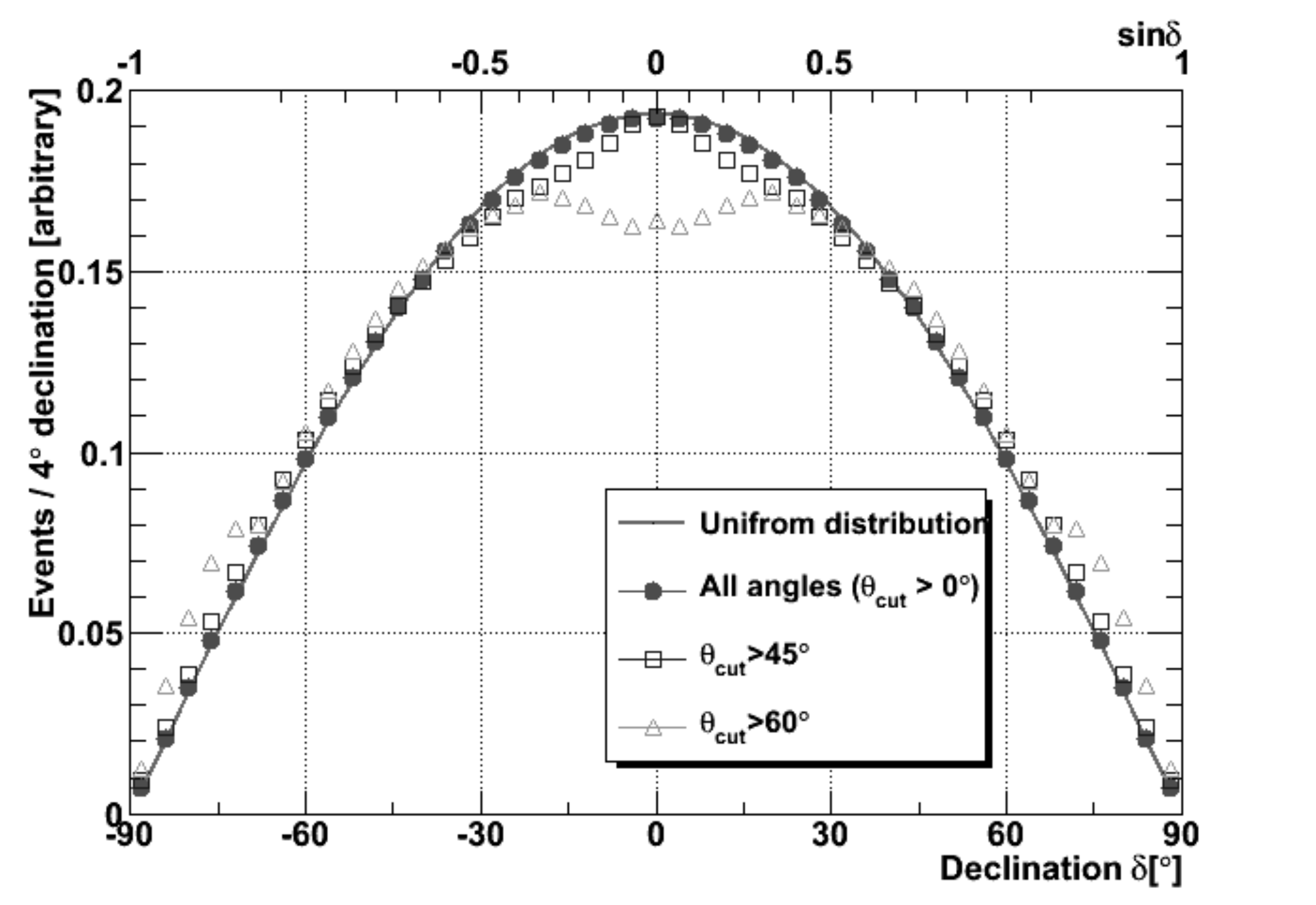}
    \caption{Declination $\delta$ distribution of triggered 
      events for different $\theta_{\rm cut} = 0^\circ$ 
      (circles), 
    $45^\circ$ (squares) and $60^\circ$ (triangles) in comparison 
      with uniform distribution (solid curve). The horizontal axis 
      on the top shows $\sin\delta$ to indicate the solid angle 
      coverage on the Celestial Sphere.}
    \label{fig3}
  \end{figure}

  \paragraph{Uniformity of exposure} 

  Unlike stationary ground-based observatories, global ISS orbit 
  and better sensitivities for large $\theta$ EAS allow to scan 
  the entire Celestial Sphere. The exposure distribution  is 
  practically flat in right accession. Apart from possible local 
  or seasonal deviation from the global average of cloud coverage 
  and BG level, the relationship between expected overall exposure 
  and declination can be analytically expressed as a function of 
  only $\theta_{\rm cut}$, knowing observable night time at a 
  given latitude. 

  Figure~\ref{fig3} shows expected distribution of triggered 
  events in declination for different $\theta_{\rm cut}=0^\circ$, 
  $45^\circ$ and $60^\circ$ cuts compared with uniform 
  distribution.
  
  For the case of $\theta_{\rm cut} = 60^\circ$ cut, minor 
  excesses and deficit may arise in very limited parts near 
  Celestial Poles and Equator, respectively.  It is because 
  sinuous variation in latitude of the orbit and JEM-EUSO stays 
  longer in high latitudes.  JEM-EUSO can achieve well constant 
  exposures for full range of $\theta$ with which arrival 
  direction analysis will be made. In the case of ground 
  observatories, first of all they are constrained in observation 
  of never-rising region below the local horizon and the 
  correction factor for non-uniform observable region may even 
  reach $\sim 3.$ 
 
  \section{Summary and discussion}

  In the present work, we simulated a large number of EAS to 
  estimate the effective aperture for present baseline 
  configuration and argued the relevant issues. 
  $A_{\rm obs}^{\rm (nadir)}$ is proportional to the square of 
  ${H_{\rm ISS}}$ which is highly dependent upon the ISS 
  operation. In the mission, the science case has assumed 
  ${H_{\rm ISS}}$ to be either $\sim$ 400 km, or 430 km following 
  the prediction at the time of EUSO mission \cite{eus}. In  
  case of lower altitudes such as 350 km, $A_{\rm obs}$ is compensated by 
  tilting $\sim 25^\circ$ to that of the nadir mode at 430 km 
  altitude without dramatic change of EAS-to-telescope distance.

  The geometrical aperture was estimated for clear sky condition. 
  It is important to mention that applying geometrical cuts helps 
  discriminate good quality events in the energy range $(2-3)\times 
  10^{19}$ eV at constant exposure with energy.  Such subset of 
  EAS data makes it possible to cross-check energy spectrum and 
  performances with ground-based experiments at equivalent 
  statistical power. Once it is carried, exposure at higher 
  energies overwhelm by removing such cuts. Taking int account 
  factors of $\eta_0$ and $\kappa_{\rm C}$, $\sim 9$ times annual 
  exposure is expected in comparison with that of Auger. 
  Particularly to increase the statistics at highest energies 
  $\sim (3-5)\times 10^{20}$ eV, we plan to operate the telescope
  in tilt mode and also with higher BG level threshold.

  The full coverage of EECR observation in  Celestial Sphere is 
  unique characteristics for the JEM-EUSO and moreover the overall 
  exposure results in almost uniform at the first order. Such an 
  advantage is more pronounced for arrival direction analysis, 
  especially against spread EECR sources.

  Some results shown herein are in progress. Further details on 
  the general performance can be also referred in \cite{san}. 
  
  \section*{Acknowledgement}
  
  KS wishes to express his gratitude to University of Alcal\'a 
  (UAH), Eberhard-Karls Universit\"at T\"ubingen and University of 
  Torino for their hospitality and excellent working conditions.  
  Computation facilities of RICC (RIKEN Integrated 
  Cluster of Clusters) System and of UAH-SPAS are acknowledged for 
  efficiently performing simulations. The present work was 
  supported in part by the Italian Ministry of Foreign Affairs, 
  General Direction for the Cultural Promotion and Cooperation.

  \clearpage
  

\newpage
\normalsize
\setcounter{section}{0}
\setcounter{figure}{0}
\setcounter{table}{0}
\setcounter{equation}{0}



\title{Precise Fluorescence Yield Measurement Using an MeV Electron Beam for JEM-EUSO Collaboration}

\shorttitle{D. Monnier Ragaigne \etal Fluorescence Yield Measurement for JEM-EUSO Collaboration}

\authors{D. Monnier Ragaigne $^{1}$, S. Dagoret-Campagne$^{1}$,P. Gorodetzky$^{2}$, J. Baret$^{1}$, M. Urban$^{1}$, C. Blaksley$^{2}$, F.
Wicek$^{1}$, T. Patzak$^{2}$ and S. Biktemerova$^{3}$ on behalf of the JEM-EUSO Collaboration}
\afiliations{$^1$Laboratoire de l'Acc\'el\'erateur Lin\'eaire, Univ Paris-Sud, CNRS/IN2P3, Orsay, France\\ $^2$Laboratoire Astroparticule et
Cosmologie, APC, Paris, France\\$^3$Joint Institute for Nuclear Research, JINR, Russia}
\email{monnier@lal.in2p3.fr}

\abstract{A new type of absolute measurement of the nitrogen fluorescence yield in the air  will be performed at LAL using 3 items which will yield 
an unprecedented precision in all conditions of pressure, temperature, and pollutants.
A 5 MeV electron beam will be provided by the new electron accelerator PHIL at LAL. As the fluorescence yield is proportional to the energy
loss of the electrons, the contribution of secondary electrons (deltas) to the signal is much more important than the contribution of the
primary electrons. It has therefore been chosen to use an integrating sphere, the basic property of which being that the probability to detect light is
independent from where the light is produced inside the sphere. An output device on this sphere will be equipped with a set of optical fibers driving the fluorescence
light to a Jobin-Yvon spectrometer equipped with an LN$_{2}$ cooled CCD. The fluorescence spectrum in the 300-430 nm range will be accurately
measured in steps of 0.1 nm resolution. A PMT equipped with a BG3 filter (the same as on JEM-EUSO) will be set on the sphere to measure
the integrated yield. The sphere will be monitored by a NIST photo-diode, and will be surrounded by a spherical envelope to create a temperature
controlled chamber (a Dewar). With this setup it will be possible to vary the temperature from $-60^{\circ} C$ to  $+40^{\circ} C$ and the pressure from 1 to 0.01 atm. The expected precision of the yield should be
better than 5\%. }

\maketitle

\section{Introduction}
A precise measurement of the energy is essential for the study of ultra-high energy
cosmic rays. Basically, two types of detectors are used for this purpose:
\begin{itemize}

\item Surface arrays which sample the shower tail: this method records the
lateral development of the shower of secondary particles using an array of particle detectors.
\item Fluorescence detectors which record the longitudinal development of the shower and
observe the atmospheric fluorescence induced by charged particles in the shower.

\end{itemize}
The second method is currently the most precise one to estimate the energy of cosmic rays,
and is used by the Fly's Eye experiment \cite{qlab1}, HiRes \cite{qlab2}, Telescope
Array \cite{qlab3}, and the Pierre Auger Observatory \cite{qlab4}. The future
JEM-EUSO telescope \cite{qlab5} will also detect extensive air showers from the International Space
Station with this method.

Fluorescence detectors provide a measurement of primary cosmic ray energy which is relatively model
independent, as the fluorescence intensity is proportional to the electromagnetic energy released 
by the shower into the atmosphere. 
For the Pierre Auger Observatory, the uncertainty in the energy using the fluorescence method is around $22\%$, and  the main source of systematic uncertainties comes from the limited accuracy
in the measurement of the air-fluorescence yield.
In the Pierre Auger Observatory\cite{qlab4} the uncertainty in the fluorescence yield
contributes $14\%$ to the total systematic error of the energy calibration.
This parameter is thus a key for determining the energy of
ultra-high energy cosmic rays detected by a fluorescence telescope.
We will measure the fluorescence yield using a 5 MeV electron beam and calibrated detectors in order to improve the
accuracy of this value to a precision of $5\%$.

\section{Fluorescence Yield}
Air-fluorescence photons are produced by the de-excitation of atmospheric
nitrogen molecules excited by the shower electrons.
Excited molecules can also decay by
colliding with other molecules, using the process of collisional quenching. 
This effect increases with
pressure, reducing fluorescence intensity.

Atmospheric effects, including pressure, temperature, and composition, must also be reproduced and studied in order to
understand the real conditions present during the production of fluorescence photons within an extensive air shower.
As the excitation cross sections show a fast decrease with energy, secondary electrons from ionization processes are the main source of fluorescence light.
For this reason, it is necessary to simulate the production of fluorescence photons in order to evaluate the fiducial volume needed for interaction.
The fluorescence spectrum
consists of a set of molecular bands represented by a set of discrete wavelengths $\lambda$. The
range of this spectrum is the near UV between 300 to 430 nm.

The fluorescence yield for a line, $Y_{\gamma}$, is defined as the number of photons emitted by the primary charged particle per meter of path.
The deposited energy of an electron per unit of length is defined as:

\begin{equation}
\rho \frac{dE}{dX}
\end{equation}
The number of photons produced with this energy depends on the fluorescence efficiency of the line, $\phi_{\gamma}$:

\begin{equation}
Y_{\gamma}(photons/e/cm)=\phi_{\gamma}\frac{\rho}{h\nu} \frac{dE}{dX}.
\end{equation}

This efficiency, $\phi_{\gamma}$, depends on the lifetime of the level (de-excitation) and also on the effect of
pressure, temperature, and composition \cite{qlab6}.

The total fluorescence yield $Y_{tot}$ is thus the sum of all $Y_{\gamma}$:
\begin{equation}
Y_{tot}=\sum\limits_{\gamma} Y_{\gamma}.
\end{equation}

Knowing both the fluorescence yield and its dependence on atmospheric properties accurately
is essential in order to obtain a reliable measurement of the energy of cosmic rays in experiments using the fluorescence method
\cite{qlab7}, \cite{qlab8} and \cite{qlab11}.
Studying the total spectrum of fluorescence emission is also fundamental for JEM-EUSO in order to optimize data analysis. 

\section{Principle of the experiment}

\subsection{Experimental Set-up}
The aim of this experiment is to measure the fluorescence yield of each line with a $5\%$ accuracy using an electron beam as a source of electrons (reproducing the electrons of
an extensive air shower), an integrating sphere with control of pressure, temperature, and composition in order to measure atmospheric effects, and calibrated detectors.

The electron beam will interact with gas inside an integrating sphere. A fraction of the emitted fluorescence light will be detected and measured with both
a Jobin-Yvon spectrometer equipped with an LN$_{2}$ cooled CCD, in order to study each spectral line separately, and also a photo-multiplier tube
equipped with a BG3 filter (the same filter as the JEM-EUSO
project).

The integrating sphere must be vacuum-tight and part of a dewar to allow studying the yield at low temperatures (down to
$-60^{\circ} C$).The basic property of the integrating sphere being that the probability to detect light is
independent from where the light is produced inside the sphere
The size of the sphere depends on pressure (due to the pressure dependence of the distance of ionization of secondary electrons and multiple
scattering) from a few centimeters at 1 atm to a few decimeters at very low pressure (0.01 atm).
The exact size of the sphere is determined using Geant4 simulations to reproduce multiple scattering and the mean free
path of secondary electrons.

The source of electrons is an electron accelerator (PHIL) developed at the Laboratoire de l'Acc\'el\'erateur Lin\'eaire (LAL) and presented in the next section.

The calibration of the detectors is fundamental in order to obtain an accurate measurement of the fluorescence
yield.

\begin{figure*}[th]
  \centering
   \includegraphics[width=7in,height=3.5in]{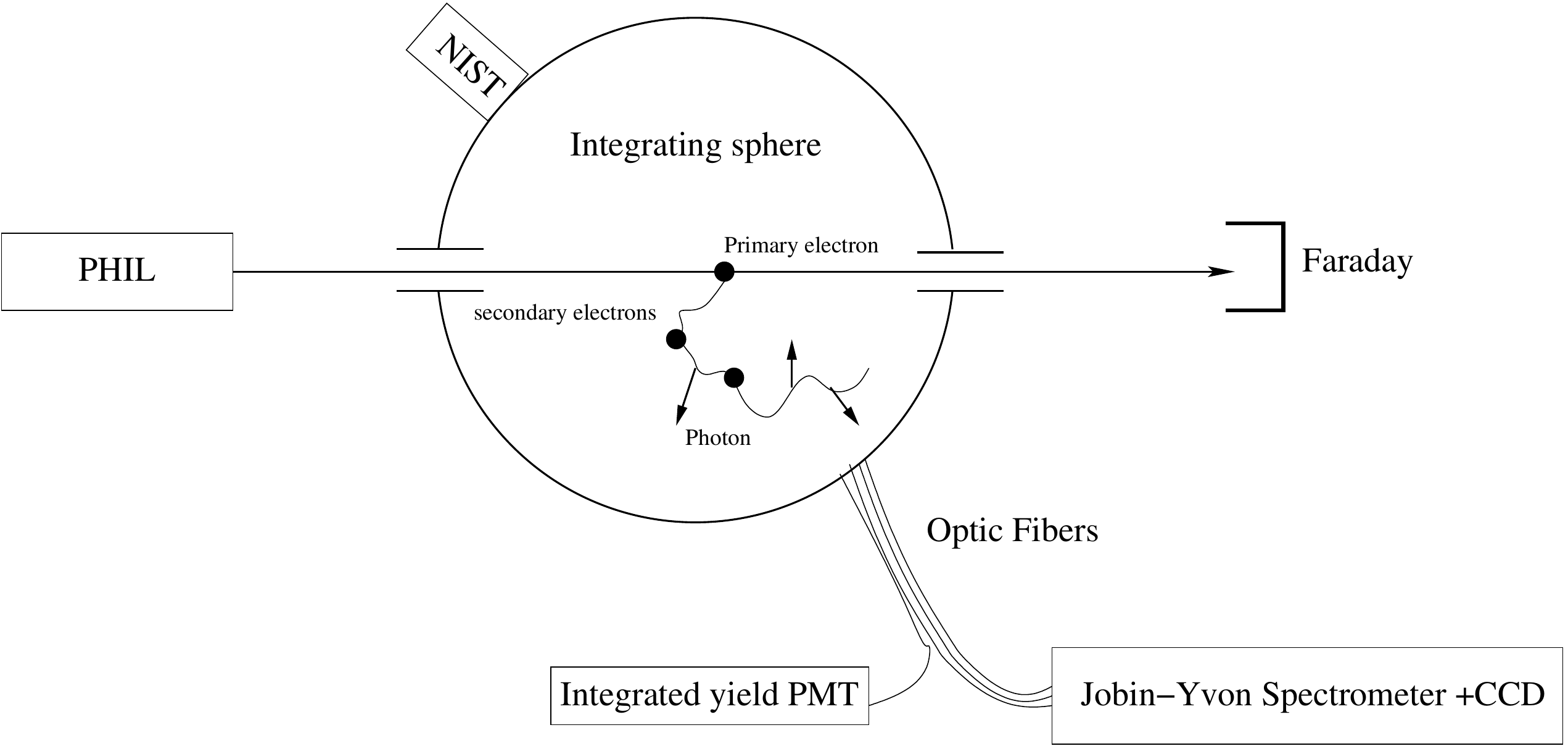}
  \caption{Design of experiment}
  \label{wide_fig}
 \end{figure*}

\subsection{PHIL: the electron Beam}
The ``PHoto-Injector at LAL''  ( \cite{qlab9} and \cite{qlab10}) is an electron
beam accelerator at LAL.
This accelerator, which is primarily dedicated to the testing and characterization of electron photo-guns
and high-frequency structures for future accelerator projects, can also be used
to simulate the electrons emitted by an extensive air shower.

PHIL is currently a 6-meter-long accelerator with 2 diagnostic beam lines. The
direct beam line will be used to inject electrons into an integrating sphere.
An Integrating Current Transformer (ICT) will provide the estimated beam
charge, beam size, and beam position measurement with high accuracy.
The main characteristics of PHIL, for our configuration, have been summed up in the
table \ref{table_PHIL}.
For the measurement of the fluorescence yield, precise knowledge of the source (energy, position, charge...) is an important part of the total
accuracy. Using the PHIL accelerator, these parameters will be available with an accuracy of $\sim2\%$.

\begin{table*}
\begin{center}
\begin{tabular}{|l||c|}
\hline
Characteristics &Values\\
\hline
Charge per bunch & between 50 pC to 300 pC      \\
Energy & 3-5 MeV \\
Energy spread  & less than 10$\%$  \\
Bunch length & a few ps\\
Beam transverse dimension & 0.5 mm\\
\hline
\end{tabular}
\caption{Characteristics of PHIL}\label{table_PHIL}
 \end{center}
\end{table*}

 \begin{figure*}[th]
  \centering
    \includegraphics[width=6in,height=3.5in]{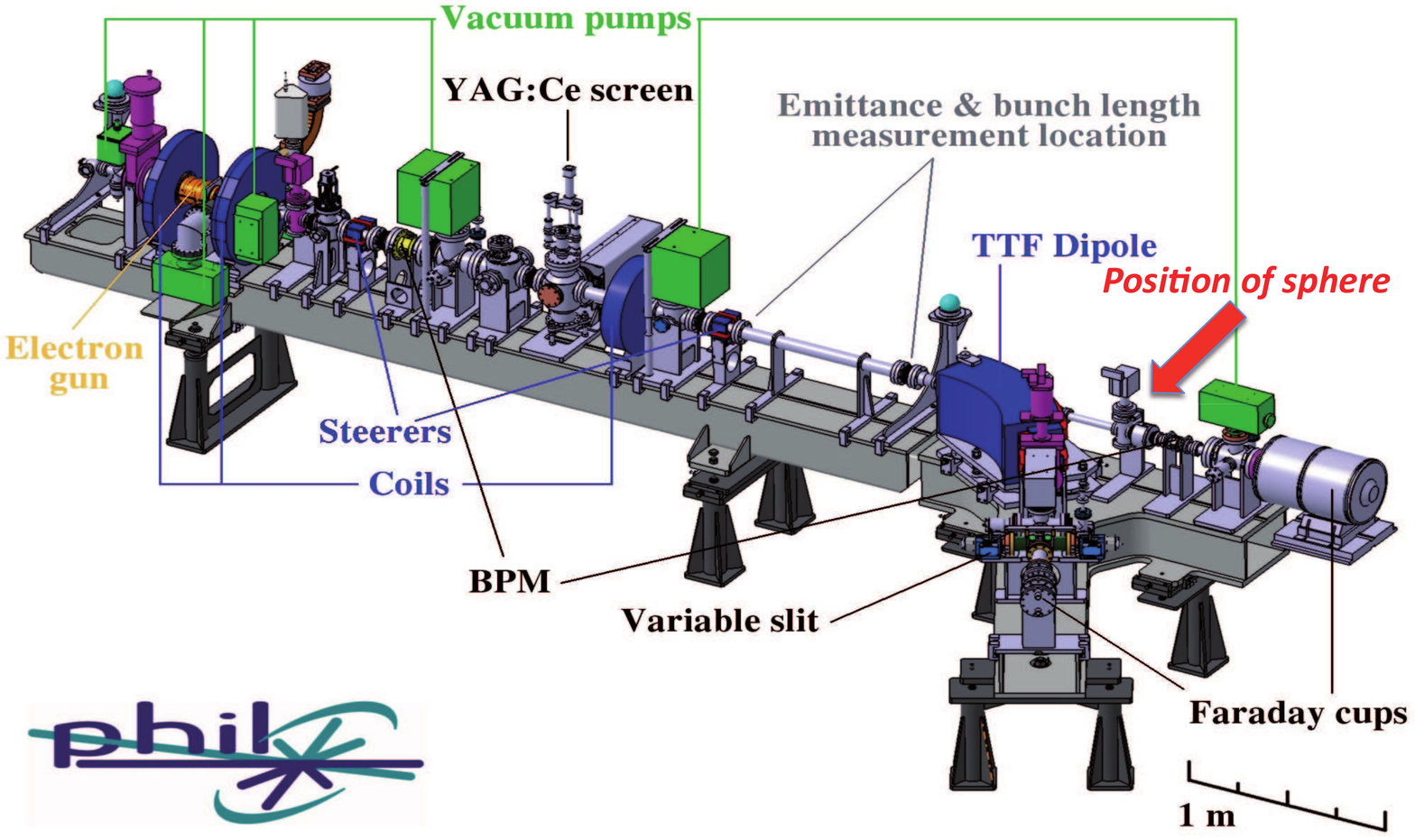}
  \caption{The PHIL accelerator: Futur position of the sphere is indicated with
  the red arrow}
  \label{wide_fig1}
 \end{figure*}

\subsection{Detectors}
 For the integrated measurement, fluorescence photons will be detected and counted by a photo-multiplier tube (PMT) with the same filter as
 in the JEM-EUSO project. The calibration of the detector is a key parameter in this kind of experiment.
  The overall PMT efficiency will be measured using a NIST photo-diode, accurate to 1.5$\%$.
 
 Spectral measurements are interesting because the effect of temperature, pressure, and composition are not the same for
 each spectral line.  These effects are also interesting for the future JEM-EUSO project in order to study the signal to noise ratio, which
 changes with the wavelength.
 
 The fluorescence lines will be measured using a Jobin-Yvon spectrometer equipped with a LN$_{2}$ cooled CCD. The CCD will be calibrated using the calibrated photo-multiplier tube
 at the second output of the spectrometer.
 
 The patented method of calibration has been developed and used with success by G. Lefeuvre, P. Gorodetsky, and their
 collaborators, and is explained in the thesis of G. Lefeuvre (see
 \cite{qlab6} and \cite{qlab12}).

 The expected accuracy of the detectors (PMT and CCD camera) should be around 2 $\%$.

\section{Summary}
Th experiment will provide both the ``integrated'' measurement and ``spectral'' measurement of the fluorescence yield with high
accuracy under a wide range of atmospheric conditions.
The first step of the experiment will debug the measurement at 1atm.
It will be performed during the next months and the study of atmospheric effects
(temperature/pressure/composition) will be made during the year 2012.

A combined $2\%$ accuracy for the detector and 2$\sim$3 $\%$ accuracy for the charge of the electron beam will allow measurement of the fluorescence yield
with an accuracy of up to $5\%$.

\section{Acknowledgements}

This work has been financially supported by the GDR PCHE in France, APC laboratory, and LAL.
We also thank the mechanics, PHIL, and vacuum team at LAL for the construction of the fluorescence bench.


\clearpage


\newpage
\normalsize
\setcounter{section}{0}
\setcounter{figure}{0}
\setcounter{table}{0}
\setcounter{equation}{0}



\title{Fluorescence yields
 by electron in moist air and its application
to the observation of ultra high energy cosmic rays from space}

\shorttitle{N. Sakaki \etal Fluorescence yield by electron in moist air}

\authors{N.~Sakaki$^{1}$,
A.~Zindo$^{1}$,M.~Nagano$^{2}$,K.~Kobayakawa$^{3}$ 
for the JEM-EUSO collaboration}
\afiliations{$^1$Department of Physics and Mathematics,
 Aoyama Gakuin University, Sagamihara 252-5258, Japan\\
$^2$Saitama-ken, 
Hasuda-shi, Higashi 5-8, 11-401, 349-0111, Japan\\
 $^3$Graduate School of Human Development and Environment,
   Kobe University, Kobe 657-8501, Japan}
\email{sakaki@phys.aoyama.ac.jp}

\abstract{ In order to explore the ultra high energy 
cosmic rays
above $10^{20}$~eV (UHECRs),
huge detection area is crucial. In the near future, 
UHECRs will be
 observed from space in projects such as
JEM-EUSO, 
to cover huge area, and fluorescent and Cherenkov light
will be detected from extensive air showers (EASs) induced by UHECRs.
Since those space-based experiments will observe most of EASs above sea,
it is necessary to take the effect of humidity into account
to obtain their longitudinal developments from the fluorescence
yields along their trajectories.
 We have measured humidity dependence of life time and of 
fluorescence yields in
air fluorescence for 10 lines between 300nm and 430nm
with Sr90 source. The fluorescence yields
 decreased with higher humidity:
for example, $\sim$20\% decrease 
was observed for 
$\sim$100\% relative humidity at 1000hPa.
The reference pressures determined from the fluorescence
yields and
the life time were consistent with each other for each line. 
If our results are applied to the UHECR observation
from space above sea,  
fluorescence yields will be reduced about 25\%
 near the sea surface
at low latitude in summer of US standard atmosphere 1966. 
Most of the observed EASs by JEM-EUSO will be inclined 
(the typical zenith angle is 60 deg.), so that the shower maximum will be far
from the sea surface. Therefore, the decrease of the yield
by humidity at shower maximum 
might be small but not negligible.}


\maketitle

\section{Introduction}
Ultra high energy cosmic ray enters the atmosphere and induces a cascade
shower. The main component is electrons, which excite 
nitrogen and
produces fluorescence 
photons
in near ultra-violet region. So called
air fluorescence method was proposed in 1960's to observe UHECRs.
The fluorescence yields 
 are nearly proportional to the deposited energy in the
atmosphere.
This method
 has been used by experiments such as Fly's eye\cite{flyseye},
High resolution fly's eye (HiRes)\cite{hires},
 Pierre Auger Observatory (Auger)\cite{auger} and Telescope array
experiment (TA)\cite{ta}. It will be also used in future experiments
 from space like TUS\cite{tus}, JEM-EUSO\cite{jemeuso},
 KLYPVE\cite{klypve}, S-EUSO\cite{seuso}.
 The principle of the air fluorescence method is
simple,
however, it is not 
straightforward
 when we apply it to the real measurement.
Because we 
need to
 understand a lot of 
factors, such as the
fluorescence yields
 in various atmospheric conditions, atmospheric
transmittance, systematics of the detector and so on. 
Above all, the
knowledge of the
fluorescence yields
is fundamental. 

We have started the
measurement of the 
fluorescence yields
in dry air and published the
results\cite{nagano03,nagano04},
because the 
experiments on ground 
so far have been performed
in dry area like a desert. However, an observation from a satellite orbit will
be main stream in the future because a huge exposure is required for the UHECR
observation. Therefore, most of showers will be observed above sea and
the fluorescence yield in moist air must be examined.

\section{Fluorescence yields in moist air}
When an electron passes through air, an excited state of N$_2$ or
N$^+_2$ will be produced and then fluorescence 
photons
will be emitted with a certain
probability. The 
fluorescence yields ($\epsilon_i$)
 for wavelength ($i$) per unit length
by an electron is expressed as a function of pressure $p$:
\begin{equation}
\epsilon _i (p) =
\rho\frac{\mathrm{d}E}{\mathrm{d}x} \left( \frac{1}
{h\nu_i} \right)\cdot \varphi_i (p) \ ,
\label{eq-eps}
\end{equation}
where $\rho$ is the gas density, $h\nu_i$ is the photon energy,
 $\mathrm{d}E/\mathrm{d}x$ is the
total energy loss of the electron. $\varphi_i(p)$ is the fraction of the
energy emitted as photons to total energy loss\cite{sakaki05}.
Hereafter we omit the suffix $i$ sometimes.
 
The reciprocal of the lifetime $\tau$ consists of three terms.
\begin{equation}
\frac{1}{\tau} = \frac{1}{\tau_{r}} + \frac{1}{\tau_{q}} +
 \frac{1}{\tau_{c}} \equiv \frac{1}{\tau_{0}} + \frac{1}{\tau_{c}} \ ,
\label{eq-1}
\end{equation}
where $\tau_r$ is the lifetime of transition with radiation from an
excited
state to a lower state, $\tau_q$ is that of internal quenching (internal
conversion plus inter-system crossing) and $\tau_c$ is that of collision
de-excitation. The reciprocal of $\tau_c$ is expressed by
\begin{equation}
\frac{1}{\tau_c} = p \sigma \sqrt{\frac{8}{\pi \mu k_B T}} \ ,
\label{eq-coltau}
\end{equation}
where
 $\sigma$ is the cross-section of
 collision de-excitation between molecules, $k_B$ is the Boltzmann
 constant, $T$ is temperature, and $\mu$ is the reduced mass of the two
 molecules. Here, the reference pressure, 
$p'$, is defined as the pressure when $\tau_c$ equals to $\tau_0$ and
\begin{equation}
\frac{1}{p'}= \tau_0 \sigma \sqrt{\frac{8}{\pi \mu k_B T}} \ .
\end{equation}

Let us consider 
the effect of water vapor.
Then $p'$ is related to $\tau_0$ with 
\begin{eqnarray}
 \frac{1}{p'}&=&(f_\mathrm{n} q_\mathrm{nn}+f_\mathrm{o} q_\mathrm{no} +
  f_\mathrm{w}q_\mathrm{nw})\tau_o \nonumber\\
 &=&
  \left(1-\frac{p_\mathrm{w}}{p}\right)\frac{1}{p'_\mathrm{dryair}}+\frac{p_\mathrm{w}}{p}\frac{1}{p'_\mathrm{H2O}}
  \label{eq-pdash},
\end{eqnarray}
where $f_n$,$f_o$ and $f_w$ are proportional to partial pressures
 of N$_2$, O$_2$ and
H$_2$O, respectively and normalized to  $f_n + f_o + f_w =1$.
$q_\mathrm{nn}$, $q_\mathrm{no}$ and $q_\mathrm{nw}$ are the quenching
rate constants of the collisional de-excitation 
between N$_2^*$ (or N$_2^{+*}$) and N$_2$, O$_2$ and H$_2$O,
respectively. 
$p_\mathrm{w}$ is water vapor pressure. $p'_\mathrm{dryair}$ and
 $p'_\mathrm{H2O}$ are the reference pressures for dry air and water
 vapor, respectively.

Then the lifetime and the fluorescence yield for each wavelength band 
are expressed with $p'$ as
\begin{eqnarray}
\frac{1}{\tau} &=& \frac{1}{\tau_{o}}\left(1+\frac{p}{p'}\right) \ ,
 \quad 
\mathrm{and}
\label{eq-tau2} \\
\epsilon (p) &=& \frac{C f_n p}{1+ \frac{p}{p'}} \ ,
\label{eq-ep} 
\end{eqnarray}
where
\begin{equation}
C=\frac{1}{R_gT}\frac{\mathrm{d}E}{\mathrm{d}x} \left( \frac{1}
{h\nu} \right)\cdot \varphi(0) \ .
\end{equation}
$\varphi(0)$ corresponds to the fluorescence 
efficiency in the absence of collisional quenching\cite{sakaki05} and
$R_g$ is the specific gas constant.

\section{Experiment}
A cubic chamber of 25~cm was used to keep air in various
conditions\cite{nagano03,nagano04}.
Decay electrons (0.85MeV on the average) from $^{90}$Sr (74MBq) were
collimated
and the number of electrons which pass through the chamber was counted
by a scintillation detector. Three 2'' photomultiplier tubes (PMTs)
selected for low noise
were attached to three sides of the chamber
 to detect fluorescence photons through bandpass filters.
The central wavelengths of the filters were 313, 325, 330, 337, 358, 370,
380, 391.4, 400 and 430~nm. The band widths were about 10~nm except the
391.4~nm filter with 5~nm width. The data were taken with the photon
counting method. The charge of the signal from each PMT and the time
difference between the electron signal and the photon signal were
recorded for coincident events of an electron signal with signal from 
one of photon PMTs.

Air in the laboratory was taken into the chamber at various pressures
between 1\,hPa and 1000\,hPa to determine the 
fluorescence yields
 in dry air.
In order to study humidity dependence of the 
fluorescence yields,
 the total pressure was fixed at 30, 100 and 1000\,hPa and the humidity was
changed between 0\% and 93\% 
under the constant temperature around 20$^\circ$C.
In order to increase or decrease humidity, air was passed through water
or silica gel.
The humidity in the chamber was measured with two
hygrometers, VAISALA HMP234 and Toplas TA502 
which were
confirmed
to work also at lower pressure than 1 atmosphere by the manufacturers.
Both hygrometers showed consistent humidity with each other during the
measurement.

\section{Results}
\begin{figure}[!ht]
 \centering
 \includegraphics[width=.5\textwidth]{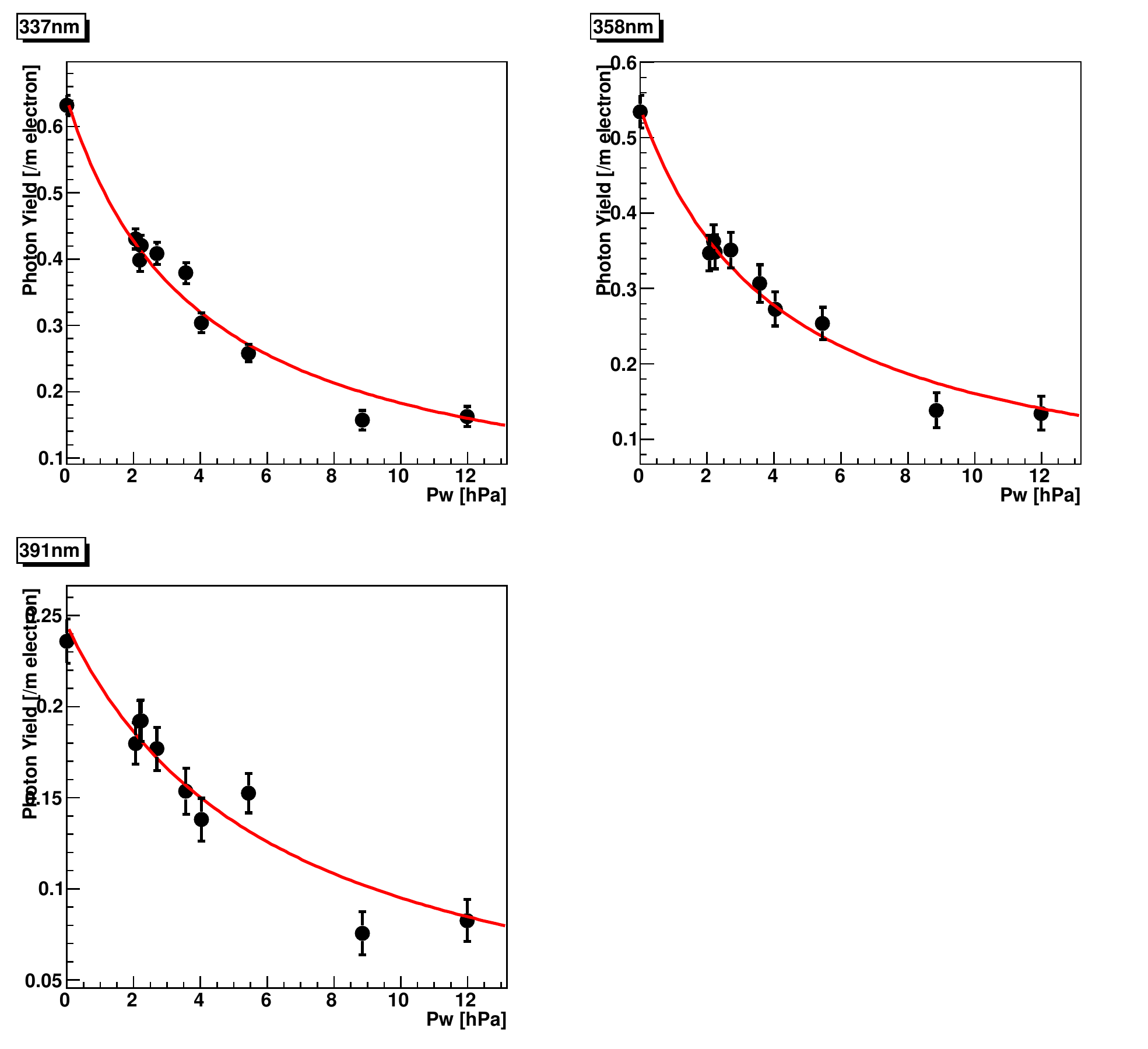}
 \caption{
 Fluorescence yields
 of 337nm, 358nm  and 391nm lines as a function of
 water vapor pressure ($p_\mathrm{w}$) at $p=$30\,hPa.
 Solid lines show the best
 fit curves by Eq.\,(\ref{eq-ep}).}
\label{fig:yieldvshum}
\end{figure}

\begin{figure}[!ht]
 \centering
 \includegraphics[width=.5\textwidth]{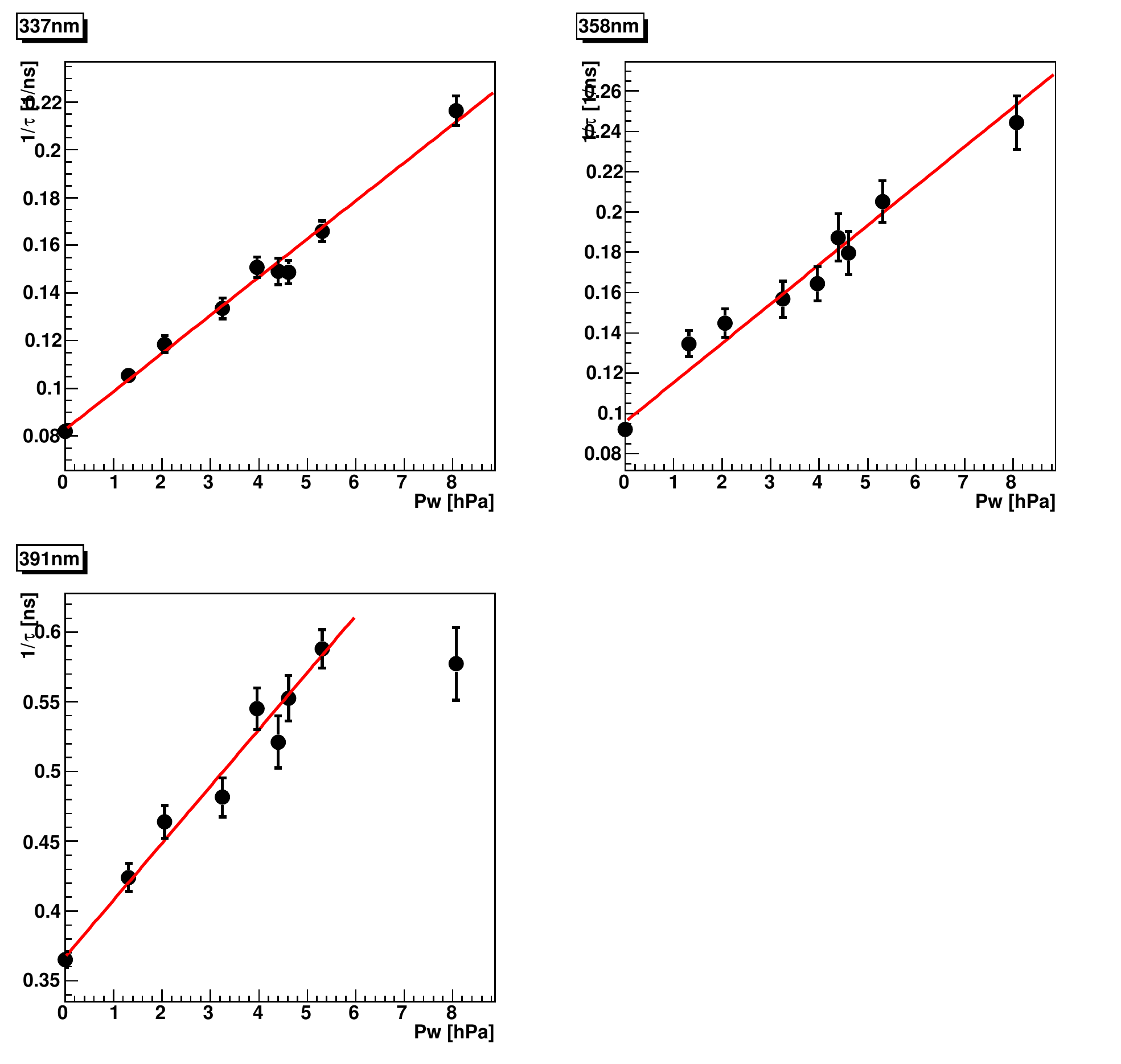}
 \caption{Reciprocal lifetime of 337nm, 358nm and 391nm lines as a function of
 water vapor pressure ($p_\mathrm{w}$)at $p=$30\,hPa.
 Solid lines show the best fit curves by
 Eq.\,(\ref{eq-tau2}).}
\label{fig:tauvshum}
\end{figure}

Fluorescence yields
per unit length per electron ($\epsilon$) was derived
 with the following equation.
\begin{equation}
\epsilon=\frac{N_\gamma}{N_e l \eta f \Omega/4\pi\mathrm{(QE)}\mathrm{(CE)}},
\end{equation}
where $N_\gamma$ is the number of detected photon signals, $N_e$ the
number of electron signals, $\eta$ the transmission of the quartz window,
$f$ the transmission of the interference filter at the wavelength of the
main nitrogen
 emission in study, $\Omega$, QE and CE the solid angle,
the quantum efficiency and the collection efficiency of the PMT,
respectively, $l$ the length of the fluorescence section.
Fluorescence yields
 and lifetime at constant total gas pressure were
measured and are shown in Figure\,\ref{fig:yieldvshum} and
Figure\,\ref{fig:tauvshum} respectively, as a function of water vapor
pressure.
Fluorescence yields
 and lifetime decrease with increasing water vapor
pressure, because N$_2$ molecules are de-excited by collision with water
molecules. 
These data are fitted by Eqs.(\ref{eq-tau2}) and (\ref{eq-ep}),
with the reference pressure in moist air expressed in
Eq.(\ref{eq-pdash}),
and then $p'_\mathrm{H2O}$ was determined. 
In this fitting process, $p'_\mathrm{dryair}$ was fixed to
that determined from the dry air data\cite{nagano04}.
$p'_\mathrm{H2O}$ 
derived from the yield data and the lifetime data
are consistent with each other within 1-2~hPa.

\begin{figure}[!ht]
 \centering
 \includegraphics[width=.5\textwidth]{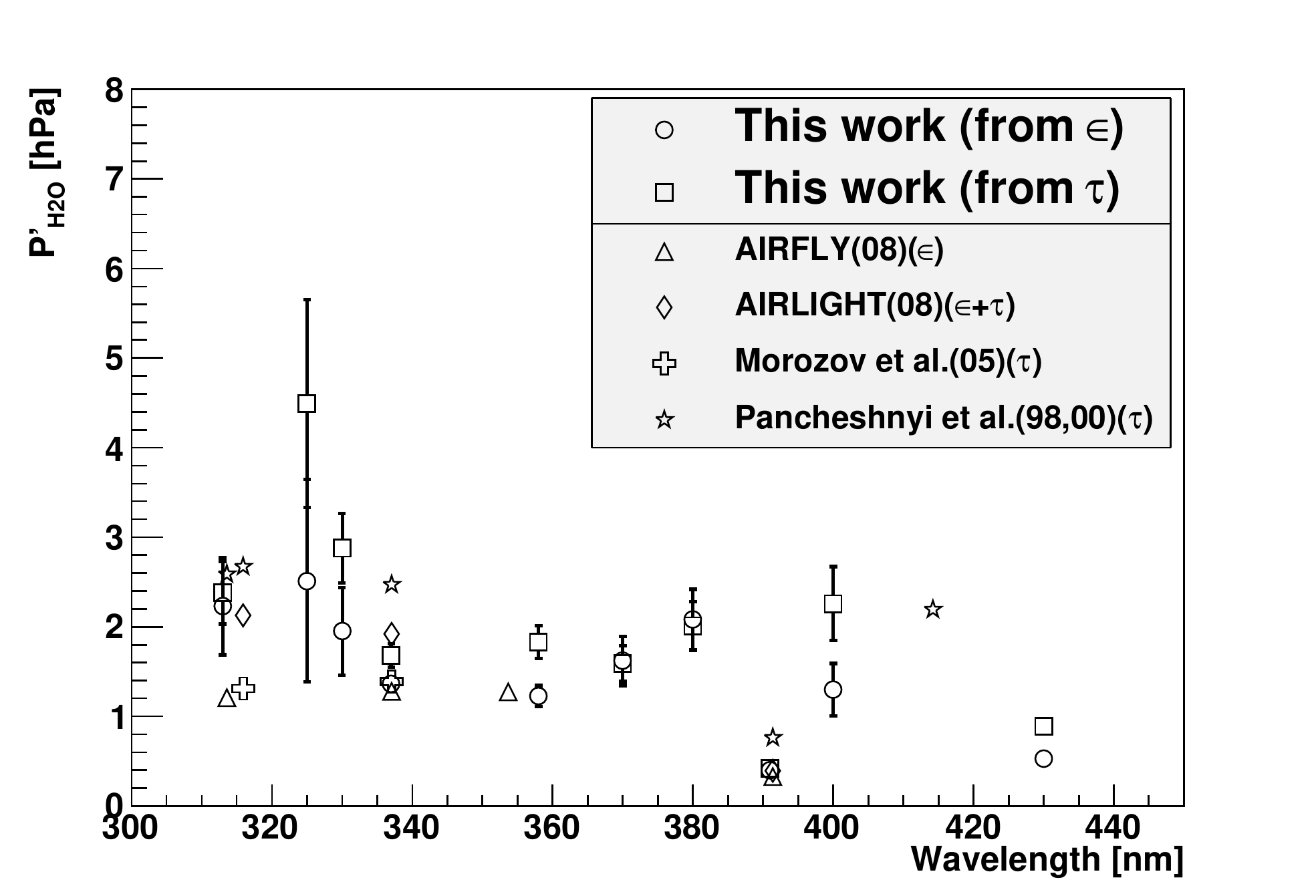}
 \caption{$p'_\mathrm{H2O}$ for 13
nitrogen lines. Our
 results are shown by circles (determined from the yield) and
 squares (from the lifetime). 
 They are compared with
 those by 
 AIRFLY\cite{airfly08} (triangles), 
 AIRLIGHT\cite{airlight08}(diamonds),
 Morozov \etal\cite{morozov05}(crosses) and
 Pancheshnyi \etal\cite{pancheshnyi98,pancheshnyi00}(stars).}
 \label{fig:pdash_h2o}
\end{figure}

Derived $p'_\mathrm{H2O}$ at $p=30$\,hPa
for 10 lines are
summarized in Figure\,\ref{fig:pdash_h2o}.  
$p'_\mathrm{H2O}$ for 1N lines (391nm and 428nm) are
about 0.4$\sim$0.8 and are
smaller than
those for 2P lines, which are around 2-3~hPa. 
$p'_\mathrm{H2O}$ for 337nm and 358nm at total
pressure 100~hPa and
1000~hPa were also determined.
$p'_\mathrm{H2O}$ determined from the yield data 
at 30~hPa, 100~hPa and 1000~hPa are 
1.36~hPa, 1.70~hPa and 1.66~hPa for 337~nm, and 1.23~hPa, 1.61~hPa and
1.27~hPa for 358~nm, respectively. Each error is 0.1-0.2~hPa.
No significant pressure dependence of $p'_\mathrm{H2O}$ is
observed.
Our results are compared with those of AIRFLY\cite{airfly08},
AIRLIGHT\cite{airlight08},  Morozov \etal\cite{morozov05} and
Pancheshnyi \etal\cite{pancheshnyi98,pancheshnyi00}
in the same figure.
They are consistent one another,
although the errors of our results 
 are relatively large
 for some lines.

\section{Application to UHECR fluorescence observation from space}
\begin{figure}[!ht]
 \centering
 \includegraphics[width=.5\textwidth]{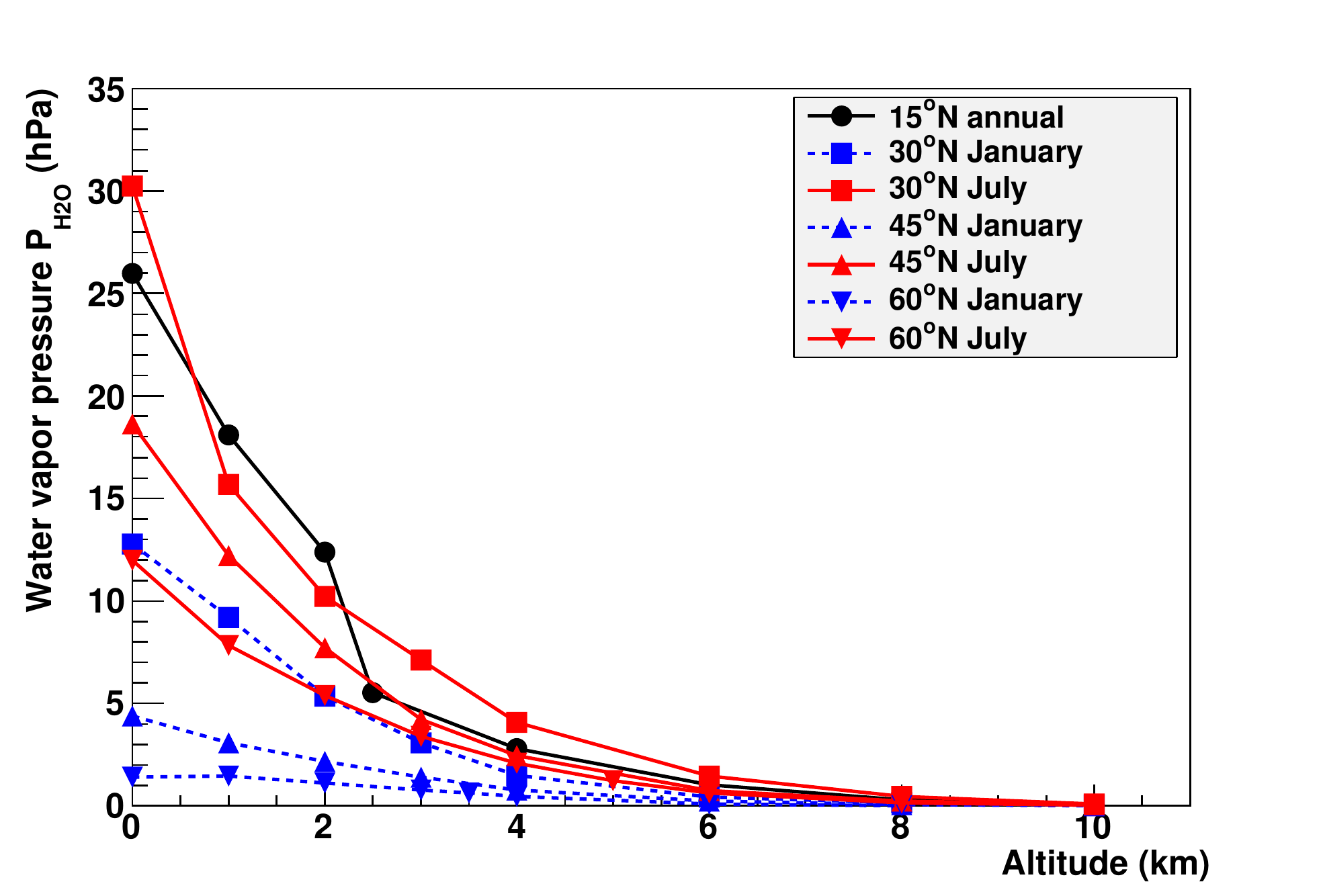}
 \caption{Water vapor pressure as a function of altitude from US
 standard atmosphere 1966 model. The data at 15$^\circ$N are shown by
 solid circles, those at 30$^\circ$N by solid squares, those at
 45$^\circ$N by solid triangles and those at 60$^\circ$N by solid
 inverted triangles. January data are connected by dashed lines and July
 data are connected by solid lines.}
 \label{fig:h2o_usstd66}
\end{figure}

US standard atmosphere 1976 model\cite{usatm76} (USstd76)
 has been used 
frequently
in the field of UHECR observation. However there is only
 dry atmosphere model in the USstd76. Therefore, we have used US
 standard atmosphere 1966 (USstd66) to see the humidity effect on fluorescence
 measurement from cosmic rays. Figure\,\ref{fig:h2o_usstd66} shows water
 vapor pressure profile as a function of altitude. In winter at high
 latitude, water vapor pressure is relatively small, however, it
 increases up to 30~hPa, which corresponds to 80\% relative humidity,
 in summer at low latitude. 

\begin{figure}[!ht]
 \centering
 \includegraphics[width=.5\textwidth]{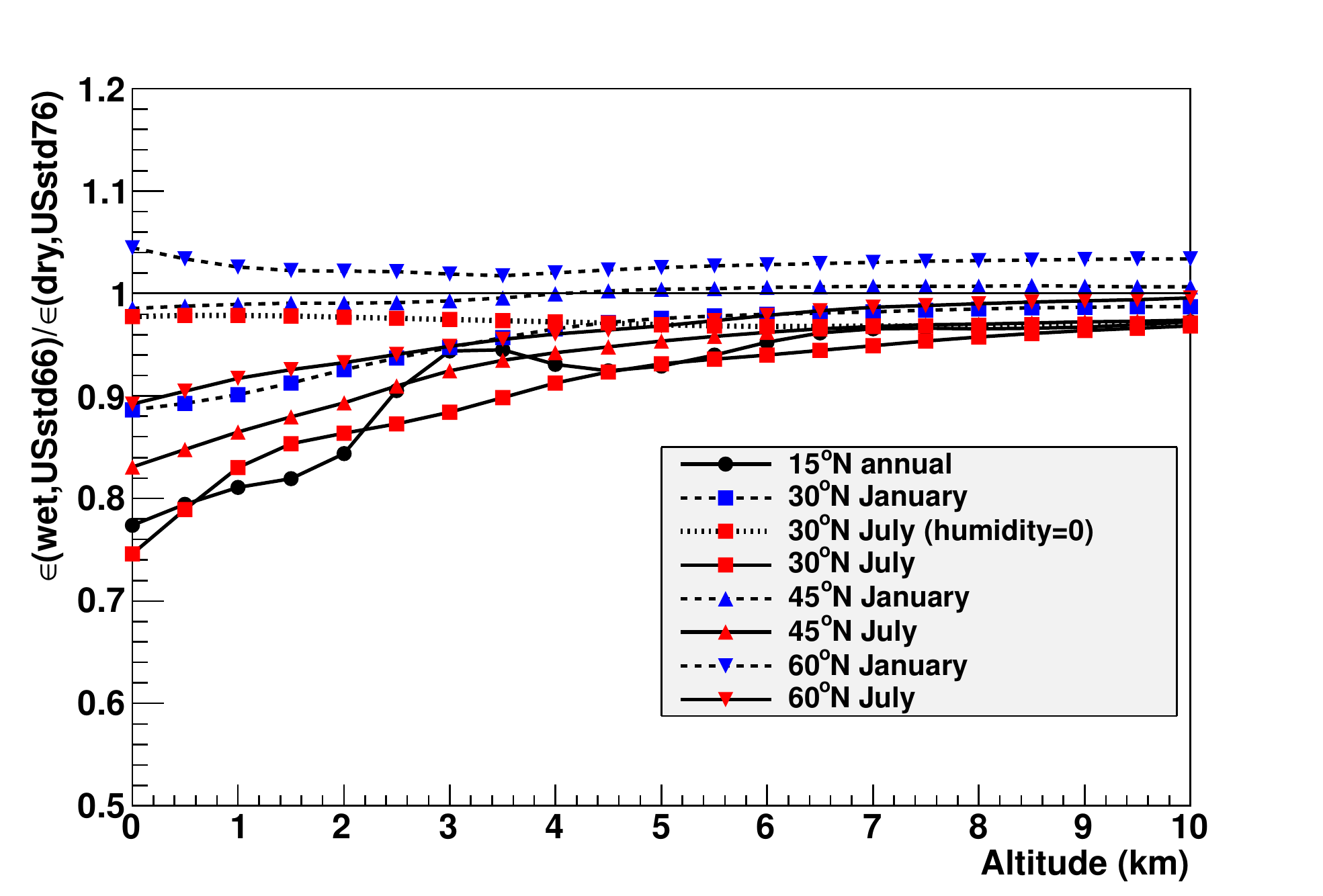}
 \caption{Ratio of total fluorescence yield between 300nm and 406nm
in moist air from US standard atmosphere 1966
to that of US standard atmosphere 1976 (dry air) as a function of
 altitude. Same markers are used as in Figure\,\ref{fig:h2o_usstd66}.}
 \label{fig:alt_yield}
\end{figure}

Using not only the humidity data but also
 the temperature and pressure data of the USstd66 model, we have
 calculated expected total 
fluorescence yields
 between 300 and 430~nm as
 a function of altitude for winter and summer at four latitudes
(15$^\circ$N, 30$^\circ$N, 45$^\circ$N and 60$^\circ$N). The
 fluorescence yields
 at each altitude was calculated with the following
 equation: 

\begin{equation}
 \epsilon=\left(\frac{\mathrm{d}E}{\mathrm{d}x}\right)_\mathrm{0.85MeV}
  \sum \frac{\varphi(0)\rho}{h\nu (1+\rho R_g\sqrt{293T}/p'_{20})},
\end{equation}
where $p'_{20}$ is the reference pressure at 20$^\circ$C, and $p'$ is
defined by Eq.(\ref{eq-pdash}). Mean $p'_\mathrm{H2O}$ from the yield data
and from the lifetime was used for each line.
The yield for USstd76 model (dry air) is
normalized to one. The decrease of the yield in summer at low
latitude is 
about 25\% at sea level (see Figure\,\ref{fig:alt_yield}).
In order to see the influence of the humidity
in USstd66 model, the ratio
 for dry air is
shown in the same figure for the 30N$^\circ$ July profile (labeled with
``(humidity=0)'' in Figure\,\ref{fig:alt_yield}). The yield agrees well
with that of USstd76 within a few \%. 
Therefore
 the decrease in
yield for 30N$^\circ$ July is
understood to be
caused by humidity, not by the difference
in temperature or pressure profile 
of both models.

For JEM-EUSO observation, median zenith angle of observed events will be
around 60 degree. EAS with $10^{20}$~eV and $\theta=60^\circ$ is
expected to reach 
its maximum development
around 6~km high. The decrease 
due to humidity in summer
is 10\% or less there. 
For vertical showers, the maximum is lower, around 2-3~km high.
The influence of humidity is larger, $\sim$15\%. 
Showers produced by neutrinos are expected to develop horizontally  
 near sea level and hence will be darker by 10\% in winter and 
25\%  in summer at low latitude.

\section{Conclusion}
In the future the UHECR observation from a satellite orbit is indispensable to
obtain huge acceptance. 
Since most of  EASs will be observed above sea,
 the influence of water vapor on the fluorescence yields must be
investigated.
We have measured the quenching of 
nitrogen fluorescence by water vapor
for ten lines and applied the result to the various atmospheric
conditions from US standard 1966 model. Fluorescence from the typical
EAS observed by JEM-EUSO (zenith angle=60$^\circ$) will be decreased by
several percent at shower maximum in summer at low latitude. For
horizontal showers near sea surface, as are induced by neutrinos, the
decrease will be larger up to 25\%. 
We have shown 
here only the decrease of the fluorescence yields
by humidity at emission point. 
Since the attenuation in atmosphere is relatively small for space-based
observations, the photon yield in moist air would be
applicable with little
modification.

The decrease in the fluorescence yields by humidity
is not negligible especially in summer at low latitude.
It is necessary to take into account 
the characteristics of the detector in each project
to estimate how much the humidity influences on the observation actually.

\clearpage


\end{document}